\documentclass[journal ]{new-aiaa}
\usepackage[utf8]{inputenc}
\usepackage{textcomp}

\usepackage{graphicx}
\usepackage{amsmath}
\usepackage[version=4]{mhchem}
\usepackage{siunitx}
\usepackage{longtable,tabularx}
\usepackage{epstopdf, epsfig}
\usepackage{bm}
\usepackage{soul}
\usepackage{float}
\usepackage{xifthen}
\usepackage[export]{adjustbox}
\usepackage[permil]{overpic}
\usepackage[auto]{contour}
\usepackage{subcaption}

\usepackage{tikz}
\usetikzlibrary{shapes,arrows}
\usepackage{verbatim}

\tikzstyle{block} = [draw, rectangle, 
    minimum height=3em, minimum width=6em]
\tikzstyle{sum} = [draw, circle, node distance=1cm]
\tikzstyle{input} = [coordinate]
\tikzstyle{output} = [coordinate]
\tikzstyle{pinstyle} = [pin edge={to-,thin,black}]

\newcommand{\blue}[1]{\textcolor{black}{#1}}
\definecolor{mygreen}{RGB}{0, 190, 0}
\newcommand{\green}[1]{\textcolor{black}{#1}}

\title{Extremum seeking control applied to \\ airfoil trailing-edge noise suppression}

\author{Tarcísio C. Déda \footnote{PhD candidate, Department of Energy.} and William R. Wolf\footnote{Assistant Professor, Department of Energy.}}
\affil{University of Campinas, Campinas, SP, Brazil, 13083-860}

\begin{document}

\maketitle

\begin{abstract}
Extremum seeking control (ESC) and its slope seeking generalization are applied in a high-fidelity flow simulation framework for reduction of acoustic noise generated by a NACA0012 airfoil. Two Reynolds numbers are studied for which different noise generation mechanisms are excited. For a low Reynolds number flow, the scattering of vortex shedding at the airfoil trailing edge produces tonal noise while, for a moderate Reynolds number case, boundary layer instabilities scatter at the trailing edge leading to noise emission at multiple tones superimposed on a broadband hump. Different control setups are investigated and they are configured to either find an optimal steady actuator intensity or an optimal position for a blowing/suction device. Implementation details are discussed regarding the control modules and design of digital filters. %considerable noise reductions and, in some cases, full noise suppcontrol setupThe best results achieved -98.8dB attenuation with lower Reynolds and -56.7dB with higher Reynolds, so the noise, in practice, can be considered suppressed.
Analyses of physical phenomena as well as of relevant behavior of the actuated plant are conducted to understand how the extremum seeking loop leverages the flow physics to control noise. Depending on the flow configuration studied and the control setup, results demonstrate that the ESC can provide considerable airfoil noise reductions.
\end{abstract}

\section*{Nomenclature}

{\renewcommand\arraystretch{1.0}
\noindent\begin{longtable*}{@{}l @{\quad=\quad} l@{}}

\green{$A$} & \green{constant blowing/suction intensity}\\
$A_c$ & blowing/suction intensity determined by ESC\\
\green{$C_f$} & \green{skin friction coefficient}\\
$J$ & ESC cost function \\
$\nabla \bar{J}$   & ESC integrator inputs \\
$m$ & ESC cost function window size\\
$\mathrm{Ma}$ & Mach number \\
$p'$  & pressure fluctuation \\
$\mathrm{Re}$ & Reynolds number \\
$T$ & ESC period\\
$T_a$ & acoustic wave period\\
$t_{c0}$ & ESC integrator turn on time\\
$\Delta t_\text{c}$ & control loop sampling time \\
$\Delta t_{\text{sim}}$ & simulation time step \\
%$k$ & discrete time index for control loop \\
\green{$u_c$} & \green{control input, determined by ESC, corresponding to the actuator position $x_c$ or intensity $A_c$}\\
\green{$\bar{u}_c$} & \green{ESC integrator output, which dynamically estimates the appropriate control input}\\
$\mathcal{W}(x)$ & window function for actuation \\
$x_c$ & actuator position determined by ESC \\
$\Delta x$ & actuation window width \\
%$x_{c_0}$ & actuator initial position\\ position or intensity) \\
$\alpha$  & ESC sinusoidal sweeping amplitude \\
$\beta$  & ESC multiplying sinusoidal wave amplitude \\
$\eta$  & ESC integrator gain \\
$\sigma$ & exponent parameter in cost function\\
$\phi$ & ESC phase delay compensation \\
$\omega$  & ESC angular frequency \\
$\omega_p$  & passband frequency \\
$\omega_s$  & stopband frequency \\
\end{longtable*}}

\section{Introduction}\label{sec:intro}

Control of unstable flows is sought in several engineering problems and finds applications in drag and noise reduction, lift increase, heat transfer and mixing enhancement, besides transition and separation delay. Active flow control has been implemented in experimental setups including cavities, jets, ramps and airfoils \citep{cattafesta1997active,  alvi2003control, tuck2004active, cattafesta2011actuators, cuvier2011flow, george2015control, sinha2018active, ghassemi2021control} and a recent experimental study of flow control was performed by \citet{zigunov2021empirical} for a bluff body. These authors used an array of blowing jets with periodic switching and applied a genetic algorithm to find out which units should be activated. The idea was to optimize drag reduction in a slanted cylinder by finding the best spatial distribution of jets in an offline fashion respective to the final proposed open-loop application. In this experimental work, the authors were able to achieve drag reduction of  11.6\%. 

In many situations, the high costs imposed by experiments may become prohibitive for preliminary design and optimization of active control strategies. In this context, the application of flow control together with model reduction has also been an active area of research \citep{barbagallo2009closed, semeraro2011feedback, brunton2014state, ma2011reduced, proctor2016dynamic} and a review on the topic is provided by \citet{rowley2017model}. 
As an example, by using a linear approximation of a mixing layer, \citet{sasaki2018closed} designed feedforward and feedback (closed-loop) control approaches. Although results from nonlinear simulations showed that the performance was different from the expected results of the linearized plant, relevant reductions in measured velocity fluctuations were observed.

The improvement in computational power and novel developments in control theory \citep{bewley2001flowcontrol, brunton2015closed} enabled high-fidelity simulations of complex unsteady flows together with active flow control. The combination of computational fluid dynamics (CFD) and active control strategies has been shown by several authors \citep{you2008active, avdis2009large, ramirez2015effects, yeh2018resolvent, ramos2019active, visbal2018exploration}.
\citet{yeh2018resolvent} applied resolvent analysis to airfoil large eddy simulation data in order to find the best open-loop control parameters of a periodic heat flux actuator near the leading edge. Through numerical simulations, they were able to tune the temporal frequency and spanwise wavelength to control flow separation, achieving a drag reduction of up to 49\% and a lift improvement of up to 54\% in different configurations at post-stall angles of attack. 

Large eddy simulations of dynamic stall were performed by \citet{visbal2018exploration} together with an exploration of flow control setups. These authors took advantage of the natural amplification of certain frequencies that occur at the separation bubble formed during dynamic stall to enhance the effectiveness of periodic actuators placed near the airfoil leading edge. In a numerical environment, the authors presented a study on how the controller modifies the flow in different configurations, and how it affects aerodynamic properties such as lift and pitching moment. Further studies on airfoil dynamic stall control were conducted by \citet{ramos2019active}. These authors investigated the actuator spanwise arrangement and frequency that provided drag reduction for a plunging airfoil by modifying the formation of the dynamic stall vortex and, hence, the unsteady pressure distribution around the airfoil. The controlled flow for the best case presented a significant drop in drag coefficient with much less alterations in lift.
Aiming to reduce airfoil trailing edge noise, \citet{ramirez2015effects} presented the effects of placing a blowing actuator at the trailing edge of a static airfoil. They showed that, for low Reynolds number flows, the actuation shifted the vortex shedding away from the airfoil surface, reducing the noise scattering mechanism.

As discussed by \citep{brunton2015closed}, adaptive techniques such as extremum seeking control (ESC) have also been applied to experimental unsteady flow problems. To reduce drag in a bluff body, \citet{beaudoin2006bluff} introduced a rotating cylinder with variable velocity in an experimental setup. An ESC cost function was employed with terms that penalized both the power loss due to drag and the power consumption of the actuation cylinder. The configuration was able to achieve a modest power reduction between 2 and 3\%.
\citet{becker2007adaptive} experimentally implemented a slope seeking control loop to increase lift in a high-lift configuration composed of airfoil and flap. By varying the amplitude of a pulsed flow outlet near the flap leading edge, the control loop optimized values of pressure coefficient gradients as an indirect way to increase lift. The online approach allowed for finding the best pulsation amplitude as the flap angle of attack varied. By dividing the wing into three different spanwise sections and independently applying slope seeking control, the online closed-loop approach was able to optimize the performance in real time, enhancing the total lift relative to a fixed pulsed jet amplitude by up to 6.8\% at high angles of attack.

Extremum seeking was employed by \citet{kim2009extremum} to reduce resonance effects in a cavity flow. By measuring pressure in two different points, an estimation of the limit cycle amplitude was made, and this cost value was targeted for minimization. The loop searched for the phase shift value of an oscillatory diaphragm that perturbed the flow, being able to reduce the natural oscillations by around -20dB, close to background noise levels. \citet{fan2017modified} developed a modified ESC scheme to optimize mixing in a jet flow by using a minijet with variable frequency. The decay rate of the jet center line velocity was used as a measurement to indirectly infer mixing, and an optimum value was achieved for a specific actuation frequency. 

\citet{brackston2016extremum} used ESC to find the best frequency and amplitude for an annular harmonic flow outlet that encircled the base of a bullet-shaped body. By probing pressure at the bullet base, the ESC loop was able to reduce the magnitude of pressure coefficient by 27\%, leading to drag reduction. In a very similar approach to the previous reference, \citet{pastoor2008feedback} applied harmonic flow actuation at the base of a bluff body while the pressure coefficient was estimated at the wall. From open-loop experiments, the authors showed that increasing the momentum transfer of the actuators caused a drop in the magnitude of pressure coefficient. This effect was observed until a certain point, after which increasing the injection power did not improve results. By implementing slope seeking control to the same problem, it was shown that it is possible to find an equilibrium operation near the optimal point, even considering variations in the flow conditions. This allowed for the power consumption to be bounded with real-time adaptation. %In their paper, the authors also presented a phase control technique that outperformed the open-loop and the slope-seeking amplitude adjustment by using only half of the actuation area with synchronized phase with respect to the aerodynamic wake. 

%\citet{hoeijmakers2008implementation} performed CFD calculations considering two blowing/suction devices in phase opposition to reduce the vortex street in a cylinder. The phase delay of the jets was determined by an ESC loop that tried to oppose the velocity fluctuations and attenuate the vortex shedding.

In this work, we propose the application of the classical ESC and slope seeking control to reduce noise generated by the flow past a NACA0012 airfoil. High-fidelity simulations are conducted for different types of jet actuation involving blowing or suction. In one setup, the controller seeks for an optimal actuation position with a fixed jet intensity while, in another, the optimal actuation intensity is sought for a fixed jet position. A cost function is computed to quantify the airfoil noise emission and the ESC loop is configured to minimize this function by finding the best actuation setup. 
Two different Reynolds numbers are investigated to test the control approach with different noise source mechanisms. For the lower Reynolds number case, vortex shedding is the source responsible for trailing-edge noise scattering at a single tonal frequency. On the other hand, for the higher Reynolds number analyzed, boundary layer instabilities are responsible for noise scattering at multiple tonal frequencies. By using this model-free technique, closed-loop control is implemented to optimize flow variables with robust tracking for reasonable plant variations. Results demonstrate that the ESC implementation  provides considerable noise reduction for the configurations tested.

\section{Methodology}\label{sec:method}

In the present work, extremum seeking control is applied for real-time minimization of acoustic noise emitted by an airfoil. The pertinent control modules are implemented in a  high-fidelity CFD simulation tool that has been validated for compressible flow applications \citep{wolfJFM2012, ramos2019active}. Besides the ESC approach, some investigations proposed in this work alternatively use slope seeking control, which is suitable for cost functions that exhibit \textit{plateaus}.
\begin{figure}
\centering
\includegraphics[width=.45\textwidth]{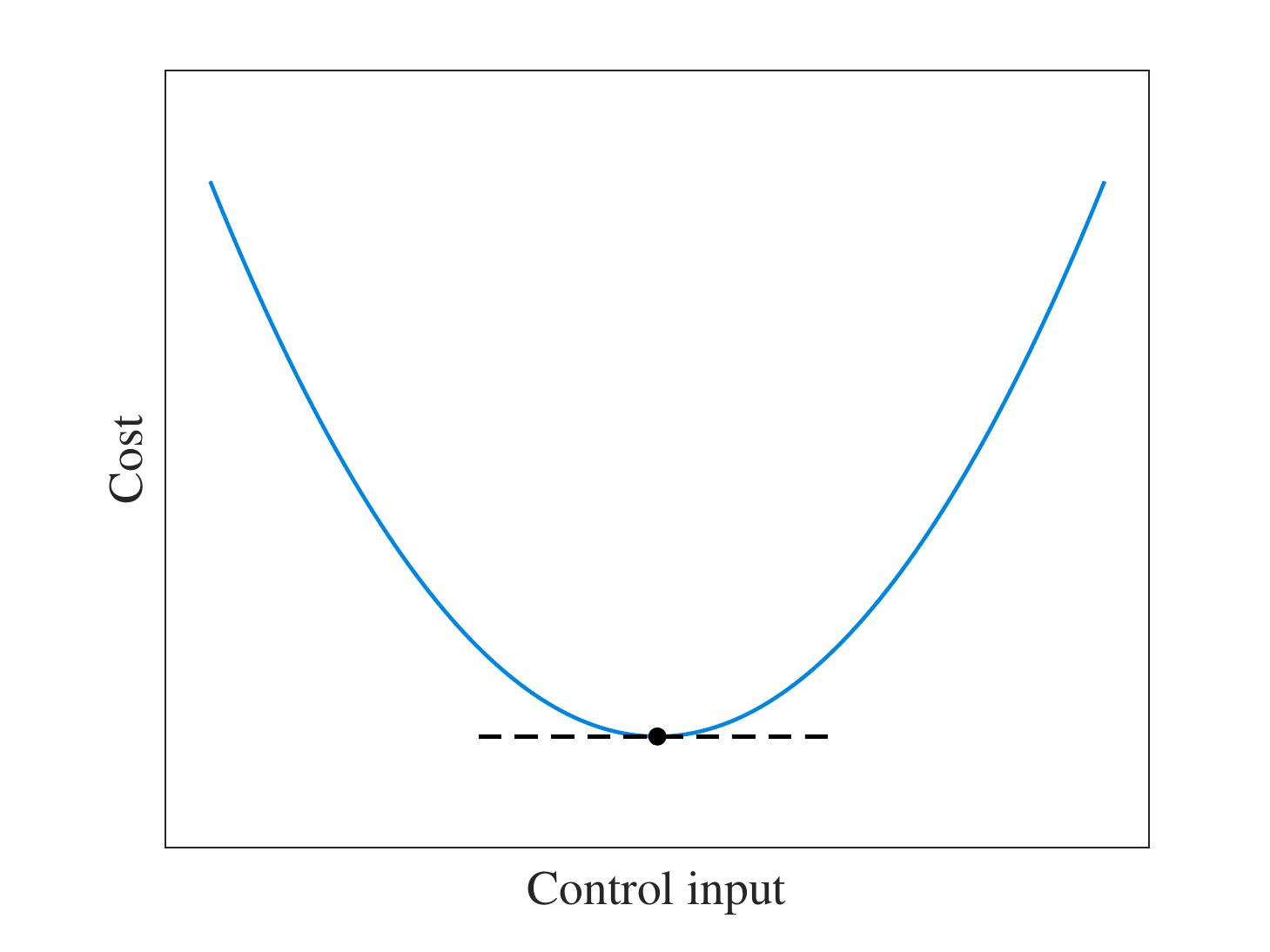}
\includegraphics[width=.45\textwidth]{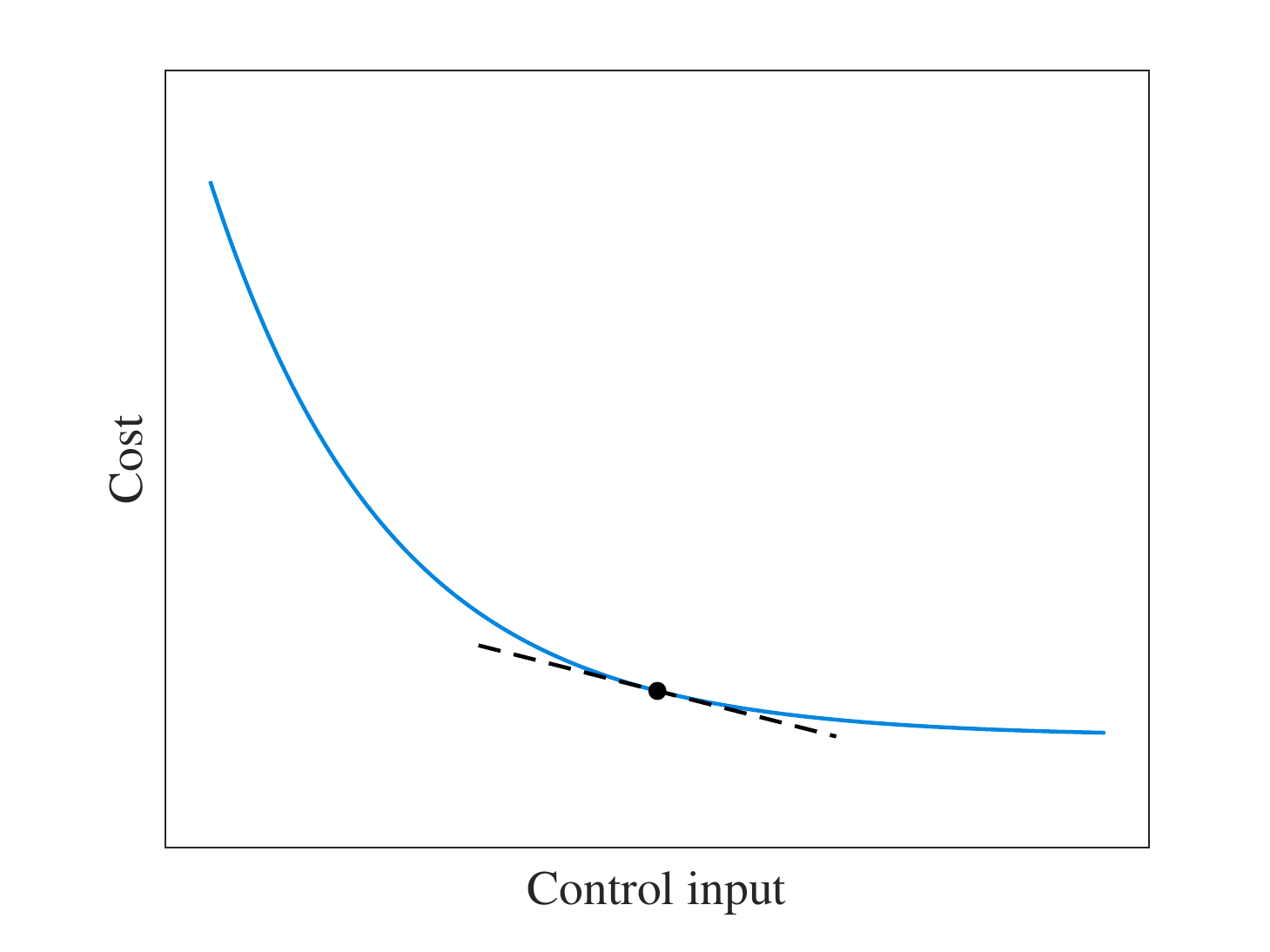}
\caption{Cost functions presenting an optimal operating point (left) and a plateau (right). The ESC approach tries to minimize a cost function while the slope seeking compensation can be applied so the control searches for a non-zero derivative, preventing excessive control effort.}
\label{f:schemes}
\end{figure}

Figure \ref{f:schemes} presents different cost functions that distinguish the applications of extremum and slope seeking control. While the former approach seeks an optimal point in the minimization of the cost function, the latter searches for a pre-defined slope value to prevent excessive control effort \citep{king2006adaptive}.
In the present applications, a cost function based on pressure fluctuations is set to quantify the airfoil noise emission. The data are sampled at a probe placed in the acoustic field to simulate a microphone and an actuator is simulated by a small jet that performs blowing or suction at the airfoil surface. The actuator can either have a variable jet intensity or a variable position along the airfoil chord.
%Permeable boundary conditions for fluid momentum are used to simulate the behavior of suction/blowing devices at the wall. The configurations presented in this work account for variable intensity or variable position of the acuators.

\subsection{Flow simulations}\label{sec:solver}

The compressible Navier-Stokes equations are solved in general curvilinear coordinates using an O-grid. Near the airfoil surface, the second-order implicit scheme from \citet{beam1978implicit} is applied to integrate the flow equations in time while, away from the wall, the flow equations are temporally advanced using a low-storage third-order Runge-Kutta scheme \citep{wray:86}. In order to exchange information between the two methods, an overlap region is employed as presented by \citep{nagarajan:04}. 
For the spatial discretization of the flow equations, the sixth-order compact scheme proposed by \citet{nagarajan:03} is applied using a staggered grid approach.  
In order to damp high wavenumber errors from the numerical solution, a sixth-order compact filter \citep{lele1992compact} is also employed.

At the airfoil wall, no-slip adiabatic boundary conditions are implemented, except where actuators are located. In these cases, blowing or suction actuation is employed and momentum values $\rho v_n$ are set according to the control strategies described in Sec. \ref{sec:act}. Here, $\rho$ is the density and $v_n$ is the velocity in the wall-normal direction. Characteristic plus sponge boundary conditions are implemented at the far field locations to minimize wave reflections. 
The flow equations are solved in non-dimensional form where length, velocity components, density and pressure are nondimensionalized by the airfoil chord $L$, freestream speed of sound $a_{\infty}$, freestream density $\rho_{\infty}$, and $\rho_{\infty}a_{\infty}^{2}$, respectively.
Further details on the solver implementation are described by \citet{nagarajan2007leading}. Despite the use of the freestream speed of sound in the nondimensionalization procedure, all results presented along this work are provided with respect to the freestream speed $U_{\infty}$.

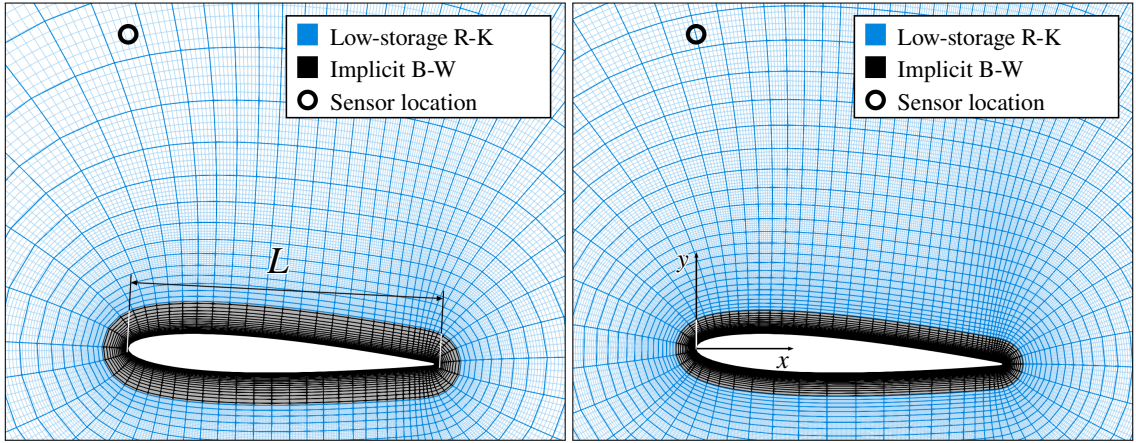
\begin{figure}
\centering

\begin{overpic}[width=0.45\textwidth, frame]{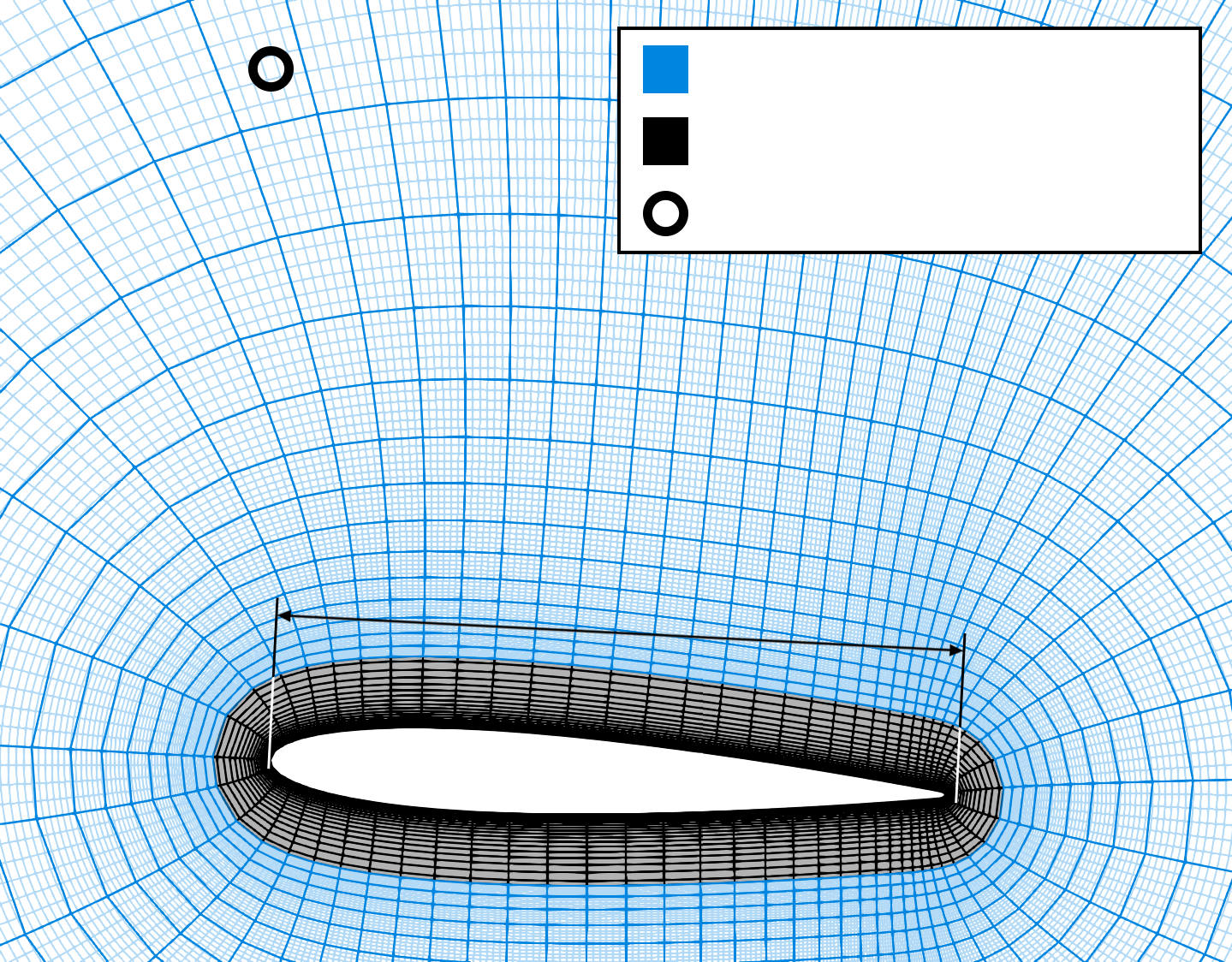}%
\put(580,707) {\small{Low-storage R-K}}%
\put(580,649) {\small{Implicit B-W}}%
\put(580,591) {\small{Sensor location}}%
\put(465,299) {\Large{\contour{white}{$L$}}}%
\end{overpic}
\begin{overpic}[width=0.45\textwidth, frame]{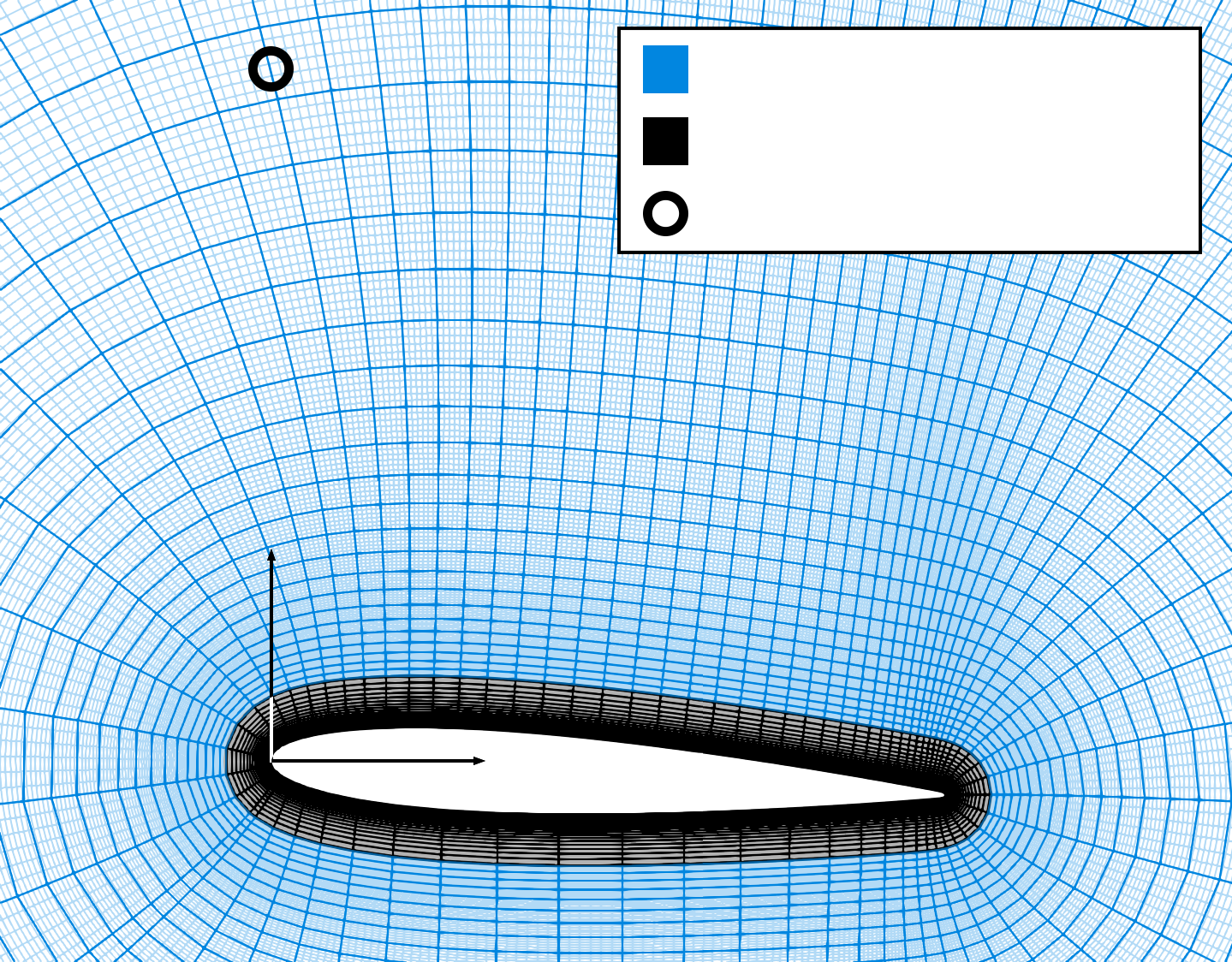}%
\put(580,707) {\small{Low-storage R-K}}%
\put(580,649) {\small{Implicit B-W}}%
\put(580,591) {\small{Sensor location}}%
\put(185,310) {{\contour{white}{$y$}}}%
\put(365,130) {{\contour{white}{$x$}}}%
\end{overpic}
\caption{\green{Computational domain used for simulations with $\mathrm{Re}=10^4$ (left) and $\mathrm{Re}=10^5$ (right). The blue mesh represents the spatial region where the explicit time integration is performed, while the black portion represents the region for the implicit method. Every 6th grid point is shown with dark lines.}}
\label{f:cdomain}
\end{figure}

The airfoil flows are investigated for two Reynolds numbers, $\text{Re}=10^4$ and $10^5$. \green{Direct numerical simulations} are conducted for Mach number $\text{Ma}=0.3$ and $3^\circ$ angle of attack and the flow conditions are chosen to allow different airfoil noise mechanisms to occur. \green{Figure \ref{f:cdomain} shows the computational grids around the NACA0012 airfoil for both Reynolds numbers and every 6th grid point is highlighted with dark lines. The origin of the Cartesian system is at the leading edge.} For the lower Reynolds number, trailing-edge noise is generated due to the presence of vortex shedding while, for the higher Reynolds number case, boundary layer instabilities are responsible for the noise generation. In the former case, the noise spectrum is composed of a single tone at the vortex shedding frequency and its harmonics. For the latter case, a main tone is expected with additional secondary tones superposed on a broadband spectrum, as shown by \cite{ricciardi2020secondary}. %In both cases, acoustic scattering by quadrupole sources at the trailing edge is a dominant feature of noise radiation \citep{hall:70}. 
Different grid setups are used for the two Reynolds numbers analyzed as shown in Table \ref{tab:grid} and they provide sufficient resolution for the present simulations according to \cite{ramirez2015effects, ricciardi2020secondary}. This table also presents the non-dimensional time steps $\Delta t_\text{sim}$ employed in each simulation, where the time scales are presented in nondimensional form with respect to $U_{\infty}$ and $L$. 
\begin{table}
  \begin{center}
\def~{\hphantom{0}}
  \begin{tabular}{lcccccc}
       \hline
       Re  & Ma & AoA & $N_x$   &   $N_y$ & Domain size & $\Delta t_\text{sim}$ \\\hline
     $10^4$ & 0.3 & $3^{\circ}$ & 400    &  250    & 12 $L$      & 4.0e-4\\
     $10^5$ & 0.3 & $3^{\circ}$ & 440    &  480    & 34 $L$    & 1.5e-4\\\hline
  \end{tabular}
  \caption{Grid and flow configurations analyzed.}
  \label{tab:grid}
  \end{center}
\end{table}

%\begin{figure}
%\centering
%\includegraphics[width=.45\textwidth, frame]{figures/rev2/10kf.png}
%\includegraphics[width=.45\textwidth, frame]{figures/rev2/100kf.png}
%\caption{\green{Computational domain used for simulations with $\mathrm{Re}=10^4$ (left) and $\mathrm{Re}=10^5$ (right). [Referenciar no texto; talvez referenciar a figura nesta seção e na seção que fala sobre sensores]}}
%\label{f:cdomain}
%\end{figure}

\subsection{Extremum seeking control}\label{sec:esc}

\begin{figure}
\centering
\begin{tikzpicture}[auto, node distance=2cm,>=latex']
    \node [input] (sweep) {};
    \node [sum, right of = sweep, node distance = 2cm] (sum) {$+$};
    \node [block, right of=sum, node distance = 2.2cm] (plant) {N-S equations};
    \node [block, right of=plant, node distance = 3.2cm] (cost) {$J(\cdot)$};
    \node [block, right of=cost, node distance = 3.2cm] (hp) {High-pass filter};
    \node [sum, below of = hp, right of = hp, node distance = 2cm] (prod) {$\times$};
    \node [right of = hp, node distance = 2cm] (rtcorner) {};
    \node [input, right of = prod, node distance = 2.5cm] (multwave) {};
    \node [block, left of=prod, node distance = 3.4cm] (lp) {Low-pass filter};
    \node [block, left of=lp, node distance =4cm] (integ) {$\eta\int_{0}^{t} (\cdot) dt$};
    \node [output, above of=cost, node distance = 5cm] (output) {};
    \draw [->] (sweep) -- node {$\alpha\sin(\omega t)$} (sum);
    \draw [->] (sum) -- node {$u_c$} (plant);
    \draw [->] (plant) -- node [name=pp, inner sep = 0, pos=0.55] {} (cost);
    
    \node [output, above of=pp, node distance = 1cm] (output) {};
    \node [input, below of=pp, node distance = 0.03cm] (input2) {};
    
    \draw [->] (cost) -- node {} (hp);
    \draw [->] (hp) -| node {} (prod);
    \draw [->] (multwave) -- node {$\beta\sin(\omega t + \phi)$} (prod);
    \draw [->] (prod) -- node {} (lp);
    \draw [->] (lp) -- node {$\nabla \bar{J}$} (integ);
    \draw [->] (integ) -| node {$\bar{u}_c$} (sum);
    \draw [->] (input2) -- node {$p'$} (output);
\end{tikzpicture}
\caption{\green{Extremum seeking control block diagram. The goal is to control (minimize) the computed output $p'$.}}
\label{f:escloop}
\end{figure}
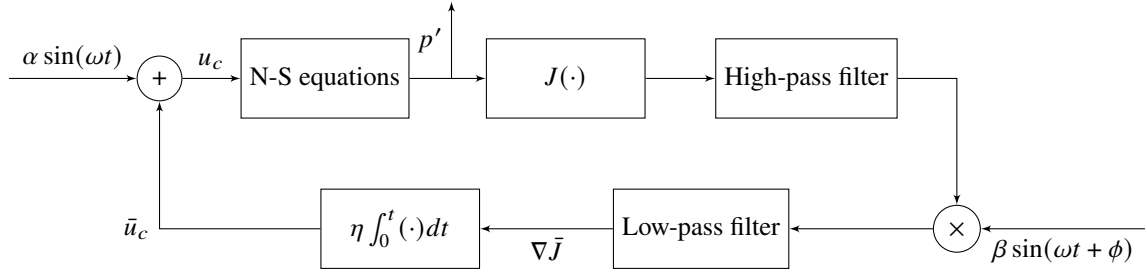

The closed-loop ESC approach implemented in this work is presented in Fig. \ref{f:escloop}. The digital control loop runs while receiving pressure fluctuation measurements $p'$ from the plant (the Navier-Stokes equations) and responds back with the pertinent interventions. \green{The goal is to minimize the computed output $p'$ via a control input $u_c$ determined by the ESC, which can be the actuator position $(u_c=x_c)$ or its intensity $(u_c=A_c)$.}

To start the search for an input $u_c^*$ that minimizes acoustic noise, an initial guess $\bar{u}_{c0}$ is set and implemented as an initial condition of a numerical integrator. An unbiased harmonic perturbation with amplitude $\alpha$ and angular frequency $\omega$ is added to the current value of $\bar{u}_c$, resulting in the actual input $u_c$ at a given time instant. Since the original plant does not change, a timescale compromise must be ensured when choosing $\omega$ so the system can respond to the periodic perturbation $\alpha \sin(\omega t)$ with reasonably low phase delay. 

\begin{figure}
\centering
\includegraphics[width=.45\textwidth, trim=0 1cm 0 0, clip]{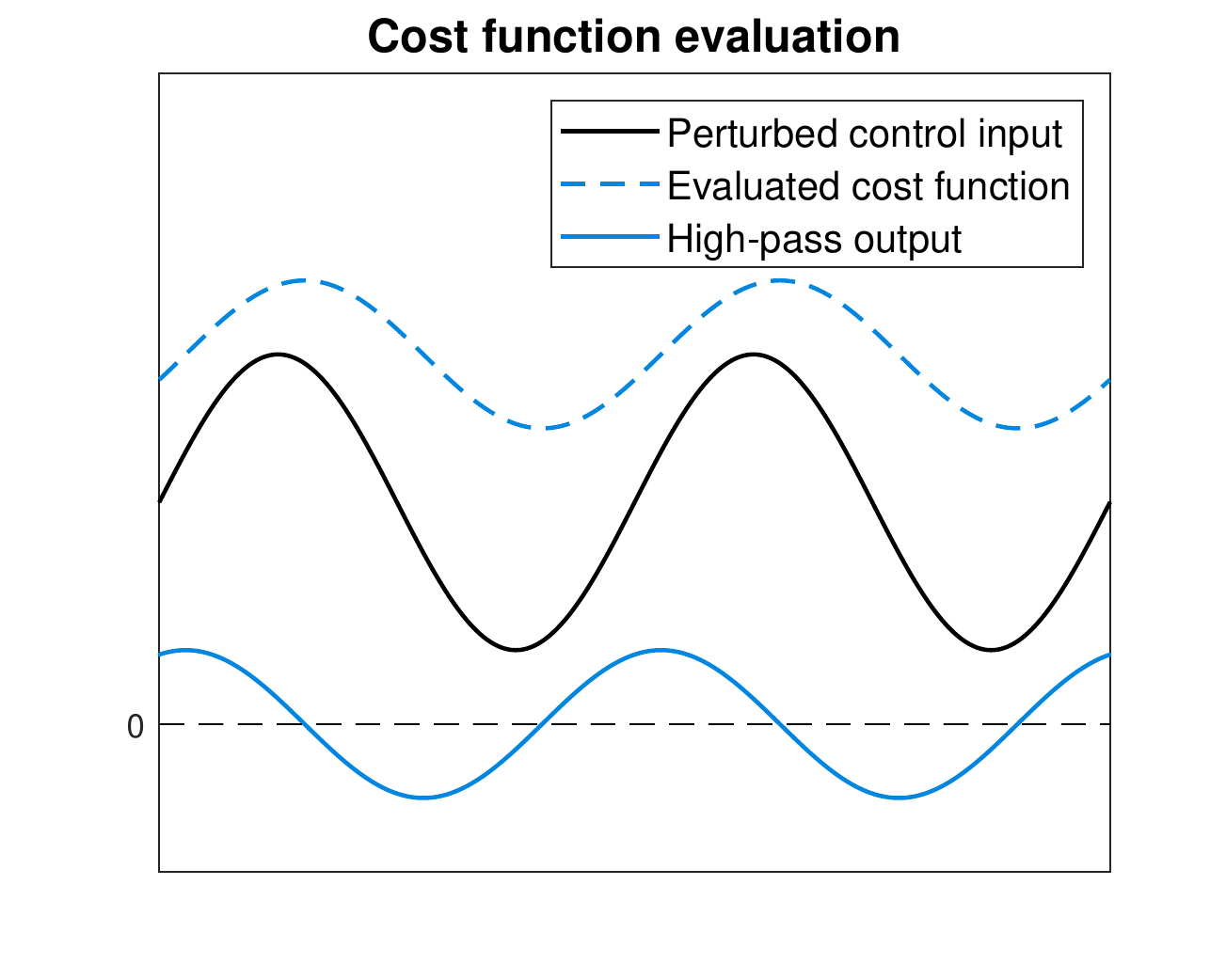}
\includegraphics[width=.45\textwidth, trim=0 1cm 0 0, clip]{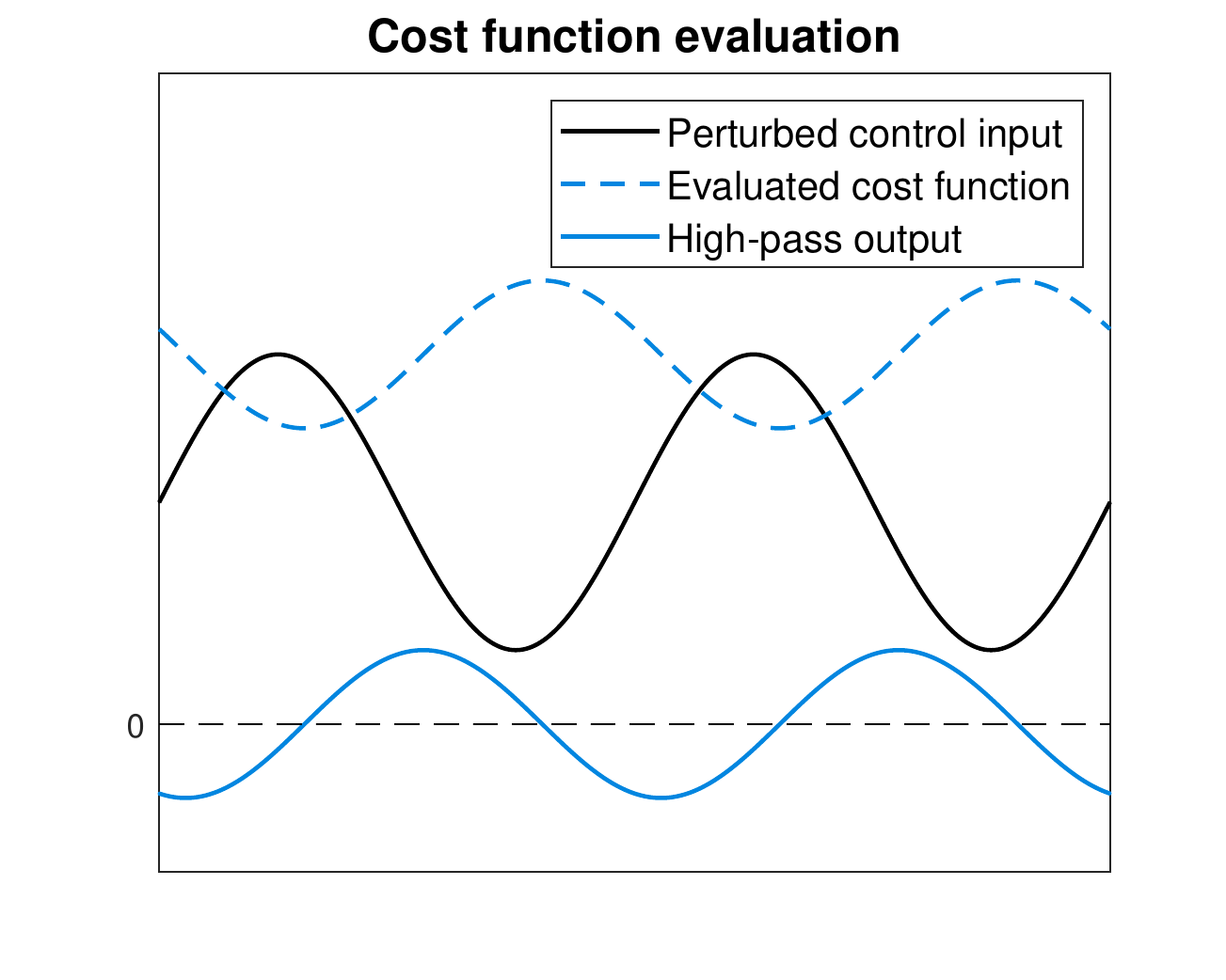}
\caption{\blue{Cost function reaction to the harmonic perturbation. The high-pass filter eliminates the DC component but the signal is shifted to the left since the filter adds phase at low frequencies. The left plot represents a case where the gradient of the cost function is positive, while the right one represents a case with a negative gradient. In general, time delay may occur.}}
\label{f:esc1}
\end{figure}

\begin{figure}
\centering
\includegraphics[width=.45\textwidth, trim=0 1cm 0 0, clip]{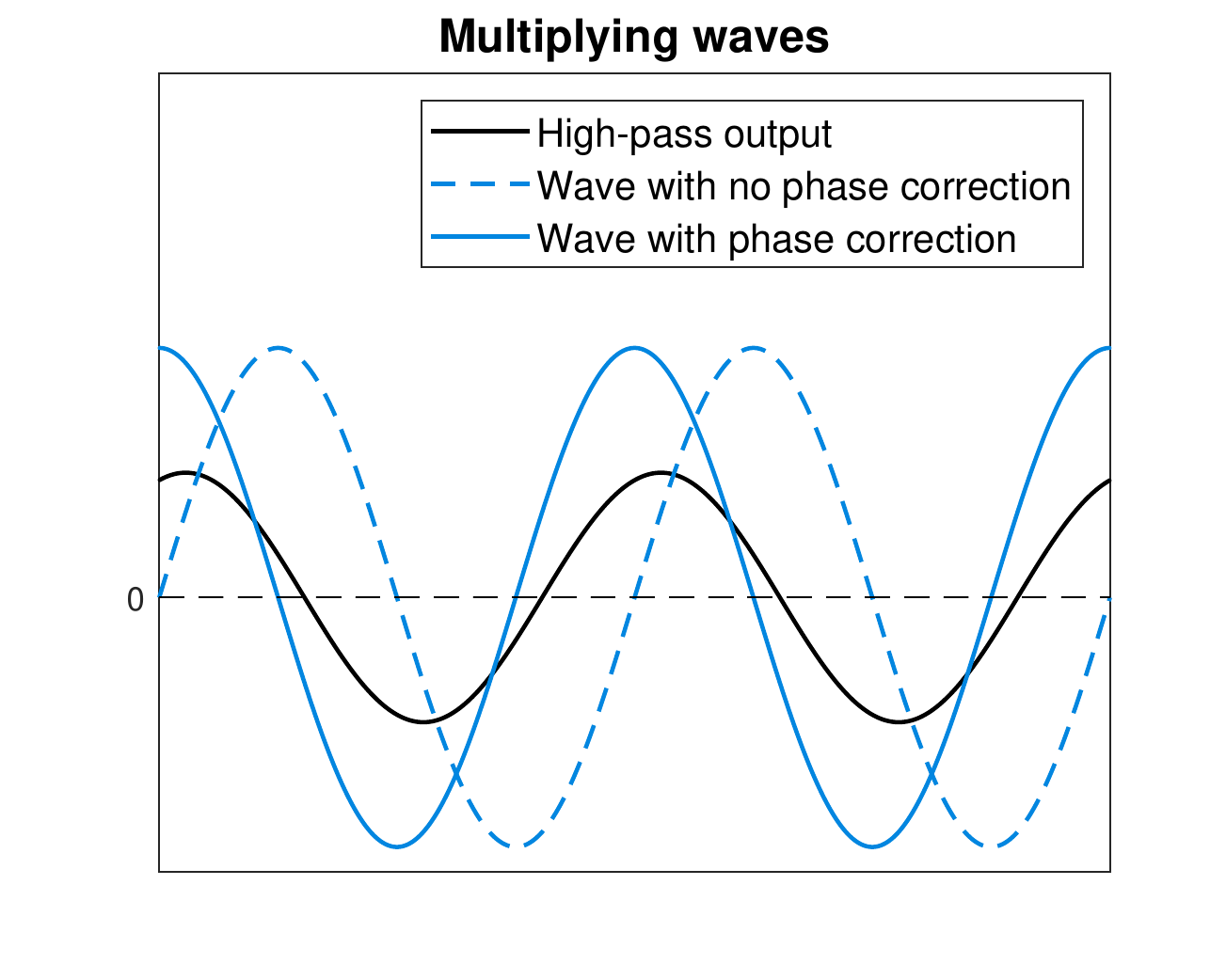}
\includegraphics[width=.45\textwidth, trim=0 1cm 0 0, clip]{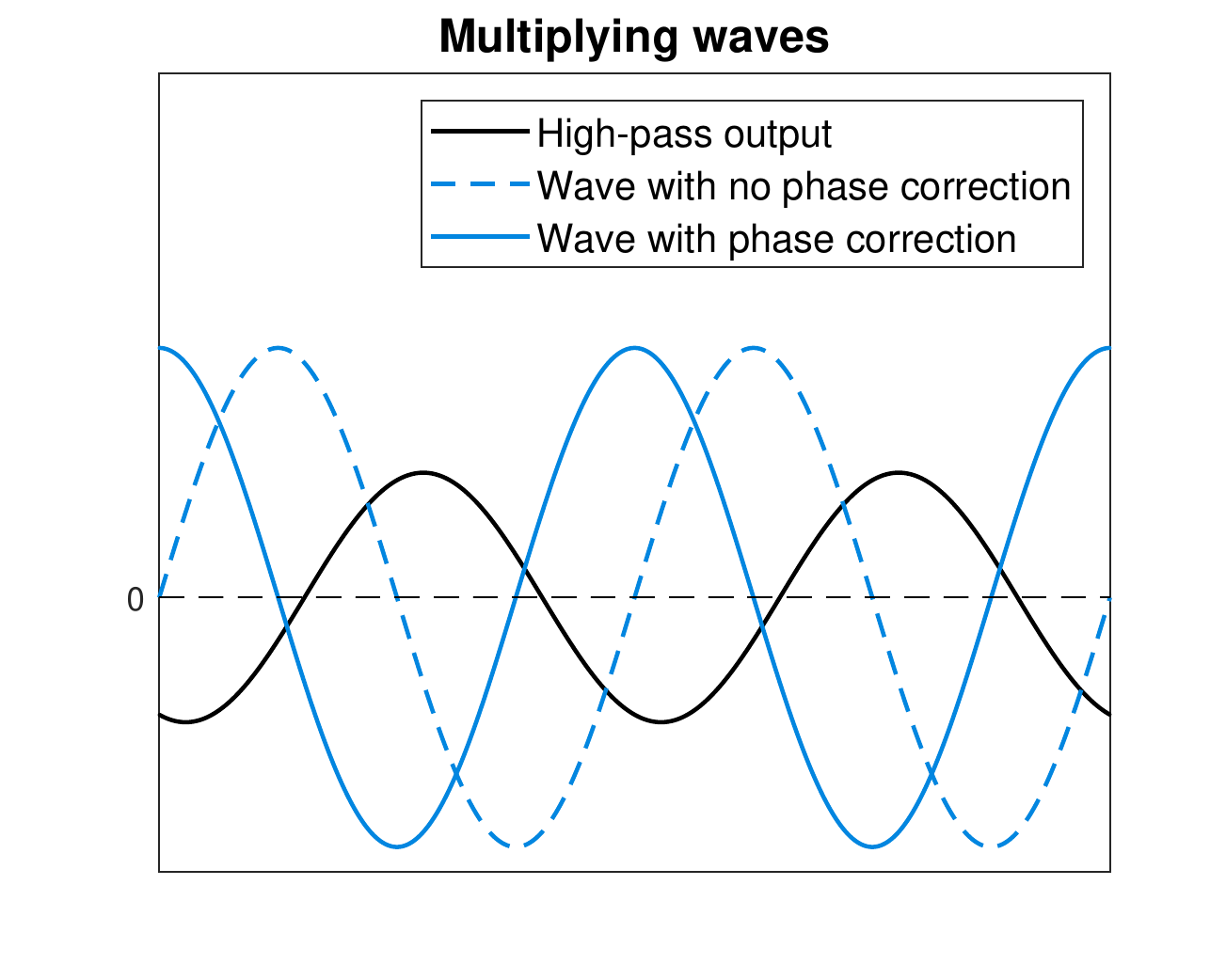}
\caption{Since the high-pass filter output will be multiplied by an oscillatory wave to extract information about the cost function gradient, the phase added by intermediate steps can be compensated to match the multiplying waves. The left plot shows a positive gradient case, while the right one shows a negative gradient case.}
\label{f:esc3}
\end{figure}

\begin{figure}
\centering
\includegraphics[width=.45\textwidth, trim=0 1cm 0 0, clip]{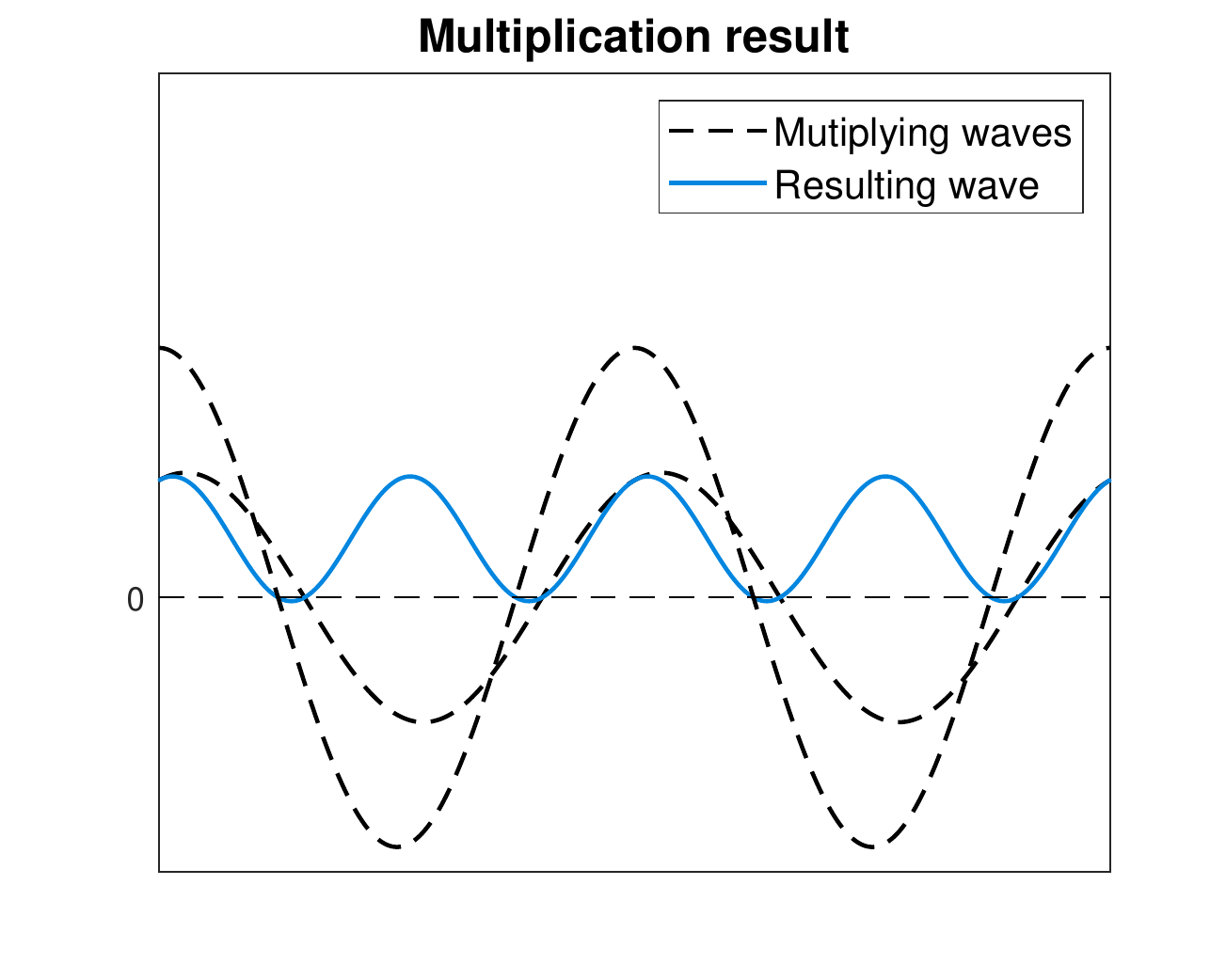}
\includegraphics[width=.45\textwidth, trim=0 1cm 0 0, clip]{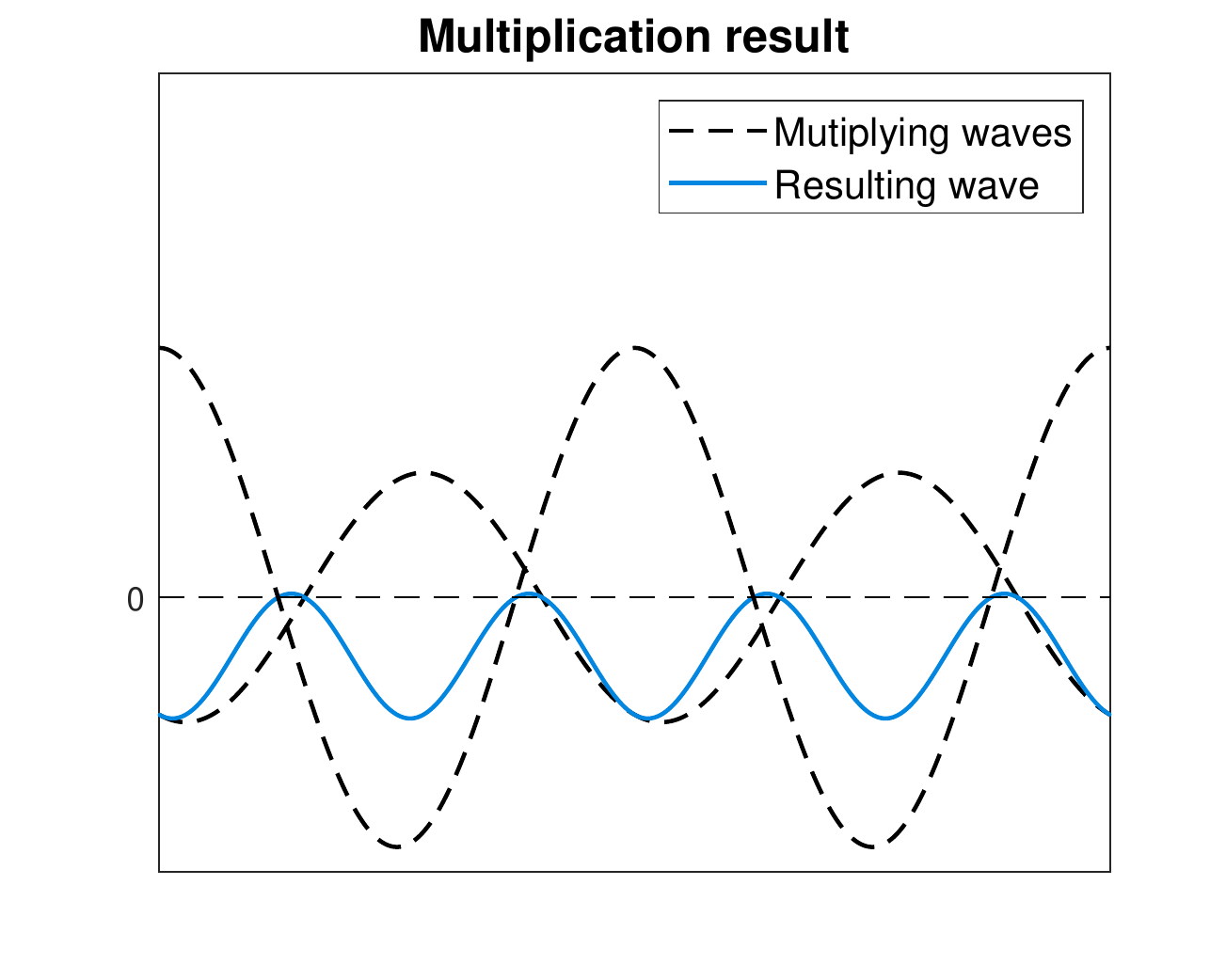}
\caption{The left plot shows a positive gradient case and the multiplication of waves results in a signal that is positive most of the time. The right one shows a negative gradient case, which produces a mostly negative signal.}
\label{f:esc4}
\end{figure}

\blue{Figures \ref{f:esc1} to \ref{f:esc4}} show a simplified example of the ESC loop behavior for $\eta$ = 0 (hence, $\bar{u}_c$ is fixed). Figure \ref{f:esc1} illustrates the expected cost function response $J$ to the periodic sweep. By ensuring proper timescale separation, the system is perturbed so the cost function responds periodically with the same frequency. The resulting phase delay is unknown and can vary with $\bar{u}_c$. 
The cost function output is high-pass filtered so its DC component is removed, as \blue{also} shown in \blue{Fig. \ref{f:esc1}}. The resulting wave is used to compute a correlation between the input and output directions.

%
%\begin{figure}
%\centering
%\includegraphics[width=.45\textwidth, trim=0 1cm 0 0, clip]{figures/esc_steps/esc1.pdf}
%\includegraphics[width=.45\textwidth, trim=0 1cm 0 0, clip]{figures/esc_steps/escN1.pdf}
%\caption{Cost function reaction to the harmonic perturbation. The left plot represents a case where the gradient of the cost function is positive, while the right one represents a case with a negative gradient. In general, time delay may occur.}
%\label{f:esc1}
%\end{figure}

%\begin{figure}
%\centering
%\includegraphics[width=.45\textwidth, trim=0 1cm 0 0, clip]{figures/esc_steps/esc2.pdf}
%\includegraphics[width=.45\textwidth, trim=0 1cm 0 0, clip]{figures/esc_steps/escN2.pdf}
%\caption{The high-pass filter eliminates the DC component but the signal is shifted to the left since the filter adds phase at low frequencies. The left plot shows a positive gradient case, while the right one shows a case with negative gradient.}
%\label{f:esc2}
%\end{figure}

High-pass filters can add positive phase $\phi$ at lower frequencies, as illustrated in \blue{Fig. \ref{f:esc1}}. Since they are implemented as digital linear dynamic systems, the exact value of $\phi$ can be easily obtained. A synthetic wave $\beta \sin(\omega t + \phi)$, which has its phase corrected, as illustrated in Fig. \ref{f:esc3}, is multiplied by the high-pass filtered signal. 
The product of the synthetic wave and the high-pass filtered output is low-pass filtered to reduce ripple, and the resulting wave $\nabla\bar{J}$ is approximately proportional to the gradient of the cost function $J$. The control law that closes the loop is given by
\begin{equation}
    \dot{\bar{u}}_c = \eta \nabla\bar{J} \mbox{ ,}
\end{equation}
where $\eta$ is a gain used to tune the convergence rate. As illustrated in Fig. \ref{f:esc4}, the direction to update $\bar{u}_c$ is given by the sign of the product and the numerical integration follows the appropriate direction; for minimization problems, a negative value is set to $\eta$.
All of the digital modules used to implement ESC have the same time step $\Delta t_c$, which is a multiple of the simulation time step $\Delta t_\text{sim}$.

\subsection{Slope seeking control} \label{sec:ssc}

\begin{figure}
\centering
\begin{tikzpicture}[auto, node distance=2cm,>=latex']
    \node [input] (sweep) {};
    \node [sum, right of = sweep, node distance = 2cm] (sum) {$+$};
    \node [block, right of=sum, node distance = 2.2cm] (plant) {N-S equations};
    \node [block, right of=plant, node distance = 3.2cm] (cost) {$J(\cdot)$};
    \node [block, right of=cost, node distance = 3.2cm] (hp) {High-pass filter};
    \node [sum, below of = hp, right of = hp, node distance = 2cm] (prod) {$\times$};
    \node [right of = hp, node distance = 2cm] (rtcorner) {};
    \node [input, right of = prod, node distance = 2.5cm] (multwave) {};
    \node [block, left of=prod, node distance = 2.5cm] (lp) {Low-pass filter};
    \node [sum, left of=lp, node distance =3.0cm] (ss) {$+$};
    \node [input, above of = ss, node distance = 1.2cm] (slope) {};
    \node [block, left of=ss, node distance =2.5cm] (integ) {$\eta\int_{0}^{t} (\cdot) dt$};
    \node [output, above of=cost, node distance = 5cm] (output) {};
    
    \draw [->] (sweep) -- node {$\alpha\sin(\omega t)$} (sum);
    \draw [->] (sum) -- node {$u_c$} (plant);
    \draw [->] (plant) -- node [name=pp, inner sep = 0, pos=0.55] {} (cost);
    
    \node [output, above of=pp, node distance = 1cm] (output) {};
    \node [input, below of=pp, node distance = 0.03cm] (input2) {};
    
    \draw [->] (cost) -- node {} (hp);
    \draw [->] (hp) -| node {} (prod);
    \draw [->] (multwave) -- node {$\beta\sin(\omega t + \phi)$} (prod);
    \draw [->] (prod) -- node {} (lp);
    \draw [->] (lp) -- node {$\nabla \bar{J}$} (ss);
    \draw [->] (ss) -- node {} (integ);
    \draw [->] (slope) -- node {$\delta_{ss}$} (ss);
    \draw [->] (integ) -| node {$\bar{u}_c$} (sum);
    \draw [->] (input2) -- node {$p'$} (output);
\end{tikzpicture}
\caption{\green{Slope seeking block diagram. The goal is to control (reduce) the measured output $p'$.}}
\label{f:ssloop}
\end{figure}
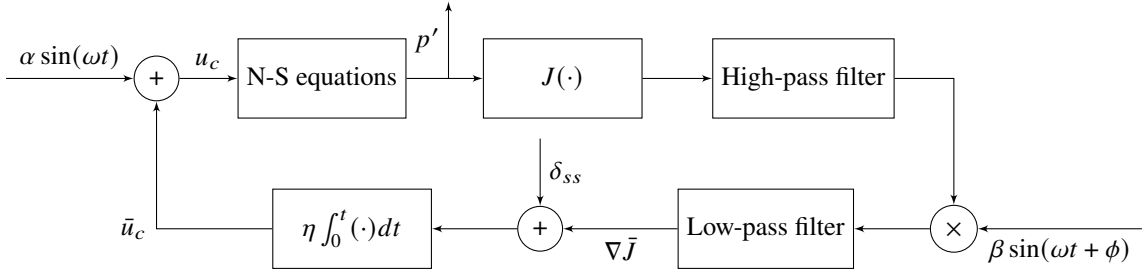

In some of the flow configurations studied in this work, the cost function presents a \textit{plateau}. When the cost function reaches this region, increasing the control effort does not change the output significantly. The slope seeking generalization can be used to find a derivative $J' \neq 0$ by setting a new offset value $\delta_{ss}$ to drift the equilibrium to a different point, as shown in Fig. \ref{f:ssloop}. 

An approximation for the value of $\delta_{ss}$ based on a target $J'_{\text{ref}}$ can be computed as

\begin{equation}
\delta_{ss} = \frac{\alpha \beta}{2} J'_{\text{ref}}\, |(H(j\omega))| \mbox{ ,}
\end{equation}
where $|(H(j\omega))|$ is the amplitude gain of the high-pass filter at frequency $\omega$. This formulation is similar to that \blue{presented} by \citet{ariyur2004slope}. Here, $\alpha$ and $\beta$ compensate the chosen harmonic amplitudes for the loop and the factor $1/2$ is used to compensate the time average of $\sin^2 (\omega t)$. The term $|(H(j\omega))|$ compensates the amplitude attenuation imposed by the high-pass filter. In this approach, the absolute value $|(H(j\omega))|$ should be used instead of the real part since the filter phase $\phi$ is already rectified when the waves are multiplied. Also, the phase delay between the input and the cost function output is considered negligible for sufficiently low frequencies $\omega$.

\subsection{Probing and cost function}\label{sec:probe}

For all configurations studied in this work, probing of instantaneous pressure is conducted in real-time at $(x,y) = (0,1)$, i.e., at one chord above the leading edge,  \green{as shown in Fig. \ref{f:cdomain}.}
The position is chosen in order to capture the acoustic waves propagating upstream. Typically, low frequency trailing-edge noise is of dipolar character \citep{curle:55} and its main direction of propagation is perpendicular to the airfoil chord. On the other hand, high frequency trailing-edge noise propagates mostly upstream following a cardioid pattern \citep{hall:70}. Due to the finite airfoil chord, secondary diffraction at the leading edge will introduce phase interference at the medium and high frequencies, leading to multiple lobes in the sound directivity. Hence, a microphone positioned above the leading edge would be able to capture the main trends of airfoil noise emission for both the low and high frequencies.
\green{It is important to mention that the sensor could also be placed on the airfoil surface, in regions absent of convective hydrodynamic disturbances. We observed that the signal computed for a sensor placed near the leading edge is highly coherent with that at the acoustic field since only acoustic disturbances propagate upstream. This setup could be more easily implemented in an experiment using surface microphones.}

The probed values are high-pass filtered to obtain the fluctuations $p'$ in the observer position. Due to the high frequencies of the acoustic emission (when compared to $\omega$ from the ESC), a very fast filter can be designed so the transient presents a very short settling time and does not interfere with the control loop. The acoustic oscillation period $T_a$ is measured from the uncontrolled plant and the following parameters are used:
\begin{itemize}
\item allowed passband ripple is of -2dB;
\item minimum stopband attenuation is set to -20dB;
\item cutoff angular frequency is set to 40\% of $2\pi/T_a$;
\item passband/stopband frequency ratio is set to 300\%.
\end{itemize}

The implementation details of a digital filter with this set of parameters are further described in Sec. \ref{sec:ldf}. With $p'(t)$ computed, the cost function is defined as
\begin{equation}
    J_{k} = \left[\sum_{i=0}^{m-1}\left(p'_{k-i}\right)^2\right]^\sigma \mbox{ ,}
    \label{e:cost}
\end{equation}
where $\sigma$ is a positive exponent used to modulate differences between the larger and lower values assumed by the function, $m$ is the number of measures (window size) used to compute the signal energy and $k$ is the discrete time iteration so that $p'_k = p'(t=k \Delta t_c)$. This penalty function increases or decreases according to the intensity of acoustic waves. An $m$ value is chosen so it is large enough to reduce high-frequency ripple in the cost function but also short enough to avoid delays due to the use of much earlier samples.

\subsection{Actuation and momentum transfer} \label{sec:act}

Permeable surface boundary conditions are applied on the airfoil wall to perform the blowing and suction actuation. Given a positional actuation window, the device can be characterized by the maximum blowing/suction momentum $A$ and the actuator horizontal position $x$, where the latter is given by the non-dimensional position with respect to the airfoil chord. In this work, two different control input approaches are employed. The first one makes use of a fixed amplitude $A$ with variable position, which is treated as a control input $u_c = x_c$; thus the controller searches for the best position for the actuator to reduce the airfoil sound emission. The second approach considers a fixed actuator position $x$ with variable momentum intensity. Hence, the control input $u_c = A_c$ is used so the controller now searches for an optimal amplitude instead of the actuator position.

Two types of window functions are used in this work. For the cases where the control input is $A_c$ and the position is fixed at the airfoil suction side, the actuator is a square window. Given a position $x$ and a window length $\Delta x$, the grid points $x_i$ at the wall from $x-\Delta x/2$ to $x+\Delta x/2$ are activated with $\rho v_n = A_c$ as
\begin{align}
    &(\rho v_n)_i = A_c \,\, \mathcal{W}_1\left[2(x_i - x)/\Delta x\right] \mbox{ ,}\\
    &\mathcal{W}_1(\theta) = \begin{cases} 
    1 ,& \text{if } -1 < \theta < 1\\
    0,              & \text{otherwise} 
\end{cases}\mbox{ ,}
\end{align}
where $\mathcal{W}_1(\theta)$ is the square window function. Hence, for those points outside the actuation region, $\rho v_n = 0$. When the actuator has its position fixed, a square window can be used since every grid point is either always activated or deactivated (although the blowing or suction intensity may vary).

%In cases where the actuator moves, $x = x_c$, the values of momentum for each point $x_i$ are given by

\begin{figure}
\centering
\includegraphics[width=.99\textwidth]{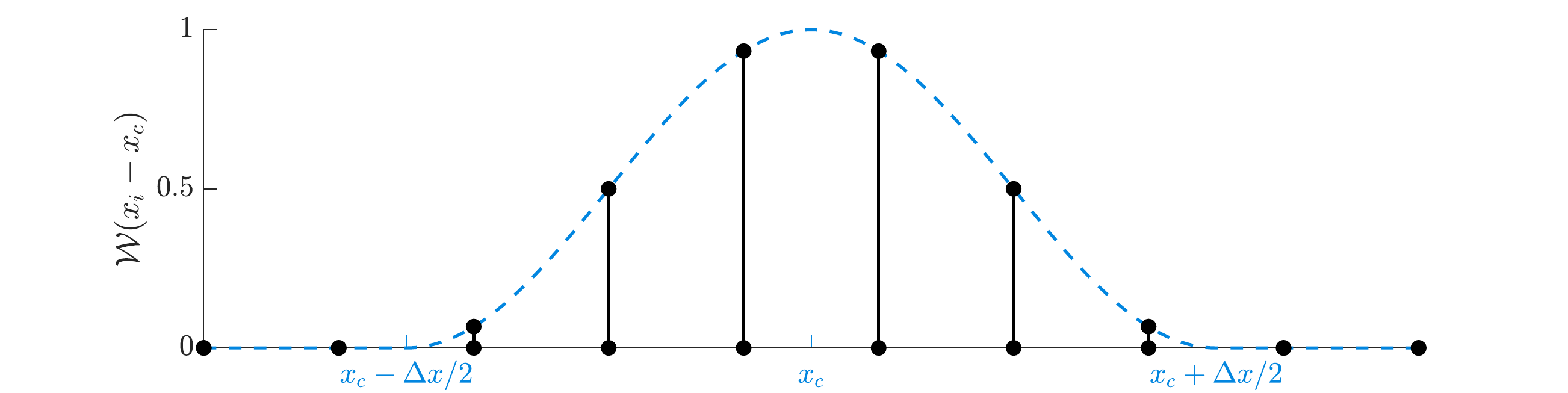}
\includegraphics[width=.99\textwidth]{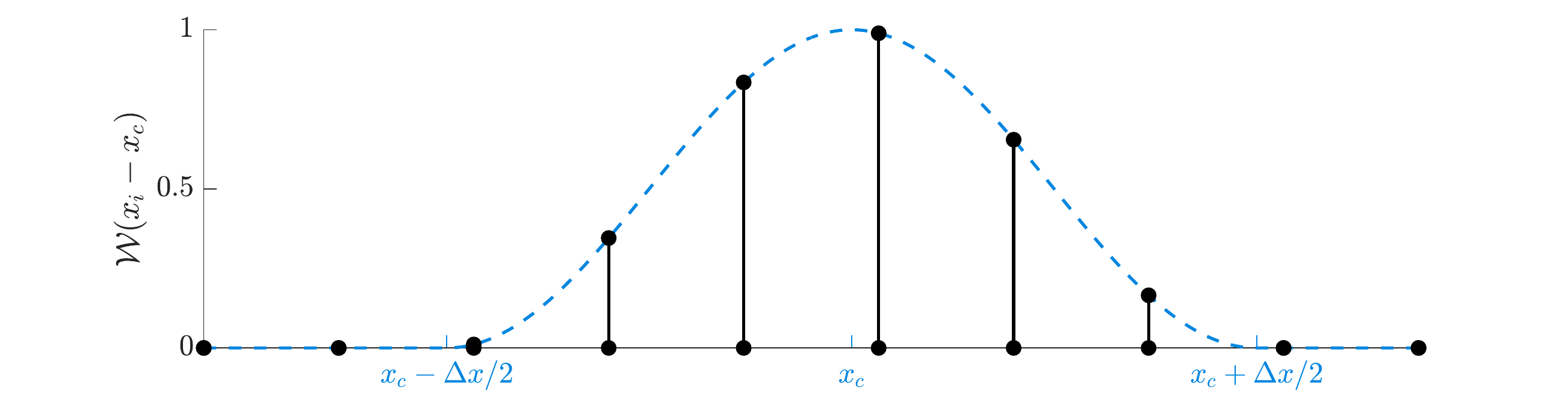}
\caption{Window used to determine the actuation region. The function is centered at $x_c$. By moving $x_c$ to the right, for example, the actuation intensity increases at the grid points to the right and decreases at points to the left.}
\label{f:act}
\end{figure}

In cases where the control input is $x_c$, or in those cases where it is $A_c$, but with the position $x$ locked at the trailing edge, a Hann window function %\citep{harris1978then}
is used. The cosine-based function is used to dampen signal discontinuities at the actuation boundaries and to enable a smoother actuator movement. The way the momentum at the actuated grid points $x_i$ behave is illustrated in Fig. \ref{f:act}
and it is also given by
\begin{align}
    &(\rho v_n)_i = A \,\, \mathcal{W}_2\left[2\pi (x_i - x)/\Delta x\right] \mbox{ ,}\\ \label{e:act} \\
    &\mathcal{W}_2(\theta) = \begin{cases}
    0.5 + 0.5\cos(\theta) \label{e:win} ,& \text{if } -\pi < \theta < \pi\\
    0,              & \text{otherwise}
\end{cases}\mbox{ ,}
\end{align}where $\mathcal{W}_2(\theta)$ is the cosine based window function. \green{In a practical application, this approach could be implemented, for example, by activating certain elements of an array of jets as presented by \citet{zigunov2021empirical}} 

\begin{figure}
\centering
\includegraphics[width=.49\textwidth]{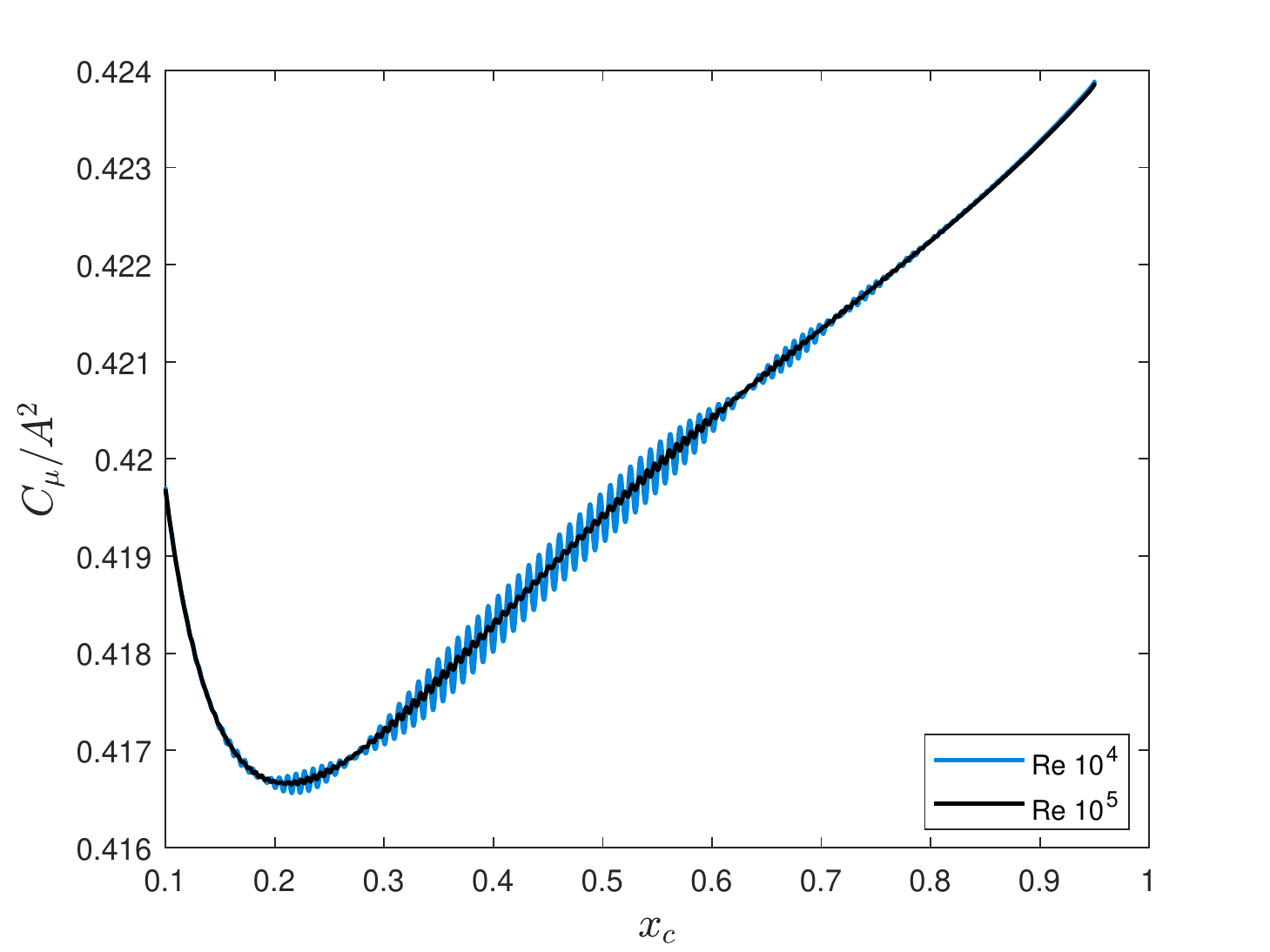}
\caption{Momentum transfer $C_\mu$ normalized by the squared maximum momentum $A$. The window function $\mathcal{W}_2$ is used and the variation of $C_\mu$ along the wall with moving actuation is under 1\% above and below the average. %The high-frequency fluctuations observed are due to the discrete form of $\mathcal{W}_2$ applied to the individual grid points, and $\Delta x = 5\%$ is chosen to keep its ripple low. The lower frequency variation in $C_\mu$ occurs because $\mathcal{W}_2$ employs the discrete $x$-coordinates as argument, which leads to different lengths of actuation along the airfoil chord.
}
\label{f:cmu_pos}
\end{figure}

An instantaneous momentum transfer coefficient $C_\mu$ is computed numerically according to the expression
\begin{equation}
    C_\mu = \frac{2\int_s \rho v_n^2 \,ds}{\rho_\infty U_\infty^2 L} \mbox{ ,}
\end{equation}
where $\rho_\infty$ and $U_\infty$ are the freestream density and velocity, respectively, and $L = 1$ is the non-dimensional chord length. \green{The momentum transfer coefficient provides a direct evaluation of energy consumption by the actuation and values of $C_\mu$ are provided for different configurations studied in Appendix B.} The integral is computed over the actuation region $s$. For simplicity, the density of the actuation jet is considered as $\rho = \rho_{\infty} = 1$. Figure \ref{f:cmu_pos} presents $C_\mu$ normalized by the squared maximum momentum $A$ as a function of the actuator position using $\mathcal{W}_2$ with $\Delta x = 5\%$. This window length was used in a set of simulations as detailed in Sec. \ref{sec:cases}, being wide enough to keep low grid-induced ripple along the wall. Some of the values for $A$ used in this work are: 0.060 ($C_\mu \approx 1.511e-3$), 0.030 ($C_\mu \approx 3.778e-4$) and 0.012 ($C_\mu \approx 6.045e-5$). These values resemble those found in literature \citep{ramos2019active, goodfellow2013momentum, benton2017high, benton2018evaluation} for flow actuation in airfoil flows. The variation of $C_\mu$ along the wall with moving actuation is under 1\% above and below the average. The high-wavenumber fluctuations observed in Fig. \ref{f:cmu_pos} are due to the discrete form of $\mathcal{W}_2$ applied to the individual grid points. The lower wavenumber variation in $C_\mu$ occurs because $\mathcal{W}_2$ employs the discrete $x$-coordinates as argument, which leads to slightly different lengths of actuation along the airfoil chord.

\begin{figure}
\centering
\includegraphics[width=.49\textwidth]{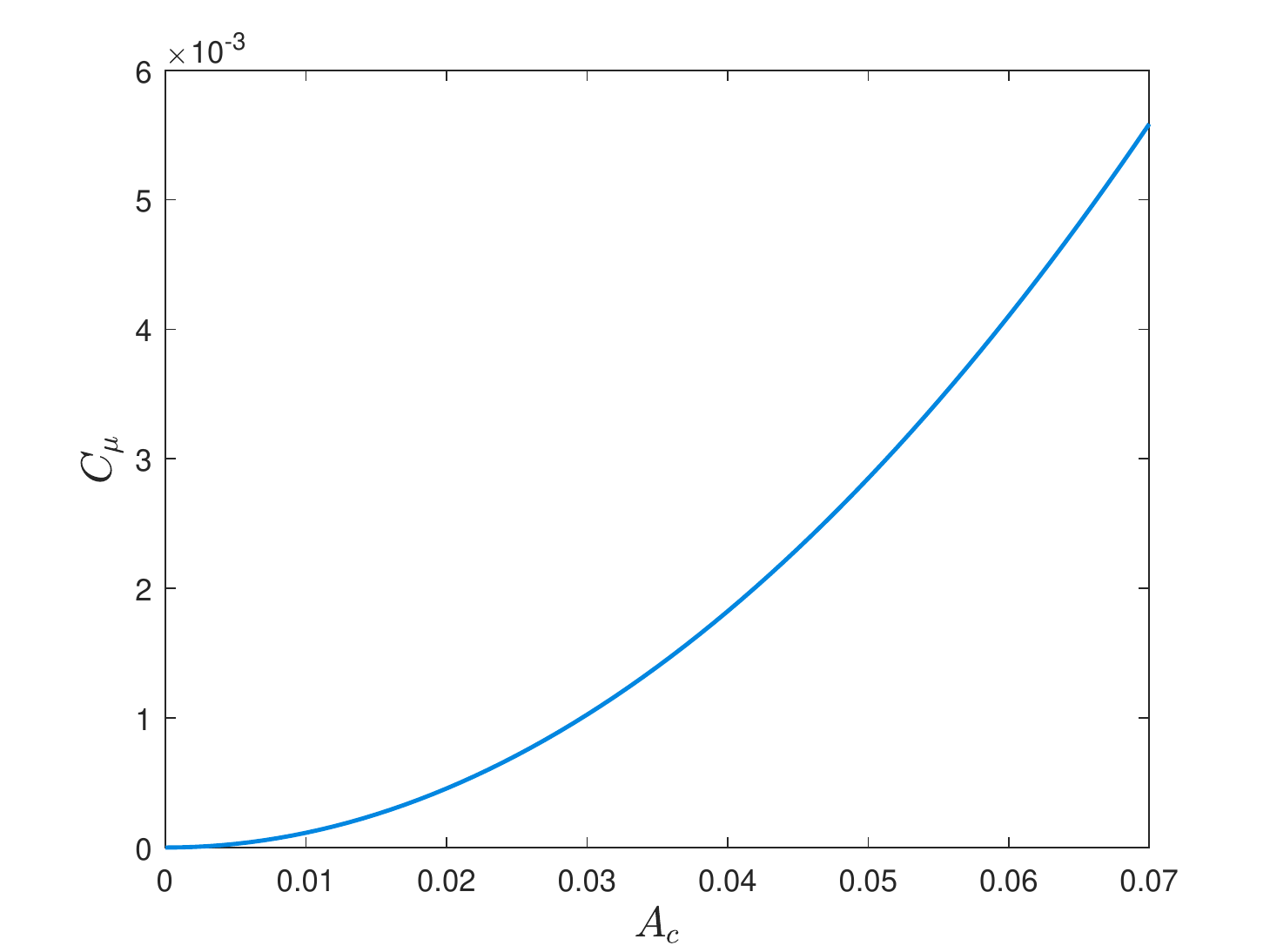}
\includegraphics[width=.49\textwidth]{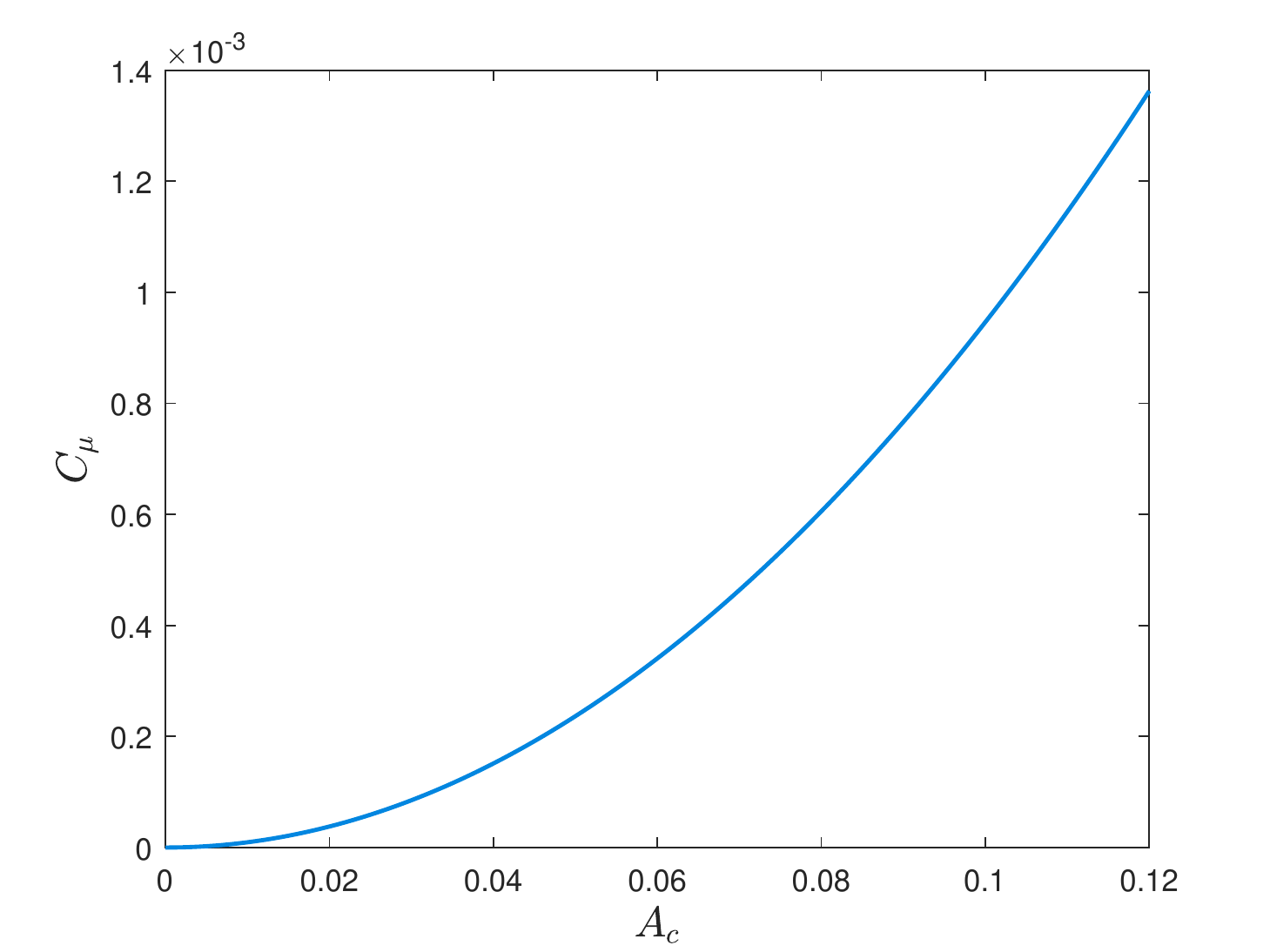}
\caption{Coefficient $C_\mu$ as a function of $A_c$ for actuators with varying intensity and fixed position. The left plot shows the case of an actuator with window function $\mathcal{W}_1$ at the suction side with $x_c = 0.925$ and $\Delta x = 5\%$. The right plot shows the case where the actuation is placed at the trailing edge, with further details provided in Sec. \ref{sec:cases}.}
\label{f:cmu_amp}
\end{figure}

The momentum transfer coefficients associated with the position-locked actuators are presented in Fig. \ref{f:cmu_amp}. The left plot is for the square window function $\mathcal{W}_1$ at the suction side with $x_c = 0.925$ and $\Delta x = 5\%$. The right plot shows $C_\mu$ for those cases where the actuator is placed at the trailing edge, and where $\mathcal{W}_2$ is used. All these configurations are studied in the present work and further details are presented in Sec. \ref{sec:cases}.

\subsection{Linear dynamic systems} \label{sec:lds}

Digital linear systems are incorporated in the original CFD numerical tool to compose the ESC and slope seeking loops. These modules are used to implement the digital filters and the integrator. The coefficients $a_i$ and $b_i$ are stored for each $n$-th order transfer function in the form
\begin{equation}
    L(z) = \frac{a_0 + a_1z^{-1} + \dots + a_nz^{-n}}{b_0 + b_1z^{-1} + \dots + b_nz^{-n}} \mbox{ ,}
\end{equation}where $z$ is the $\mathcal{Z}$-transform variable. These coefficients can be used to evaluate the difference equation
\begin{equation}
    y[k] = \frac{1}{b_0}\bigg{(}\sum_{i=0}^{n} a_ix[k-i] - \sum_{i=1}^{n} b_iy[k-i]\bigg{)} \mbox{ ,}
\end{equation}where $x[k]$ and $y[k]$ are the system input and output, respectively. A number of past input and output values must be saved according to the order $n$.

\subsubsection{Integrator}

The numerical integrator is implemented by following the trapezoidal rule so the corresponding transfer function is set as
\begin{equation}
    I(z) = \frac{\eta \Delta t_c}{2} \frac{1 + z^{-1}}{1 - z^{-1}} \mbox{ ,}
\end{equation}
where $\eta$ is the integration gain as described in Sec. \ref{sec:esc}.

\subsubsection{Digital Filters} \label{sec:ldf}

The digital filters applied in each of the simulations are: a high-pass filter for pressure fluctuation probing (Sec. \ref{sec:probe}); a high-pass filter for the control loop (Sec. \ref{sec:esc}); and a low-pass filter for the control loop (Sec. \ref{sec:esc}).
All filters designed in the present work are Chebyshev type I \citep{tan2018digital}. First, the following set of characteristcs are chosen:
\begin{itemize}
\item allowed passband ripple;
\item minimum stopband attenuation;
\item cutoff frequency as a percent of $\omega$ for the loop filters, or $2\pi/T_a$ for the probing, where $T_a$ is the acoustic wave period;
\item frequency ratio $\omega_p/\omega_s$ for high-pass filters, or $\omega_s/\omega_p$ for low pass filters, where $\omega_p$ is the limit of the passband and $\omega_s$ is the limit of the stopband.
\end{itemize}

Next, the required order $n$ of the system is estimated. A set of $n$ poles for an equivalent analog filter is obtained at $s=s_1, s=s_2, \dots, s=s_n$, where $s$ is the Laplace transform variable. These poles are initially obtained for a low-pass Chebyshev type I filter, and are converted (if needed) to a high-pass one through the relation
\begin{equation}
    s_{i,\text{hp}} = \frac{\omega_p^2}{s_{i,\text{lp}}} \mbox{ .} \label{eq:hpconv}
\end{equation}
A correspondent set of poles is obtained for the digital equivalent filter by using the bilinear (or Tustin) transform 
\begin{equation}
    z_i = \frac{1 + (s_i\Delta t_c)/2}{1 - (s_i\Delta t_c)/2} \mbox{ ,} \label{eq:tustin}
\end{equation}
which can preserve the frequency response characteristics at the range of interest. Frequency pre-warping is not applied in this work \citep{franklin2015feedback}. 

The coefficients $b_i$ for the transfer function denominator are calculated from the roots $z_i$. The polynomial $b_0 + b_1z^{-1} + \dots + b_nz^{-n}$ is then evaluated at every $z=z_i$ to check if relevant numerical errors are carried out. This test is important because distortions can occur at lower frequencies when $z_i - 1$ gets very close to zero due to the sampling time $\Delta t_c$ being much lower than $2\pi/\omega_p$. Some tests performed during this work showed that the evaluation of $\prod_{i=1}^n{(z-z_i)}$ to find the coefficients $b_i$ can produce very small errors that significantly alter the filter behavior. For all cases studied in this work, the results of this robustness check are in the order of the numerical precision.

To compute the coefficients of the numerator polynomial for low pass filters, the zeros are all placed at $z_i = 0$. From the direct zero mapping equivalence $z_i = e^{s_i\Delta t_c}$, $s_i \xrightarrow{} -\infty$ as $z_i \xrightarrow{} 0$, which is coherent with analog Chebyshev type I low-pass filters having no zeros. Alternatively, the bilinear transform (Eq. \ref{eq:tustin}) could have been used to obtain better high-frequency response characteristcs, which would lead to $z_i = -1$. Since in the present work $z_i = 0$ is considered, $a_0 = \sum b_i$ and $a_{i\neq 0}=0$ are chosen so the low frequency gain (near $z=1$) is unity (0 dB).
 
For the high-pass filter case, the equivalence in Eq. \ref{eq:hpconv} gives $s_{i,hp} \xrightarrow{} 0$ as $s_{i,lp} \xrightarrow{} -\infty$. From the analog zeros $s_{i,hp} = 0$, the discrete zeros are then placed at $z_i = 1$; this value can be obtained either from the direct zero mapping equivalence $z_i = e^{s_i\Delta t_c}$ or the bilinear transform (Eq. \ref{eq:tustin}). The coefficients $b_i^*$ are obtained by expanding the Newton binomial

\begin{equation}
    (z + 1)^n = b_0^* + b_1^*z^{-1} + \dots + b_n^*z^{-n} \mbox{ ,}
\end{equation}
and, with the transfer function computed with coefficients $b_i^*$, a high frequency gain $\gamma$ is obtained (for example, at the Nyquist frequency $\pi/{\Delta t_c}$). Since a 0 dB gain is desired at the passband, $b_i = b_i^*/\gamma$ can be calculated for the final transfer function.

The frequency response for the transfer function $F(z)$ at a desired frequency $\omega_r$ is calculated by evaluating it at $z = e^{j \omega_r \Delta t_c}$. By doing so, $\gamma$ is obtained for the Nyquist frequency through $|F(z = -1)|$. The compensation phase $\phi$ described in Sec. \ref{sec:esc} for the ESC loop high-pass filter can be obtained through the argument $\angle F(z = e^{j \omega \Delta t _c})$, where $\omega$ is the ESC sweeping frequency, also introduced in Sec. \ref{sec:esc}. More details about the filters employed in the present simulations are provided in the Appendix A. %\ref{appA}. %the Appendix.
 
\section{Results}\label{sec:cases}

\newcommand{\ifequals}[3]{\ifx{#1}{#2}{#3}\fi}

\newcommand{\cs}[1]{\ifnum#1=011{1}\fi\ifnum#1=006{2}\fi\ifnum#1=007{3}\fi\ifnum#1=012{4}\fi\ifnum#1=013{5}\fi\ifnum#1=009{6}\fi\ifnum#1=010{7}\fi\ifnum#1=015{8}\fi\ifnum#1=016{9}\fi\ifnum#1=027{10}\fi\ifnum#1=017{11}\fi\ifnum#1=018{12}\fi\ifnum#1=020{13}\fi\ifnum#1=024{14}\fi\ifnum#1=025{15}\fi\ifnum#1=026{16}\fi\ifnum#1=105{17}\fi\ifnum#1=106{18}\fi\ifnum#1=107{19}\fi\ifnum#1=108{20}\fi\ifnum#1=109{21}\fi}

In this section, studies of airfoil noise reduction are presented for controlled flows according to the seeking approach proposed. The subsections are divided according to the flow Reynolds number as well as to the adopted actuation setup. The signals generated by the control loops are displayed to present results for each case. In all simulations, the controller time step is set to $\Delta t_c = 80\Delta t_\text{sim}$.
This value is sufficient to capture the dynamics present in the flow, allowing the resolution of frequencies two times higher than the maximum resolved in the pressure fluctuations. Also, a much smaller $\Delta t_c$ would decrease the robustness of the filter behavior (at low frequencies) to the numerical errors pointed out in Sec. \ref{sec:ldf}.

% A summary of results in terms of noise attenuation is also presented in decibels for the controlled cases that achieved convergence of control inputs in the Appendix B. \ref{appB}.

\subsection{Reynolds number 10,000}

\begin{figure}
\centering
\includegraphics[width=.47\textwidth]{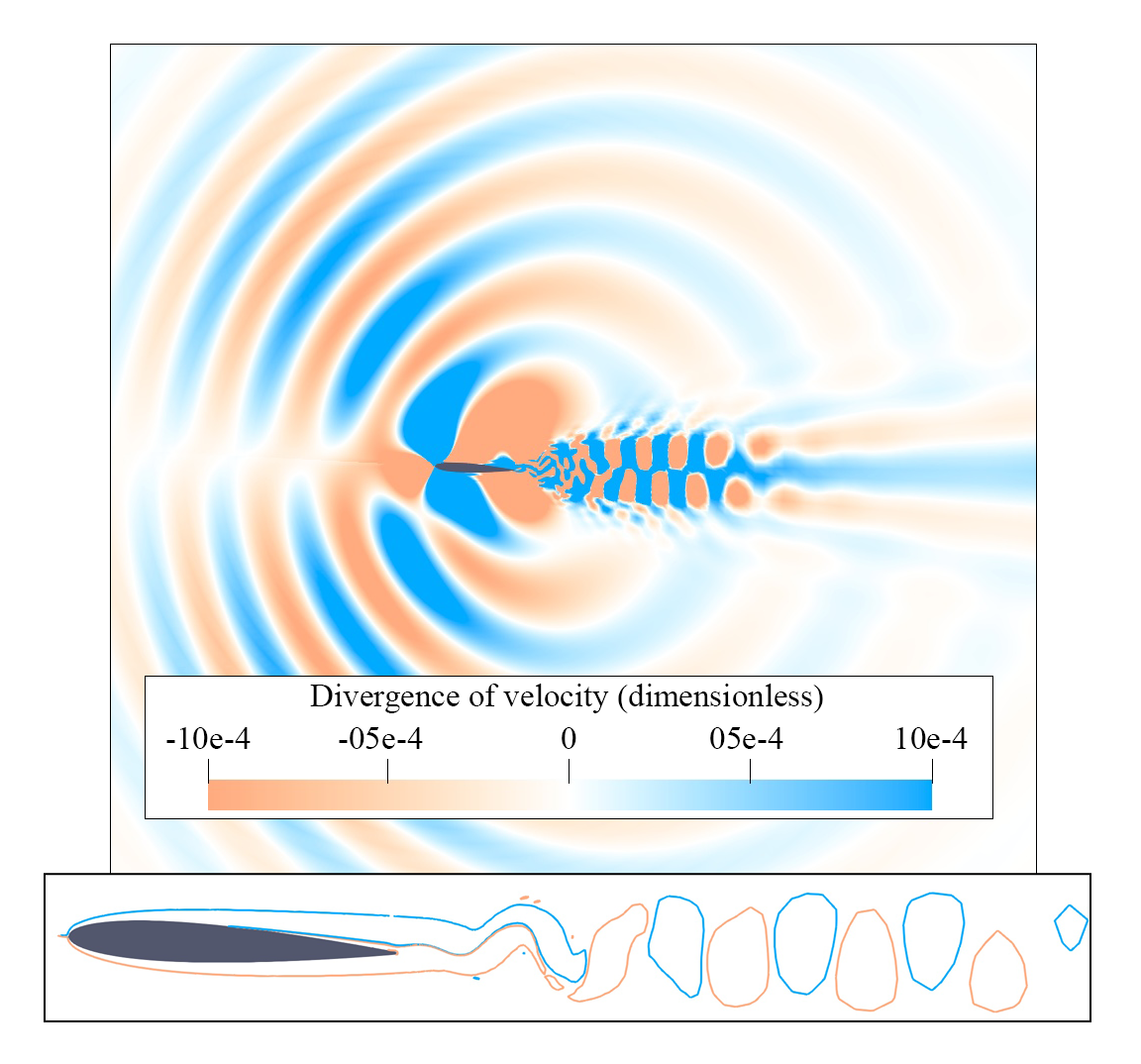}
\includegraphics[width=.45\textwidth]{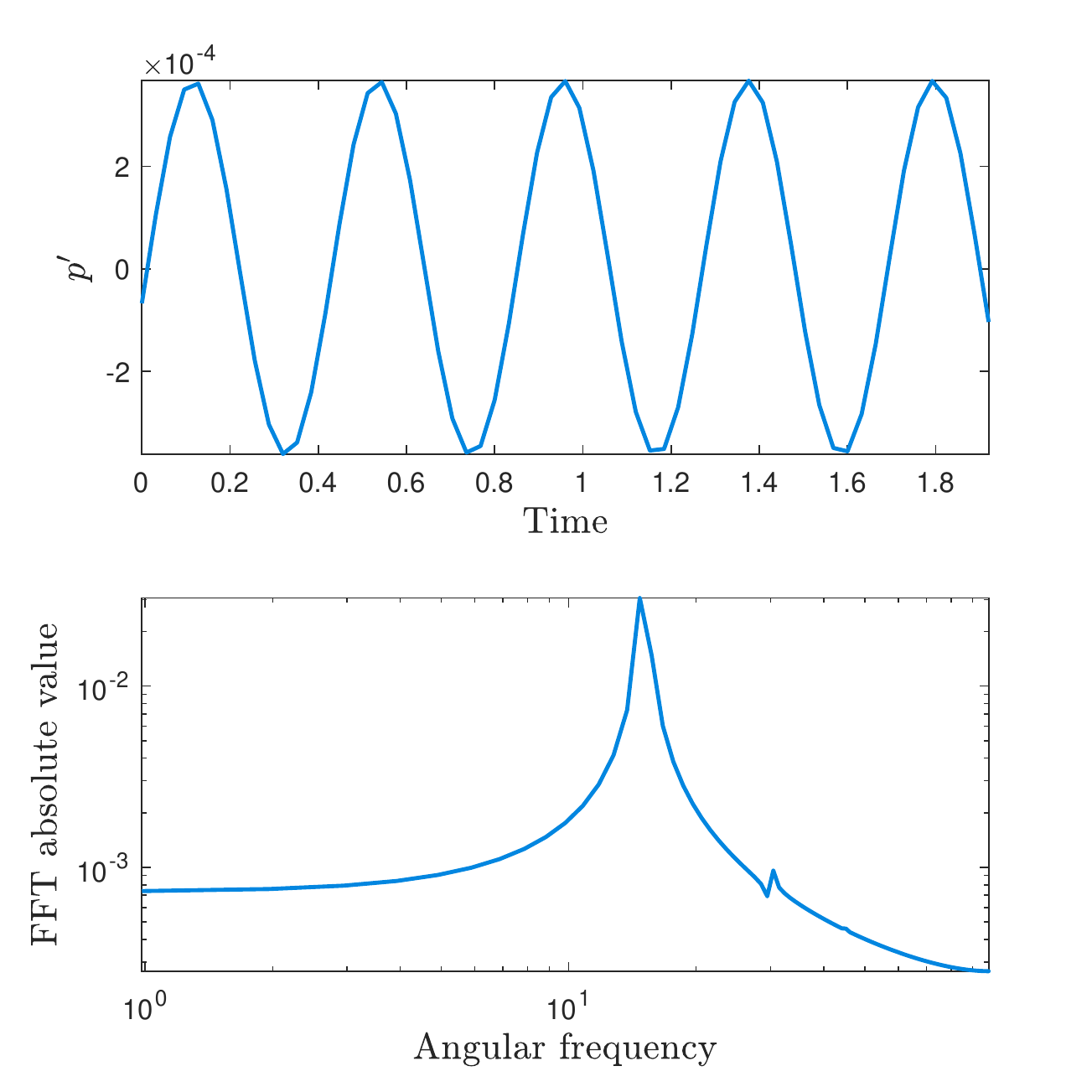}
\caption{The left image shows contours of divergence of velocity and isolines of z-vorticity for a NACA0012 airfoil at $\mathrm{Re} = 10^4$, $\mathrm{Ma} = 0.3$ and $3^\circ$ angle of attack. The right plots show the temporal signal of pressure fluctuation and its Fourier transform computed at the sensor location. No flow control is applied.}
\label{f:10k0}
\end{figure}

The flow past a NACA0012 airfoil with $\text{Re} = 10^4$, $\text{Ma} = 0.3$ and  $3^\circ$ angle of attack is studied first. This flow develops a von Kármán vortex street that generates tonal noise at the shedding frequency and its harmonics. The interaction of the vortical structures with the trailing edge produces acoustic waves that radiate in phase opposition above and below the airfoil. These waves can be seen in the contours of divergence of velocity in Fig. \ref{f:10k0} together with the vortical structures downstream the airfoil. In the same figure, the temporal signal computed at the sensor location and its Fourier transform are shown.

Three actuation approaches are presented for this case. First, a moving actuator with constant suction is applied so an optimal location can be searched at the airfoil suction side to minimize noise by manipulating the flow along the airfoil and its wake. Noise reduction is also sought in the second approach using an actuator with fixed chord position and varying intensity. Finally, a third approach introduced by \citet{ramirez2015effects} is applied, where an actuator with varying intensity is placed at the trailing edge. This previous reference shows that, at the present Reynolds number, the incident sound field is due to a volume quadrupole source distribution near the trailing edge. As discussed by \citet{curle:55}, the quadrupole source is mostly reactive and leads to acoustic scattering in the presence of the rigid airfoil surface, generating a dipolar acoustic field. \citet{ramirez2015effects} shows that when blowing is applied at the trailing edge, the quadrupolar field is displaced downstream, reducing the scattered noise component radiated to the far-field. In the present work, instead of using a fixed momentum, as in the previous reference, the blowing/suction intensity is the control input for the closed-loop system.

\subsubsection{Chordwise moving actuation with constant intensity} \label{sec:moving10}

\begin{figure}
\centering
\includegraphics[width=.32\textwidth]{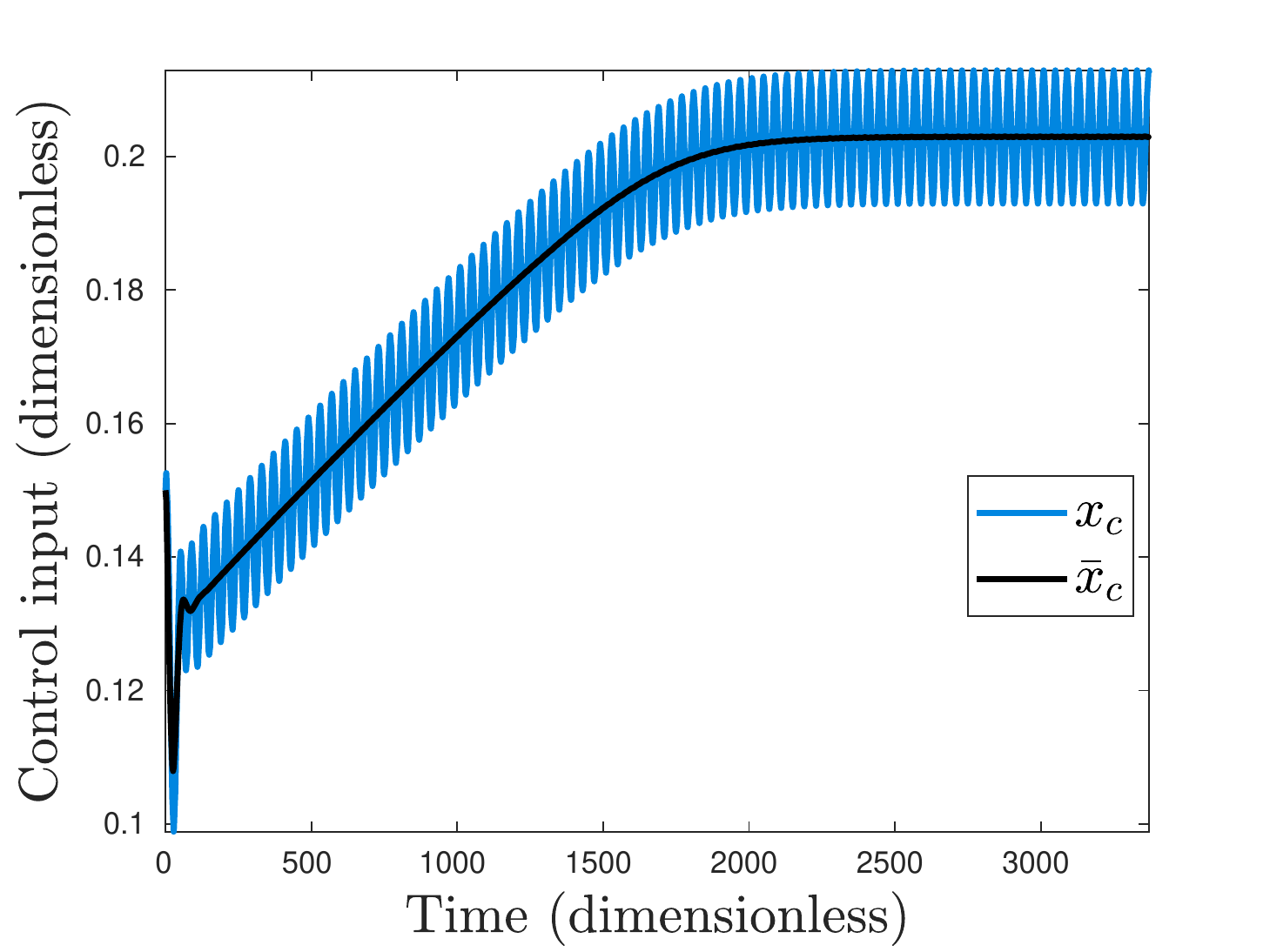}
\includegraphics[width=.32\textwidth]{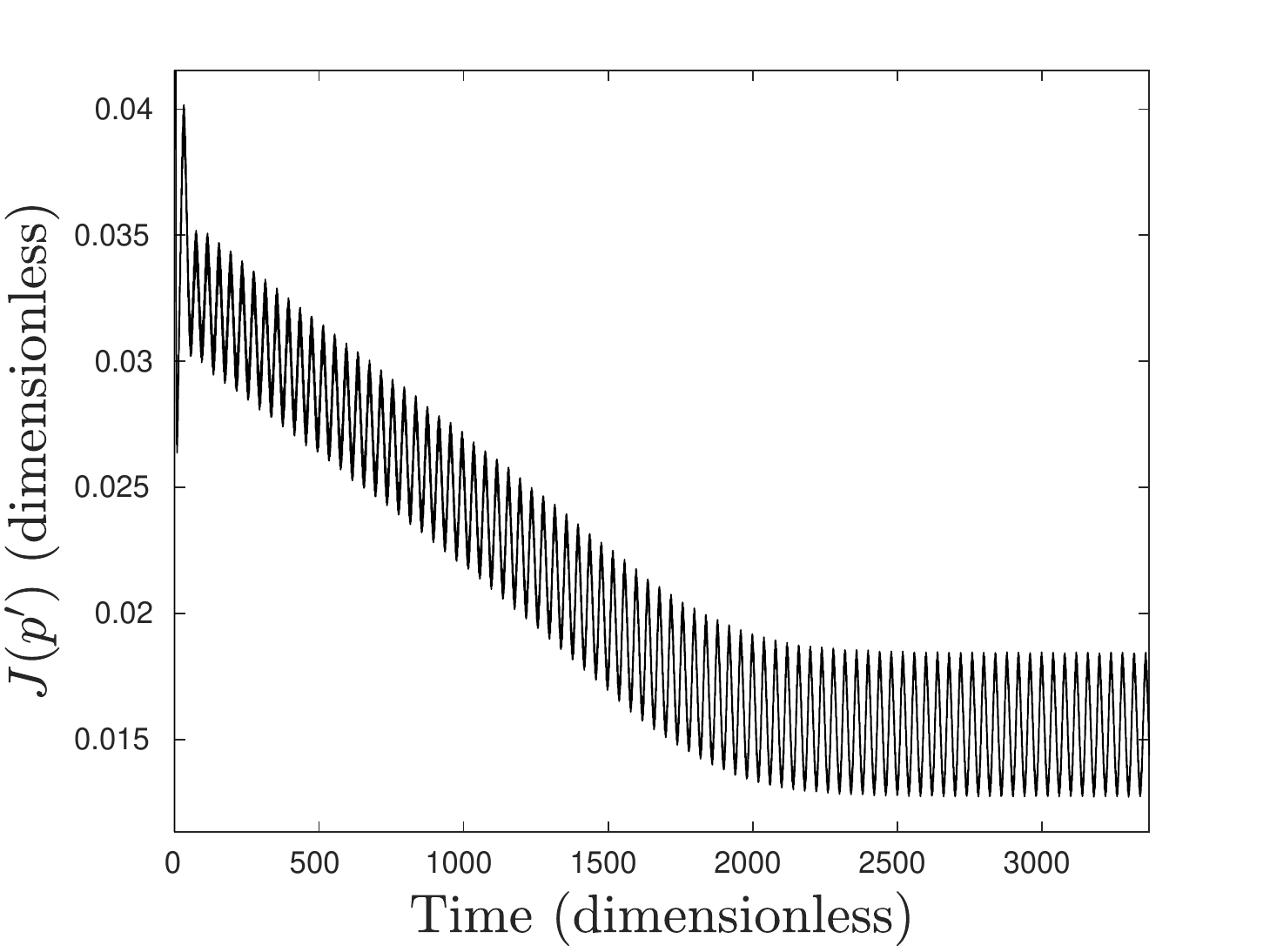}
\includegraphics[width=.32\textwidth]{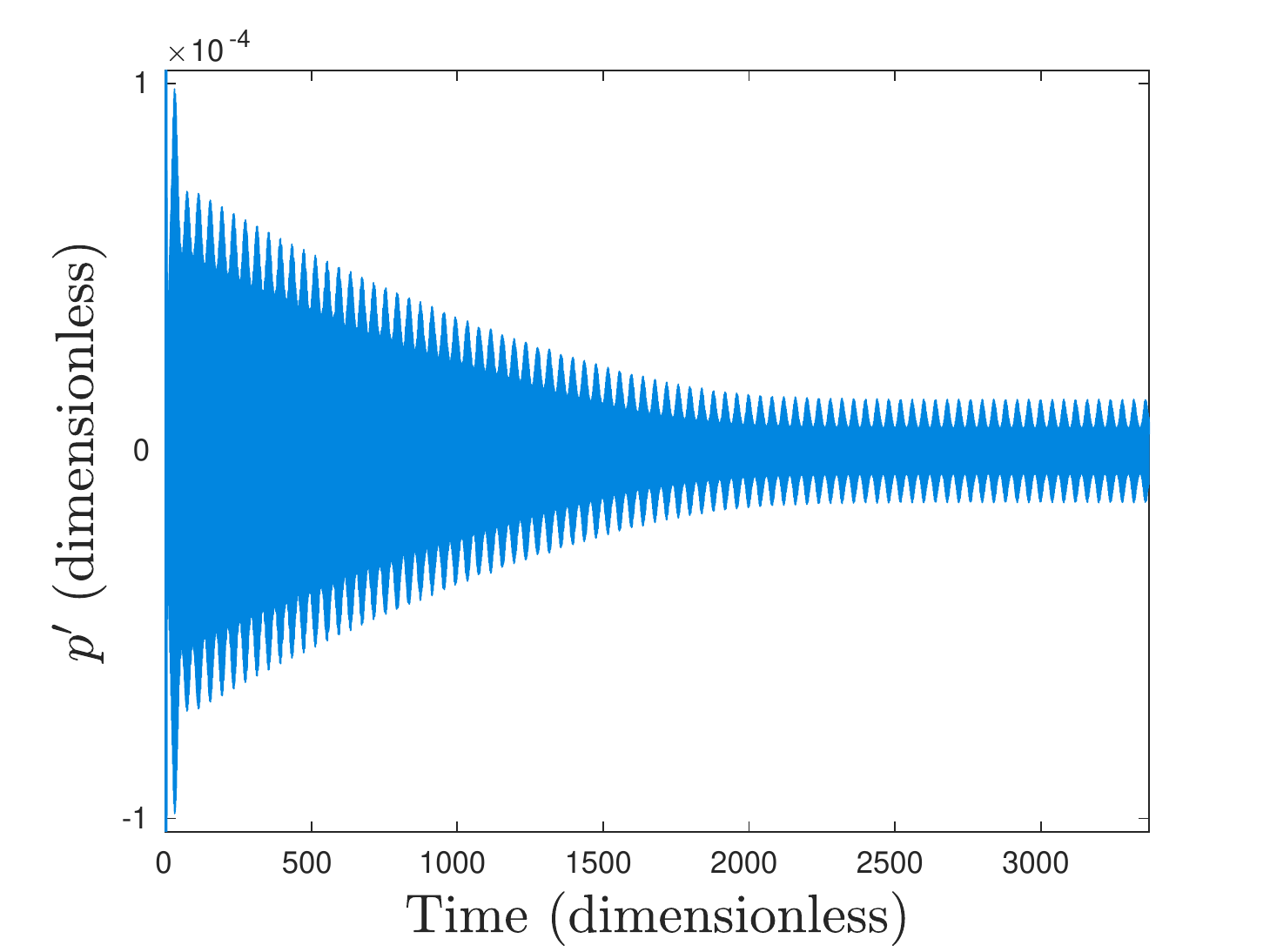}
\caption{Case \cs{11}. Instantaneous, $x_c$, and mean, $\bar{x}_c$, actuator position (left), cost function (center), and acoustic pressure (right). When operating at improper timescales, the loop may not work properly. In this case, since $\omega$ is not slow enough, phasing between the input and output waves can reach 90$^\circ$, rendering $\bar{x}_c$ not able to change.}
\label{f:e011}
\end{figure}

In the present setup, an actuator is placed at an initial position $x_{c0}$ and, at a time instant $t_{c0}$, the control integration starts. In Table \ref{tab:cases10kpos}, control parameters are presented for 10 cases studied with moving actuation. They all use the $\mathcal{W}_2$ window function with $\Delta x = 5\%$ of the chord length.
\begin{table}
  \begin{center}
\def~{\hphantom{0}}
  \begin{tabular}{lcccccccc}
     \hline
     Case    & $A$    & $x_{c0}$ & $t_{c0}$ & $\alpha=\beta$ & $m$ & $2\pi/\omega$ & $\eta$     & Results\\\hline
     \cs{11} & -6.0\% &    0.15  & 0.00e+00 & 0.01            & 20  & 4.00e+01      & -6.00e+00  & Fig. \ref{f:e011}\\
     \cs{6}  & -6.0\% &    0.15  & 5.00e+01 & 0.01            & 20  & 4.00e+01      & -5.00e+01  & Fig. \ref{f:e006}\\
     \cs{7}  & -6.0\% &    0.92  & 8.00e+01 & 0.01            & 20  & 4.00e+01      & -5.00e+01  & Fig. \ref{f:e007}\\
     \cs{12} & -6.0\% &    0.19  & 2.13e+02 & 0.01            & 20  & 1.00e+02      & -6.00e+00  & Fig. \ref{f:e012}\\
     \cs{13} & -6.0\% &    0.92  & 0.00e+00 & 0.01            & 20  & 4.00e+01      & -6.00e+00  & Fig. \ref{f:e013}\\
     \cs{9}  & -3.0\% &    0.15  & 8.00e+01 & 0.01            & 20  & 4.00e+01      & -5.00e+01  & Fig. \ref{f:e009}\\
     \cs{10} & -3.0\% &    0.92  & 1.50e+02 & 0.01            & 20  & 4.00e+01      & -1.00e+02  & Fig. \ref{f:e010}\\
     \cs{15} & -1.2\% &    0.82  & 2.13e+02 & 0.01            & 20  & 1.00e+02      & -2.00e+02  & Fig. \ref{f:e015}\\
     \cs{16} & -1.2\% &    0.20  & 2.13e+02 & 0.01            & 20  & 1.00e+02      & -2.00e+02  & Fig. \ref{f:e016}\\
     \cs{27} & -1.2\% &    0.20  & 2.00e+02 & 0.01            & 20  & 1.00e+02      & -1.30e+02  & Fig. \ref{f:e027}\\\hline
  \end{tabular}
  \caption{Control parameters for cases with moving actuation at $\text{Re} = 10^4$. The amplitude $A$ is given as a percentage of the freestream momentum %(which is equal, in value, to the Mach number $\text{Ma} = 0.3$, since $\rho_\infty=1$) 
  and the negative sign denotes suction. The integrator turn-on time $t_{c0}$ and the ESC wave period $2\pi/\omega$ are dimensionless temporal parameters relative to freestream velocity and chord. The initial actuator position $x_{c0}$ consists of the horizontal location relative to the chord length. The ESC parameter $\alpha$ is also relative to the chord length.}
  \label{tab:cases10kpos}
  \end{center}
\end{table}
Case \cs{11} shows an example when the timescales of the flow response to the actuator and those from the controller periodic perturbation are not well separated. Since $\omega$ is not slow enough, the controller stops moving before an optimal actuation position is reached. This occurs due to system response delay that renders the actuator to get stuck. When the input/output lag reaches a correspondent phase of $90^\circ$, the average of the product between the waves becomes zero. Figure \ref{f:e011} shows the temporal evolution of the actuator position, cost function and acoustic pressure. From this figure, it is possible to see that the actuator moves downstream along the airfoil reducing the cost value and, hence, the noise computed at the sensor position. However, due to the $90^\circ$ input/output lag, the actuator reaches a steady mean position. The simulation depicted as case \cs{12} uses a similar configuration but with a lower frequency $\omega$ that does not allow the phase to reach $90^\circ$. The unwanted oscillations that occur at the very beginning of the simulation are related to the settling time of the ESC high-pass filter. In some of the next cases, the integration in the ESC loop is turned on at an instant $t_{c0}\neq 0$ to attenuate this effect.

\begin{figure}
\centering
\includegraphics[width=.32\textwidth]{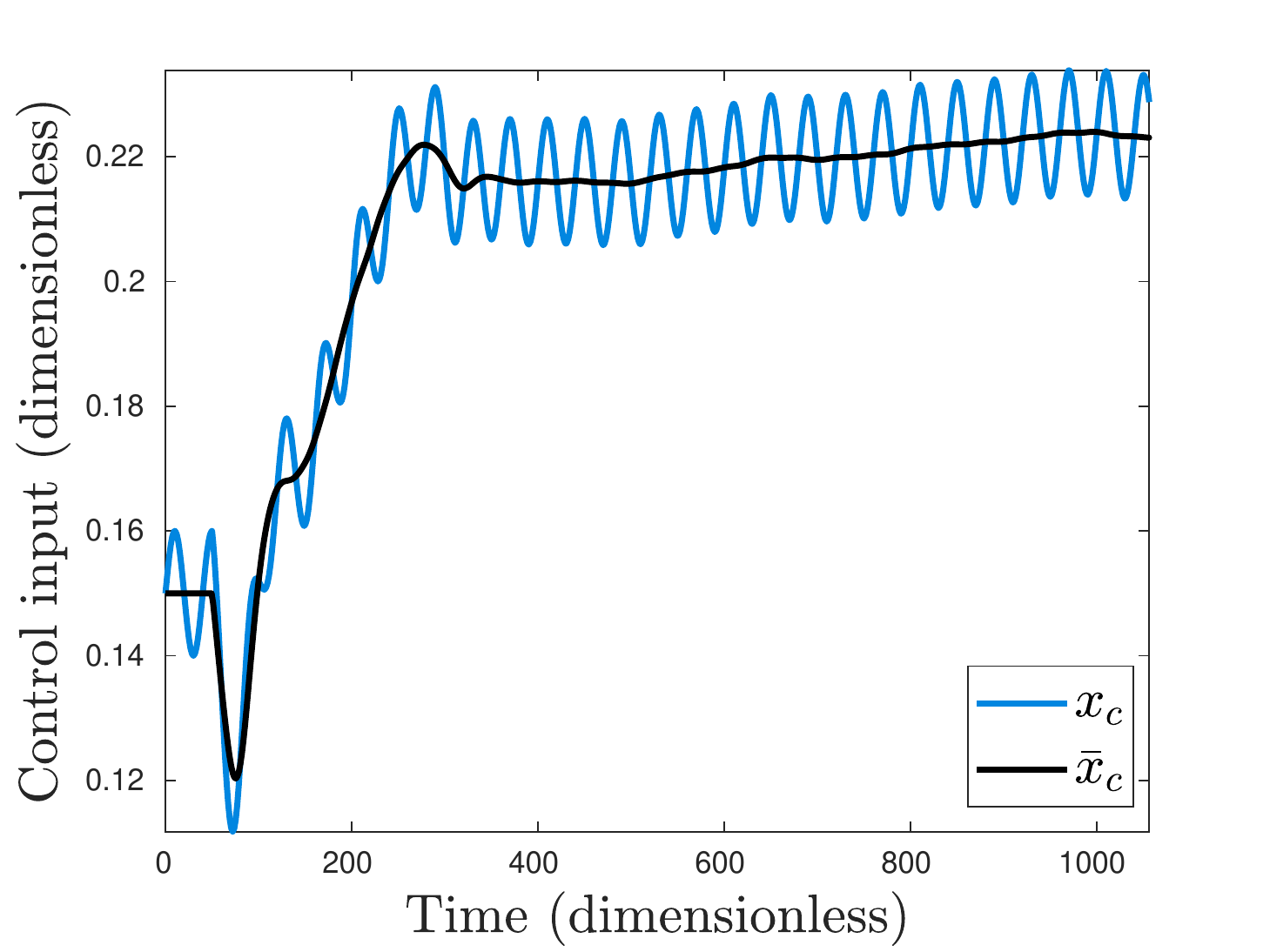}
\includegraphics[width=.32\textwidth]{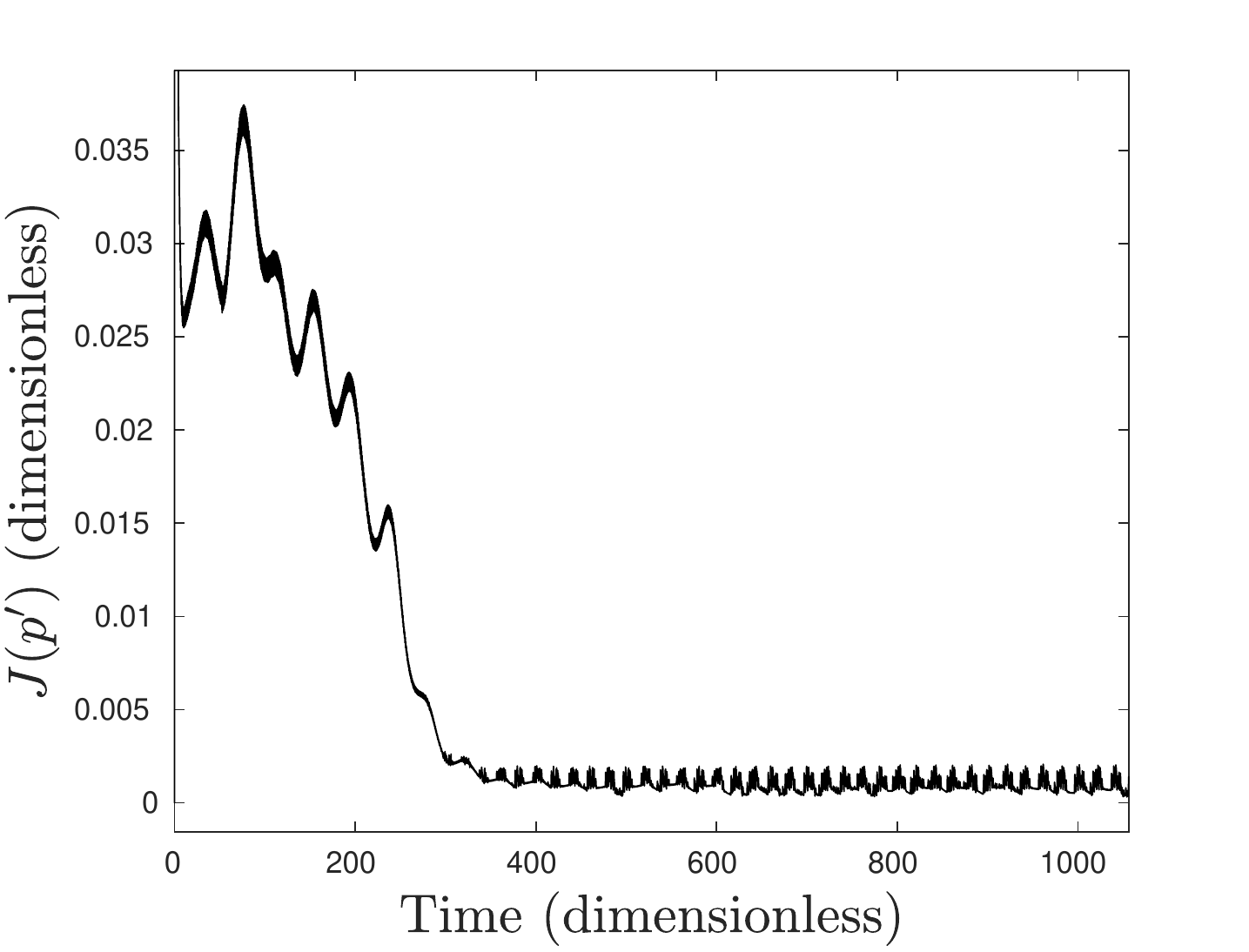}
\includegraphics[width=.32\textwidth]{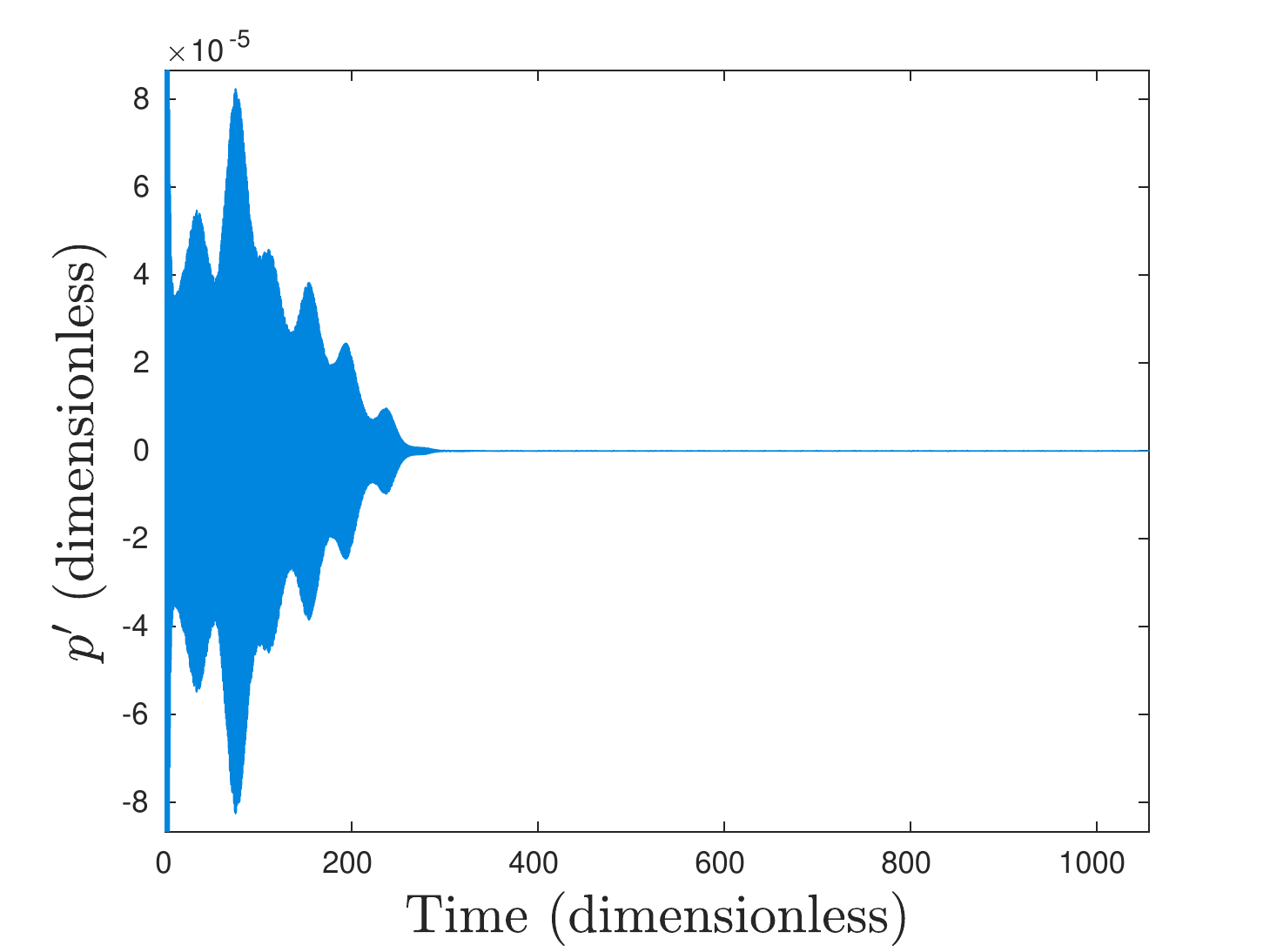}
\includegraphics[width=.32\textwidth]{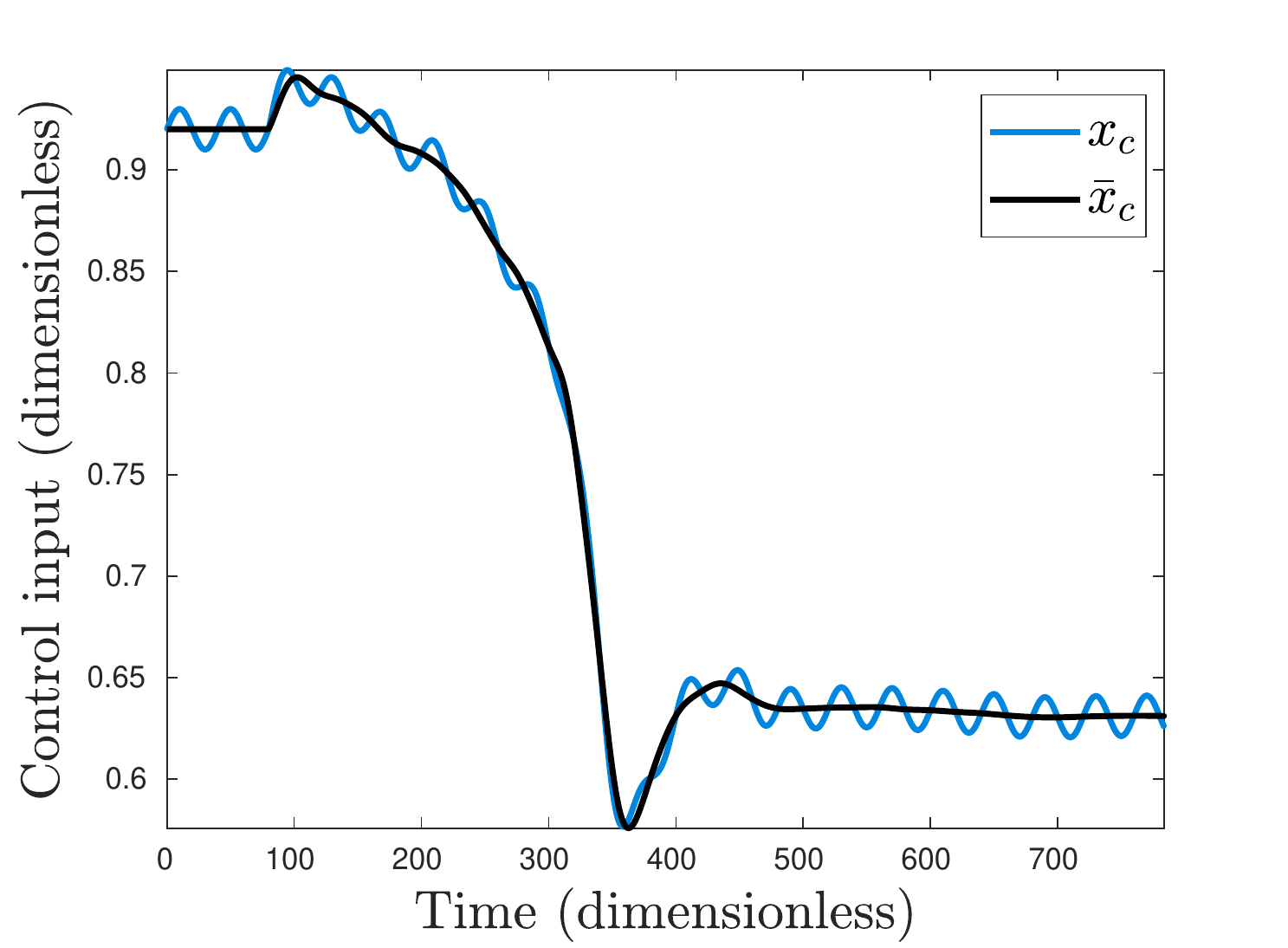}
\includegraphics[width=.32\textwidth]{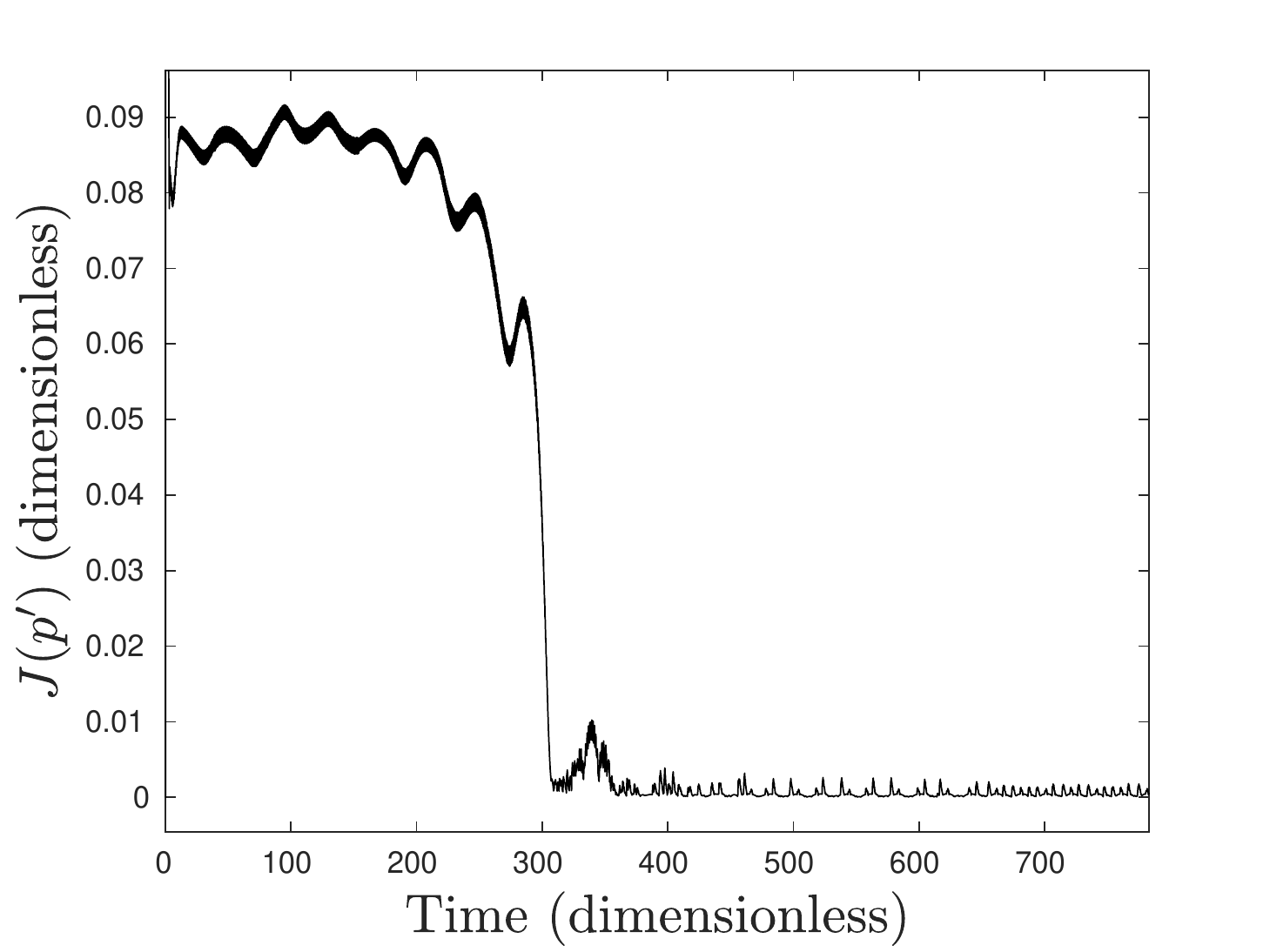}
\includegraphics[width=.32\textwidth]{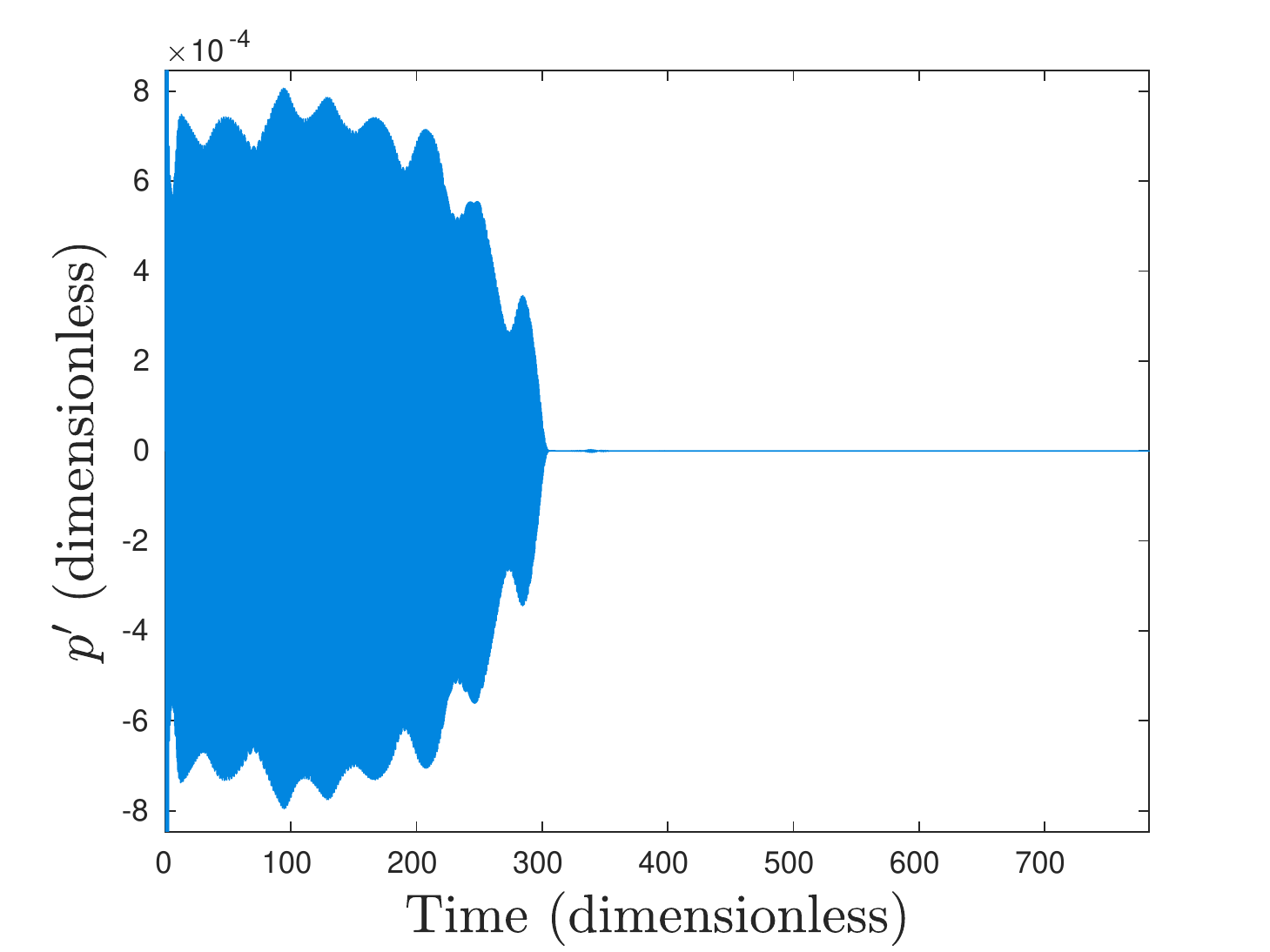}
\caption{Cases \cs{6} and \cs{7}. For the present cases, the flow is reattached when the actuator reaches a specific region. Here, a higher integrator gain $\eta$ is applied compared to case \cs{11} while the suction value $A=-6\%$ is kept. Comparing a starting actuation position near the leading edge (top row) or the trailing edge (bottom row), it is possible to find that the region of reattachment ranges about $0.22 \lesssim x_c \lesssim  0.80$. The overshoots before convergence occur due to integration of fast transients that are not high-pass filtered, and a reduction in $\eta$ dampens the overshoots, as will be shown in the next simulations.}
\label{f:e006}\label{f:e007}
\end{figure}

By fixing a maximum suction momentum with $A = -6.0\%$ (relative to that of the freestream), the controller drives the actuator position towards the center of the airfoil since this minimizes the cost function. It can be observed in Fig. \ref{f:e006} that there is a region between $0.22 \lesssim x_c \lesssim 0.80$ at which the flow reattaches, suppressing the vortex shedding and, hence, the acoustic noise generation. This figure presents the results for cases \cs{6} and \cs{7}, where the actuator is placed before or after the re-attachment region. Despite being similar to case \cs{11}, case \cs{6} has a higher integrator gain $\eta$  which allows for convergence by overshooting the actuator past the sticking point. The impact of the higher gain is observed in Fig. \ref{f:e006} by the faster rise time. When flow re-attachment occurs, the computed cost function behaves discontinuously and fast transients occur due to abrupt variations in this function that are not high-pass filtered. The discontinuity introduces signal transients that move the actuator when integrated. Normally, the high-pass filter allows the components that are due to the input oscillation to pass but, with the discontinuity in the cost function, it is unable to separate the central and the fluctuation values of $J$. From the cost function presented in the bottom row of Fig. \ref{f:e006}, it is possible to notice that at $t = 300$ the minimum is achieved. At this time instant, the actuator position is at $\bar{x}_c \approx 0.80$, after which it has a sudden drop and overshoot, which finally brings the actuator to $\bar{x}_c \approx 0.64$. Indeed, for the present setup, we observed through an open loop study that any actuator with $A=-6.0\%$ positioned between $\bar{x}_c \approx 0.22$ and $0.80$ is able to reattach the flow. We also noticed that the present results show hysteresis, which means that the actuation region for the flow to remain attached is larger than that needed to first reach the reattachment.

\begin{figure}
\centering
\includegraphics[width=.32\textwidth]{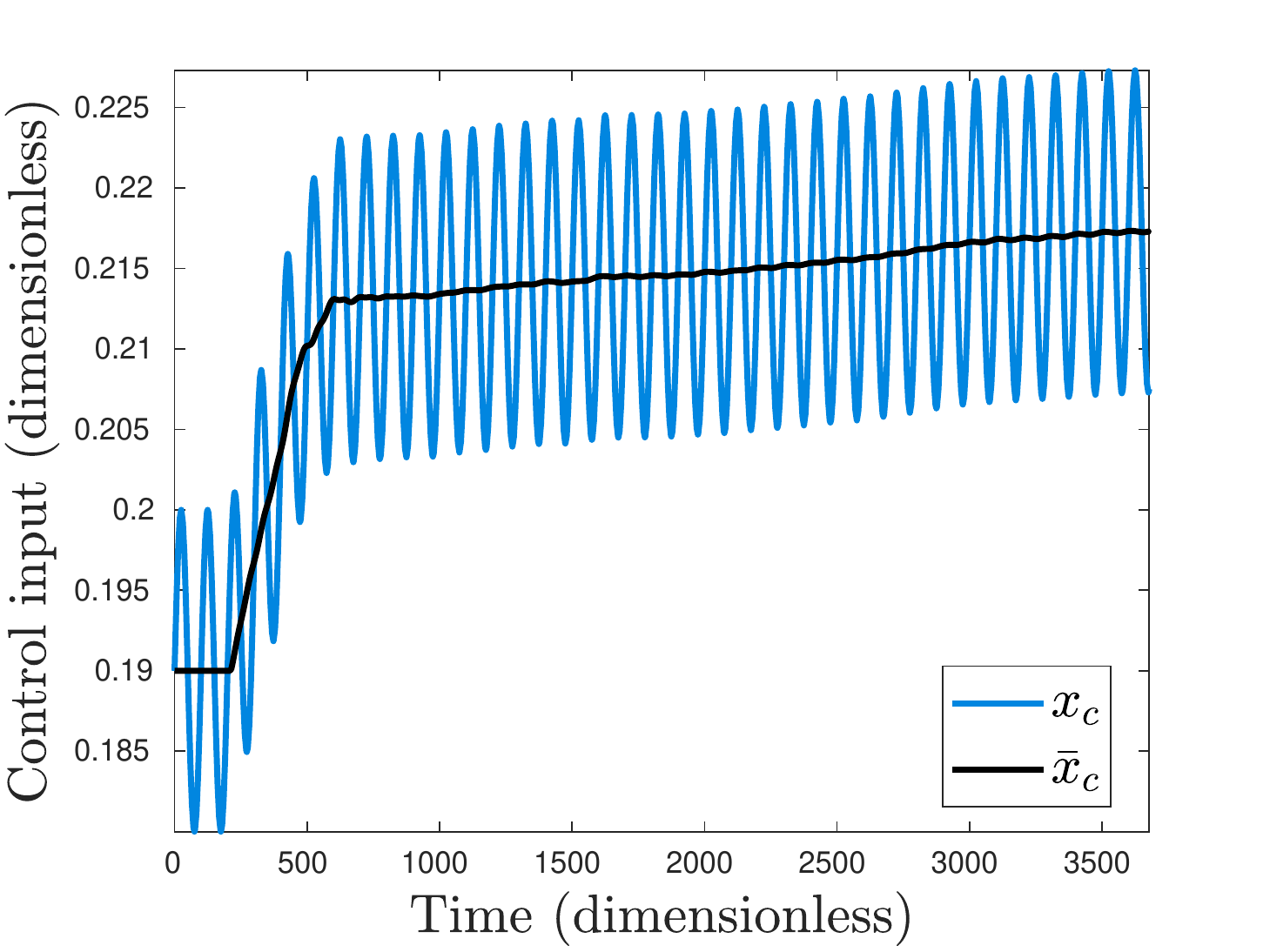}
\includegraphics[width=.32\textwidth]{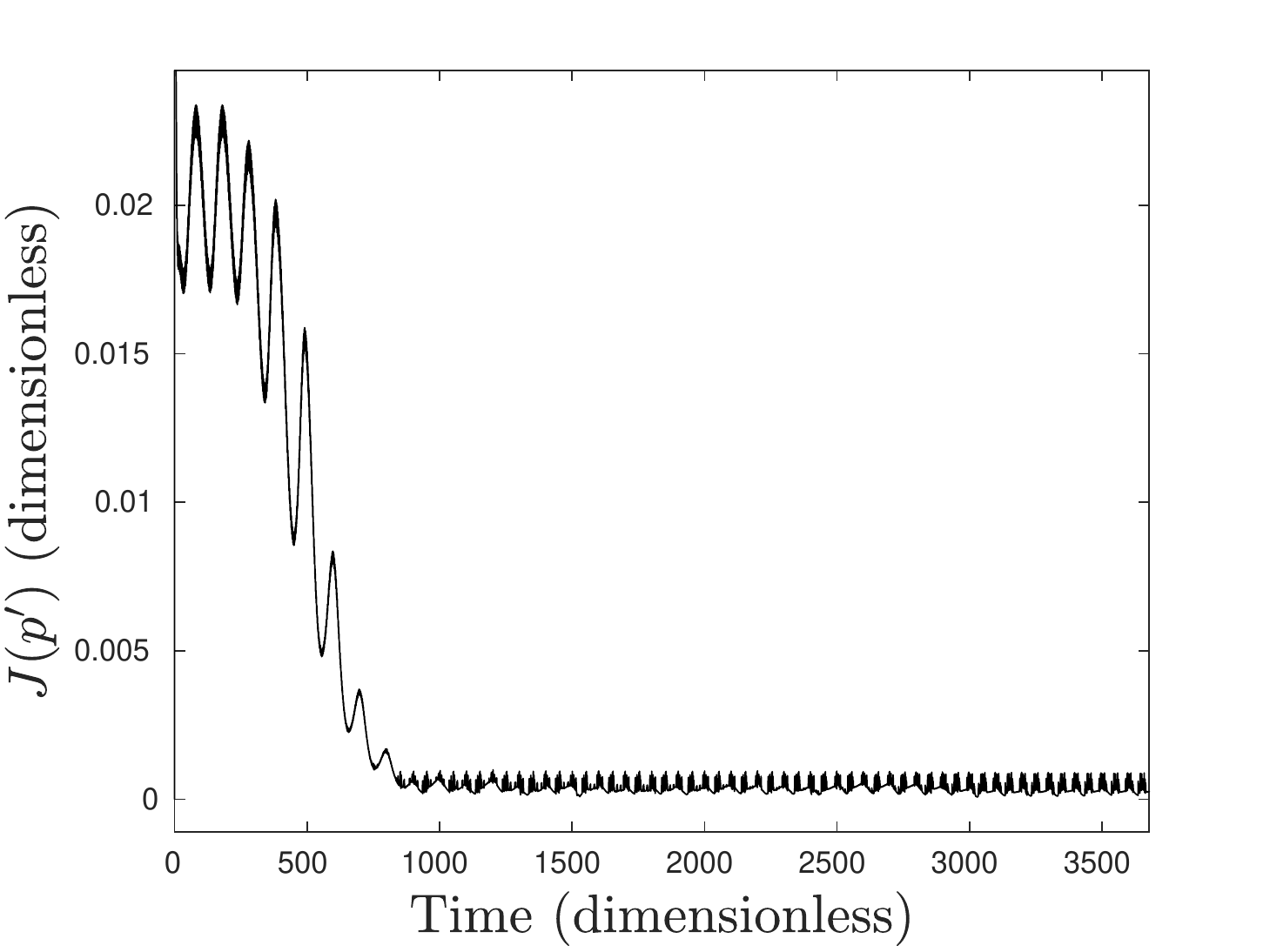}
\includegraphics[width=.32\textwidth]{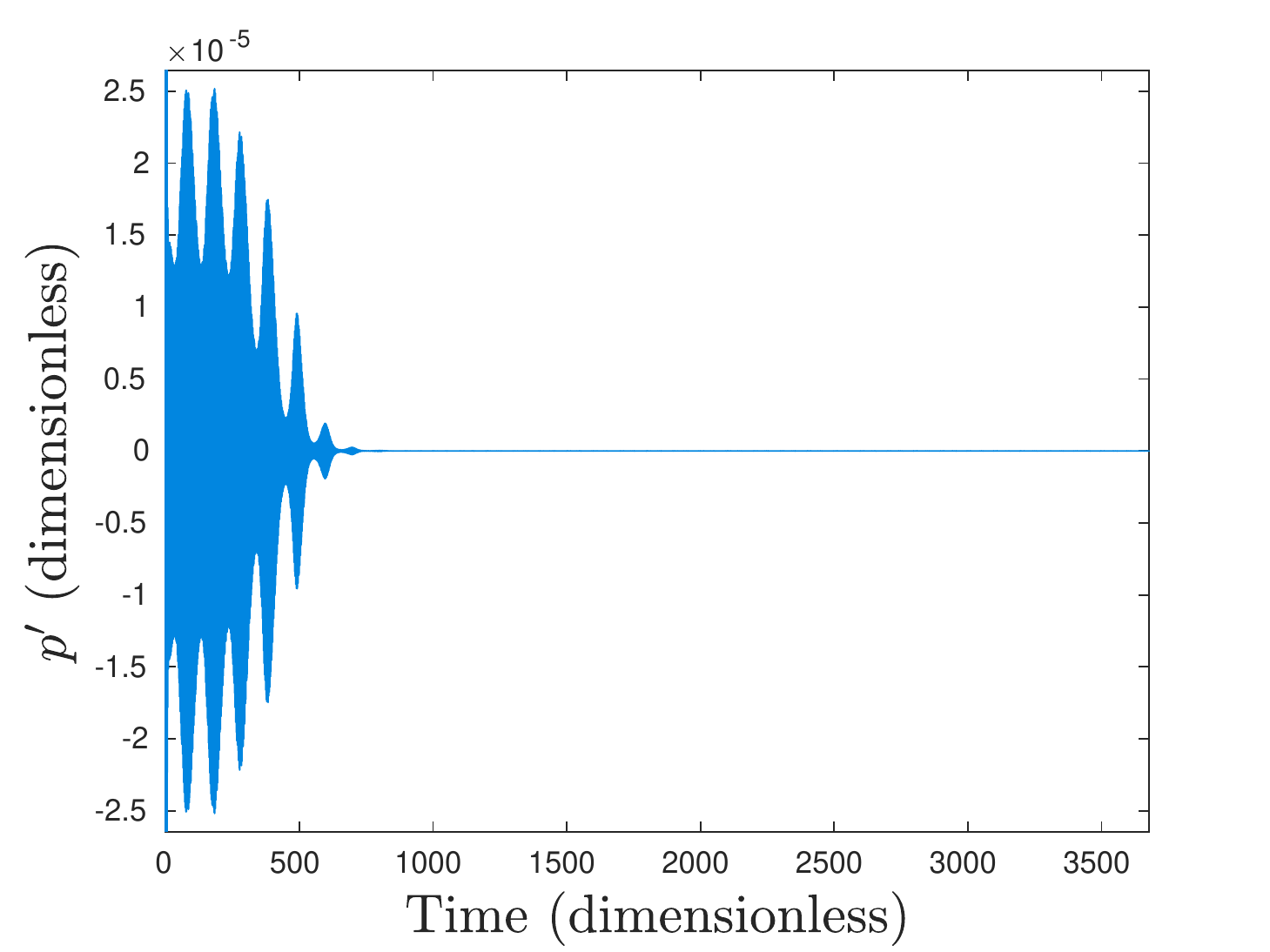}
\includegraphics[width=.32\textwidth]{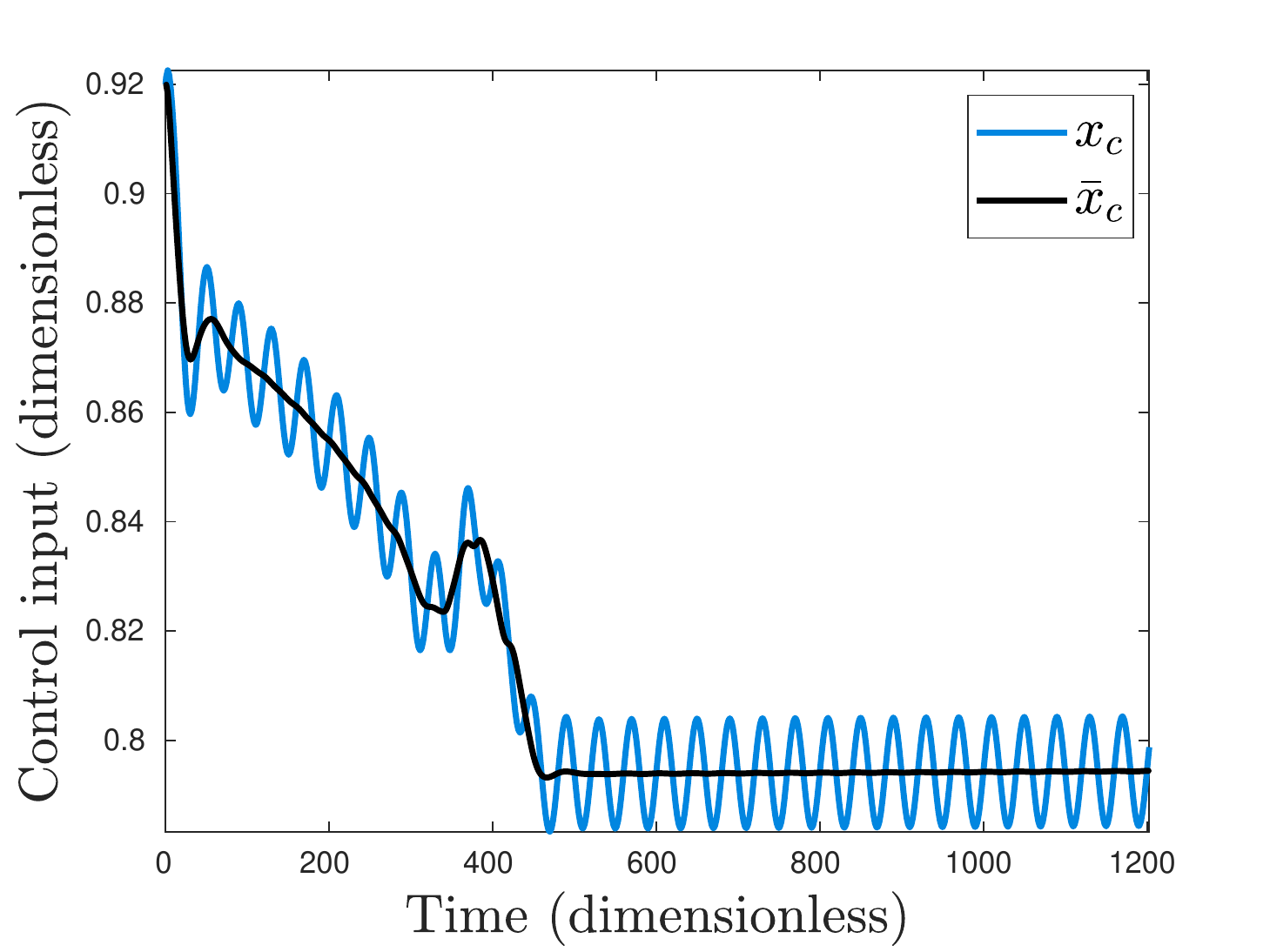}
\includegraphics[width=.32\textwidth]{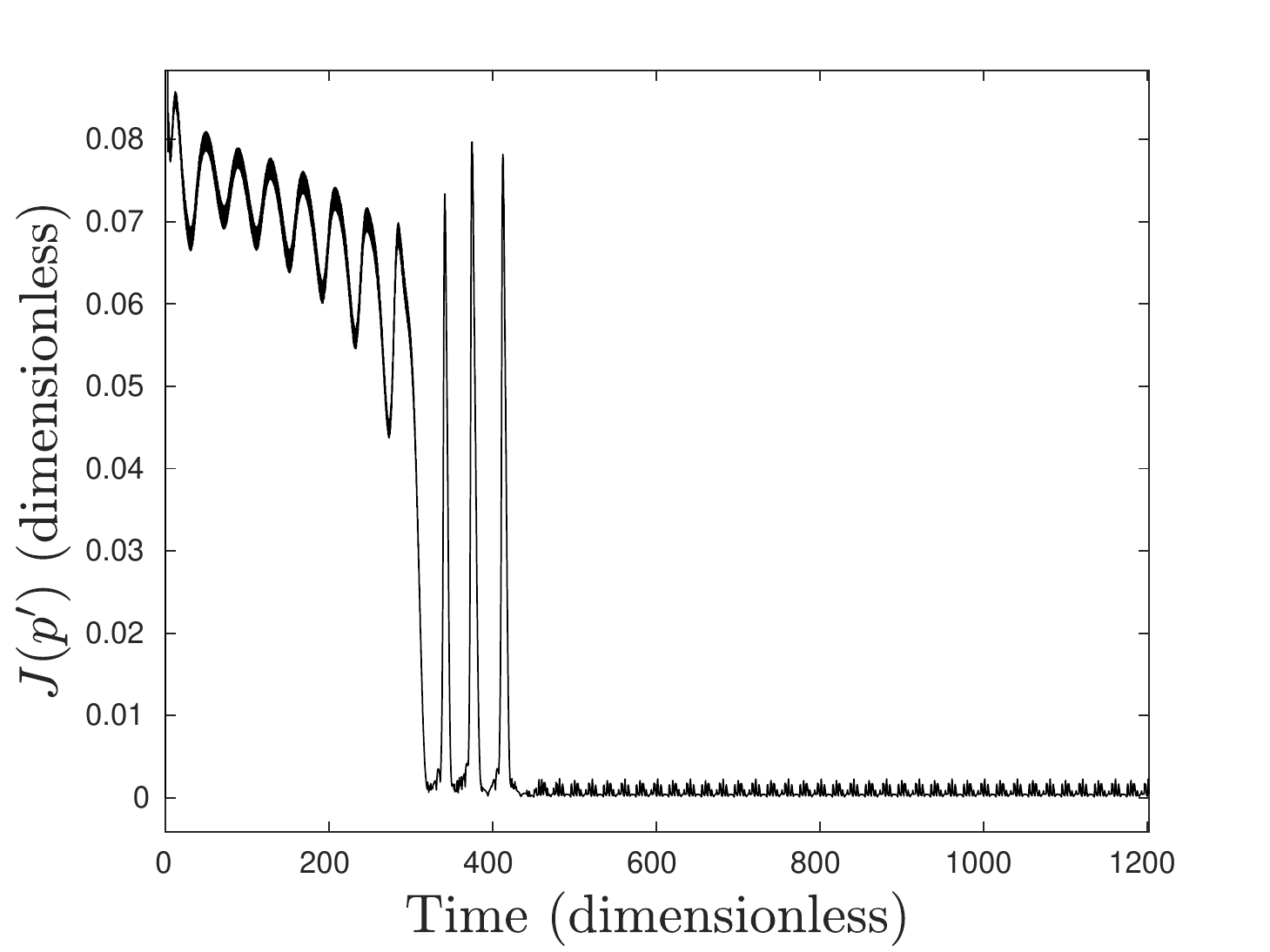}
\includegraphics[width=.32\textwidth]{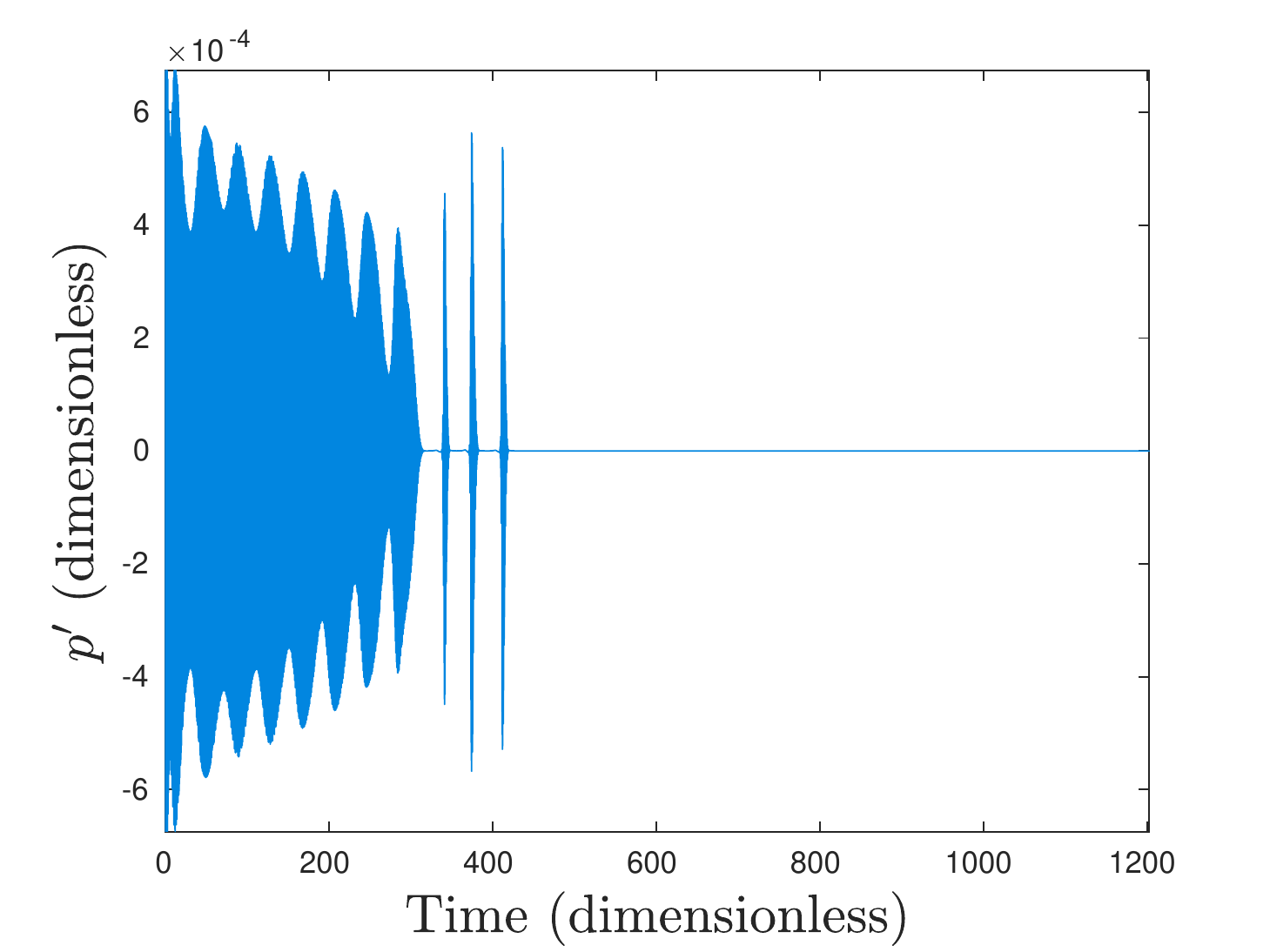}
\caption{Cases \cs{12} and \cs{13}. Overshoots can be reduced by setting lower magnitude values for $\eta$, as the observation of final values of $x_c$ suggests. After high-frequency transients, when the flow reattaches, acoustic waves are suppressed and the actuator motion occurs only due to residual noise. Due to lower integrator gains $\eta$, the actuator position stays near the reattachment boundary.}
\label{f:e012}\label{f:e013}
\end{figure}

Results obtained for cases \cs{12} and \cs{13} are presented in Fig. \ref{f:e012}. These setups are obtained for lower values of the integrator gain $\eta$. In case \cs{12}, the ESC frequency $\omega$ is lower than that for case \cs{13}. While the actuator is positioned near the leading edge for the former setup, it is positioned close to the trailing edge for the latter one. Since the noise emission as a function of $x_c$ presents a discontinuity when reattachment occurs, rapid variations can be seen in the actuator motion, as shown in the bottom row of Fig. \ref{f:e007}. A reduced integration gain makes the actuation motion slower, preventing or attenuating this phenomenon. As a trade-off, a slower settling time is observed. In case \cs{12}, the frequency $\omega$ is reduced to rectify the issue seen in case \cs{11}. Since the gain $\eta$ is as low as in that case, the same value for $\omega$ would lead to the $90^\circ$ phase previously described. This also means that the control in case \cs{6} only converges due to intense integration of the transient signals that appear during reattachment.

\begin{figure}
\centering
\includegraphics[width=.32\textwidth]{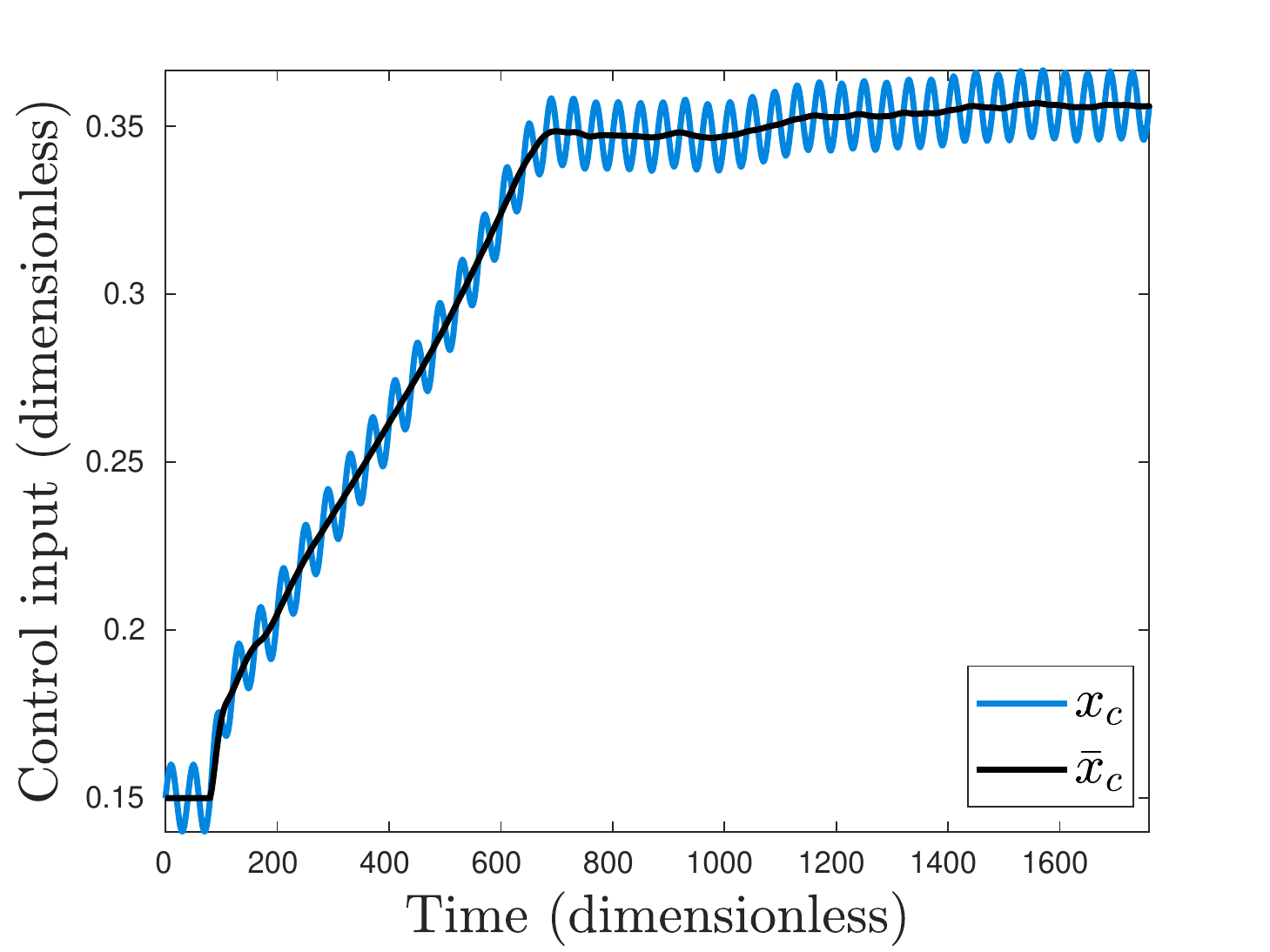}
\includegraphics[width=.32\textwidth]{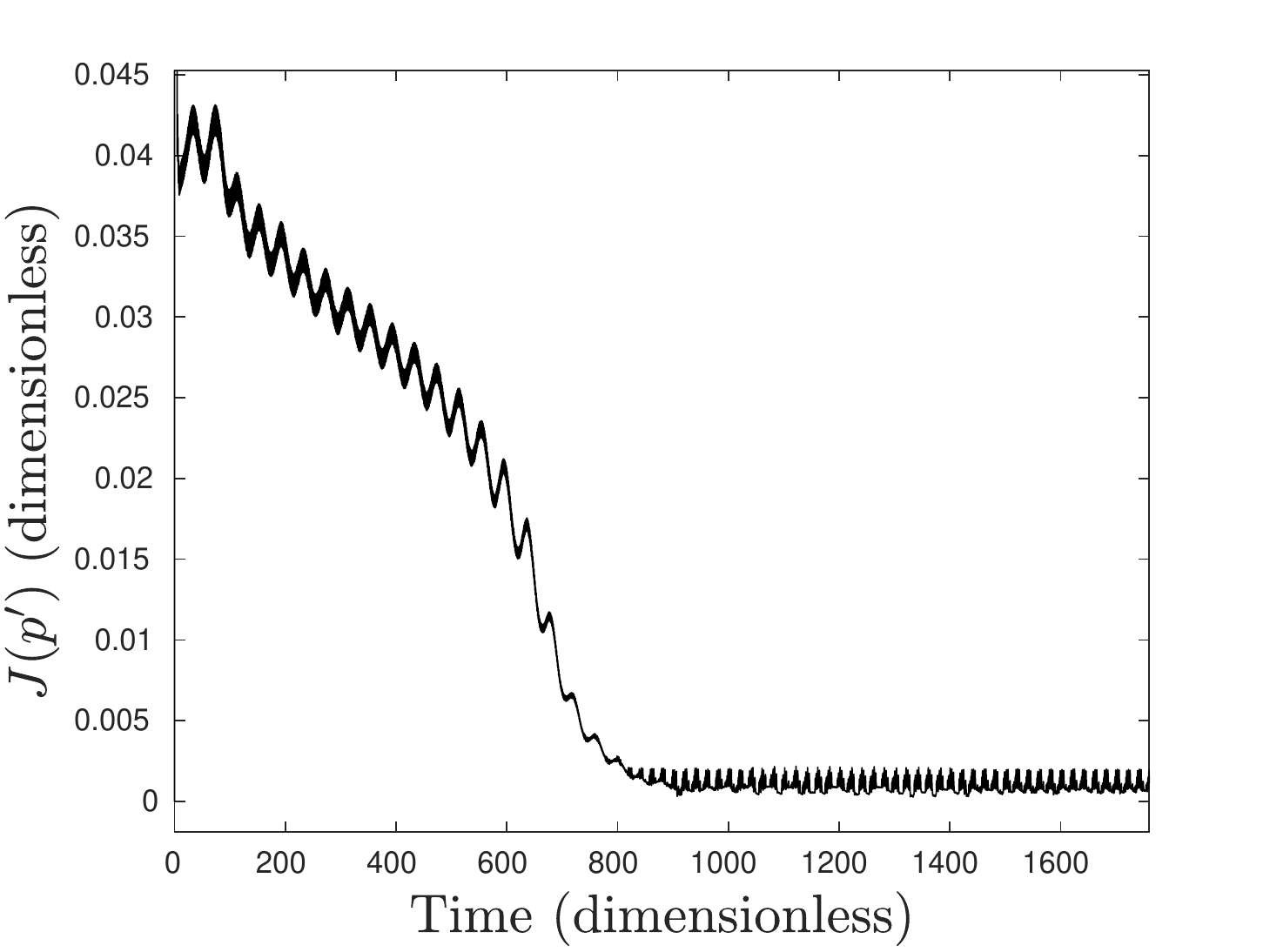}
\includegraphics[width=.32\textwidth]{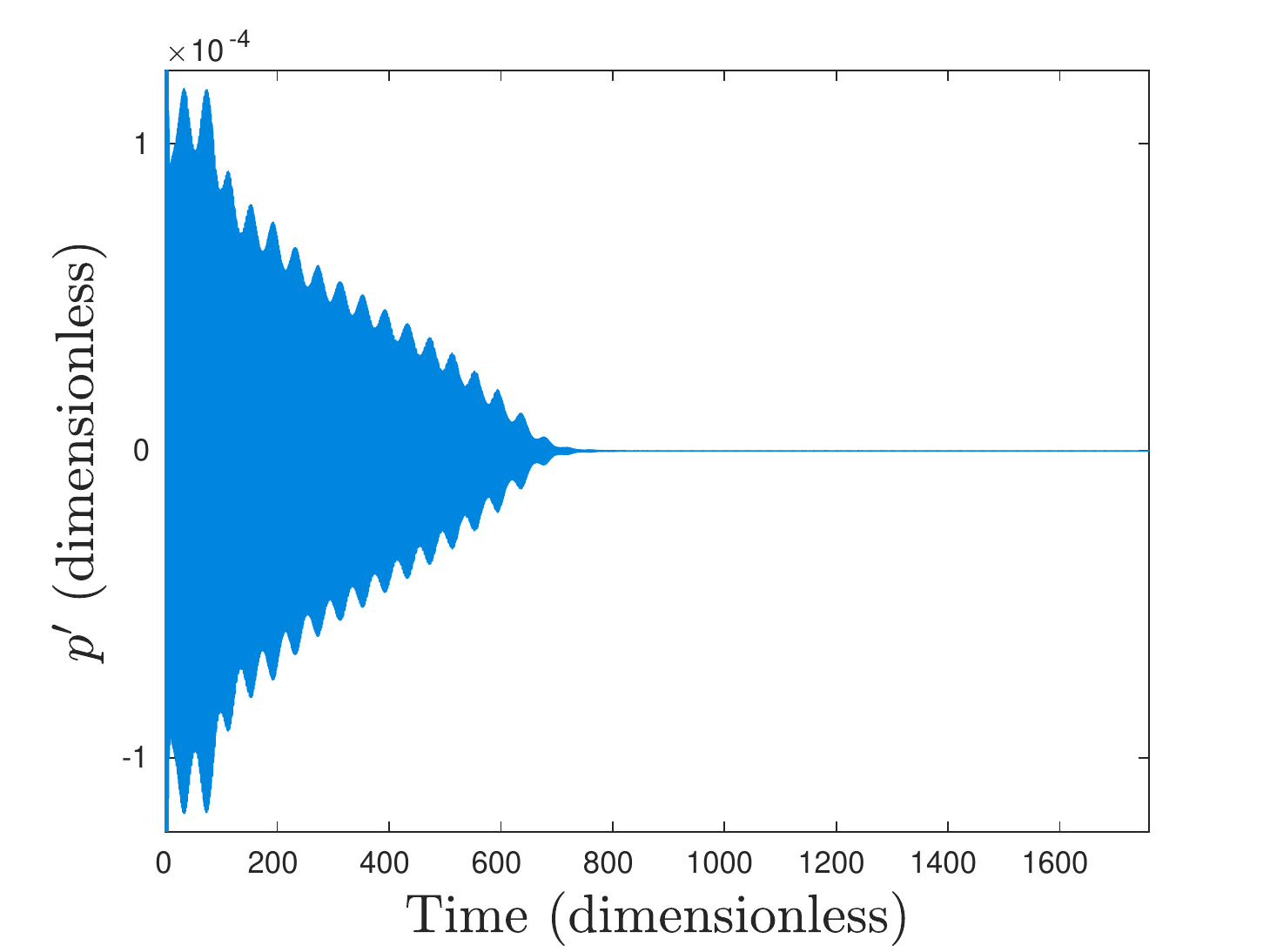}
\includegraphics[width=.32\textwidth]{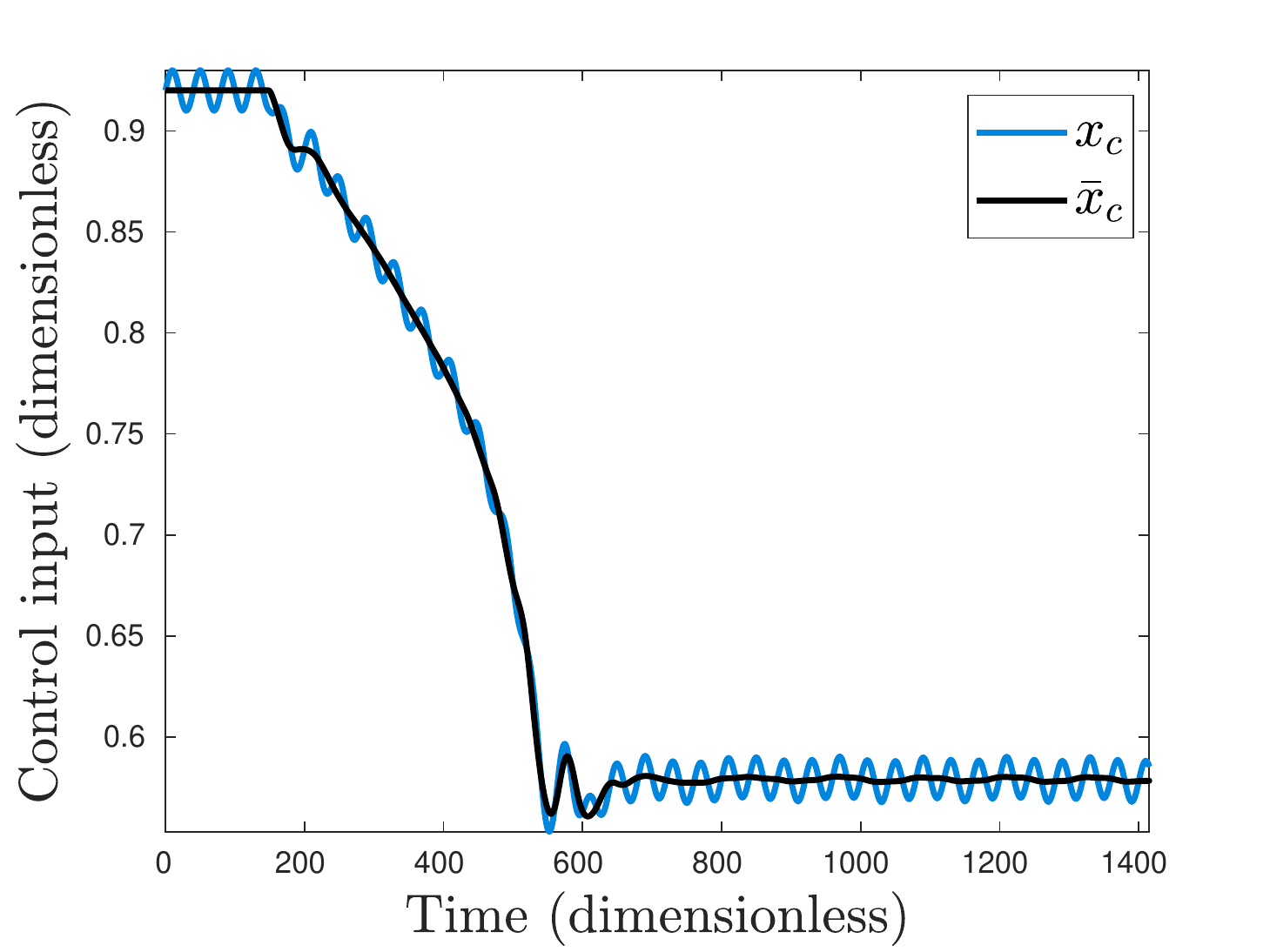}
\includegraphics[width=.32\textwidth]{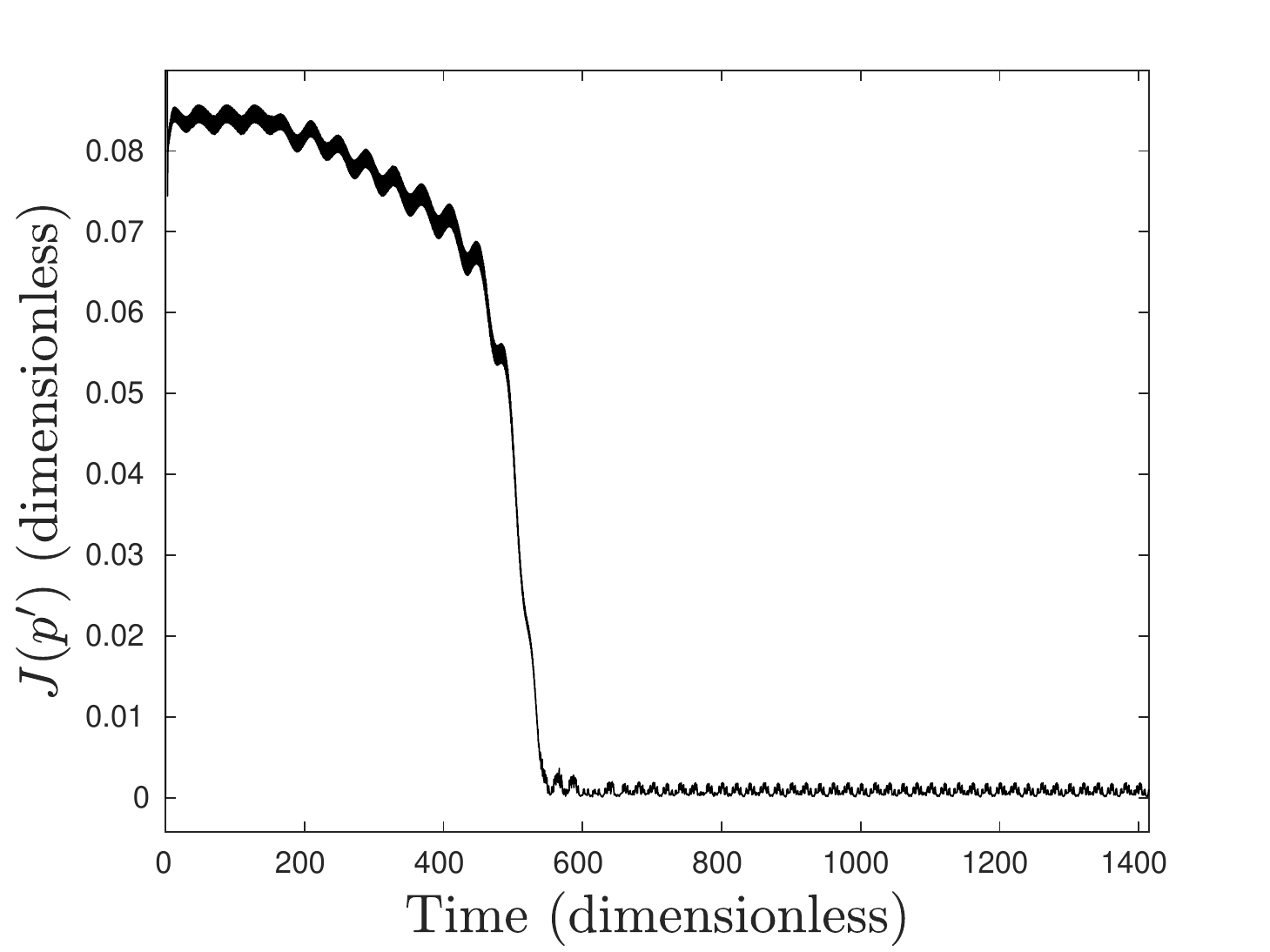}
\includegraphics[width=.32\textwidth]{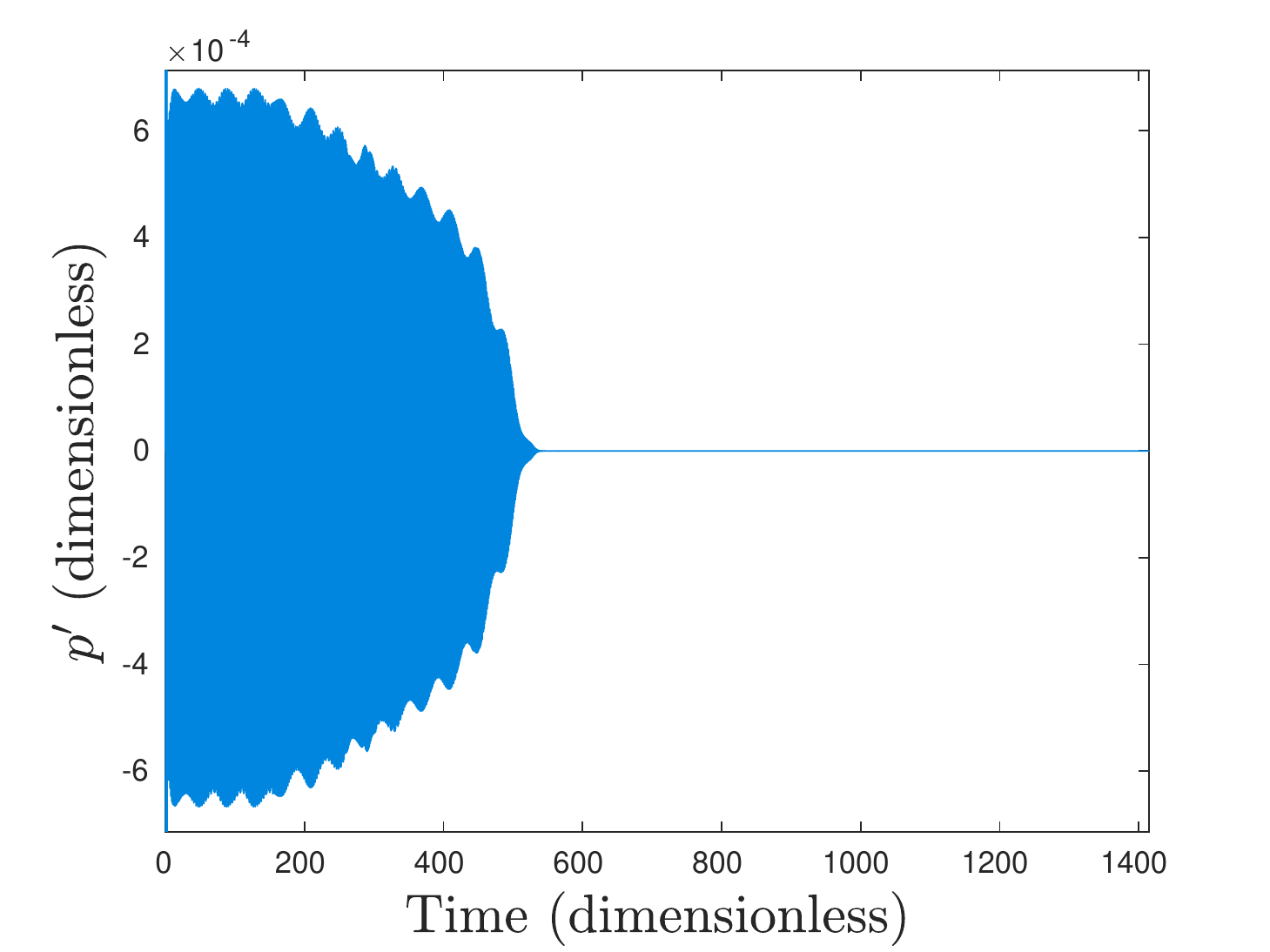}
\caption{Cases \cs{9} and \cs{10}. Lower values of $A$ result in narrower reattachment regions ($0.35 \lesssim x_c \lesssim 0.65$), while the same results are obtained in terms of acoustic noise reduction after convergence.}
\label{f:e009}\label{f:e010}
\end{figure}

A lower suction intensity of $A=-3.0\%$ is tested in cases \cs{9} and \cs{10}. Hence, an assessment is performed to verify if the flow can still be reattached with less power consumption to suppress the airfoil noise. 
While the frequencies $\omega$ are the same for both these cases, the integrator gains are different, with case \cs{10} having a higher value.
Results shown in Fig. \ref{f:e009} demonstrate that noise can be suppressed, but the region at which the flow reattaches is reduced, becoming delimited by $0.35 \lesssim x_c \lesssim 0.65$. The higher gain of case \cs{10} leads to a more abrupt drop of the cost function, similar to those observed in previous cases. %However, differently from case \cs{7}, where the actuator is brought to a position in the middle of the reattachment region, the lower suction intensity keeps the actuator near the boundary for which reattachment occurs \red{[essa frase parece estranha]}.

\begin{figure}
\centering
\includegraphics[width=.32\textwidth]{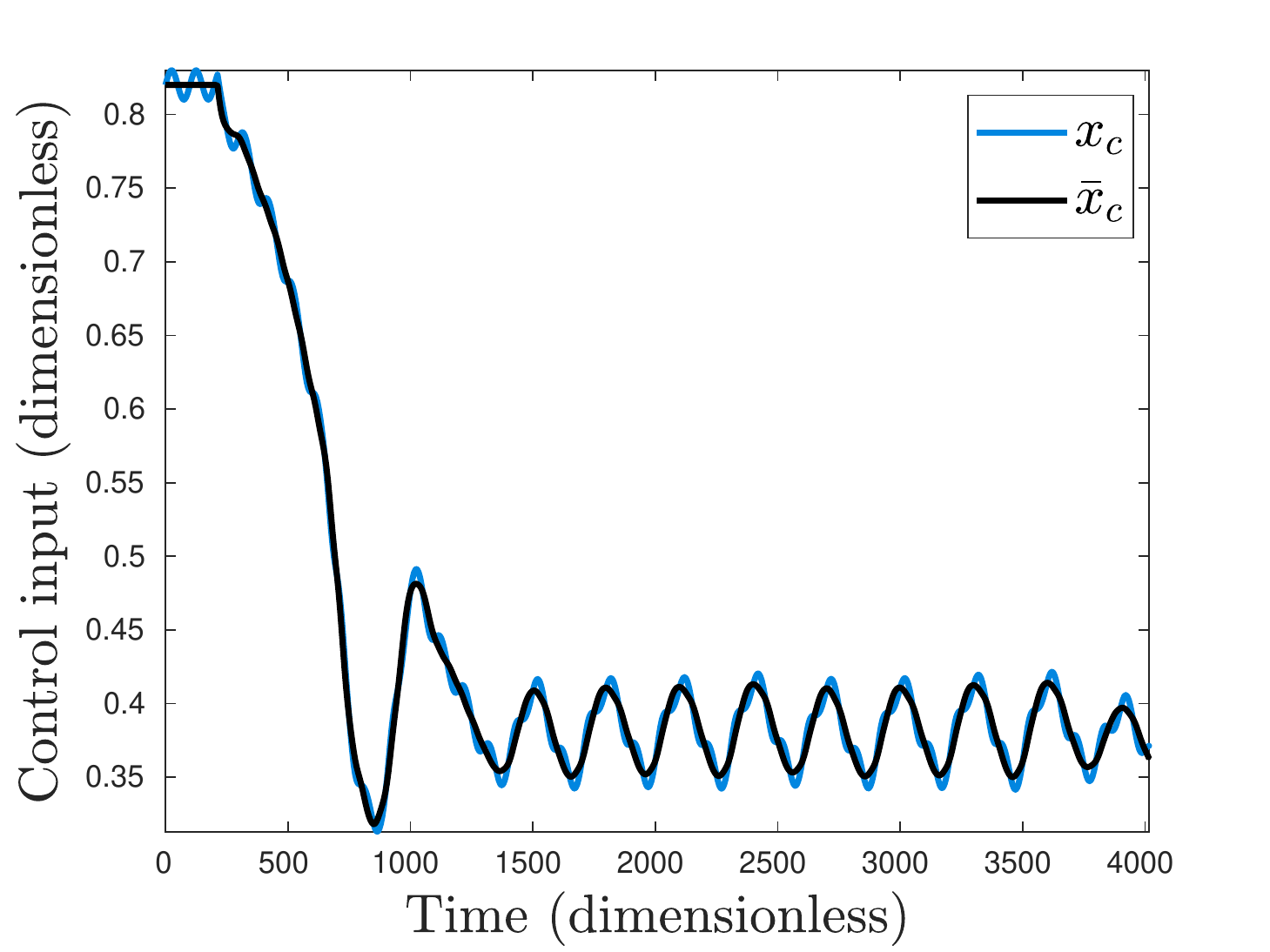}
\includegraphics[width=.32\textwidth]{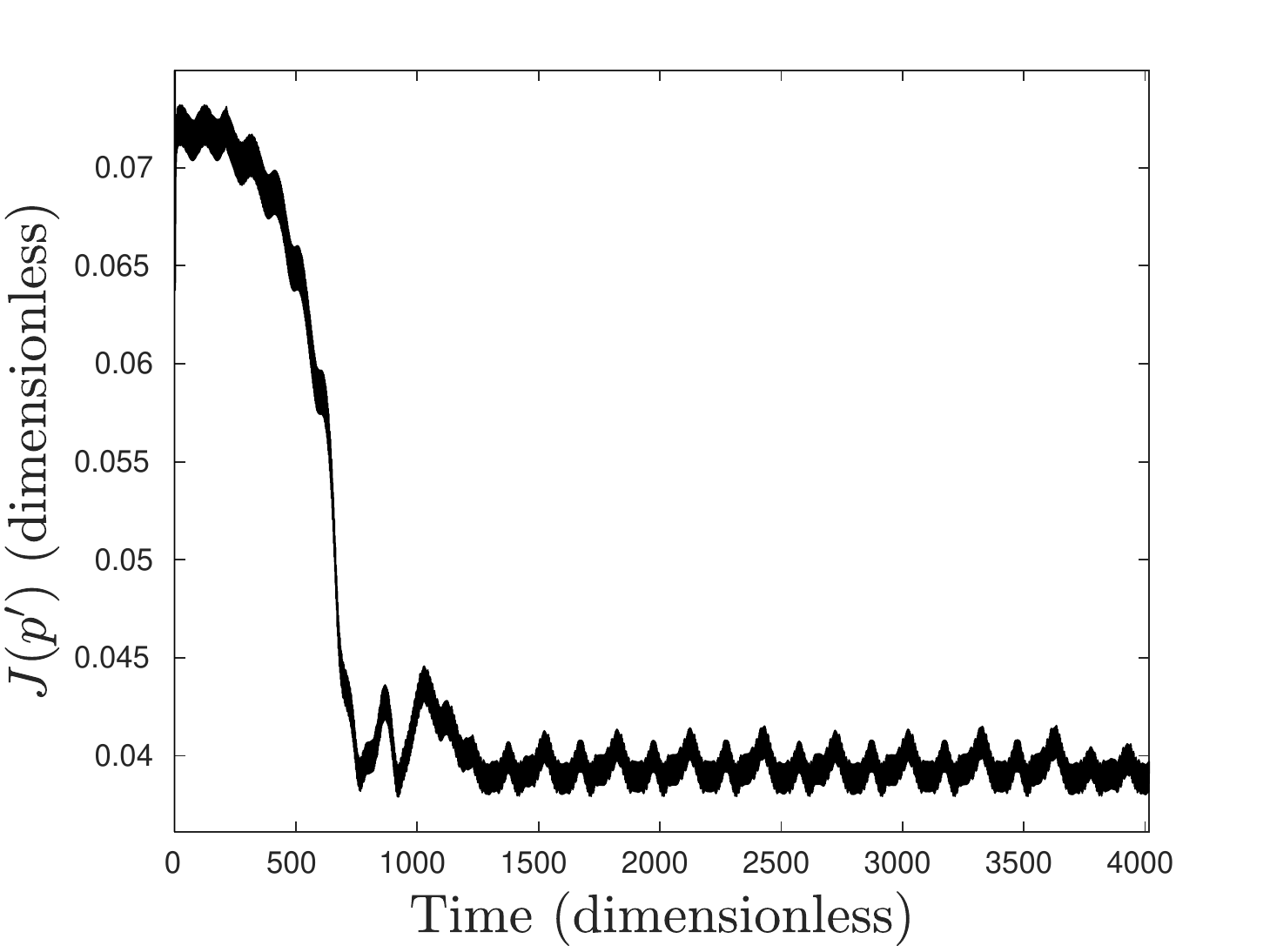}
\includegraphics[width=.32\textwidth]{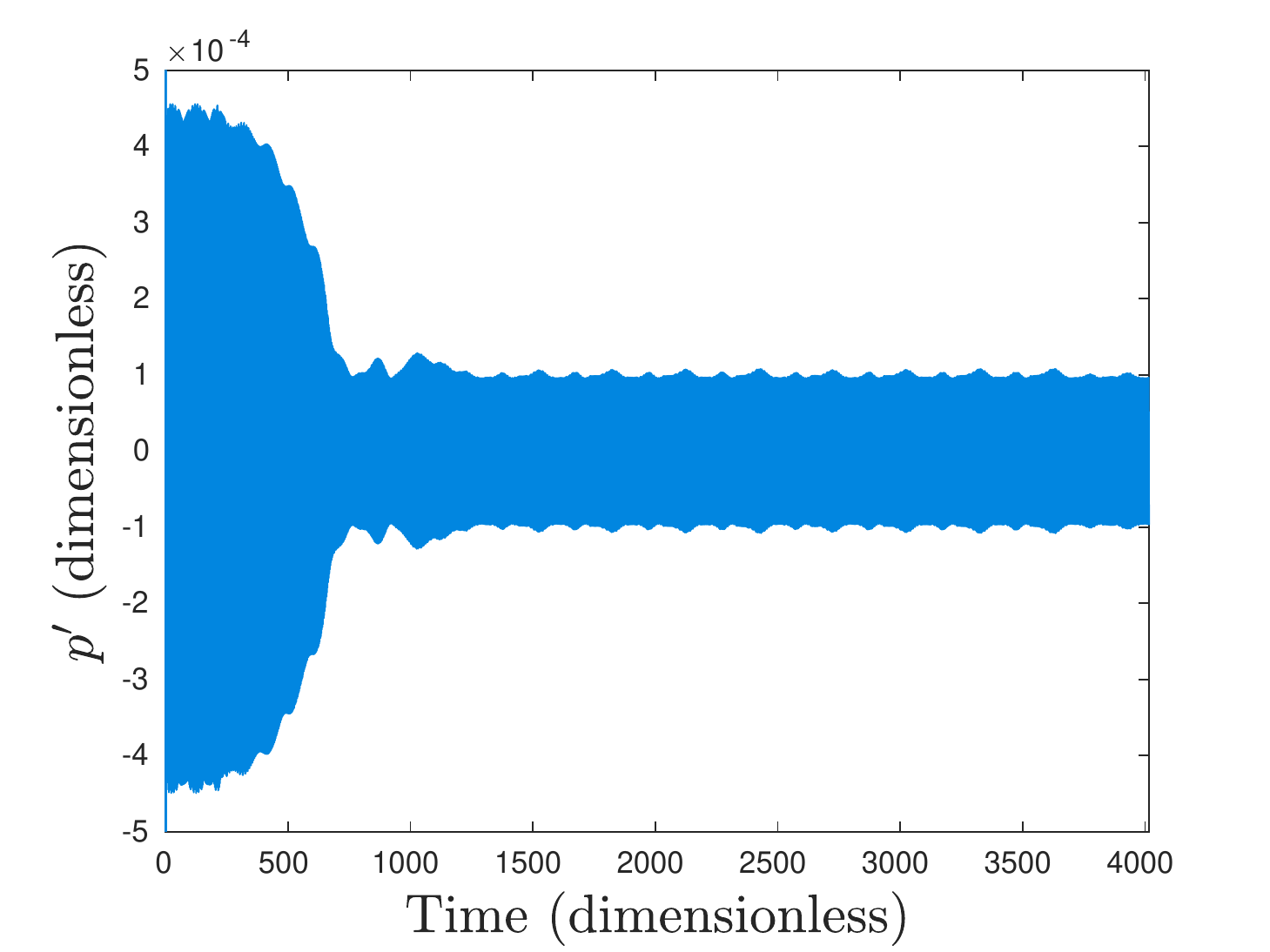}
\includegraphics[width=.32\textwidth]{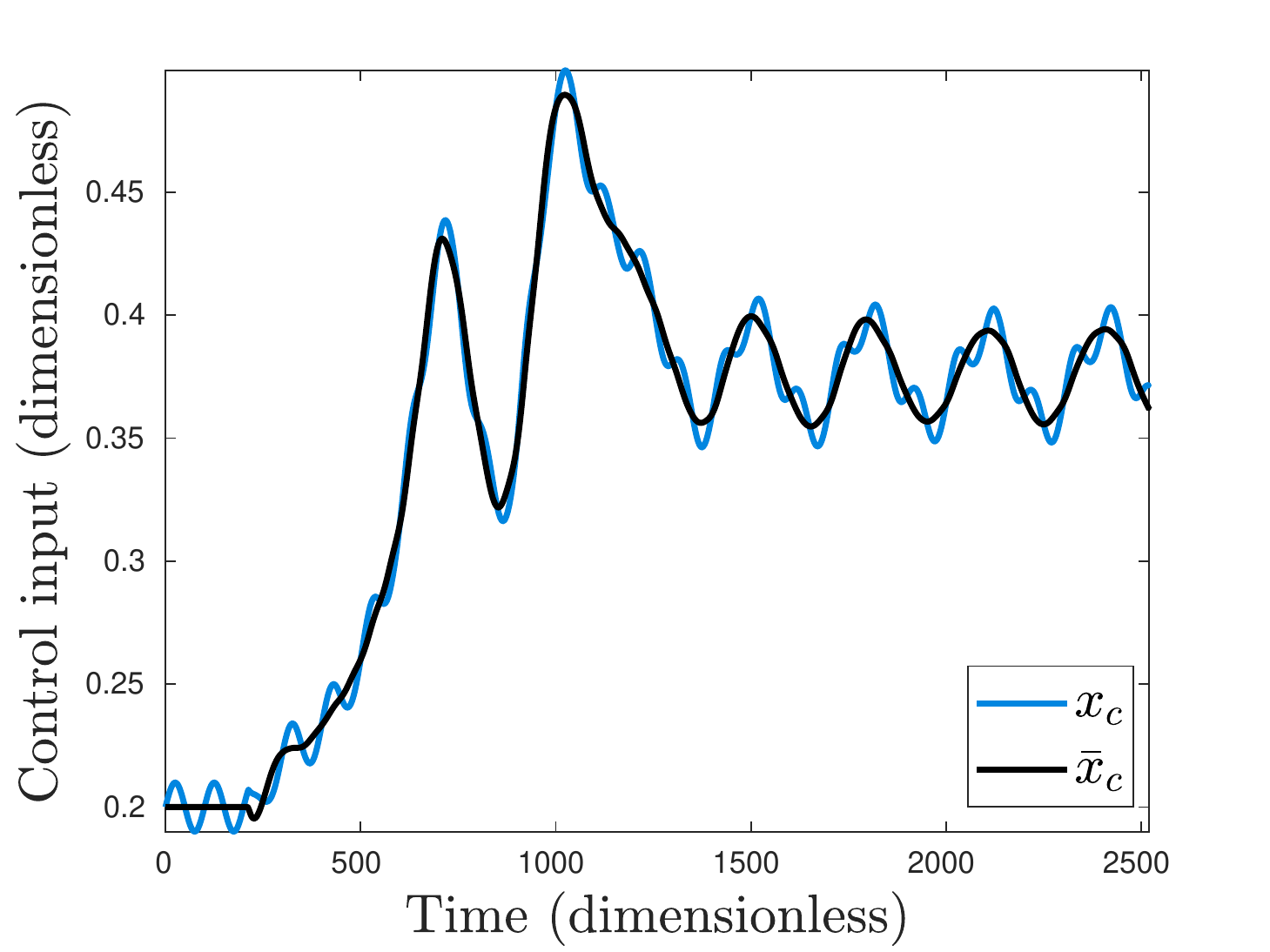}
\includegraphics[width=.32\textwidth]{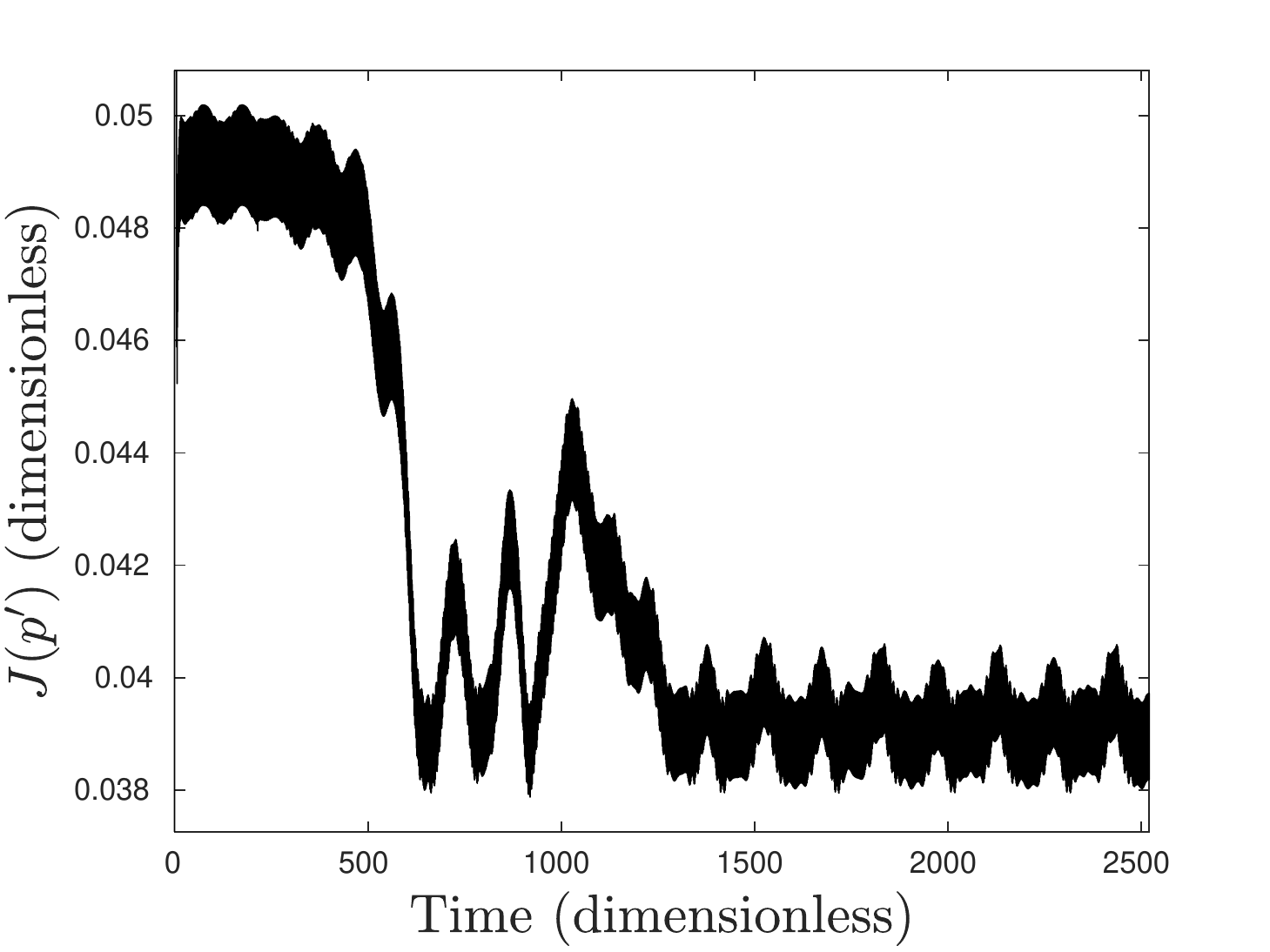}
\includegraphics[width=.32\textwidth]{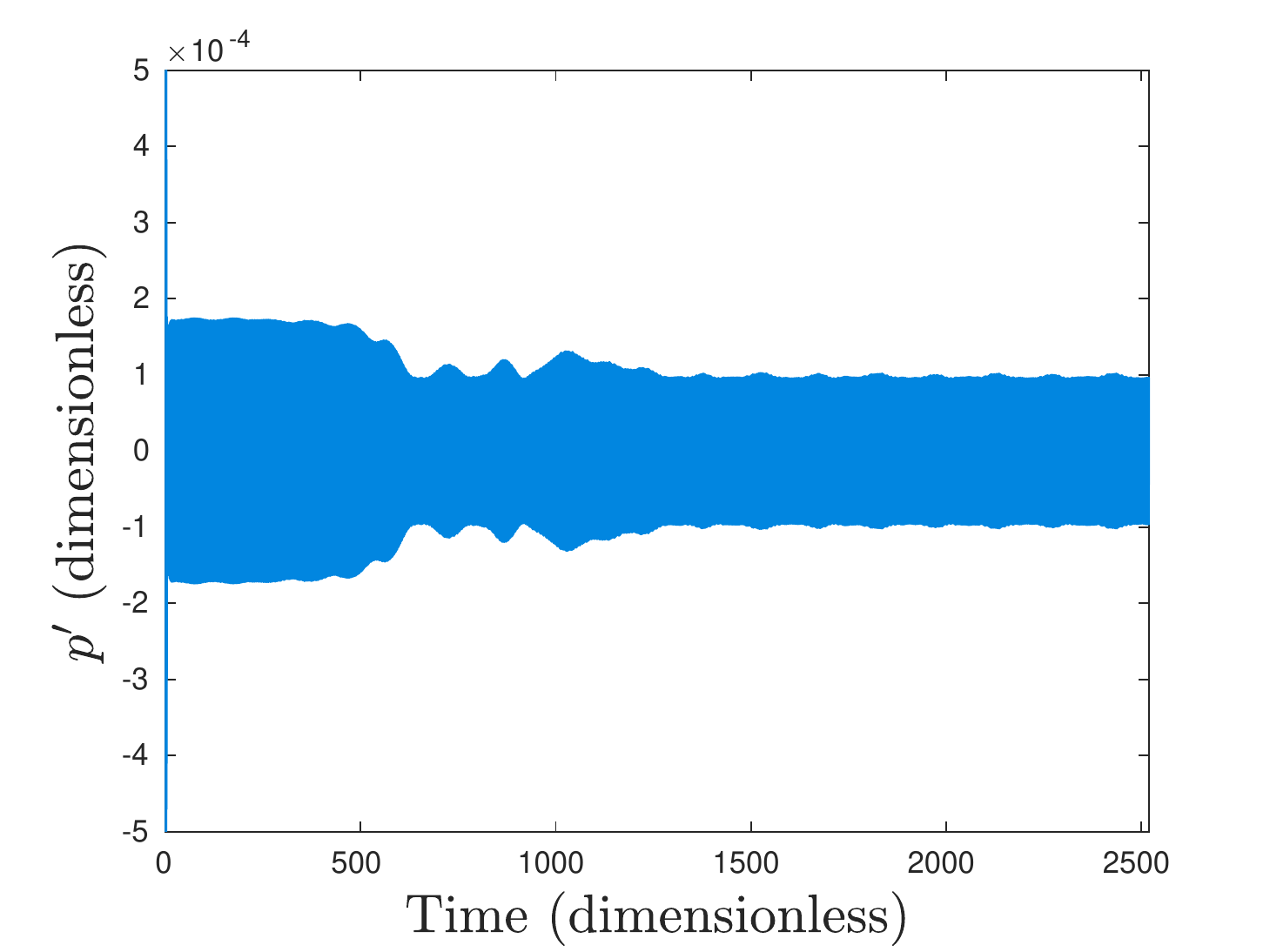}
\caption{Cases \cs{15} and \cs{16}. By further reducing the actuator intensity to $A=-1.2\%$, flow reattachment is not observed and the loop drives $x_c$ to an optimal position independently from the starting point. Some oscillations occur around the optimal position at a frequency lower than $\omega$ due to the high integration gain $\eta$.}
\label{f:e015}\label{f:e016}
\end{figure}

\begin{figure}
\centering
\includegraphics[width=.32\textwidth]{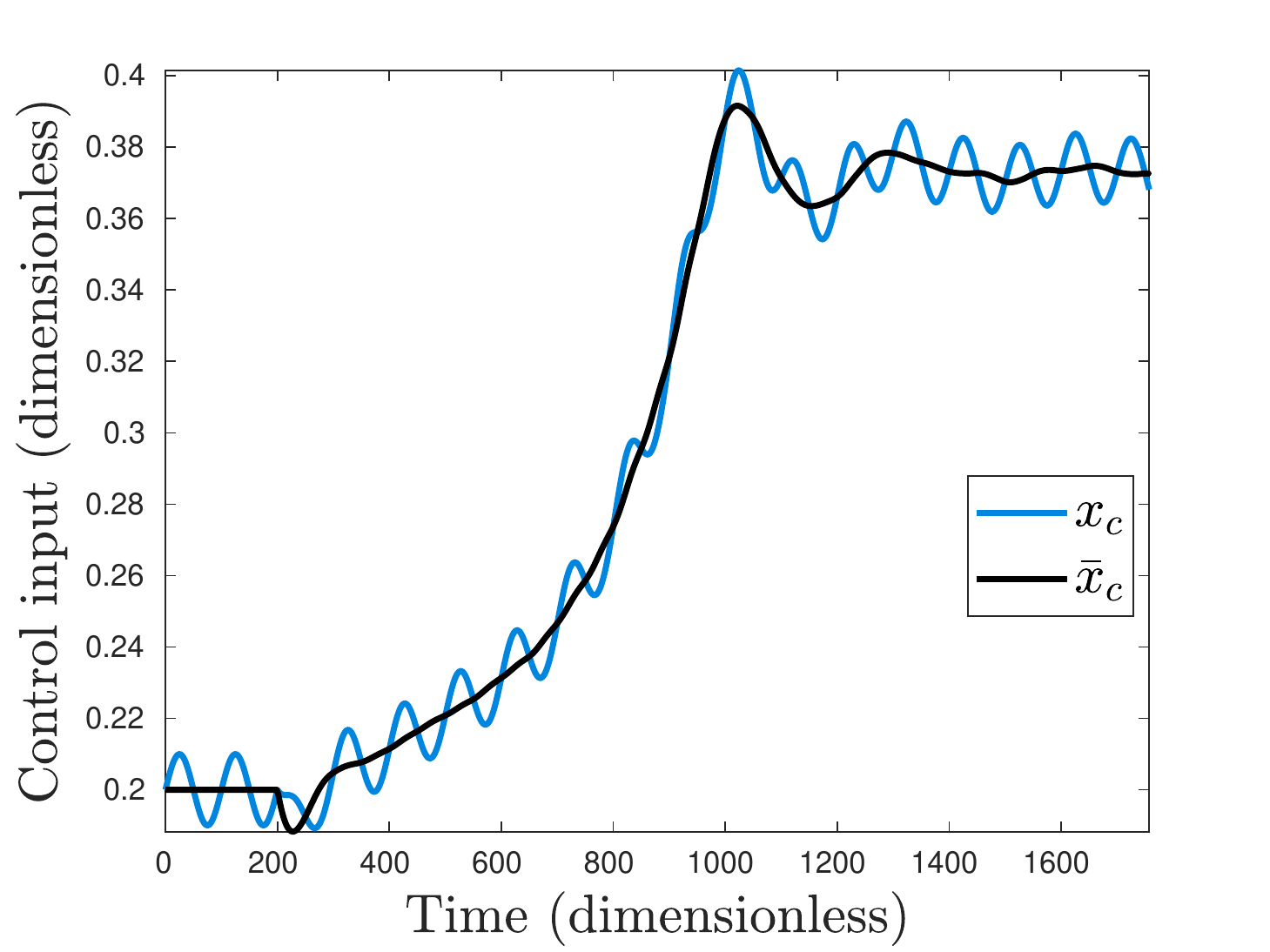}
\includegraphics[width=.32\textwidth]{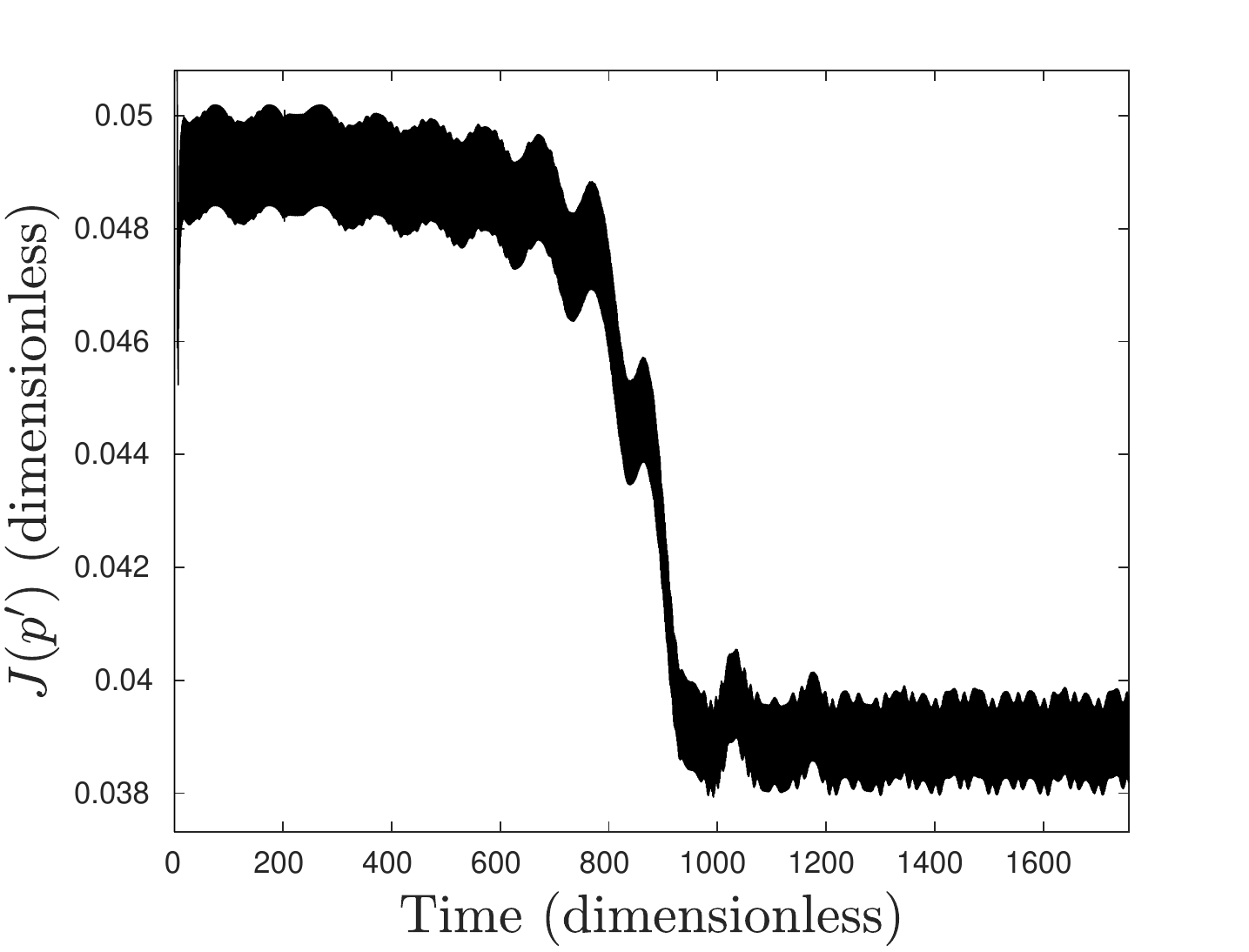}
\includegraphics[width=.32\textwidth]{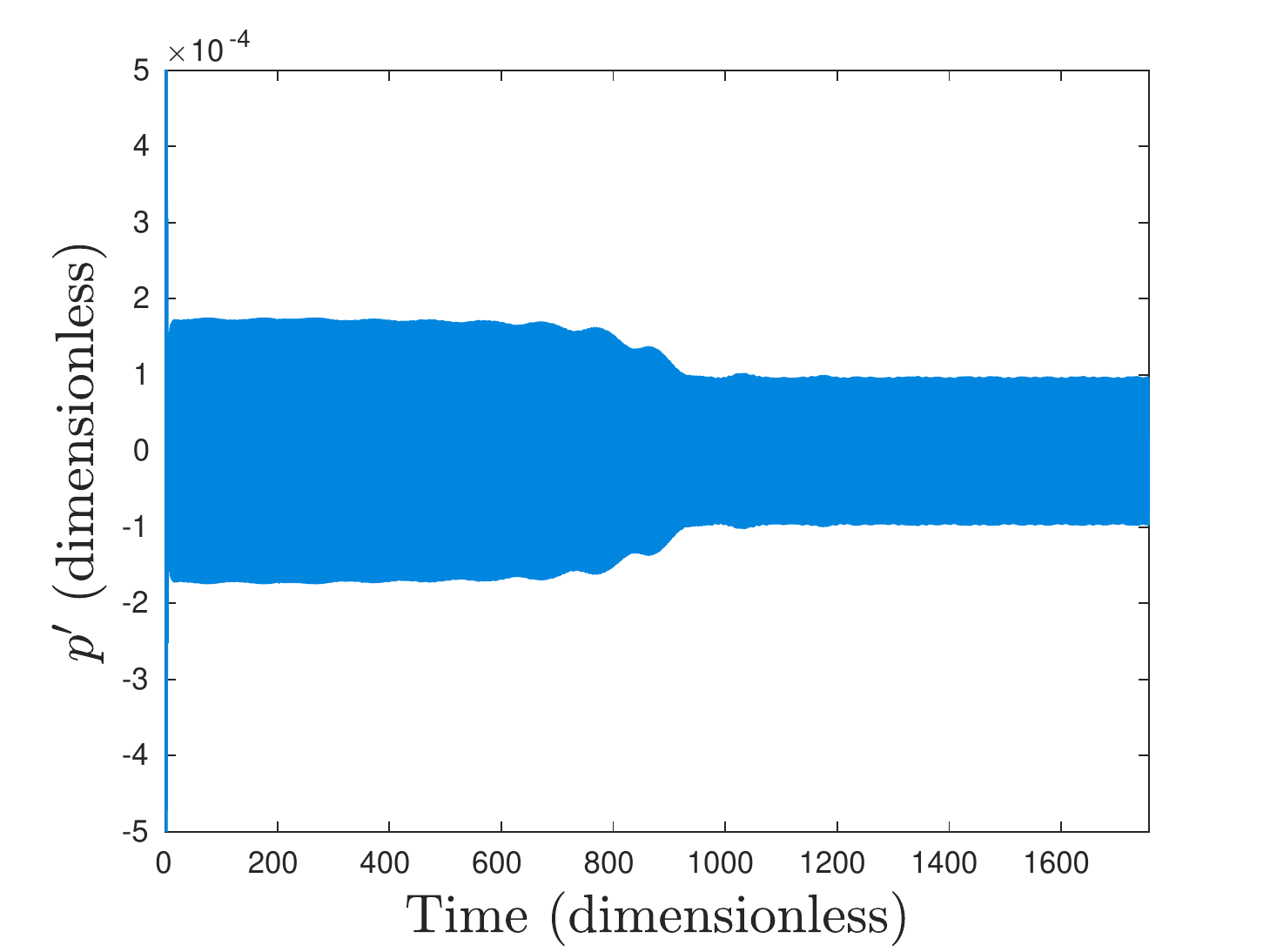}
\caption{Case \cs{27}. A lower value of $\eta$, compared to those employed in cases \cs{15} and \cs{16}, increases the rise time but reduces overshoots and undesired oscillations around the optimal position.}
\label{f:e027}
\end{figure}

An even lower suction intensity of $A=-1.2\%$ is tested in cases \cs{15} -- \cs{27}. As can be seen in Figs. \ref{f:e015} and \ref{f:e027}, for these cases, the suction device is not able to reattach the flow and, therefore, noise generation is minimized but not suppressed. As a result, the actuator converges to a specific location at $\bar{x}_c \approx 0.37$ independently from the initial position being near the leading or trailing edge. Results are shown in Fig. \ref{f:e015} for cases \cs{15} and \cs{16} and it is possible to see oscillations of the mean actuator position $\bar{x}_c$ around the equilibrium location. This oscillatory motion occurs due to the high integrator gain $\eta$ which results in a fast displacement of the jet actuator that, in turn, exceeds the optimal target. As shown in Fig. \ref{f:e027}, case \cs{27} rectifies this issue by reducing the magnitude of $\eta$. However, the lower integration gain leads to a more pronounced rise time.

\begin{figure}
\centering
\includegraphics[width=.32\textwidth]{figures/acc_vort/10k/0000.png}
\includegraphics[width=.32\textwidth]{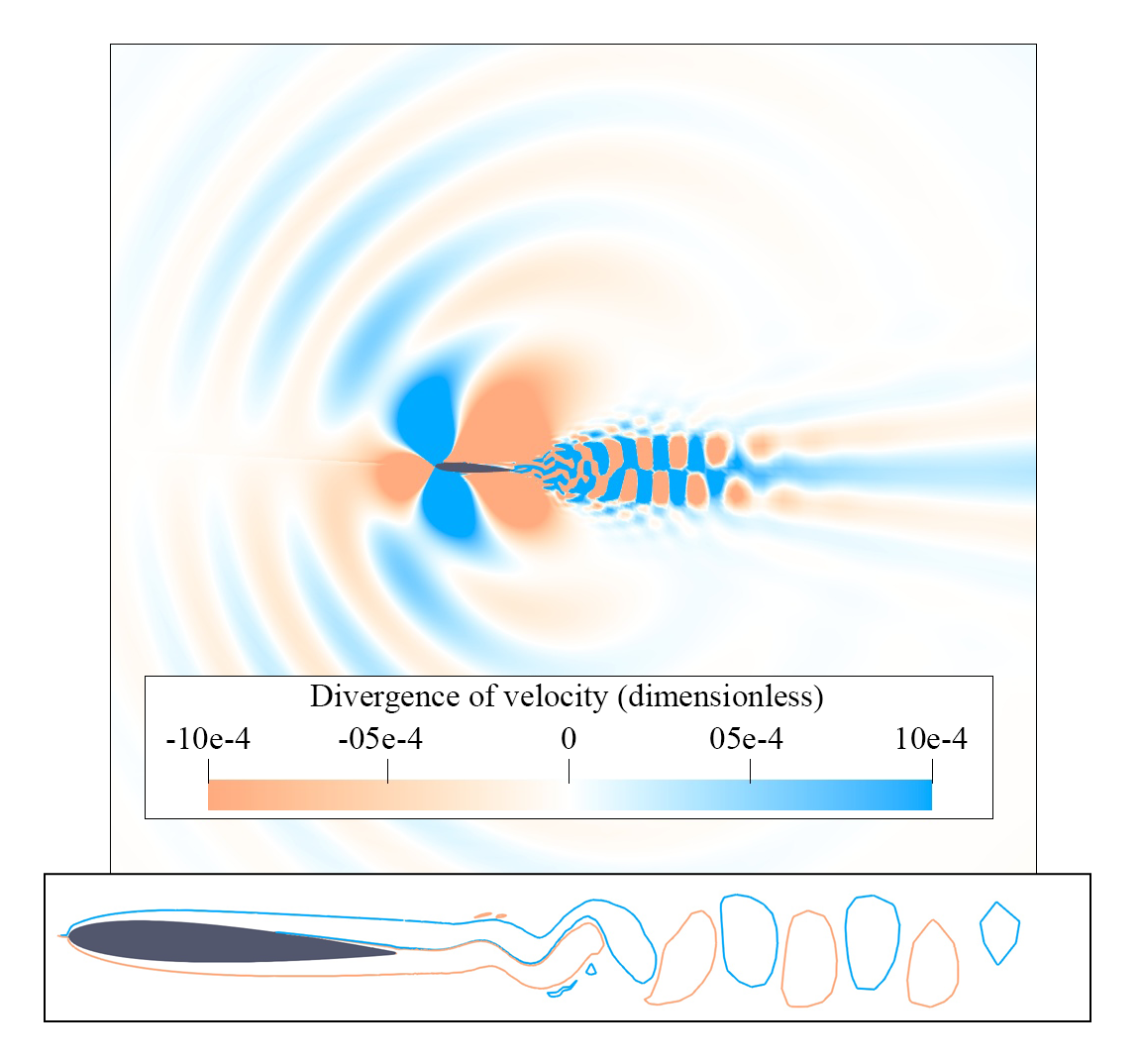}
\includegraphics[width=.32\textwidth]{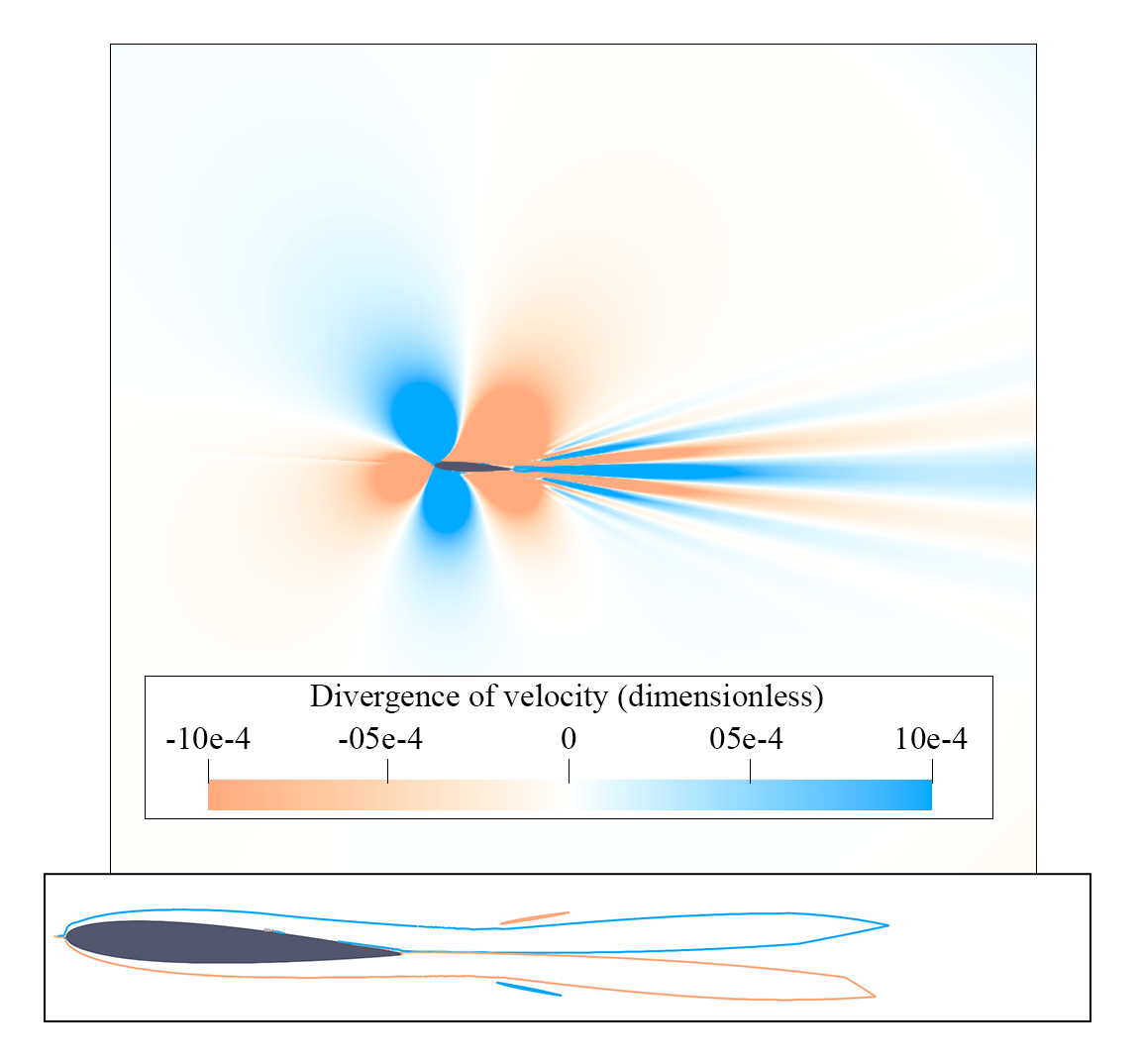}
\caption{Comparison between different actuation intensities. The images depict divergence of velocity (top) and isolines of z-vorticity (bottom). The passive flow (left) develops vortex shedding that generates noise when interacting with the trailing edge. With a weak suction jet of intensity $A=-1.2\%$ (center), the flow remains unsteady, but noise generation is reduced. With a stronger actuation of $A=-3.0\%$ (right), the flow becomes steady and noise is suppressed.} 
\label{f:10kall}
\end{figure}

A comparison between different flow fields, with and without actuation, is presented in Fig. \ref{f:10kall}. The divergence of velocity is shown as an indicative of the acoustic field for the passive flow in the left plot. The center and right figures show the divergence of velocity for controlled cases with $A=-1.2\%$ and $A=-3.0\%$, respectively. In the lower actuation amplitude case, the flow is still unsteady and displays vortex shedding but, in the other case, the flow is reattached and steady, resulting in noise suppression. The vorticity field is also shown for all three simulations in the same figure.

\subsubsection{Actuation with varying intensity and fixed position on the suction side}

In this section, a new control configuration is proposed where suction (or blowing) actuation is placed at a fixed position $x = 0.925$ on the airfoil suction side. The control parameter is the intensity of the jet $A_c$ and Table \ref{tab:cases10susi} shows the simulation setups investigated for cases \cs{17} -- \cs{20}. The square window function $\mathcal{W}_1$ is employed in the present setups with $\Delta x = 5.0\%$. For all cases, the noise generation is gradually reduced when suction is intensified, so the control loop automatically increases the magnitude of $A_c$ in the negative direction. After a critical intensity value, the flow reattaches and the airfoil noise is suppressed. %
%The flow Reynolds and Mach numbers are kept as $\text{Re} = 10^4$ and $\text{Ma} = 0.3$, respectively.
%
\begin{table}
  \begin{center}
\def~{\hphantom{0}}
  \begin{tabular}{lccccccccc}
     \hline
     Case    & $x$    & $A_{c0}$   & $t_{c0}$ & $\alpha=\beta$ & $m$ & $2\pi/\omega$ & $\eta$     & $w$ (Eq. \ref{e:cost2}) & Results\\\hline
     \cs{17} & 0.925 &    -3.0\%  & 8.50e+01 & 0.007          & 20  & 4.00e+01      & -4.00e+00  & 0.00                    & Fig. \ref{f:e017}\\
     \cs{18} & 0.925 &    -3.0\%  & 8.50e+01 & 0.004          & 20  & 4.00e+01      & -8.00e+00  & 0.25                    & Fig. \ref{f:e018}\\
     \cs{20} & 0.925 &    -5.5\%  & 3.50e+02 & 0.004          & 20  & 1.60e+02      & -2.00e+01  & 0.25                    & Fig. \ref{f:e020}\\\hline
  \end{tabular}
  \caption{Control parameters for actuation with varying intensity and fixed position on the suction side. %$\text{Re} = 10^4$.  The initial intensity $A_{c0}$ is given as a percentage of the freestream momentum and the negative sign denotes suction. %(which is equal, in value, to the Mach number $\text{Ma} = 0.3$, since $\rho_\infty=1$).
  %The integrator turn-on time $t_{c0}$ and the ESC wave period $2\pi/\omega$ are dimensionless temporal parameters computed relative to the freestream speed and airfoil chord. The fixed position $x$ represents the nondimensional actuator position. The ESC parameter $\alpha$ is relative to the freestream momentum.
  The term $w$ is a weight used to penalize the cost function based on the actuation intensity.
  }
  \label{tab:cases10susi}
  \end{center}
\end{table}

\begin{figure}
\centering
\includegraphics[width=.32\textwidth]{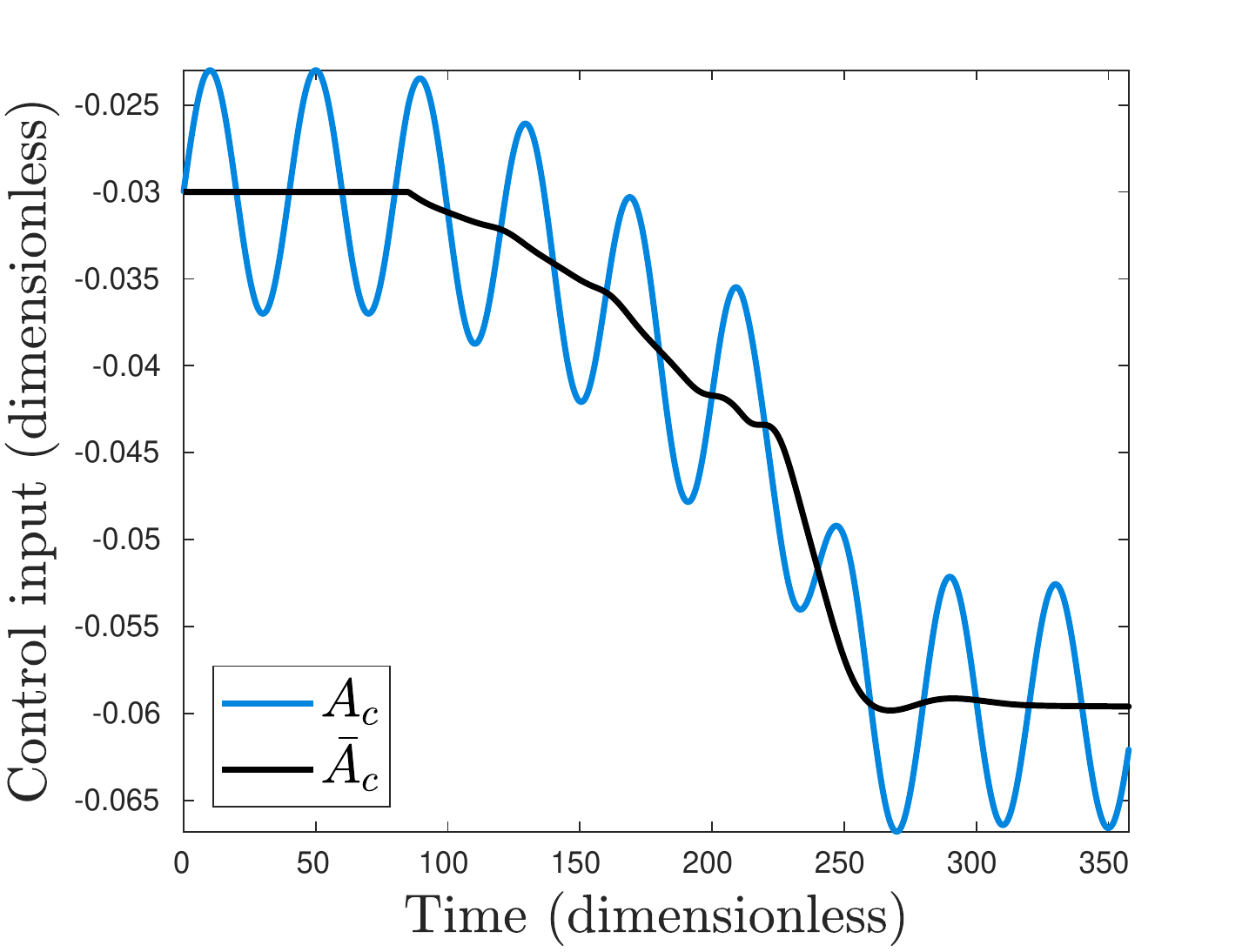}
\includegraphics[width=.32\textwidth]{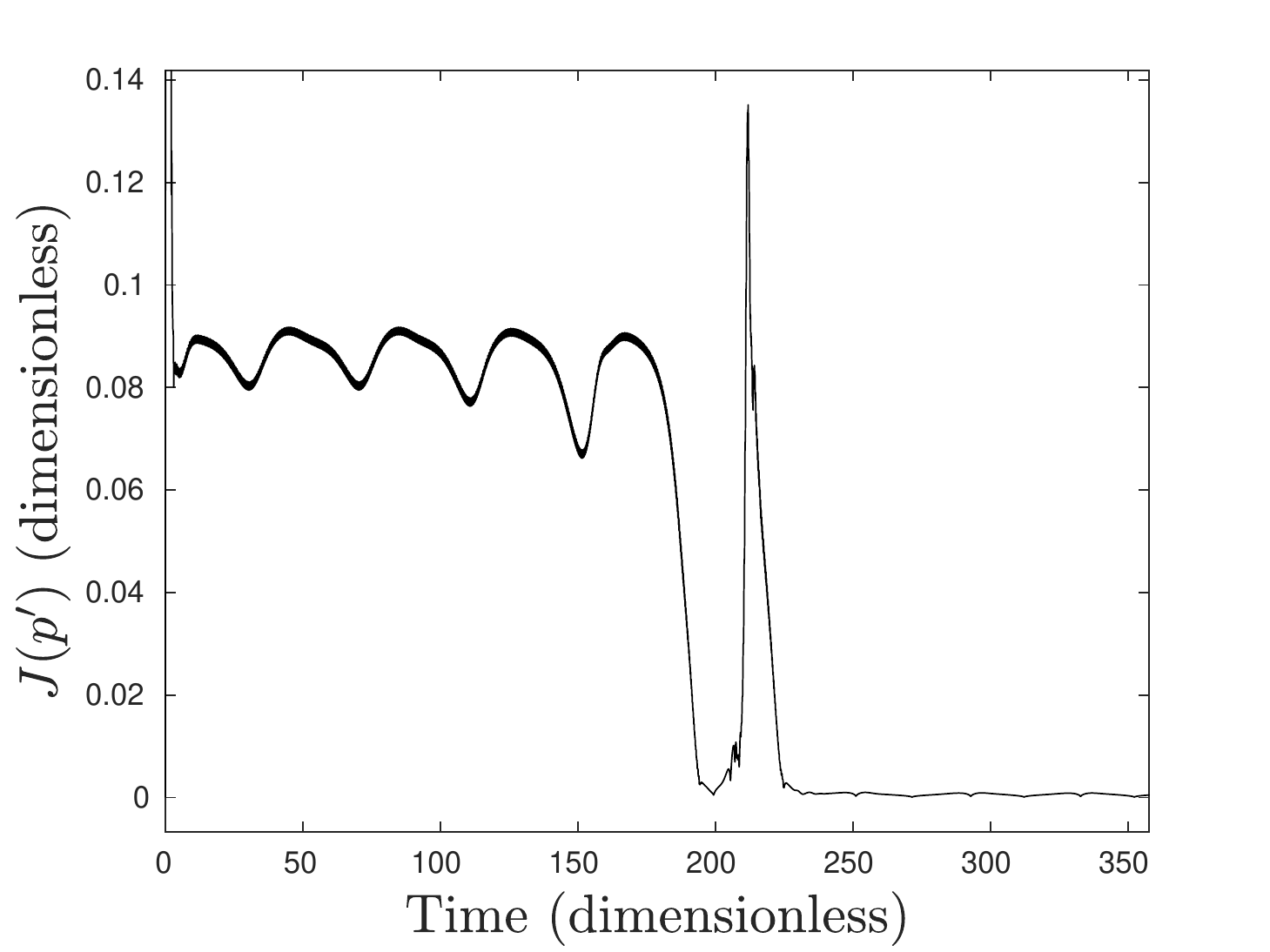}
\includegraphics[width=.32\textwidth]{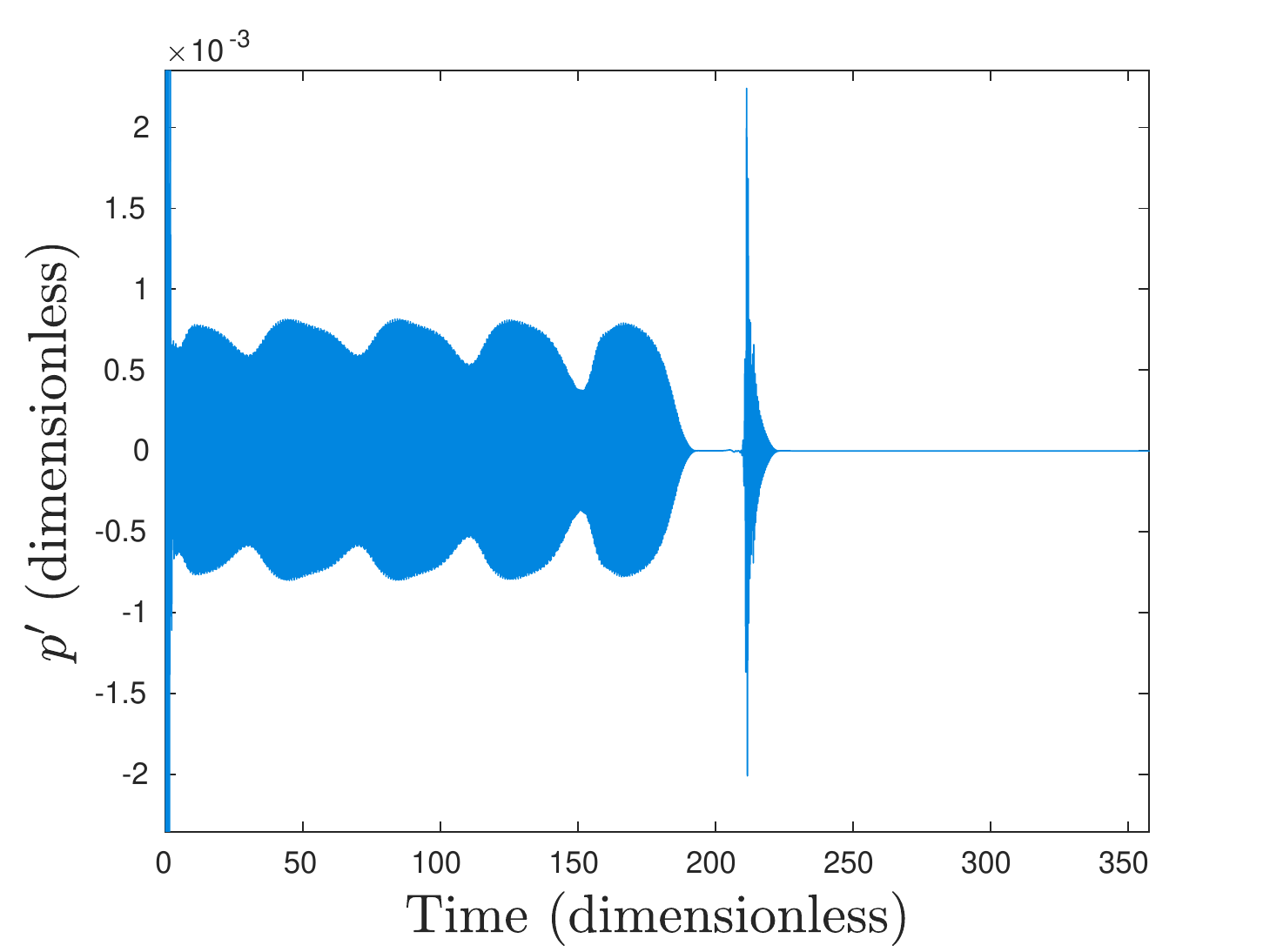}
\caption{Case \cs{17}. This setup seeks an optimal suction intensity $A_c$ to reduce noise at a fixed actuator position $x = 0.925$. The ESC is able to attach the flow with $A_c \approx -5\%$, but during the ESC sweep the flow detaches again leading to a peak in the cost function $J$. Due to this peak and the filter dynamics, the ESC finds a higher actuation magnitude $|A_c|$ to maitain the flow attached and suppress noise.}
\label{f:e017}
\end{figure}

Results for simulation \cs{17} are presented in Fig. \ref{f:e017}, where it is possible to see that the initial suction actuation $A_{c0}$ is not sufficient to reattach the flow. The actuation intensity from the ESC can be observed in the blue line of the left plot in Fig. \ref{f:e017}. When the critical value of $A_c \approx -5.0\%$ is reached, the flow becomes attached and the noise is suppressed at a time just before $t \approx 200$. However, due to the harmonic actuation, the control input is raised above a minimum for which the flow detaches again. This behavior can be observed in the center plot of Fig. \ref{f:e017}, which shows the cost function given by Eq. \ref{e:cost2}. The brief noise spike at $t\approx 210$ occurs due to the sudden flow detachment that results from the ESC perturbation. This occurs when the absolute value of $A_c$ decreases due to the harmonic sweep, which leads to a discontinuous behavior in the transition between the attached and detached flows.

\blue{After the strong variation in the cost function the actuation magnitude $A_c$ increases, as shown in Fig. \ref{f:e017}. In this case, the suction magnitude for reattachment is higher than that for which the flow first became attached.}
This behavior in the control input is related to the filter dynamics and the sharp peak observed in the cost function. Hence, the suction magnitude first needed to suppress noise is lower than that achieved by the ESC at later times. Moreover, since significant noise is not generated after the cost function peak, the control loop no longer has information about a direction to change $A_c$. Similarly, in a more realistic application, control effort drifting due to measurement noise could lead to unnecessarily high power consumption. In order to overcome this issue, a modified cost function is proposed as 
\begin{equation}
    J_{k} = \left[\sum_{i=0}^{m-1}\left(p'_{k-i}\right)^2\right]^\sigma + w |A_c| \mbox{ ,}
    \label{e:cost2}
\end{equation}
where $w$ is a constant weight used to penalize the controller when the actuator intensity  grows. The new cost function is applied to cases \cs{18} and \cs{20} and it balances the noise reduction and the control effort. 

\begin{figure}
\centering
\includegraphics[width=.32\textwidth]{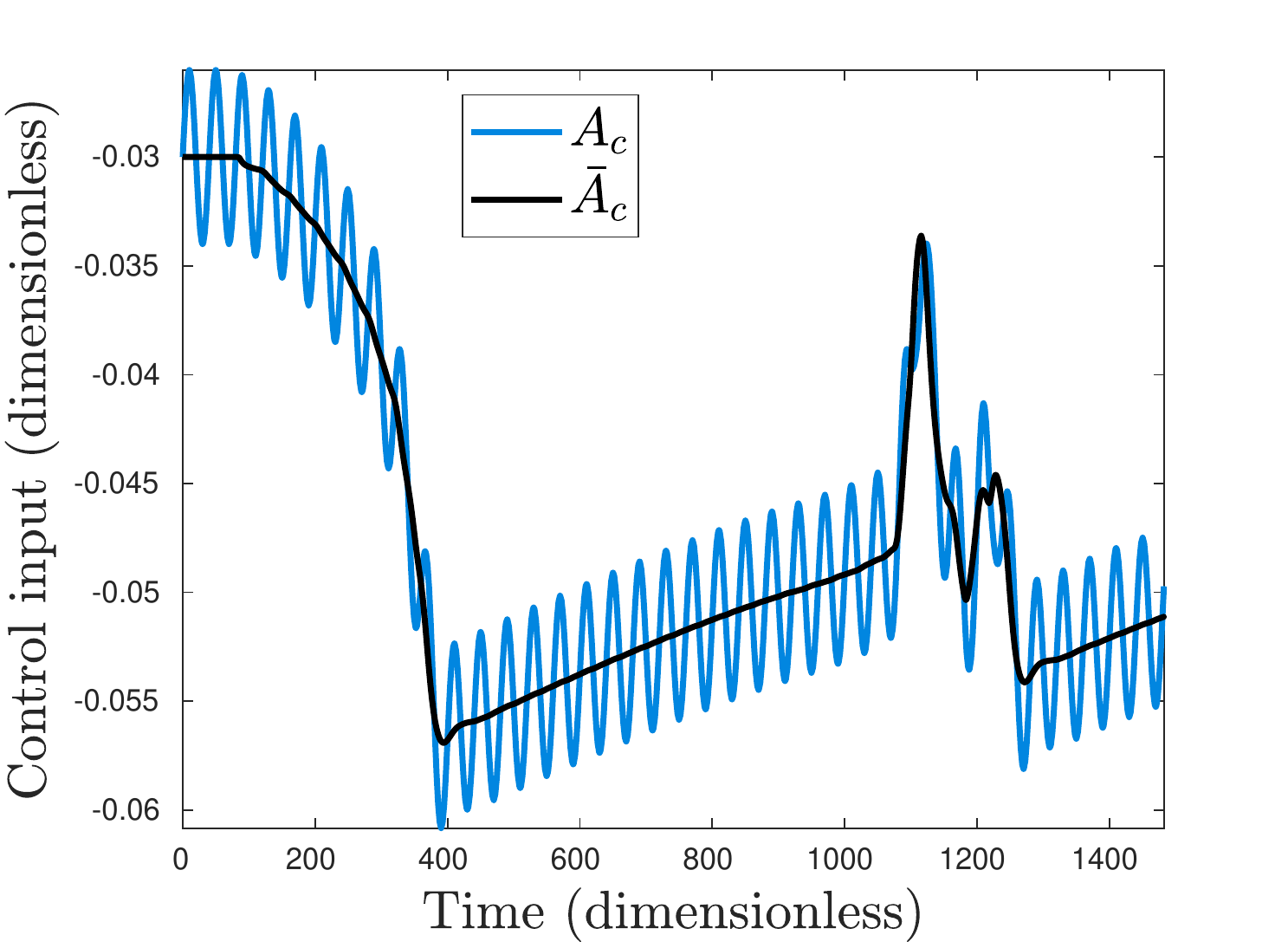}
\includegraphics[width=.32\textwidth]{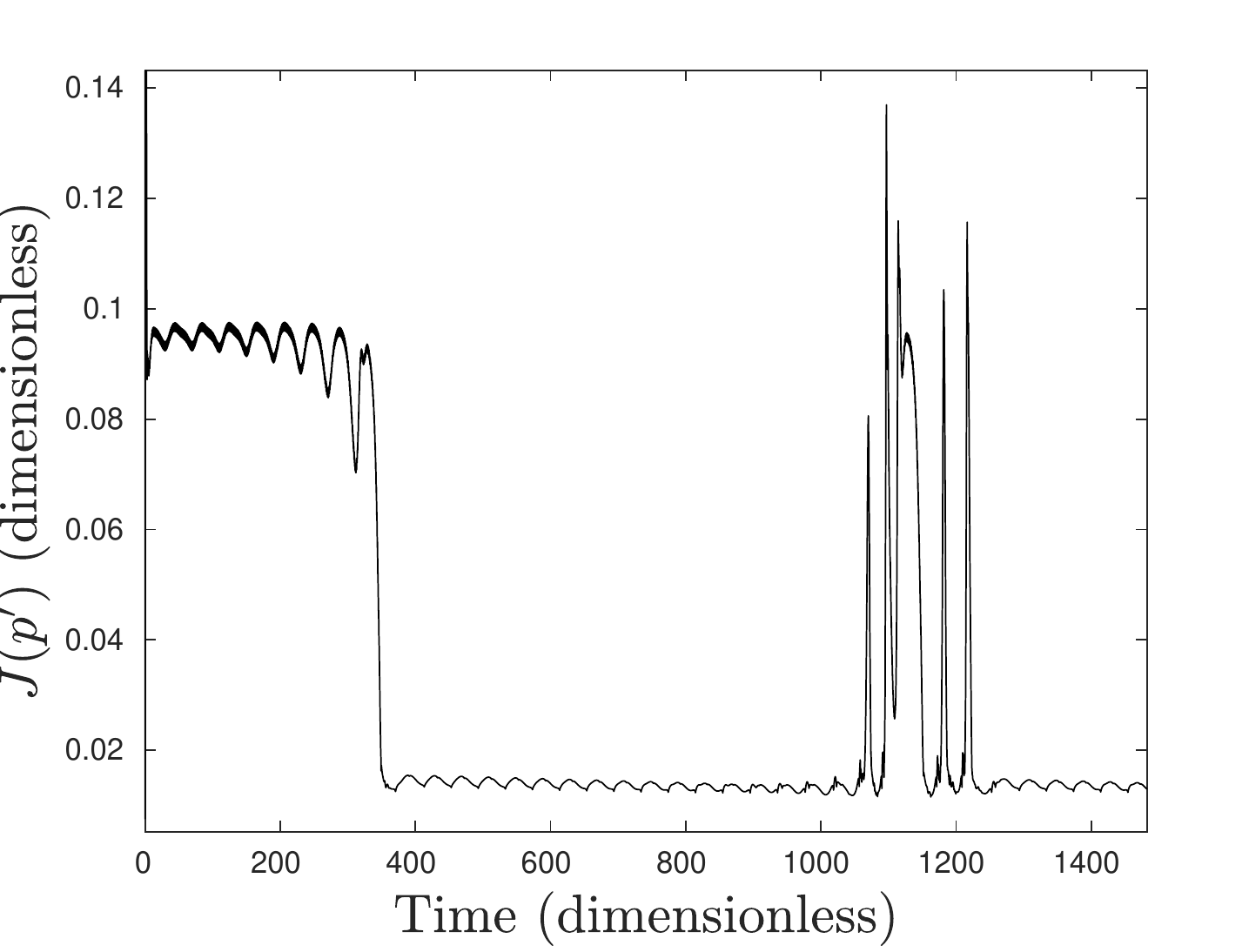}
\includegraphics[width=.32\textwidth]{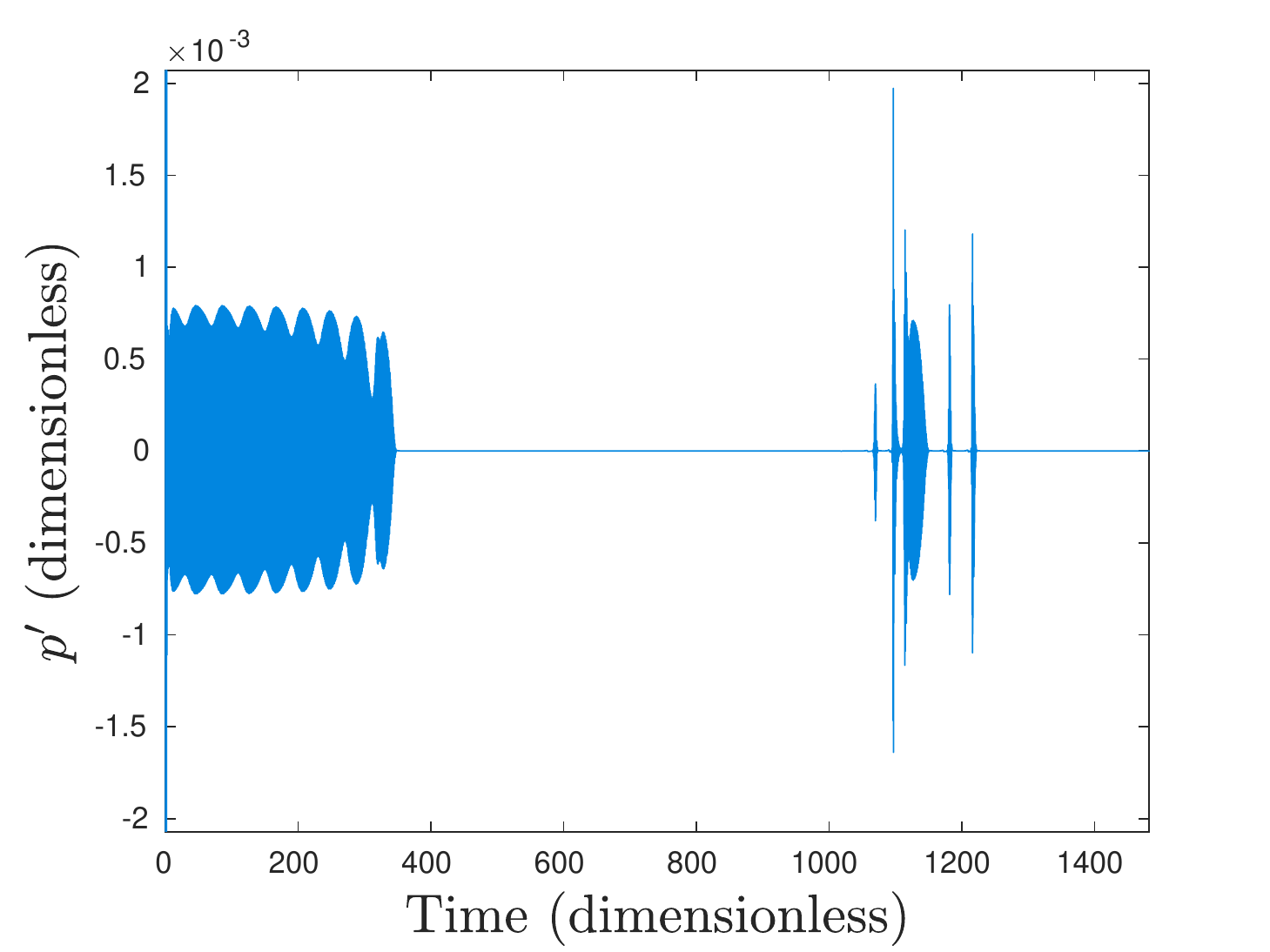}
\caption{Case \cs{18}. In order to avoid the hysteresis observed in Fig. \ref{f:e017}, a new term is added to the cost function including a penalty $w$ to the control effort (Eq. \ref{e:cost2}). With this term, when noise is suppressed at $t\approx 400$, the magnitude of the actuation intensity is reduced. At $t\approx 1050$, the flow detaches again leading to a series of reattachments and detachments observed in the peaks and valleys of the cost function $J$, after which the ESC is able to \blue{find an actuation where noise is again suppressed.}}%find equilibrium.}
\label{f:e018}
\end{figure}

Through an inspection of the results presented in Fig. \ref{f:e018}, it is possible to visualize the reduction in the absolute value of $A_c$ after flow attachment at $t \approx 400$. In this case, when the acoustic pressure computed by the sensor is suppressed, the term used to penalize the control input dominates in Eq. \ref{e:cost2}. Then, the magnitude of the actuation intensity $|A_c|$ is continuously reduced until the flow becomes detached again at $t\approx 1050$, which leads to a discontinuity in the cost function. In this case, the controller is unable to keep the flow at an equilibrium point and the term related to the acoustic noise in the cost function dominates again over the actuator intensity penalty. The suction intensity grows again to reduce noise and this leads to a series of flow reattachments and detachments due to the filter dynamics as can be seen in the cost function for $1000 \lesssim t \lesssim 1200$. At some point, the ESC is able to reach an optimal actuation intensity that suppresses noise again and the loop is restarted from a point similar to that observed for $t \approx 400$, where the penalty weight plays again an important role in the cost function. %but at a critical point, the flow rapidly reattaches and the term related to the control input starts dominating. The controller then makes the control effort decrease, but it detaches after another lower critical point, never reaching an equilibrium.

\begin{figure}
\centering
\includegraphics[width=.32\textwidth]{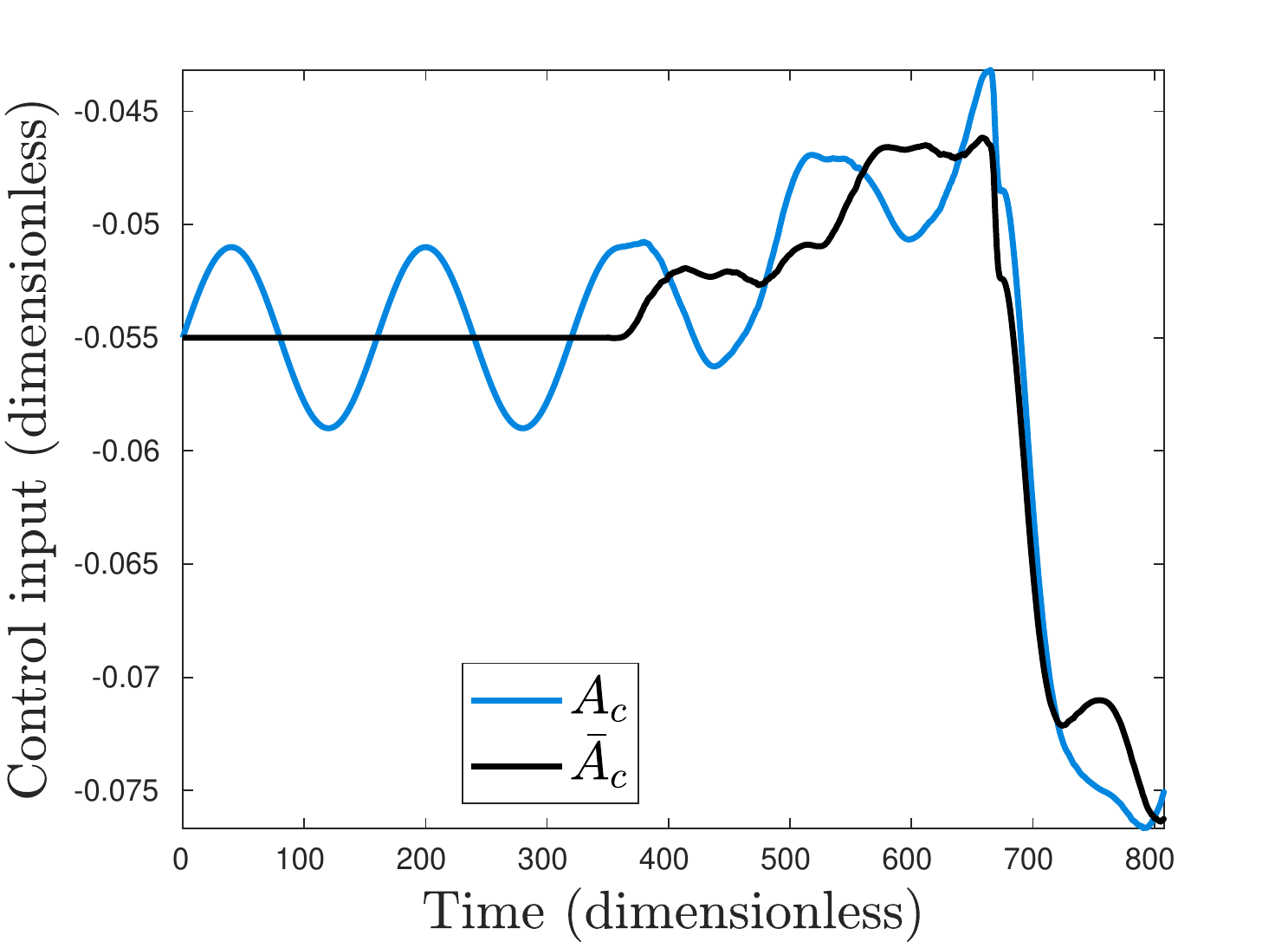}
\includegraphics[width=.32\textwidth]{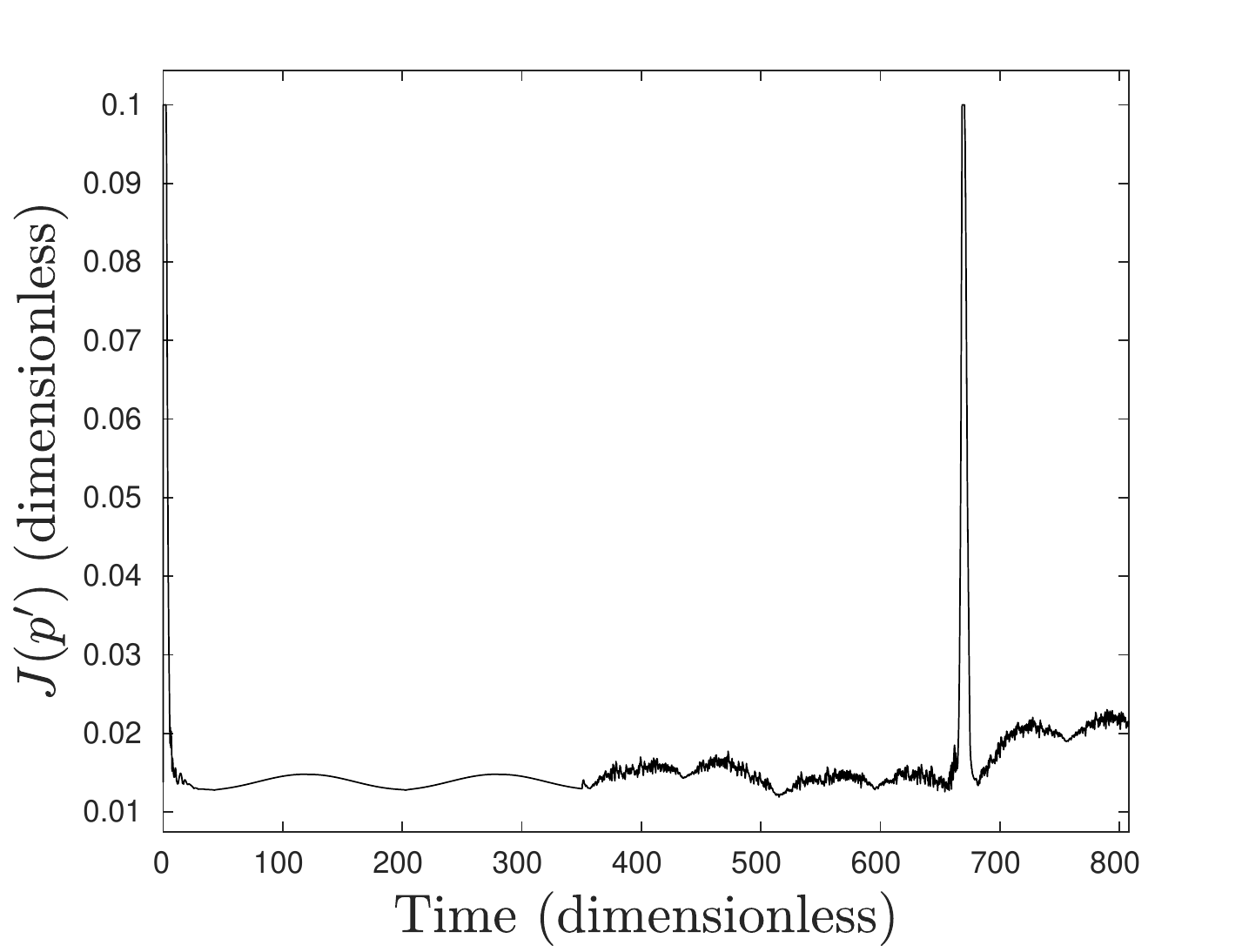}
\includegraphics[width=.32\textwidth]{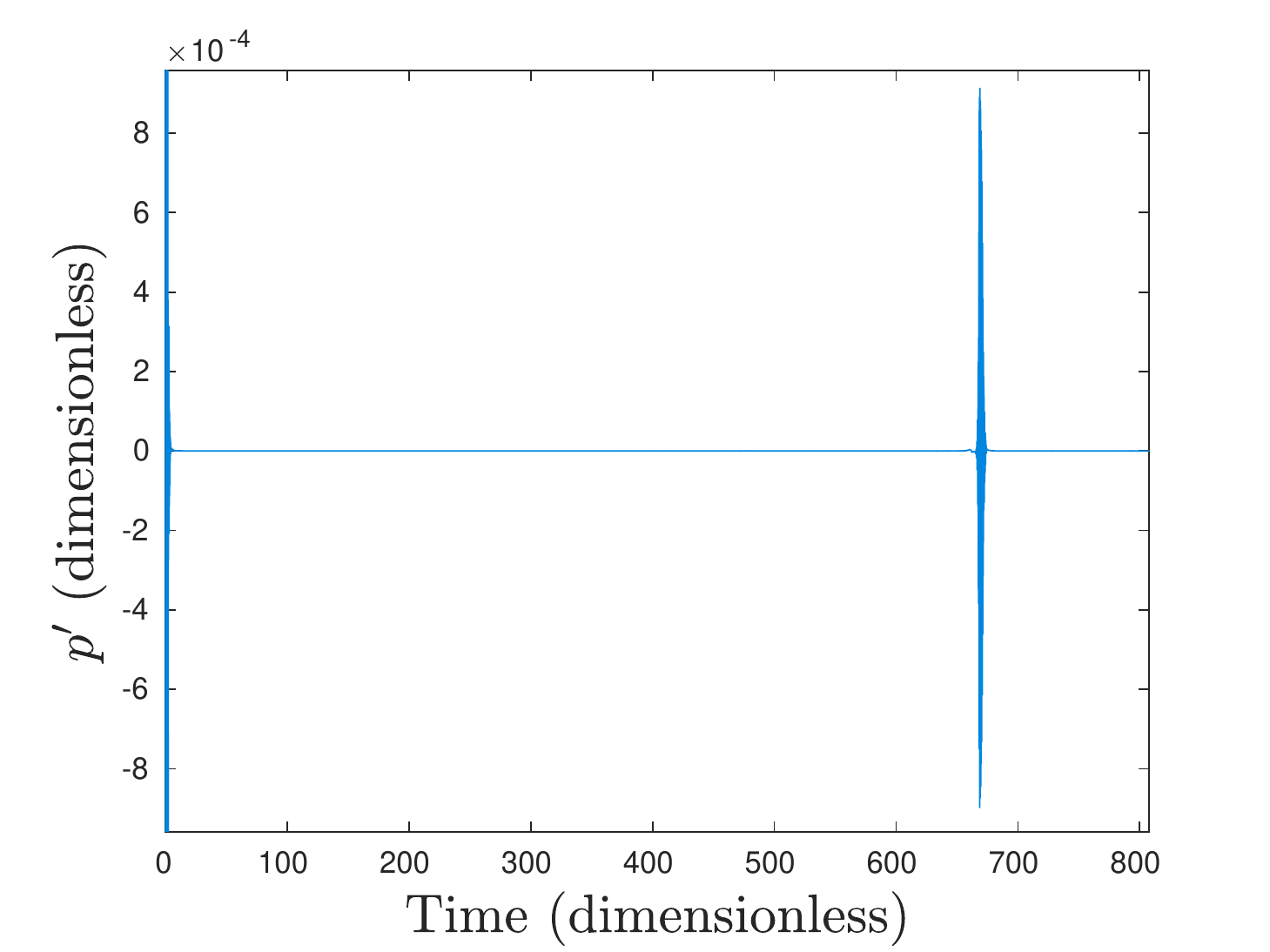}
\caption{Case \cs{20}. In order to improve the response of the ESC to flow detachments, the low-pass filter is removed and the magnitude of $\eta$ is increased.}
\label{f:e020}
\end{figure}

Some attempts to tune the ESC parameters were conducted and, in case \cs{20}, the integration gain is increased and the low-pass filter bypassed to reduce the controller response time to measurements. 
Figure \ref{f:e020} presents the results for this simulation which has a high initial actuation intensity, sufficient to maintain the flow attached. As can be seen from the figure, when the control loop is started at $t_0=350$, the ESC leads to a reduction in the magnitude of $A_c$. Even considering the changes in the controller setup, the loop is unable to find an equilibrium suction intensity. Hence, this approach presents issues due to the discontinuous physics of the controlled plant that renders the control system unable to operate as desired. Despite this issue, the response to detachment is faster than in the previous cases as can be seen from the pressure fluctuations in the right plot.

\subsubsection{Actuation with varying intensity and fixed position at the trailing edge} \label{sec:10ktred}

\begin{figure}
\centering
\includegraphics[width=.50\textwidth, trim=0 1cm 0 0, clip]{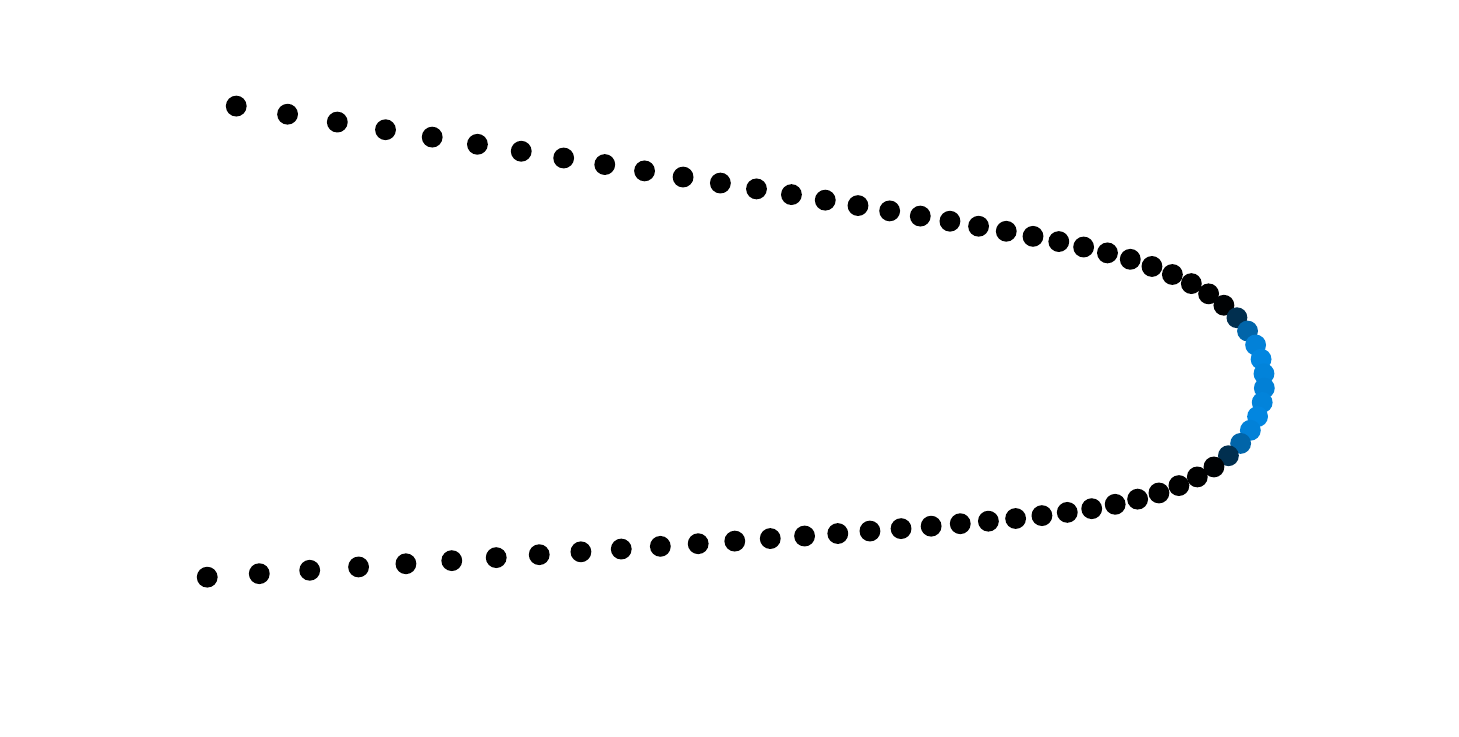}
\caption{The blue points indicate the actuator position where flow injection is applied. The intensity $A_c$ of the actuation is set by the controller.}
\label{f:walter_act}
\end{figure}

In this control setup, an actuator with varying intensity is applied fixed at the trailing edge. The actuator is placed at the grid points highlighted in blue in Fig. \ref{f:walter_act} and the window function $\mathcal{W}_2$ is employed for the present setup. This configuration was proposed by \citet{ramirez2015effects}, who observed that trailing edge blowing reduced the scattered noise field from low Reynolds number airfoil flows. These authors verified that flow actuation in the trailing edge displaced the vortex shedding further downstream the airfoil wake. Hence, the peak \green{fluctuations} of the Lighthill stress terms, which represent the incident quadrupolar sources, were also displaced. In proximity to a rigid surface, the quadrupole sources lead to an efficient acoustic scattering which, in this case, is reduced due to source displacement. \green{This approach was also studied experimentally by \citet{massarotti2019passive} for active noise control, providing attenuation of a main tone produced by the flow past an elliptical crossbar installed on a car roof rack}. Three setups are analyzed in this section with the parameters shown in Table \ref{tab:cases10tred}.
\begin{table}
  \begin{center}
\def~{\hphantom{0}}
  \begin{tabular}{lcccccccc}
     \hline
     Case    & $\delta_{ss}$ & $A_{c0}$  & $t_{c0}$ & $\alpha=\beta$ & $m$ & $2\pi/\omega$ & $\eta$    & Results\\\hline
     \cs{24} & 0.00e-00      &   0.0\%  & 200.0     & 0.007          & 20  & 80.0          & -1000.0   & Fig. \ref{f:e024}\\
     \cs{25} & 3.64e-08      &   0.0\%  & 200.0     & 0.004          & 20  & 80.0          & -2000.0   & Fig. \ref{f:e025}\\
     \cs{26} & 3.64e-08      &   13.0\%  & 200.0     & 0.004          & 20  & 80.0          & -2000.0   & Fig. \ref{f:e026}\\\hline
  \end{tabular}
  \caption{Control parameters for cases with fixed actuator at the trailing edge. The parameter $\delta_{ss}$ corresponds to the slope seeking compensation value. %The initial actuation intensity $A_{c0}$ is given as a  percentage of the freestream momentum and a positive value denotes blowing. The integrator starting time $t_{c0}$ and the ESC wave period $2\pi/\omega$ are dimensionless temporal parameters relative to the freestream velocity and airfoil chord. The ESC parameter $\alpha$ is relative to the freestream momentum.
  }
  \label{tab:cases10tred}
  \end{center}
\end{table}

\begin{figure}
\centering
\includegraphics[width=.32\textwidth]{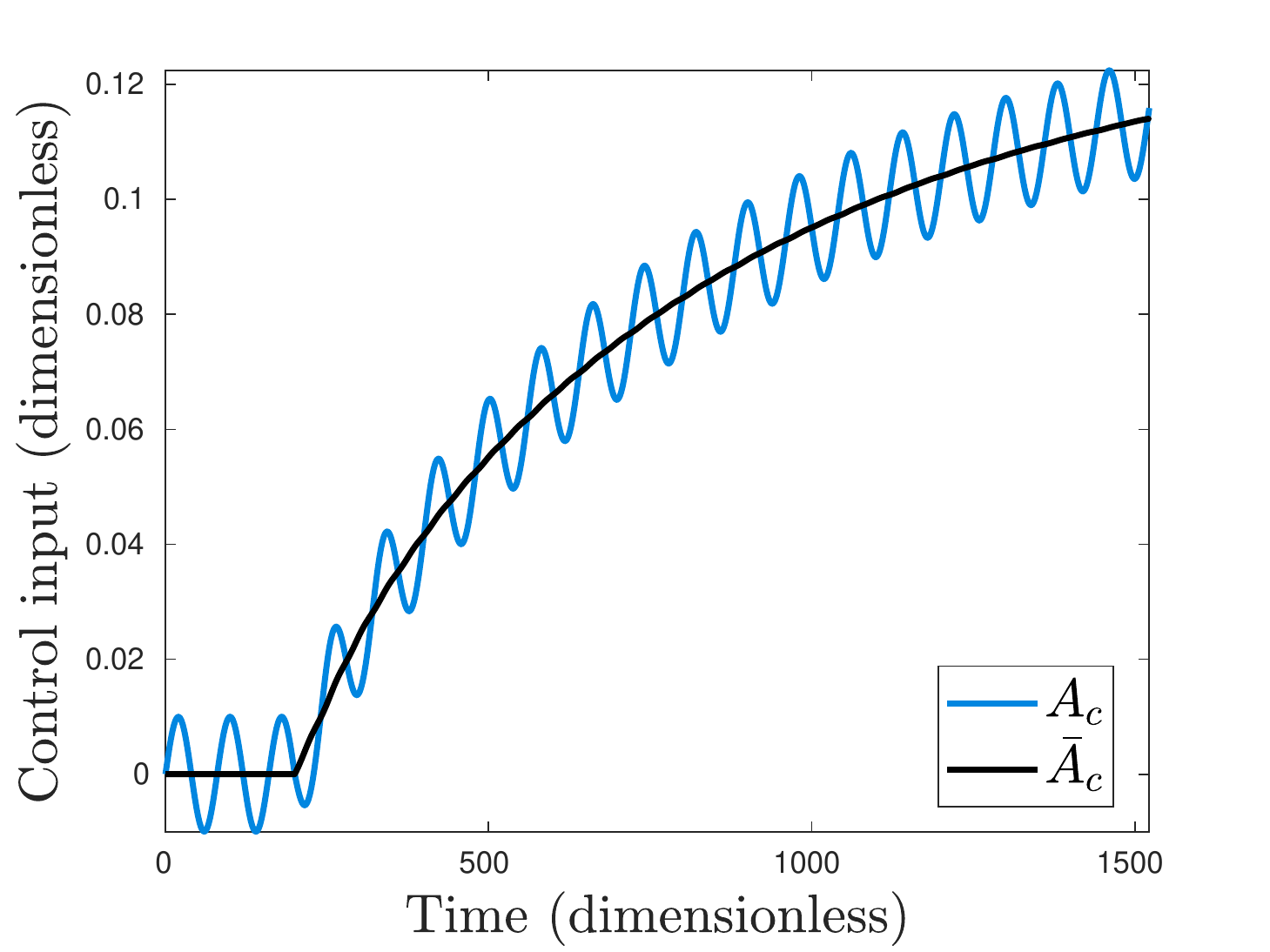}
\includegraphics[width=.32\textwidth]{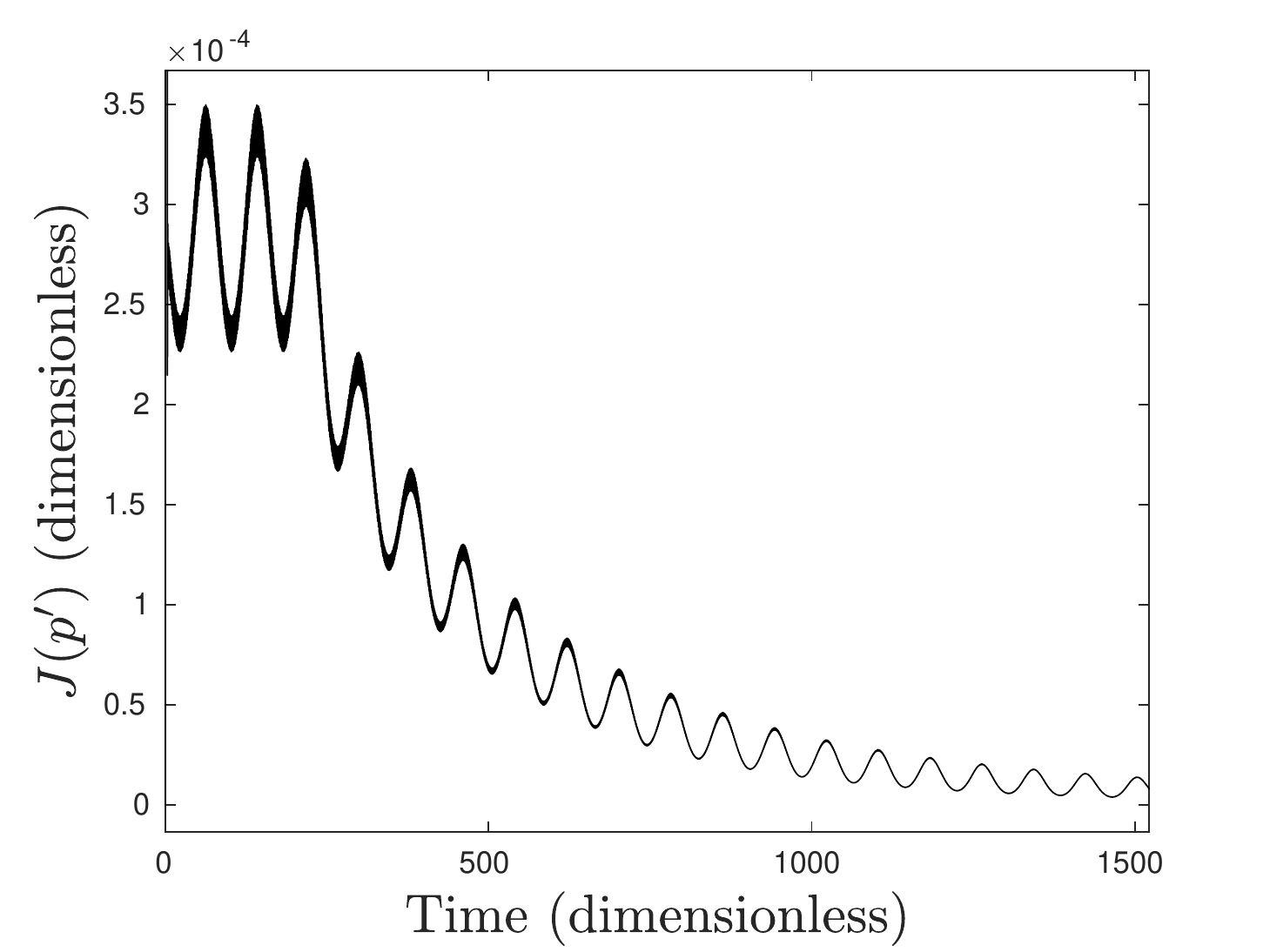}
\includegraphics[width=.32\textwidth]{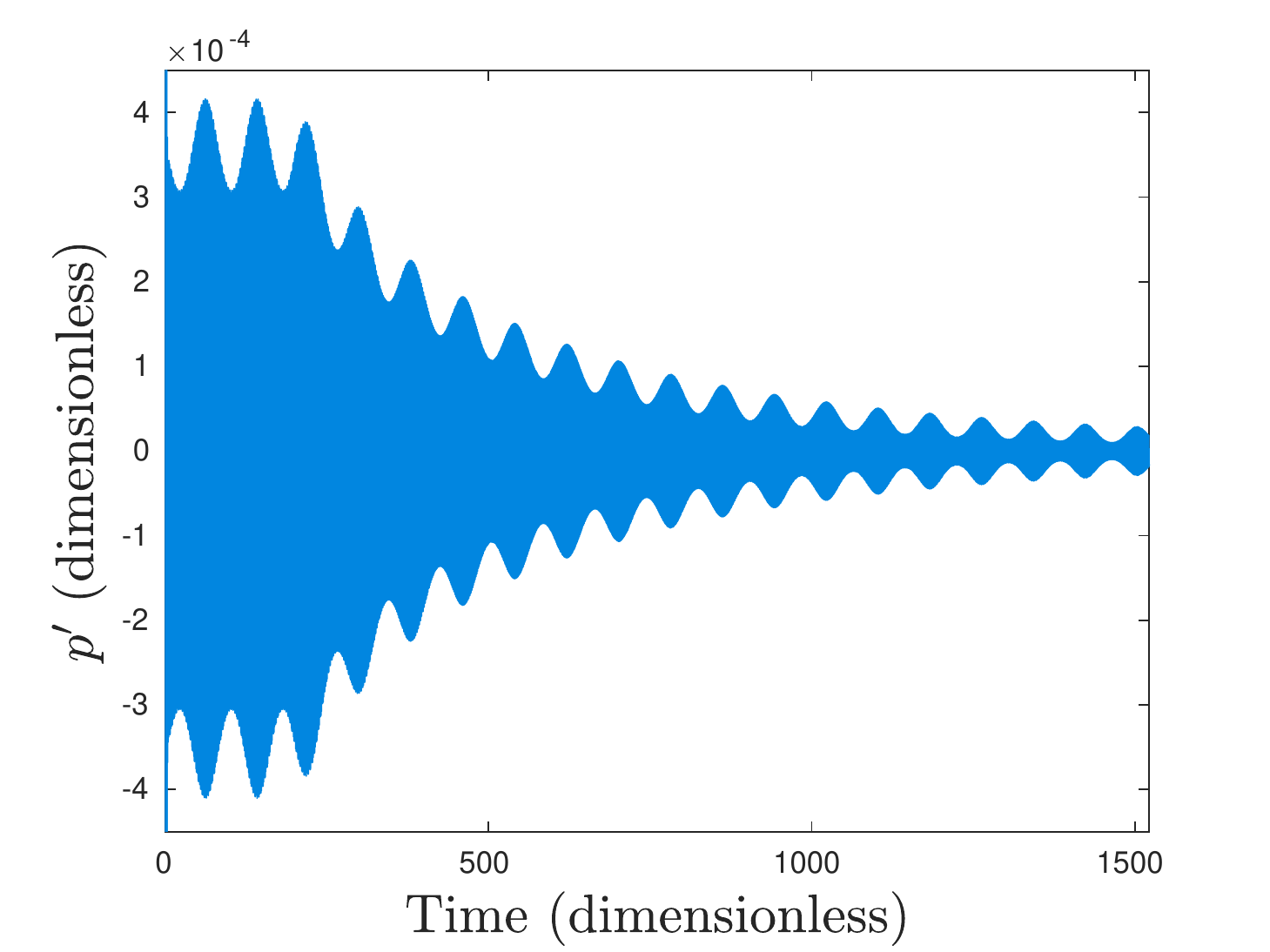}
\caption{Case \cs{24}. Application of the ESC to find an optimal actuation intensity at the trailing edge. In this case, the magnitude of $A_c$ tends to grow indefinitely, increasing the control effort and reducing the noise emission.}
\label{f:e024}
\end{figure}

The parameter $\delta_{ss}$ (see Sec. \ref{sec:ssc}) is used to introduce slope seeking compensation since the cost function for the present actuation setups presents a \textit{plateau}.
Figure \ref{f:e024} shows results obtained for the ESC implementation, i.e., $\delta_{ss} = 0$. In this case, the control system searches for an optimal actuation intensity and blowing is automatically chosen instead of suction, since displacing the vortex shedding away from the trailing edge reduces noise scattering. As can be seen from the figure, the jet magnitude keeps increasing in time and leads to more power consumption from the control. On the other hand, a higher blowing intensity reduces the acoustic pressure computed by the sensor. 

\begin{figure}
\centering
\includegraphics[width=.32\textwidth]{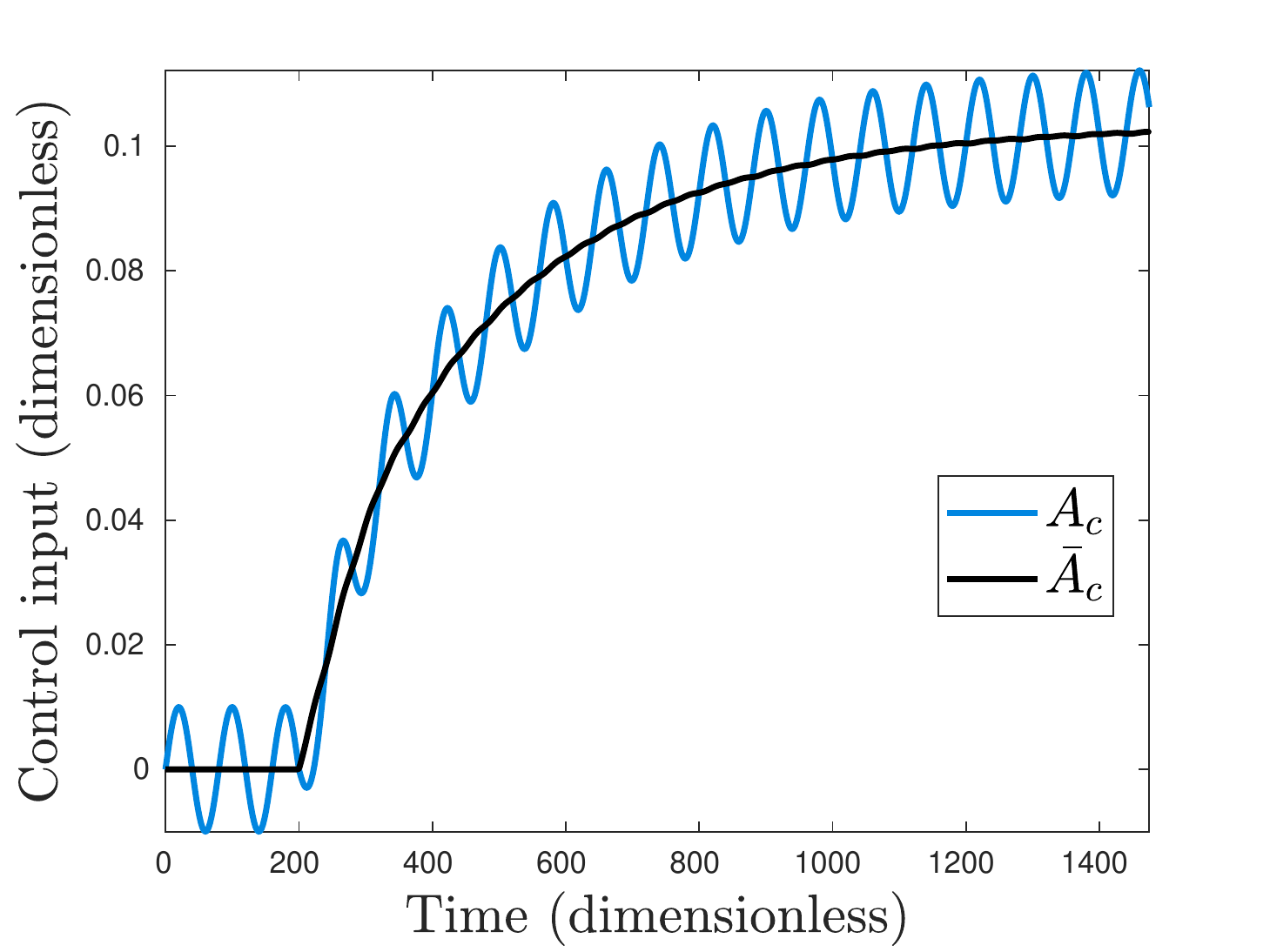}
\includegraphics[width=.32\textwidth]{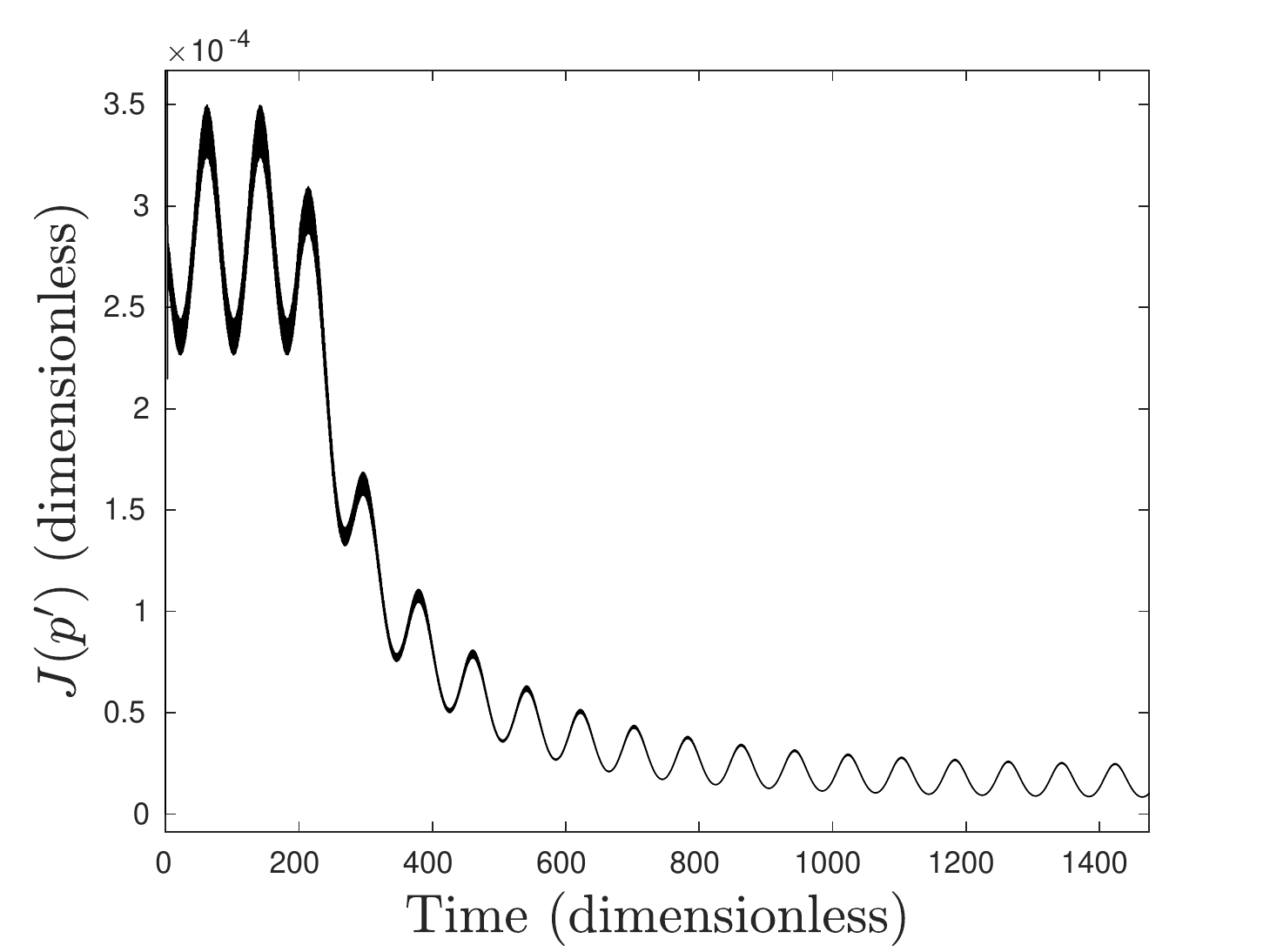}
\includegraphics[width=.32\textwidth]{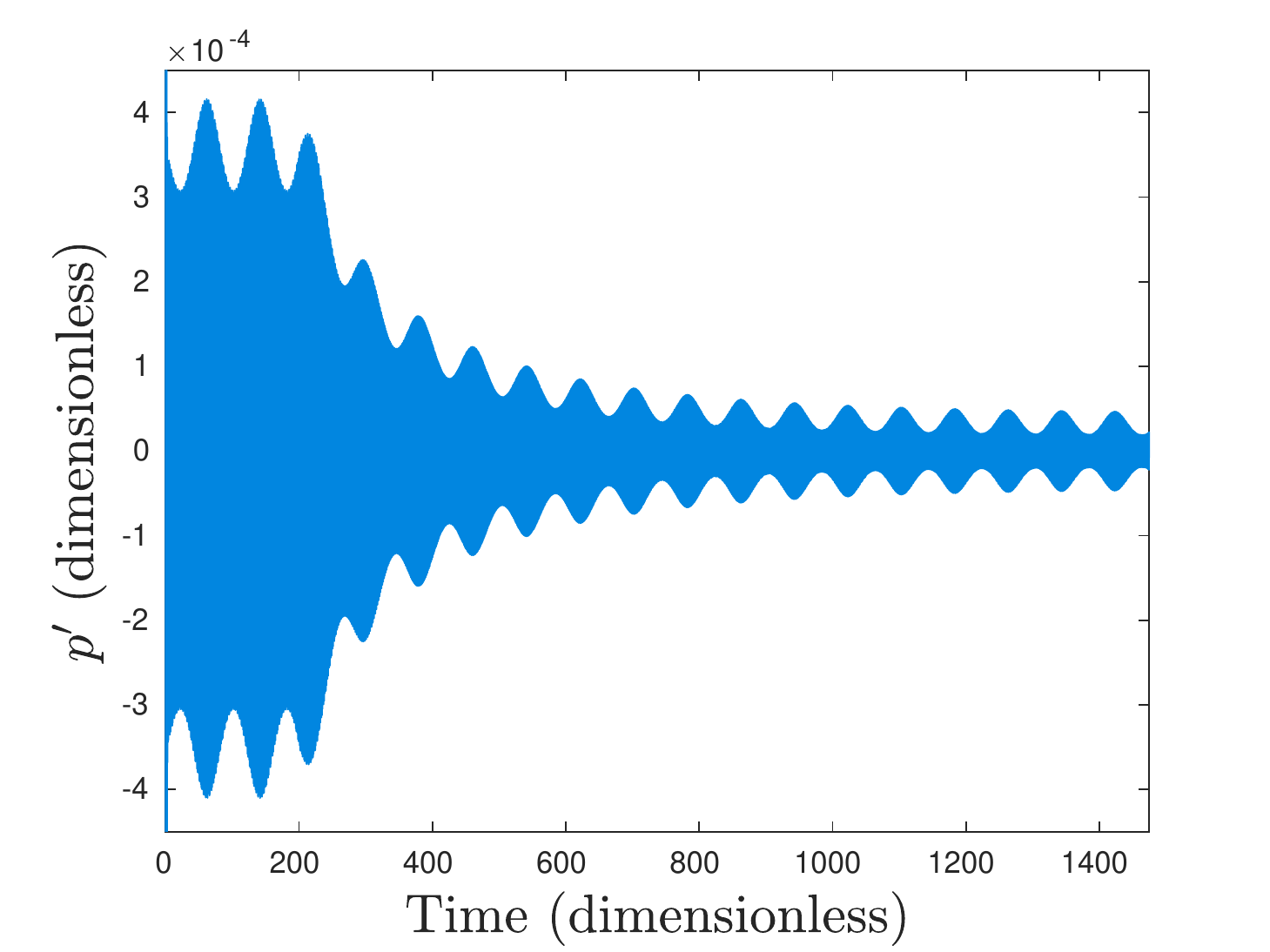}
\includegraphics[width=.32\textwidth]{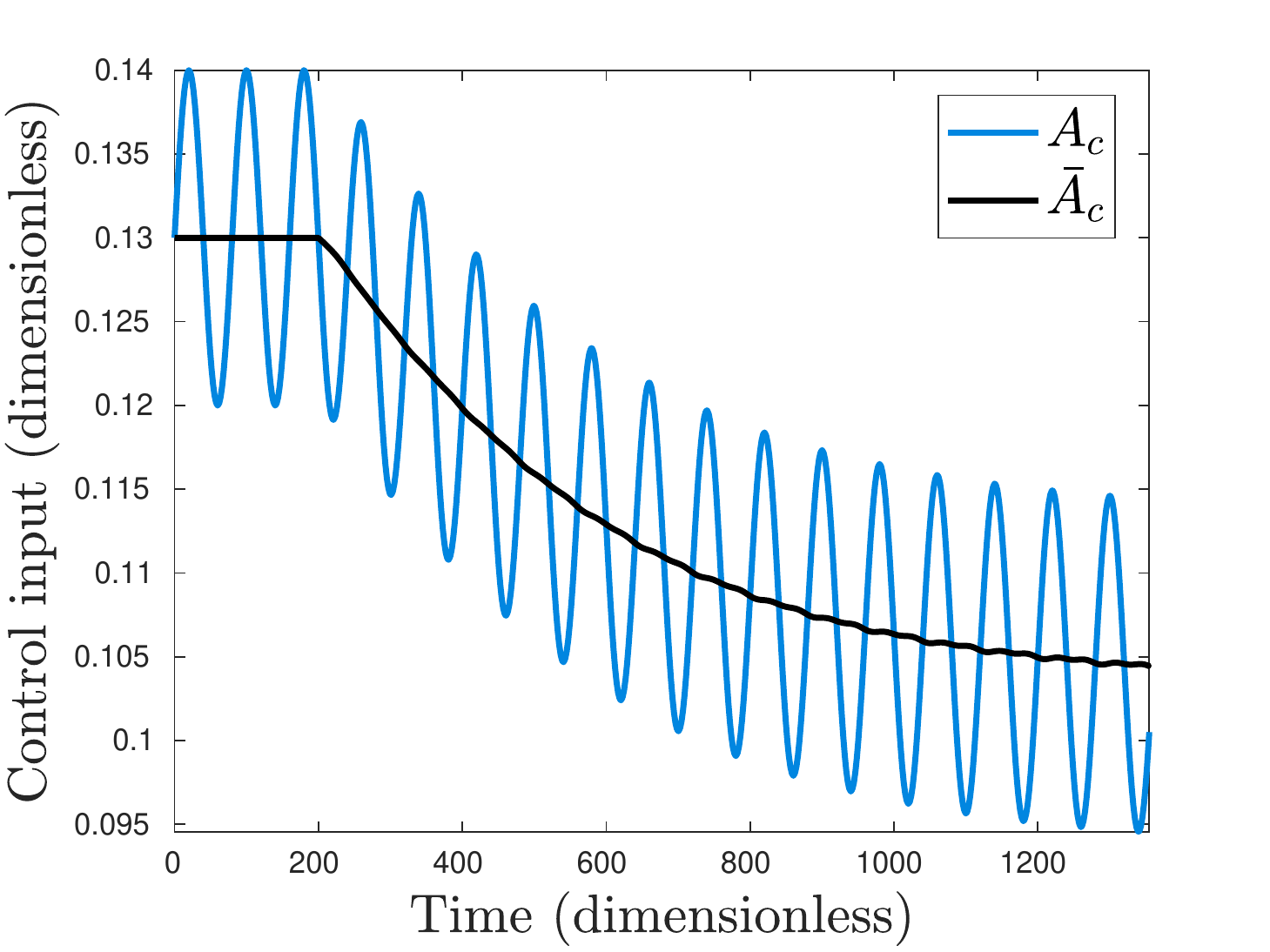}
\includegraphics[width=.32\textwidth]{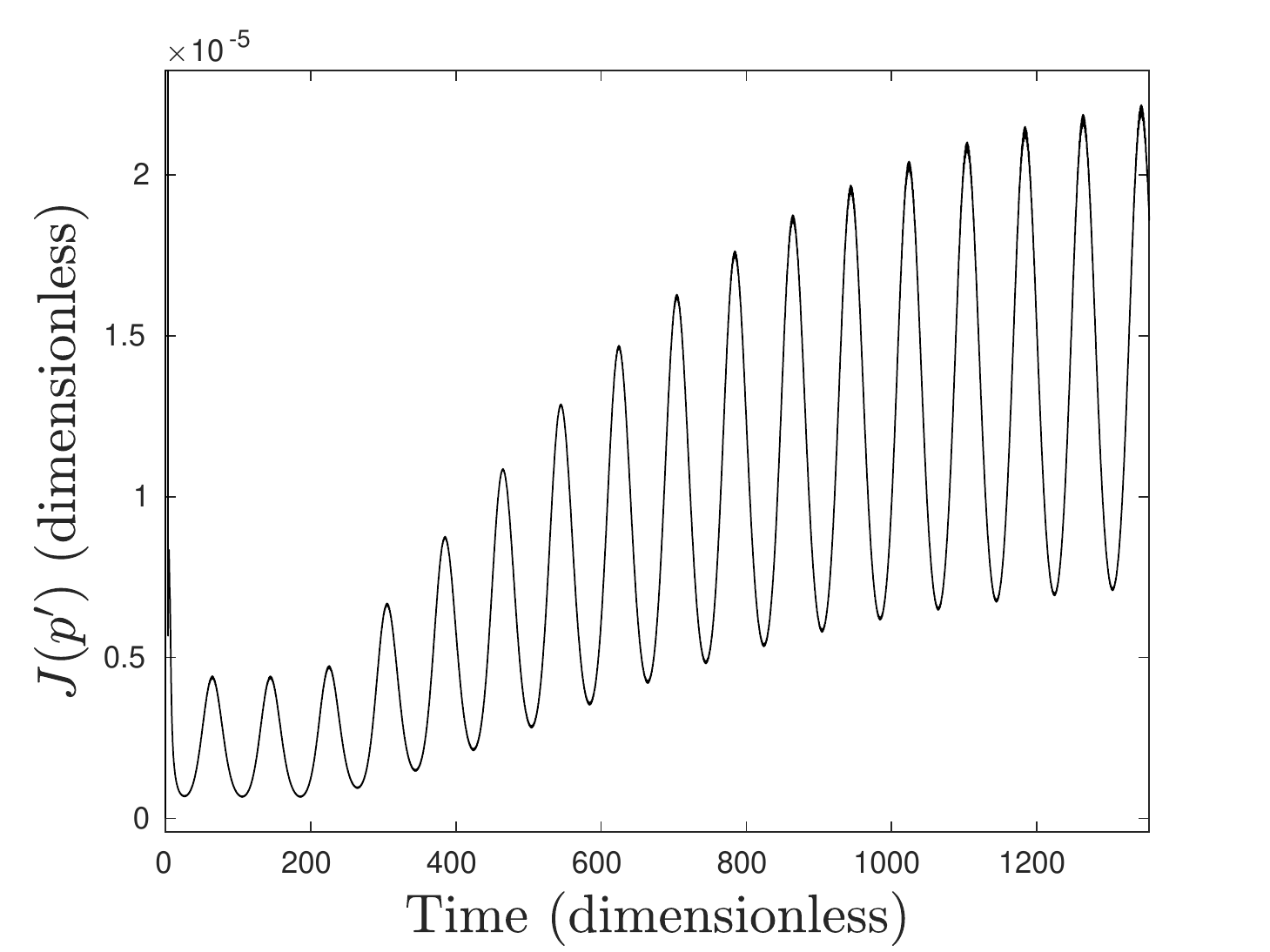}
\includegraphics[width=.32\textwidth]{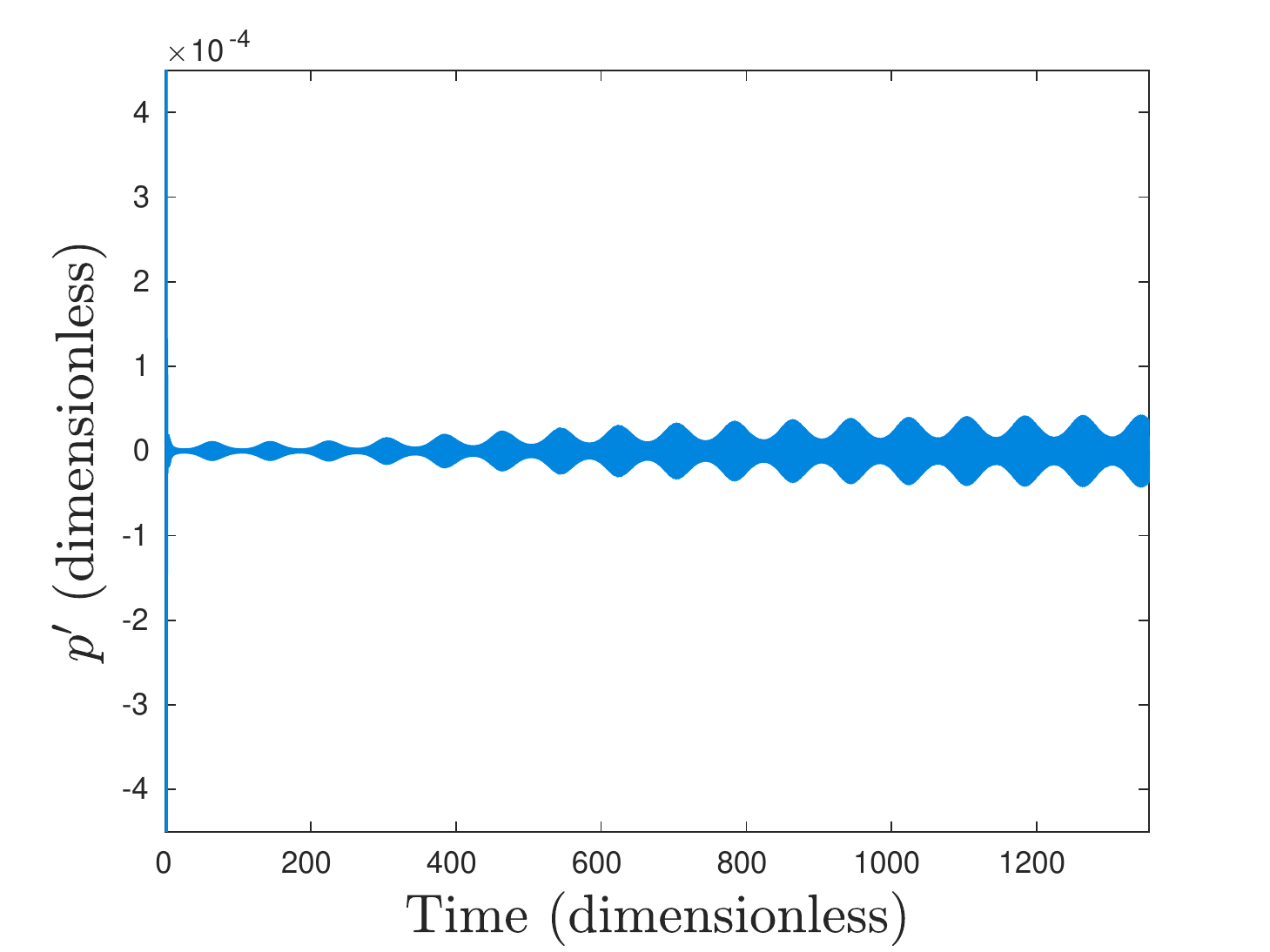}
\caption{Cases \cs{25} and \cs{26}. Application of slope seeking control to find the actuation intensity at the trailing edge. Two initial jet amplitudes are tested: the first is null (top row) and the second considers a high amplitude initial blowing (bottom row). Both cases converge to the same control actuation regardless of the initial conditions.}
\label{f:e025}\label{f:e026}
\end{figure}

\begin{figure}
\centering
\includegraphics[width=.32\textwidth]{figures/acc_vort/10k/0000.png}
\includegraphics[width=.32\textwidth]{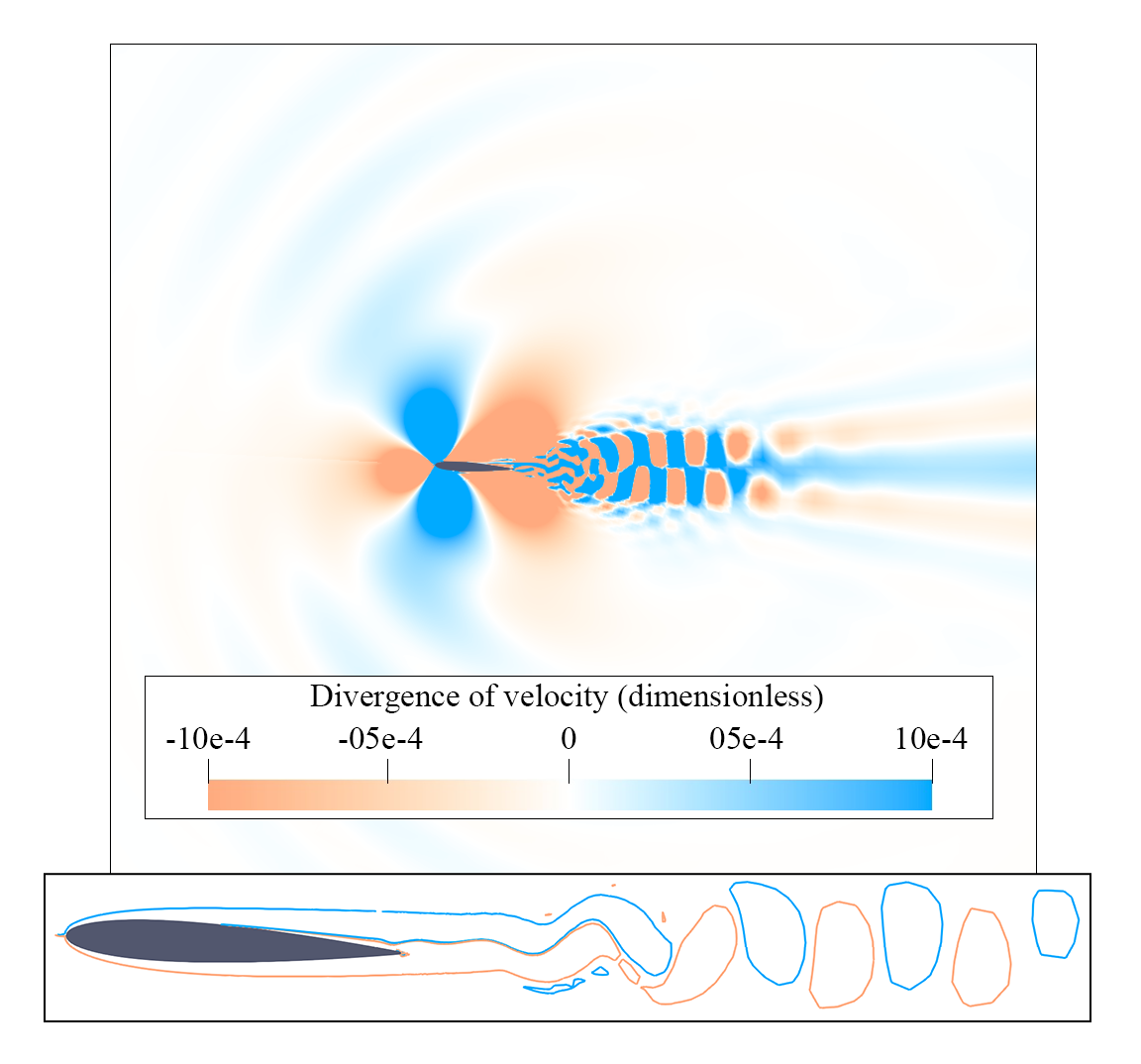}
\caption{Comparison between passive (left) and controlled (right) flows. In the controlled simulation (case \cs{25}), the vortex street forms further downstream diminishing the acoustic scattering and, hence, noise radiation.}
\label{f:10kwalter}
\end{figure}

To avoid the permanent increase in the control effort, slope seeking compensation with $\delta _{ss} = 3.64e-08$ is applied. Figure \ref{f:e025} shows the results for two cases: in the first, the initial jet intensity $A_{c0}$ is null while, in the second, it has a high initial blowing amplitude. With slope seeking control, the loop is able to find an equilibrium point that can be reached independently of the initial guess $A_{c0}$. This makes the system more robust to drifting driven by noise measurements. One can see in the figure that the same actuation intensity $A_c$ is found for both cases and that the magnitude of the pressure fluctuations probed by the sensor are the same. A comparison between the passive and the controlled flows is displayed in Fig. \ref{f:10kwalter}. As expected, the vortex street is displaced further downstream from the airfoil surface in the controlled case, reducing the noise emission.

\subsection{Reynolds number 100,000}

\begin{figure}
\centering
\includegraphics[width=.47\textwidth]{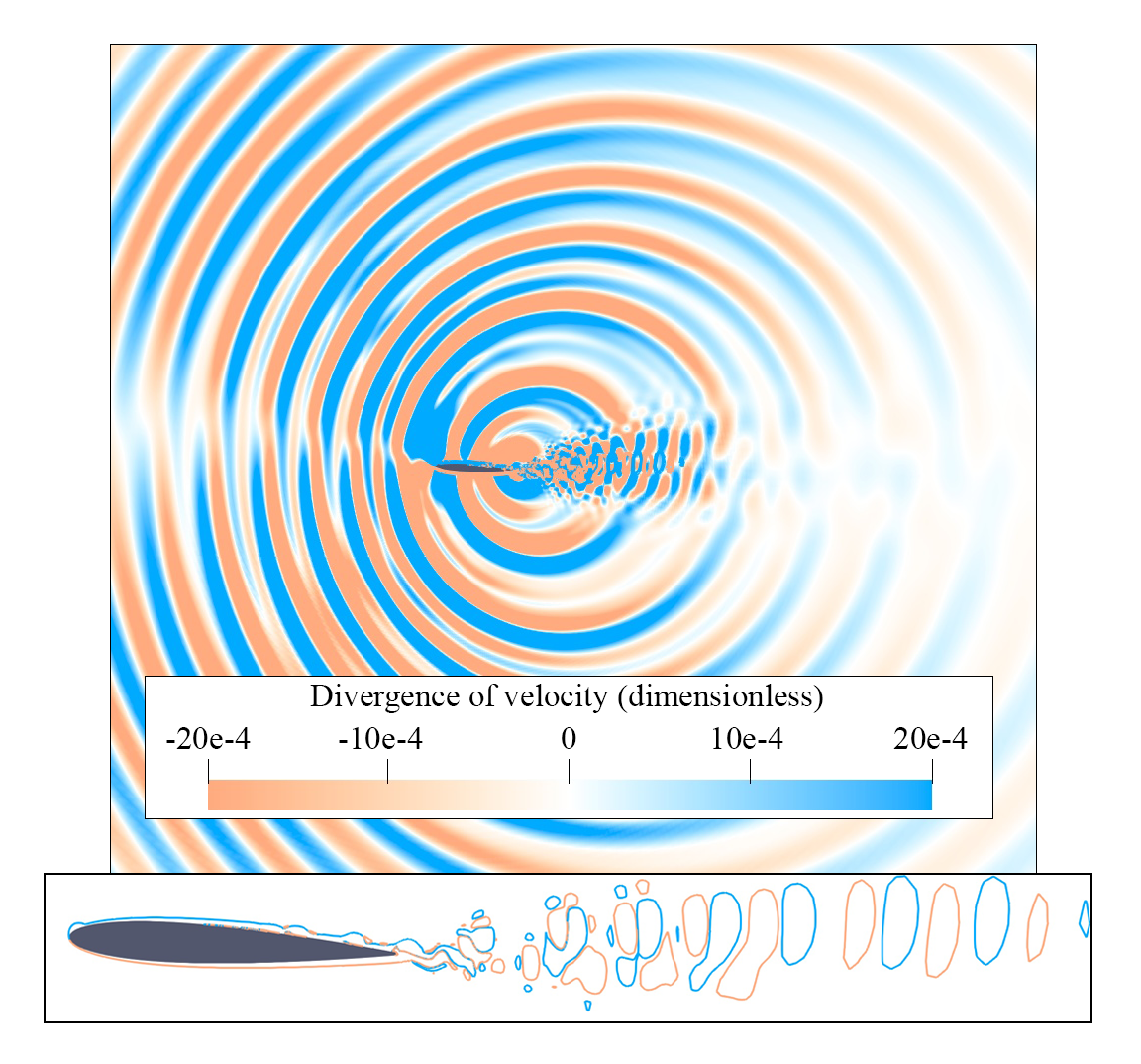}
\includegraphics[width=.45\textwidth]{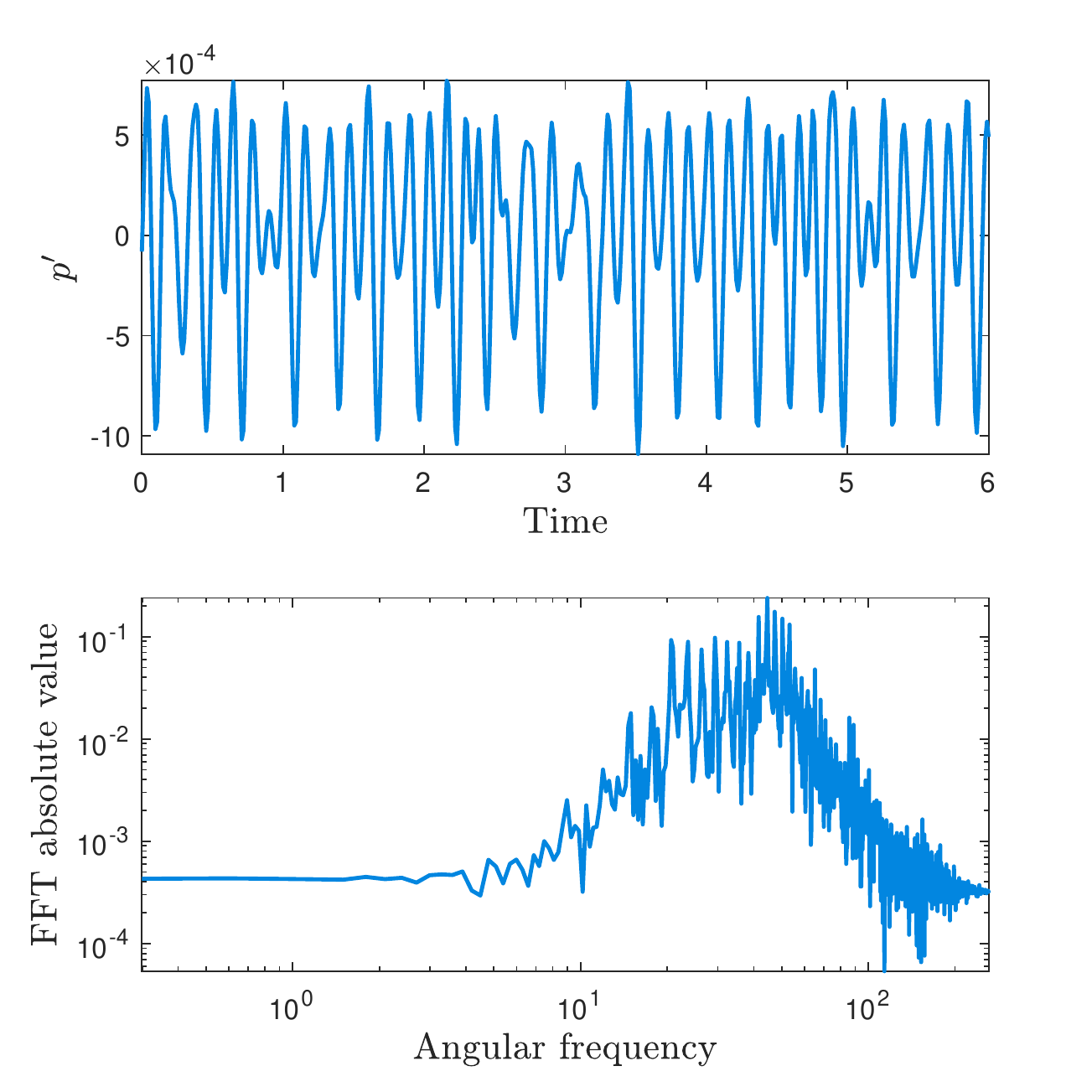}
\caption{Contours of divergence of velocity and isolines of z-vorticity without actuation for the NACA0012 airfoil at $\mathrm{Re} = 10^5$, $\mathrm{Ma} = 0.3$ and $3^\circ$ angle of attack (left). Pressure signal at the sensor location and its Fourier transform (right).}
\label{f:100k0}
\end{figure}

Flow instabilities develop over the airfoil suction side at $\text{Re}=10^5$ as discussed by \citet{ricciardi2020secondary}. These flow structures can be seen in \green{Fig. \ref{f:100k0}} that presents contours of divergence of velocity with a detail view of the z-vorticity. As shown by the previous authors, a thin separation bubble over the airfoil suction side promotes a frequency modulation of the boundary layer flow instabilities which, in turn, lead to the presence of multiple tones superimposed on a broadband noise signal. \green{Figure \ref{f:100k0}} shows both the temporal signal computed by the sensor and its Fourier transform. As can be seen from the figure, the simulation for this plant results in a non-periodic acoustic pressure signal composed of several tonal frequencies.

For this Reynolds number, two different control approaches are tested. First, a moving actuator is placed at the suction side to find an optimal position with the ESC loop. It will be shown that this control setup is able to lock the flow trajectory into a periodic one, where the flow becomes more organized with vortex shedding. A second control setup is tested where an actuator is placed at the trailing edge. This approach is similar to that presented in Sec. \ref{sec:10ktred}; however, the suction side actuator is kept turned on at the optimal position found from the moving actuation setup. In all of these cases, the adjustable actuator is turned on at $t=50$ instead of $t=0$. A window length $m=40$ is employed in all simulations with $\mathrm{Re}=10^5$.

\subsubsection{Chordwise moving actuator with constant intensity} \label{sec:susi}

Two simulations are presented for a control setup consisting of an actuator placed at the airfoil suction side. Similarly to the simulations presented in Sec. \ref{sec:moving10}, the ESC loop seeks the optimal position to reduce noise. The studies conducted with this approach using the parameters indicated in Table \ref{tab:cases100kpos} suggest that there is a region of actuation between $x=0.35$ and $x=0.80$ at which convergence is achievable. For the present suction actuator intensity $A=-6.0\%$, the ESC could not find an optimal region if the actuator position was placed outside this interval. %Outside this region, since the flow is very unorganized, the controller moved the actuator randomly, sometimes trying to find a position outside of the wall limits. 
However, we observed that through an increase in the suction intensity, the ESC was able to converge for a wider region along the airfoil chord.
\begin{table}
  \begin{center}
\def~{\hphantom{0}}
  \begin{tabular}{lcccccccc}
     \hline
     Case     & $A$    & $x_{c0}$ & $t_{c0}$ & $\alpha=\beta$ & $m$ & $2\pi/\omega$ & $\eta$ & Results\\\hline
     \cs{105} & -6.0\% &    0.80  & 130.0    & 0.01            & 40  & 5.00e+01      & -1000  & Fig. \ref{f:e105}\\
     \cs{106} & -6.0\% &    0.35  & 130.0    & 0.01            & 40  & 5.00e+01      & -1000  & Fig. \ref{f:e106}\\\hline
  \end{tabular}
  \caption{Control parameters for cases with moving actuation at $\text{Re} = 10^5$. %The jet amplitude $A$ is given as a percent of the freestream momentum (which is equal, in value, to the Mach number $\text{Ma} = 0.3$, since $\rho_\infty=1$).and the negative sign denotes suction. The control integrator turn on time $t_{c0}$ and the ESC wave period $2\pi/\omega$ are dimensionless temporal parameters relative to the freestream velocity and airfoil chord. The initial actuation position $x_{c0}$ consists of the horizontal coordinate relative to the chord length, being zero at the leading edge and one at trailing edge. The ESC parameter $\alpha$ is also provided relative to the chord.
  }
  \label{tab:cases100kpos}
  \end{center}
\end{table}

\begin{figure}
\centering
\includegraphics[width=.32\textwidth]{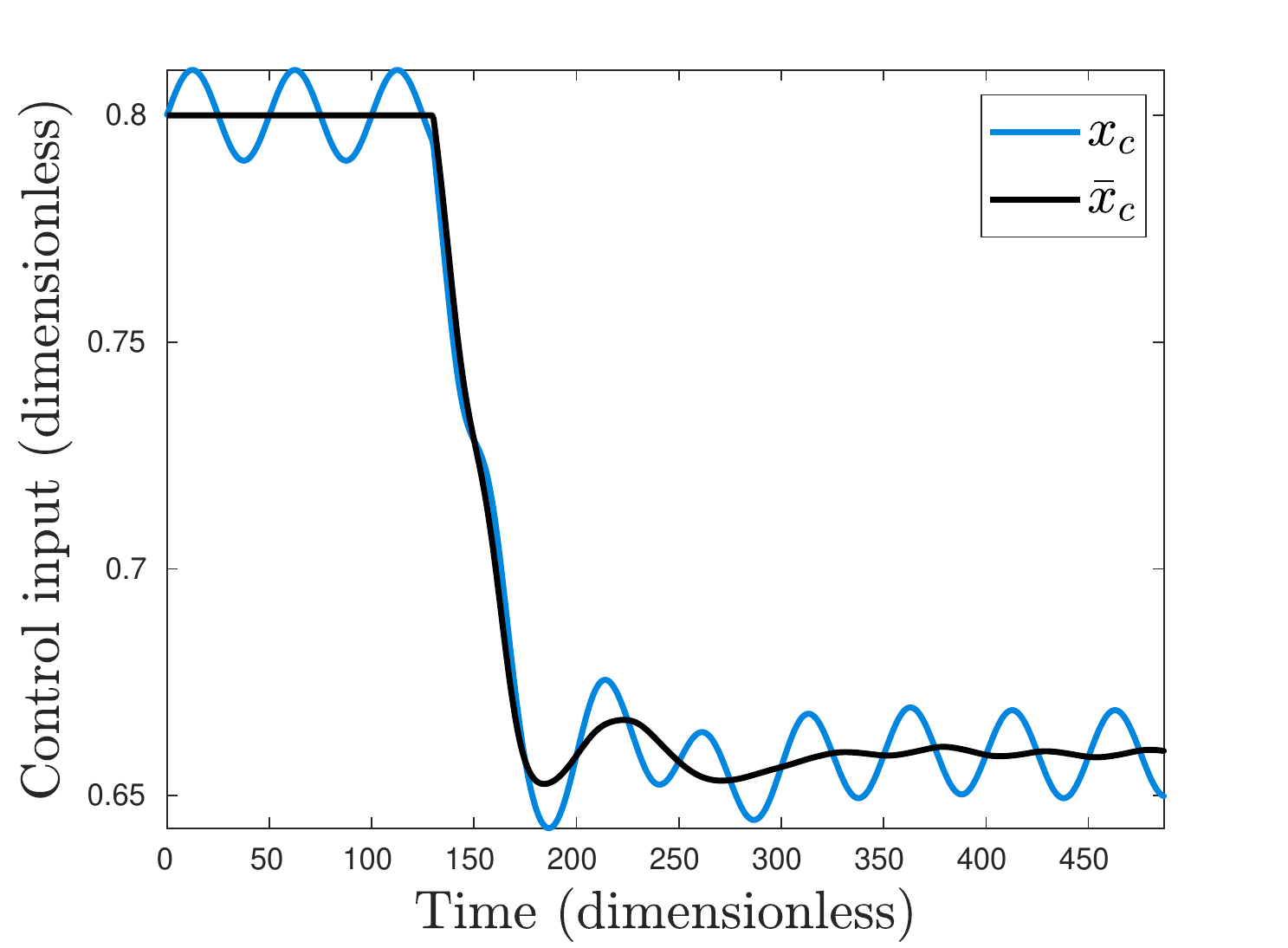}
\includegraphics[width=.32\textwidth]{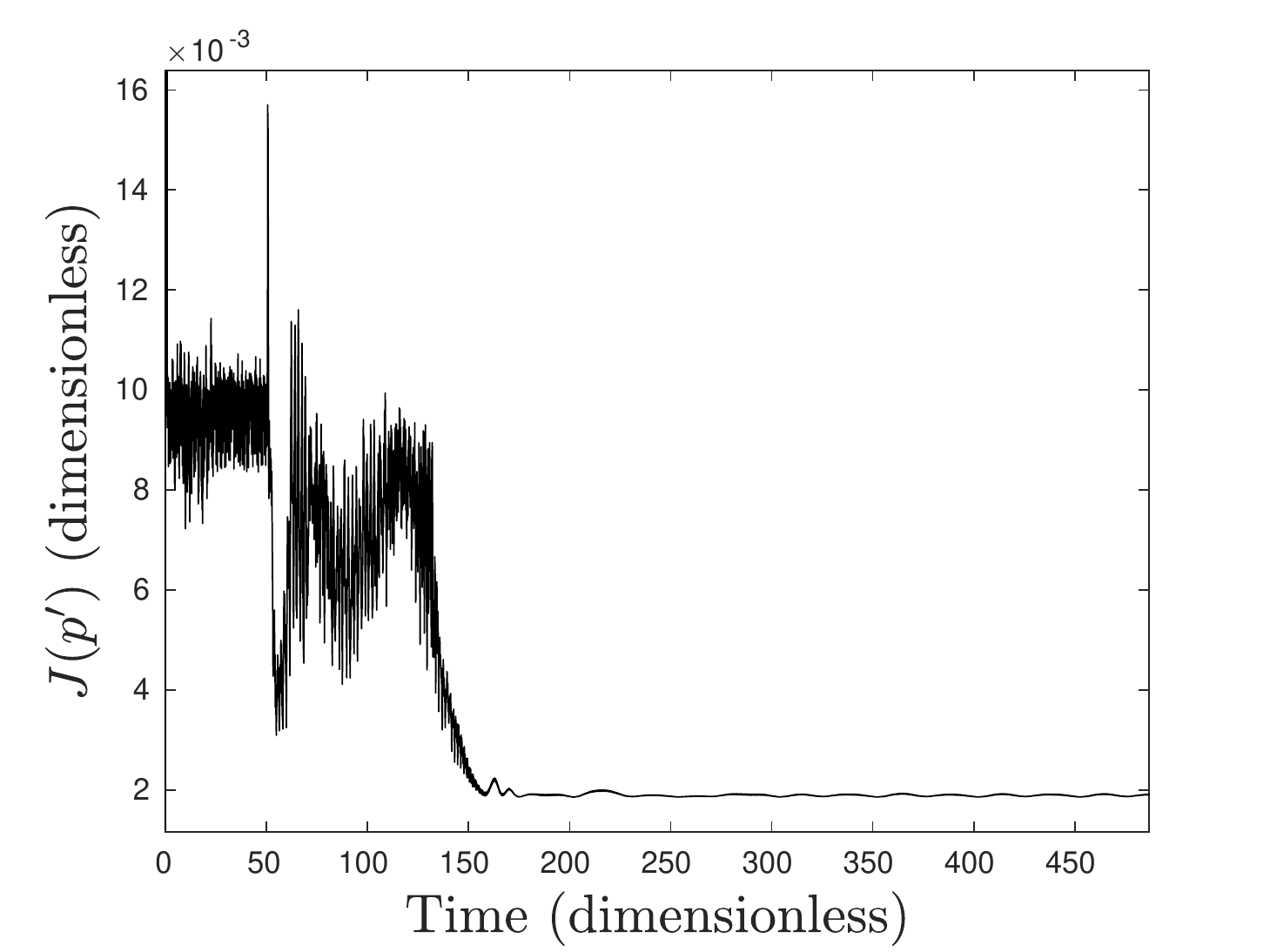}
\includegraphics[width=.32\textwidth]{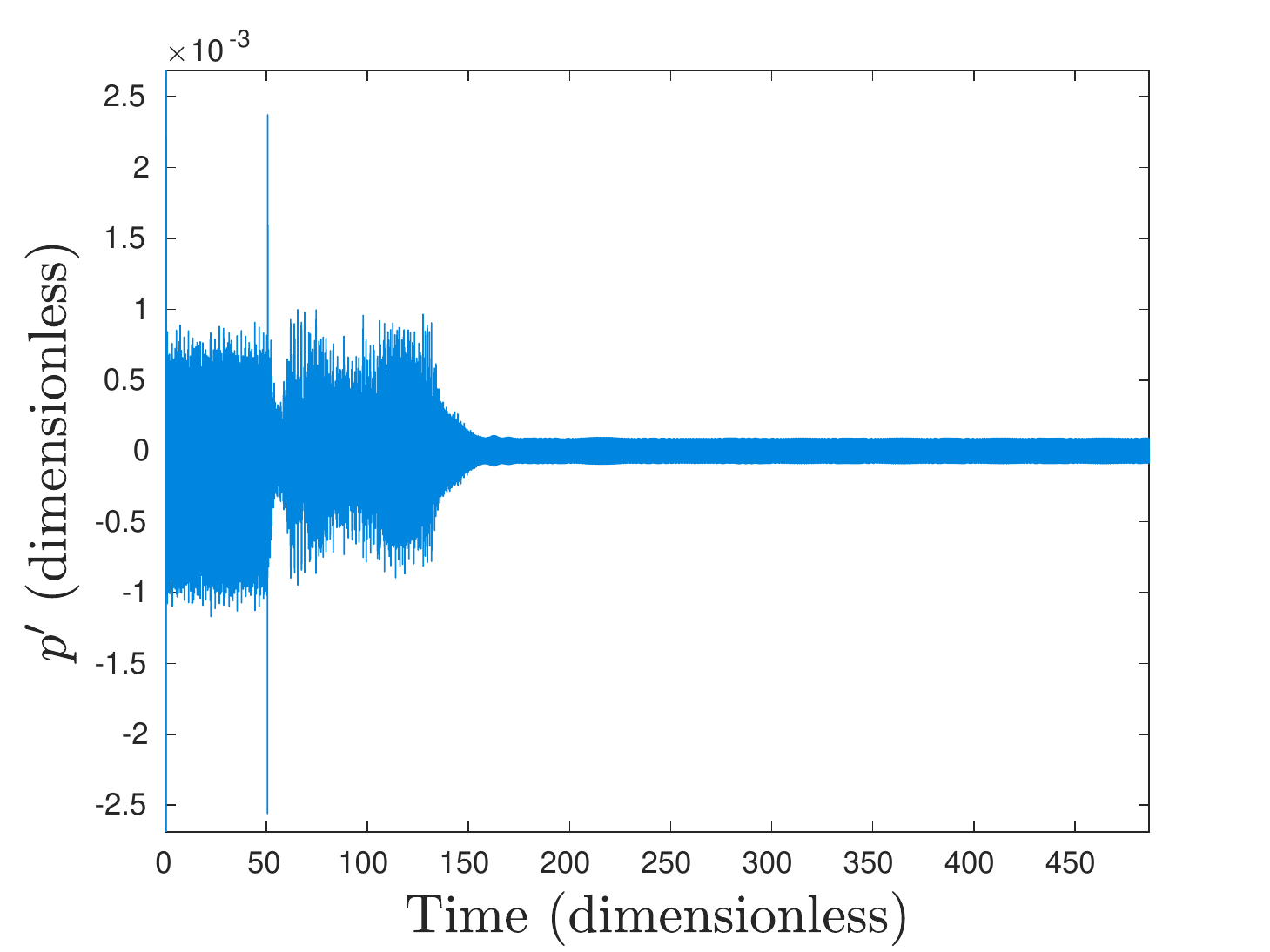}
\includegraphics[width=.32\textwidth]{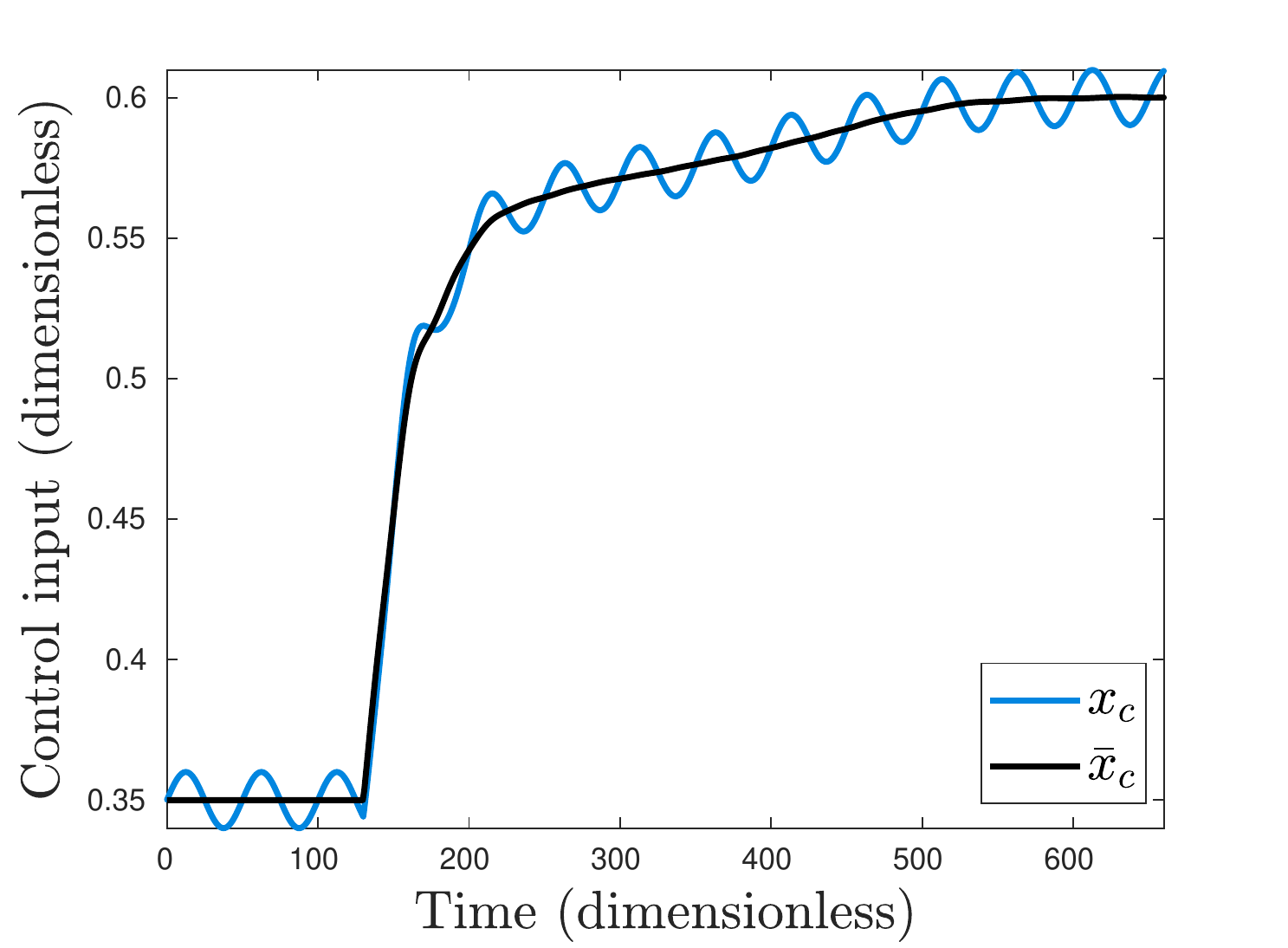}
\includegraphics[width=.32\textwidth]{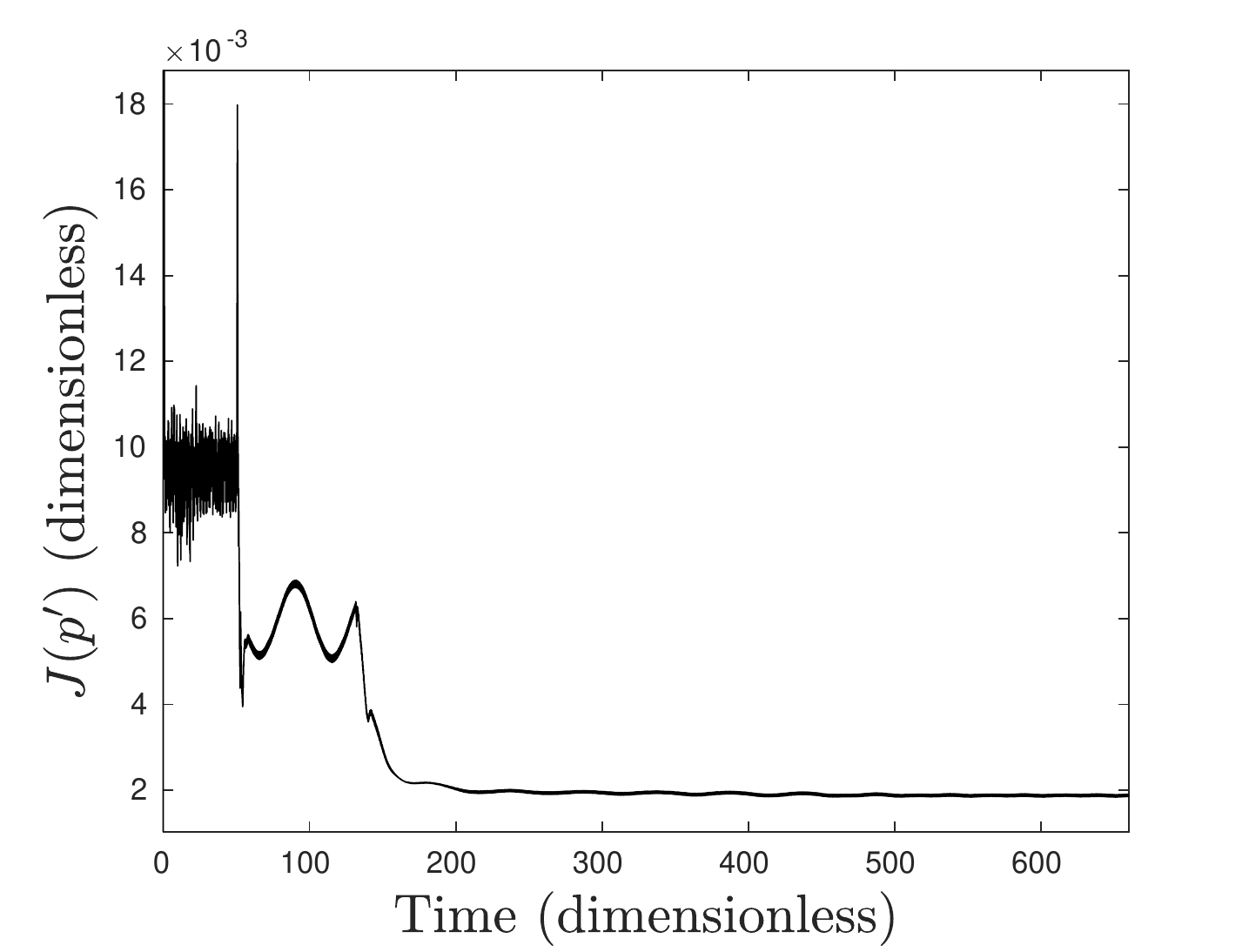}
\includegraphics[width=.32\textwidth]{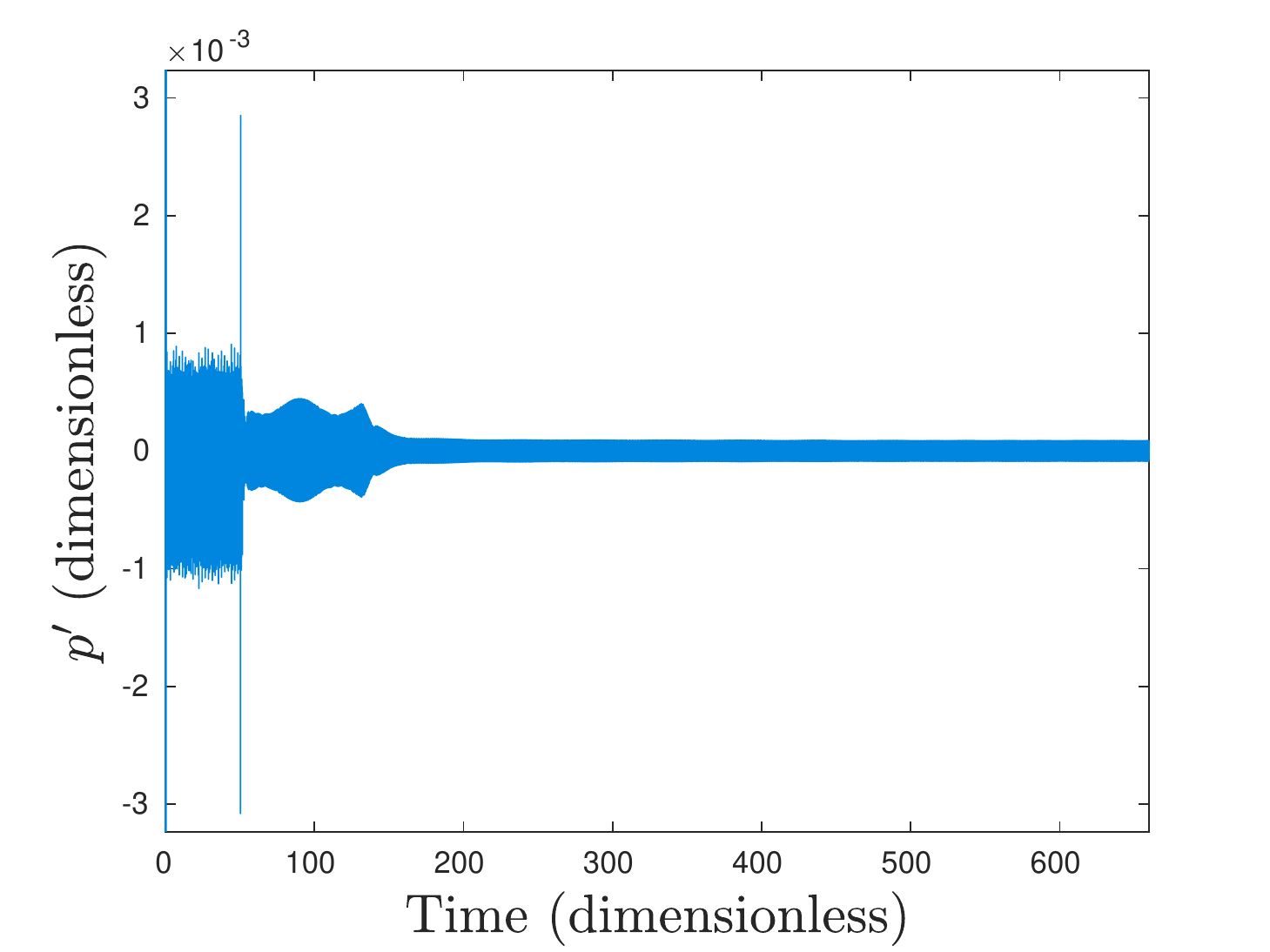}
\caption{Cases \cs{105} and \cs{106}. The ESC loop is able to find an optimal region of actuation if the initial position of the actuator with $A=-6.0\%$ is placed at $0.35 \leq x_c \leq 0.80$. It was observed that an increase in the suction actuation intensity $A$ widens this optimal region.}
\label{f:e105}\label{f:e106}
\end{figure}

Figure \ref{f:e105} presents the results for simulations \cs{105} and \cs{106}. The ESC parameters are identical for both cases, except for the initial actuator position. In the first case, the actuator starts near the trailing edge while, for the second one, it is placed closer to the leading edge. As can be observed from the plots of $x_c$ in the left column, the controller moves the actuators to an optimal region between $0.55\lesssim x_c \lesssim 0.70$. At these positions, the cost function reaches a plateau after significantly attenuating noise. With the actuator position converged, the noise generation mechanism becomes similar to that from a lower Reynolds airfoil and a single tone is observed in the noise spectrum. In this case, the boundary layer flow instabilities are suppressed and the flow unsteadiness comes exclusively from a von Kármán vortex street. %At a certain interval of $0.68 \lesssim x_c \lesssim 0.8$, for case \cs{105}, and $0.35 \lesssim x_c \lesssim 0.55$ for case \cs{106}, the cost function drops significantly. After that, a slight movement of the actuator occurs until convergence and it contributes on reducing the wave amplitude by further 4\%.

\subsubsection{Actuation with varying intensity and fixed position at the trailing edge} \label{sec:100ktred}

In the previous setup, the ESC was able to find an optimal actuation that suppressed the formation of boundary layer instabilities on the airfoil suction side. In that case, vortex shedding became the sole mechanism for noise generation, similarly to the low Reynolds number flows. Hence, in the present setup, the actuator is placed at the trailing edge in a similar fashion to the approach presented in Sec. \ref{sec:10ktred}. However, an open-loop actuator with $A=-6.0\%$ is kept fixed at the suction side at $x=0.66$ to eliminate the boundary layer instabilities as observed in case \cs{105}.

\begin{figure}
\centering
\includegraphics[width=.32\textwidth]{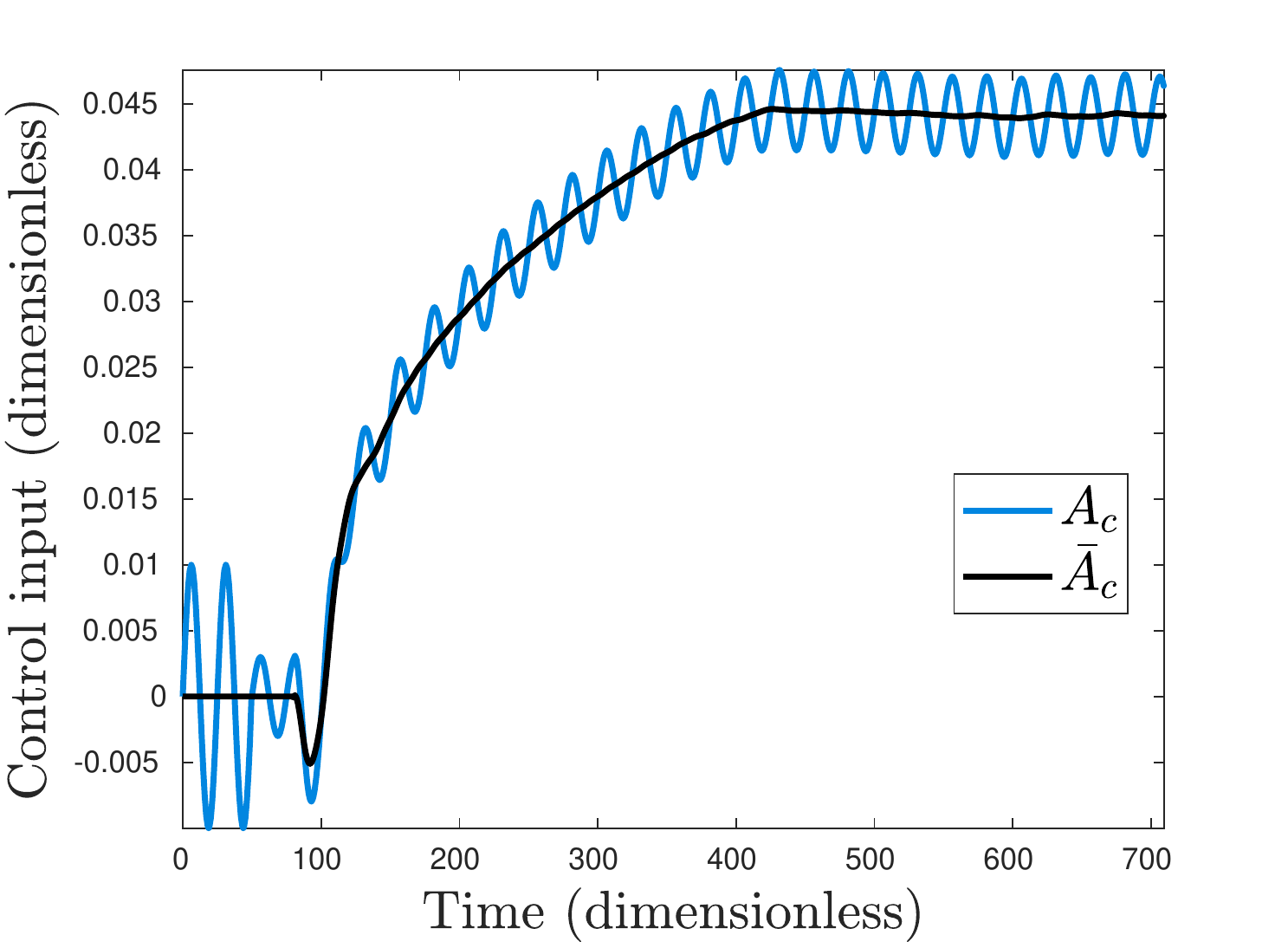}
\includegraphics[width=.32\textwidth]{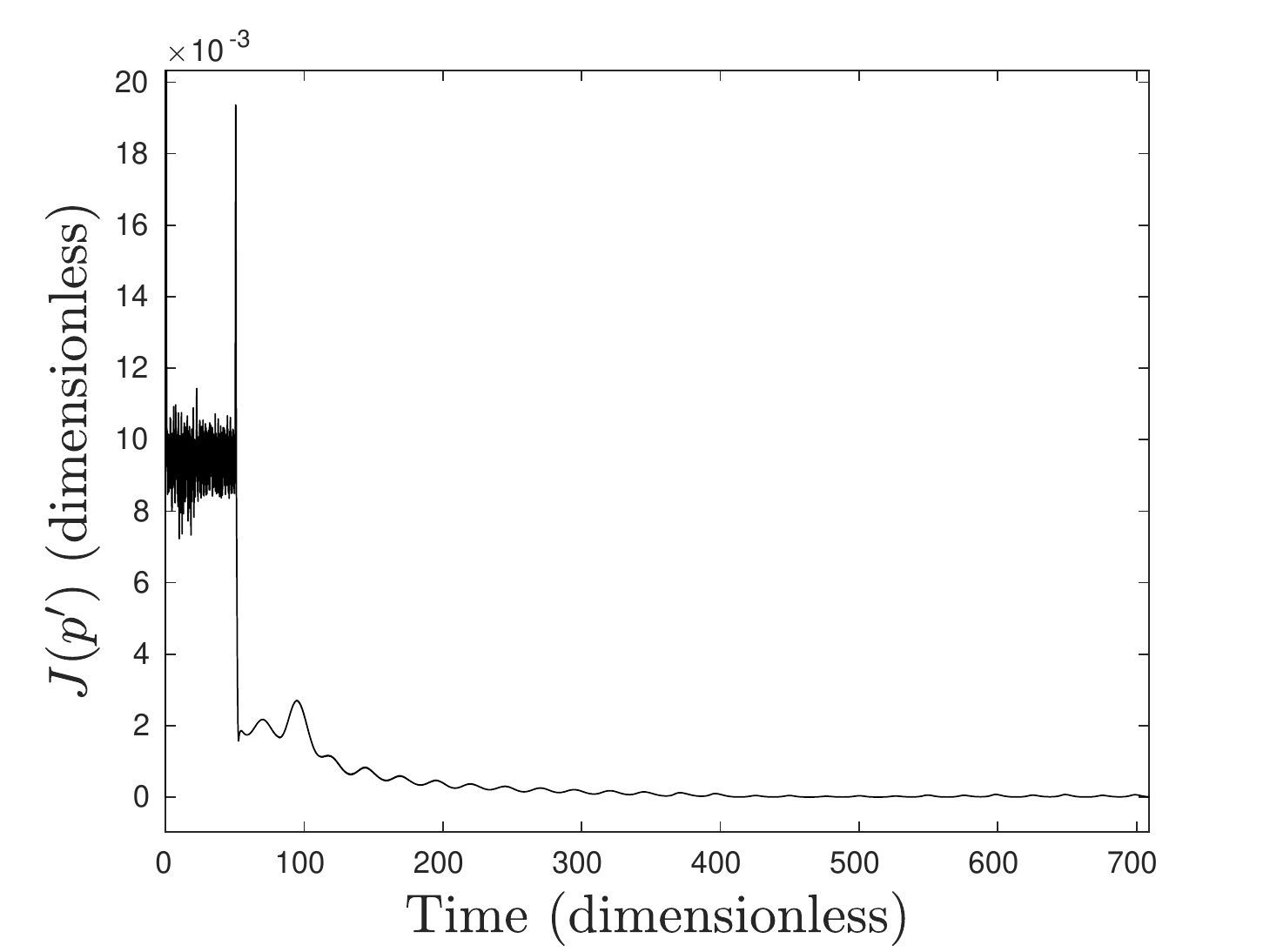}
\includegraphics[width=.32\textwidth]{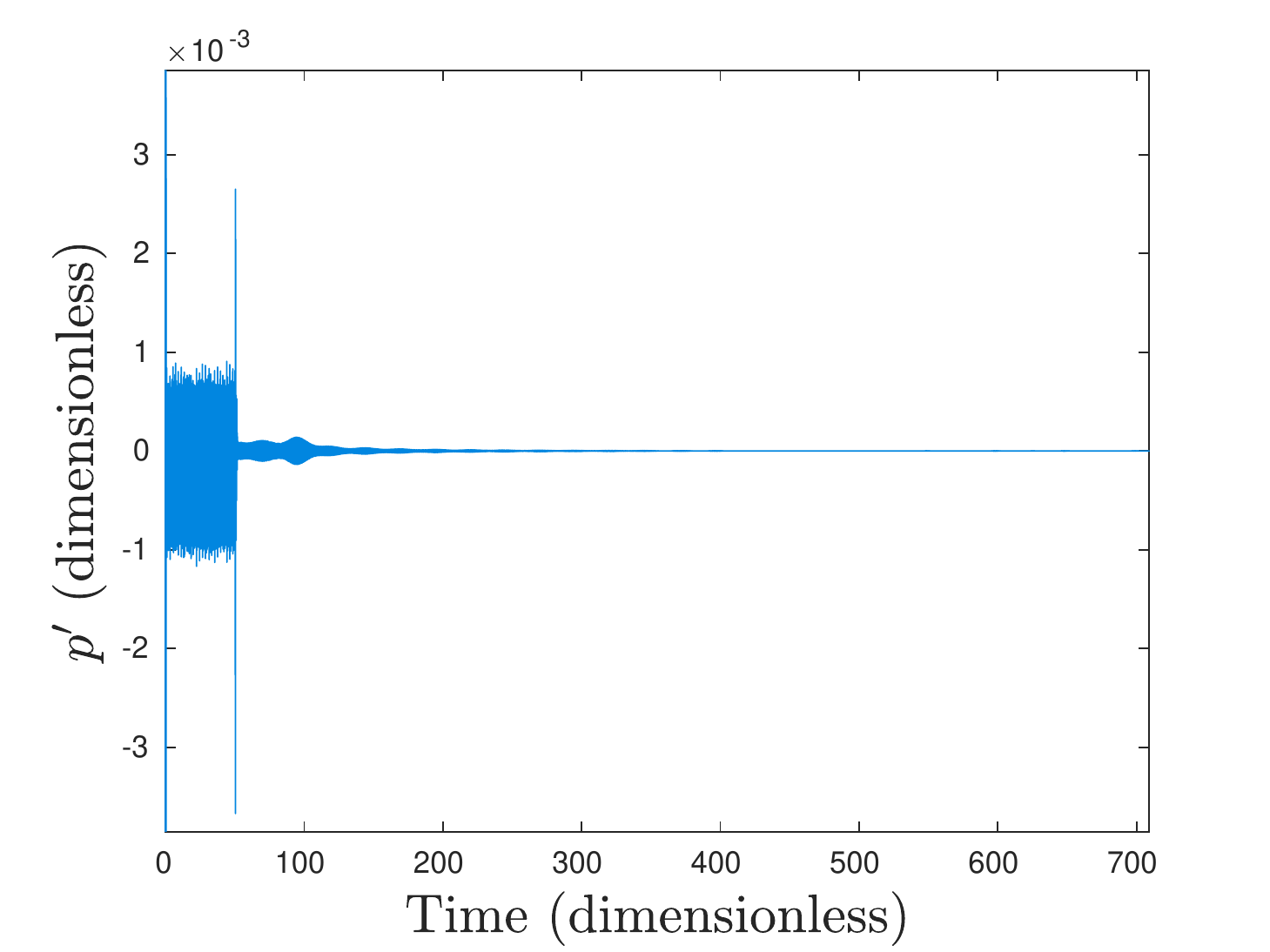}
\caption{Case \cs{107}. This case shows that it is possible to find a trailing edge blowing intensity $A_c$ to reduce noise in the ESC loop by also keeping an actuator on the suction side in the position obtained in case \cs{105}.%  After a critical jet amplitude, noise is suppressed as the flow becomes nearly steady.\red{[este ponto crítico não é alcançado aqui]}
}
\label{f:e107}
\end{figure}

\begin{figure}
\centering
\includegraphics[width=.32\textwidth]{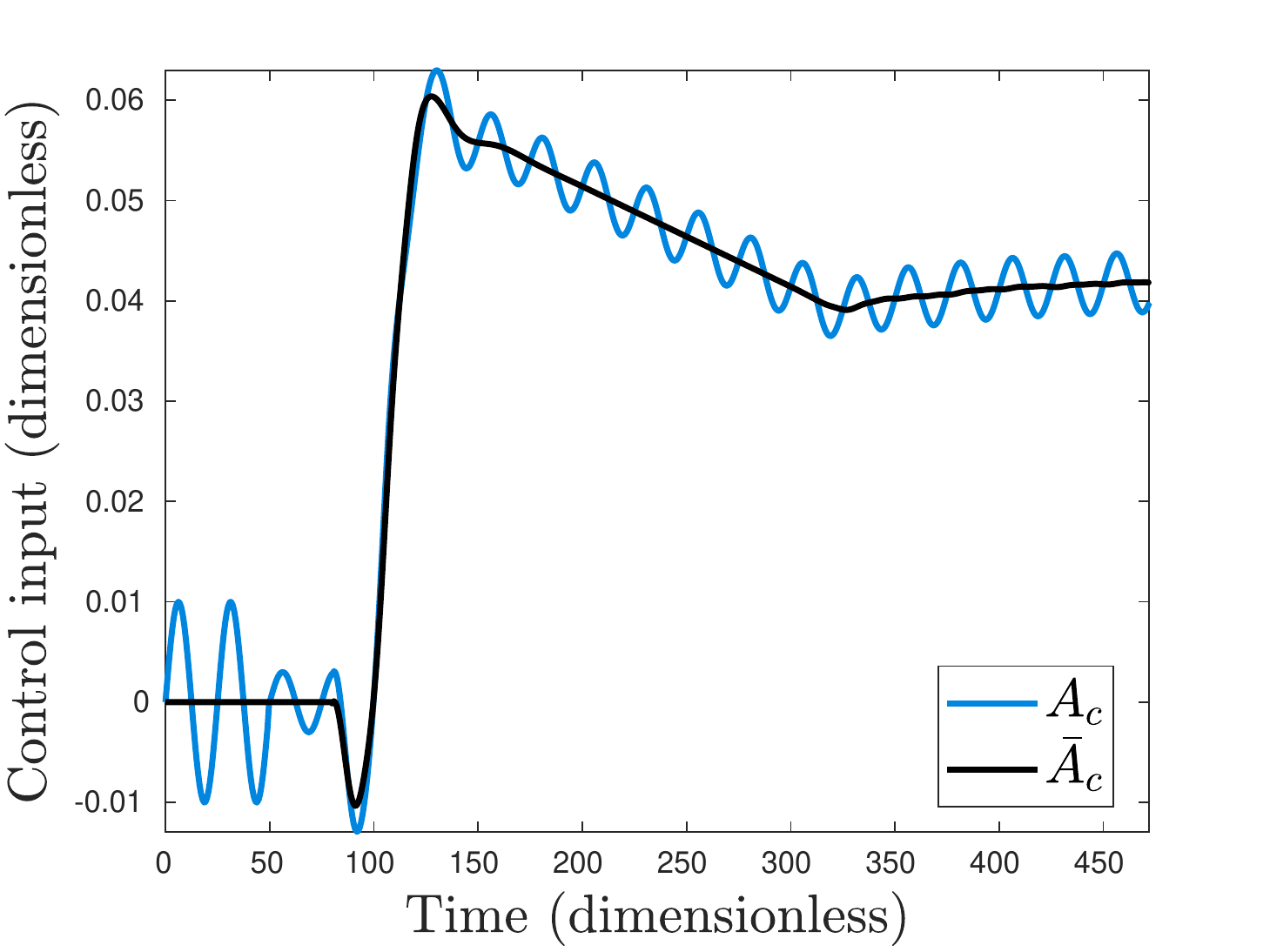}
\includegraphics[width=.32\textwidth]{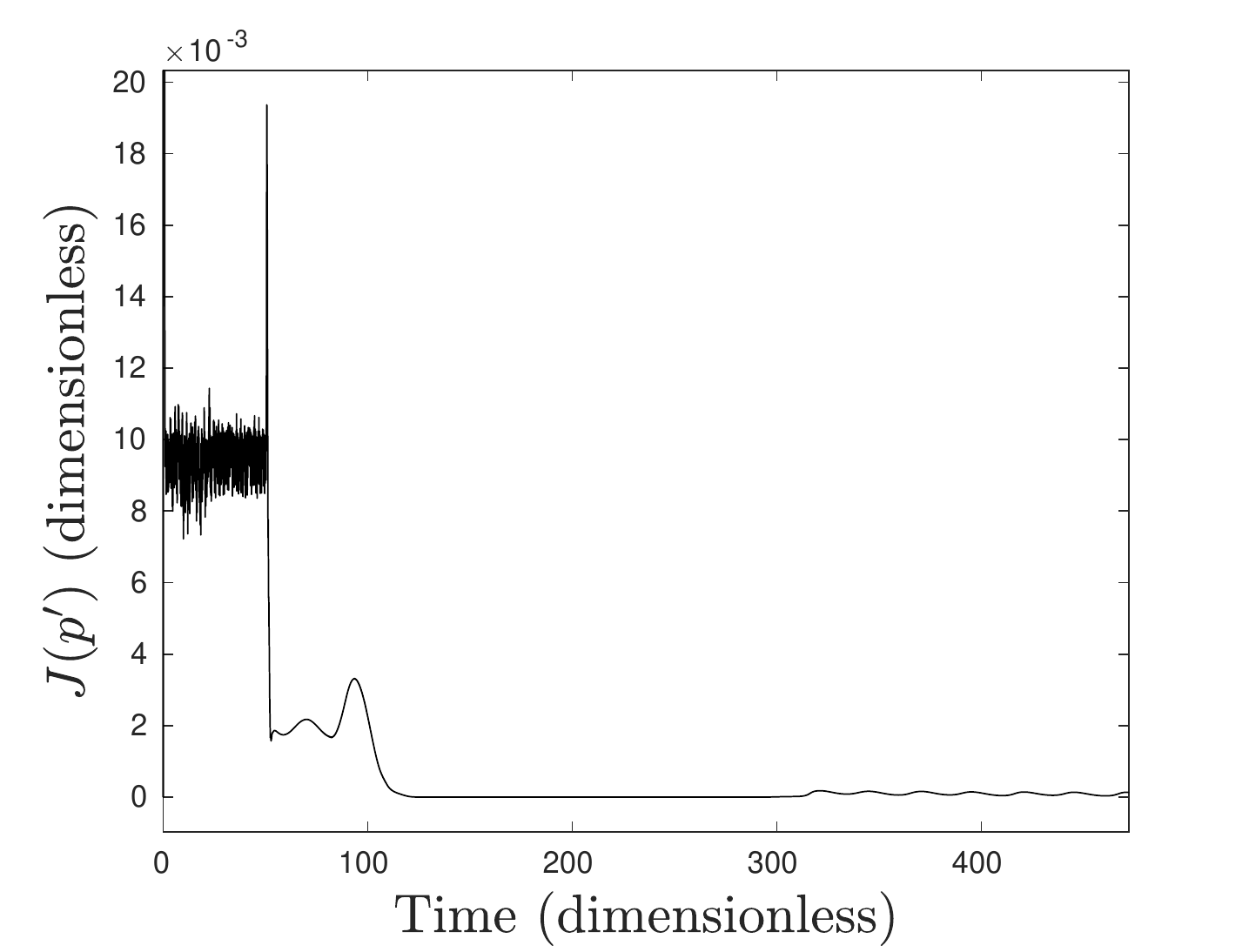}
\includegraphics[width=.32\textwidth]{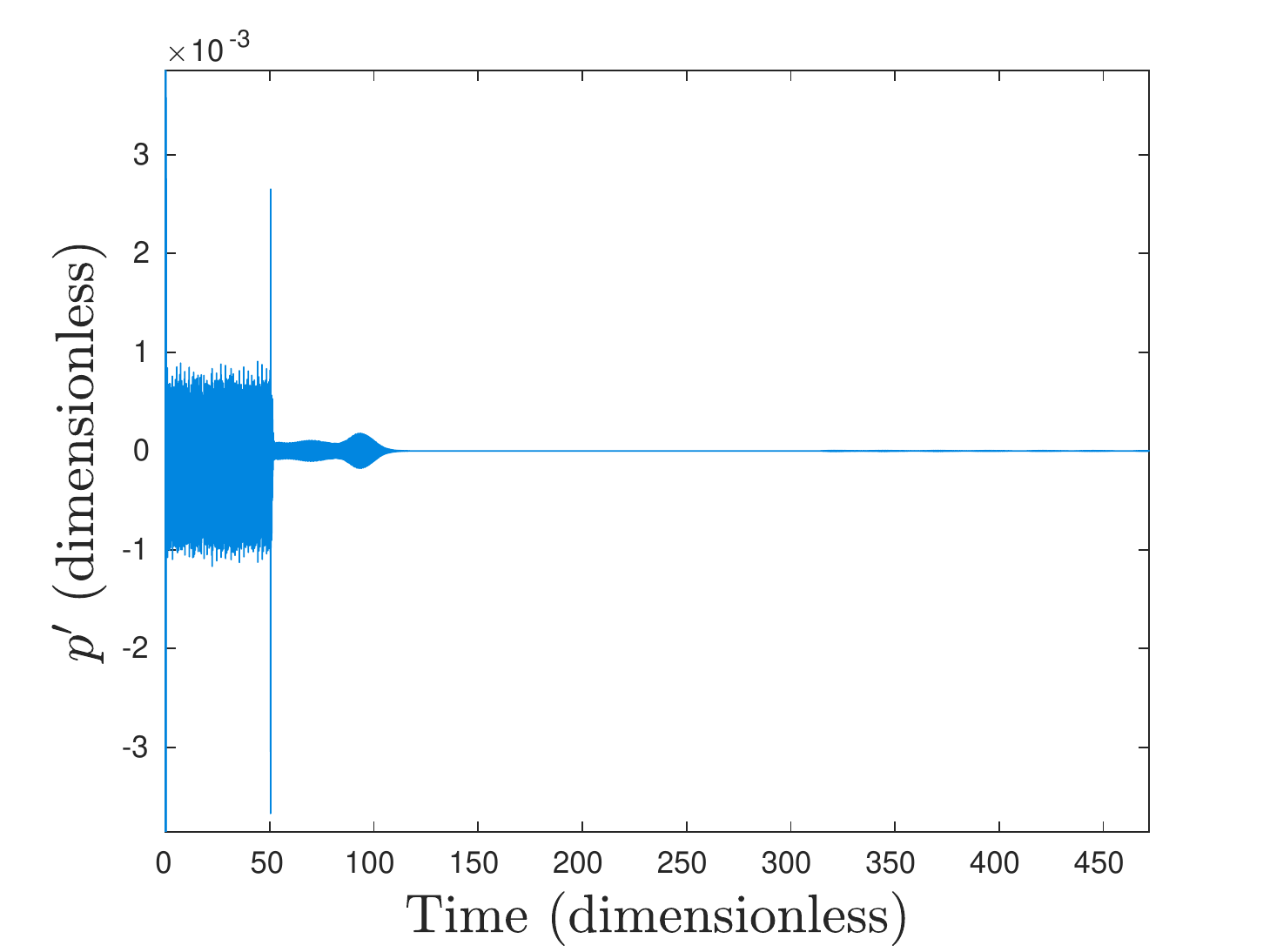}
\caption{Case \cs{108}. With slope seeking compensation, it is possible to keep the trailing edge blowing intensity $A_c$ below the critical stabilization point. This is useful to avoid drifting to unnecessarily high values of $A_c$ since, after the flow is stabilized, no meaningful information is measured by the sensor and the controller cannot make proper decisions.}
\label{f:e108}
\end{figure}

\begin{figure}
\centering
\includegraphics[width=.32\textwidth]{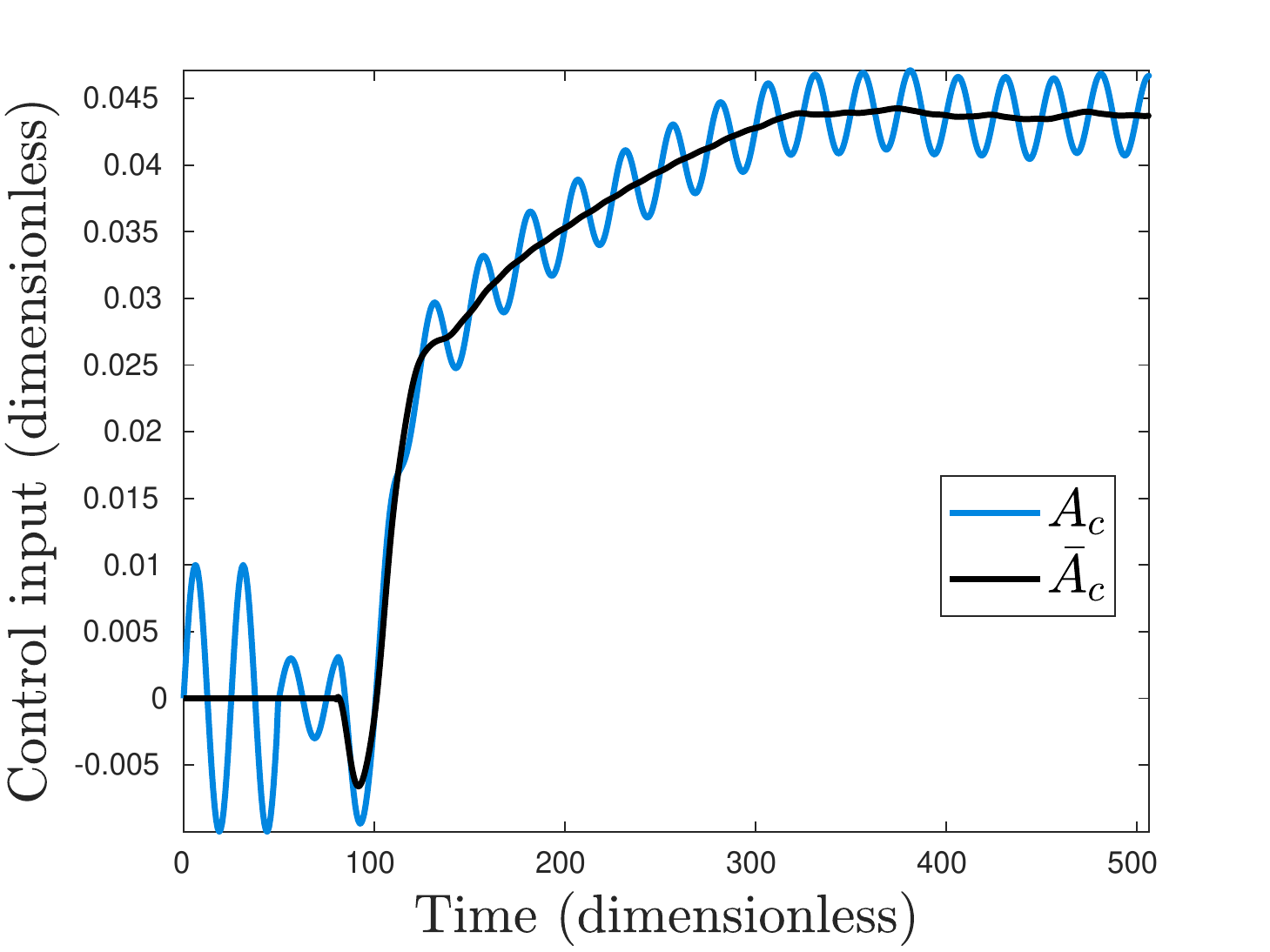}
\includegraphics[width=.32\textwidth]{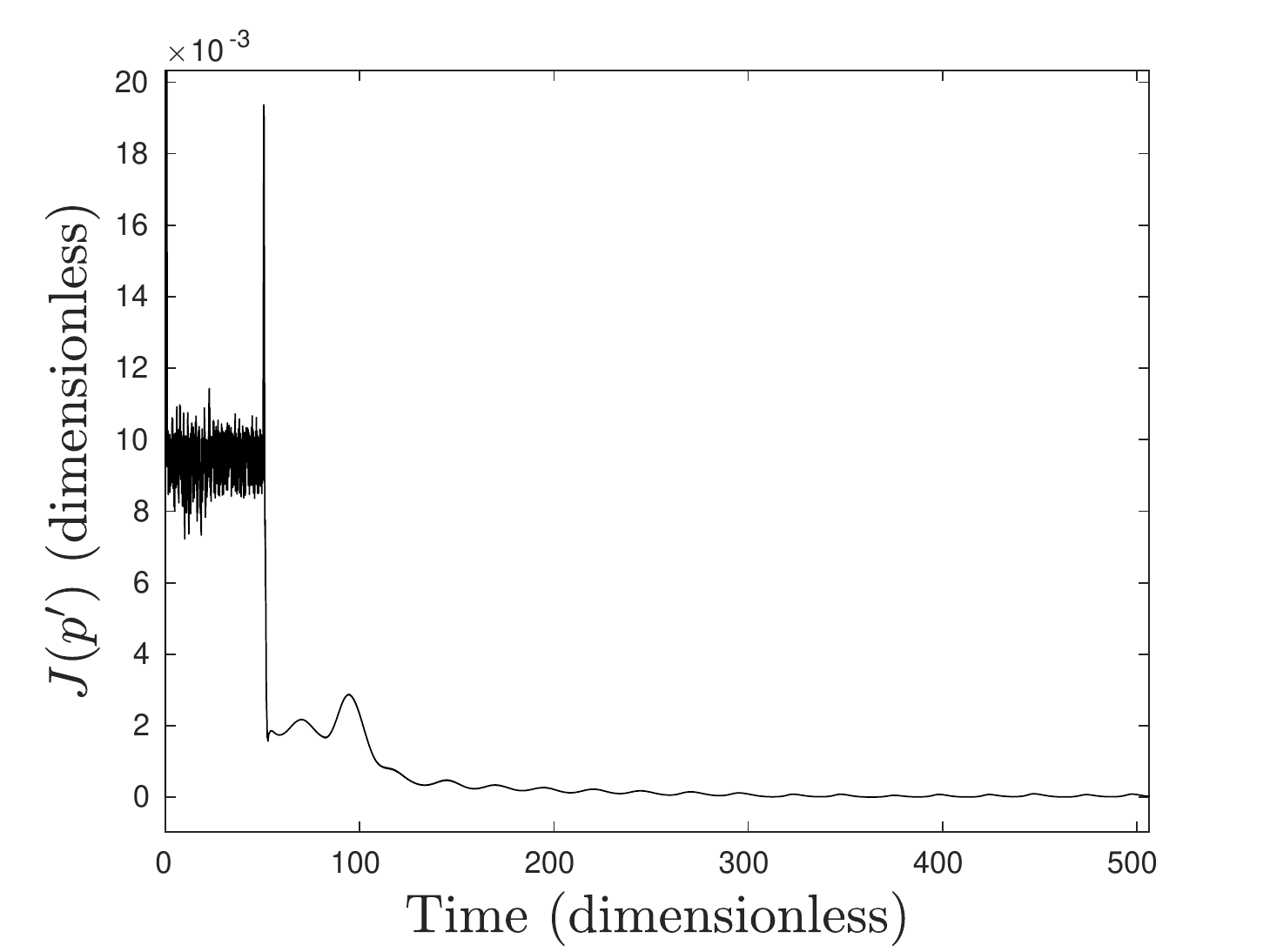}
\includegraphics[width=.32\textwidth]{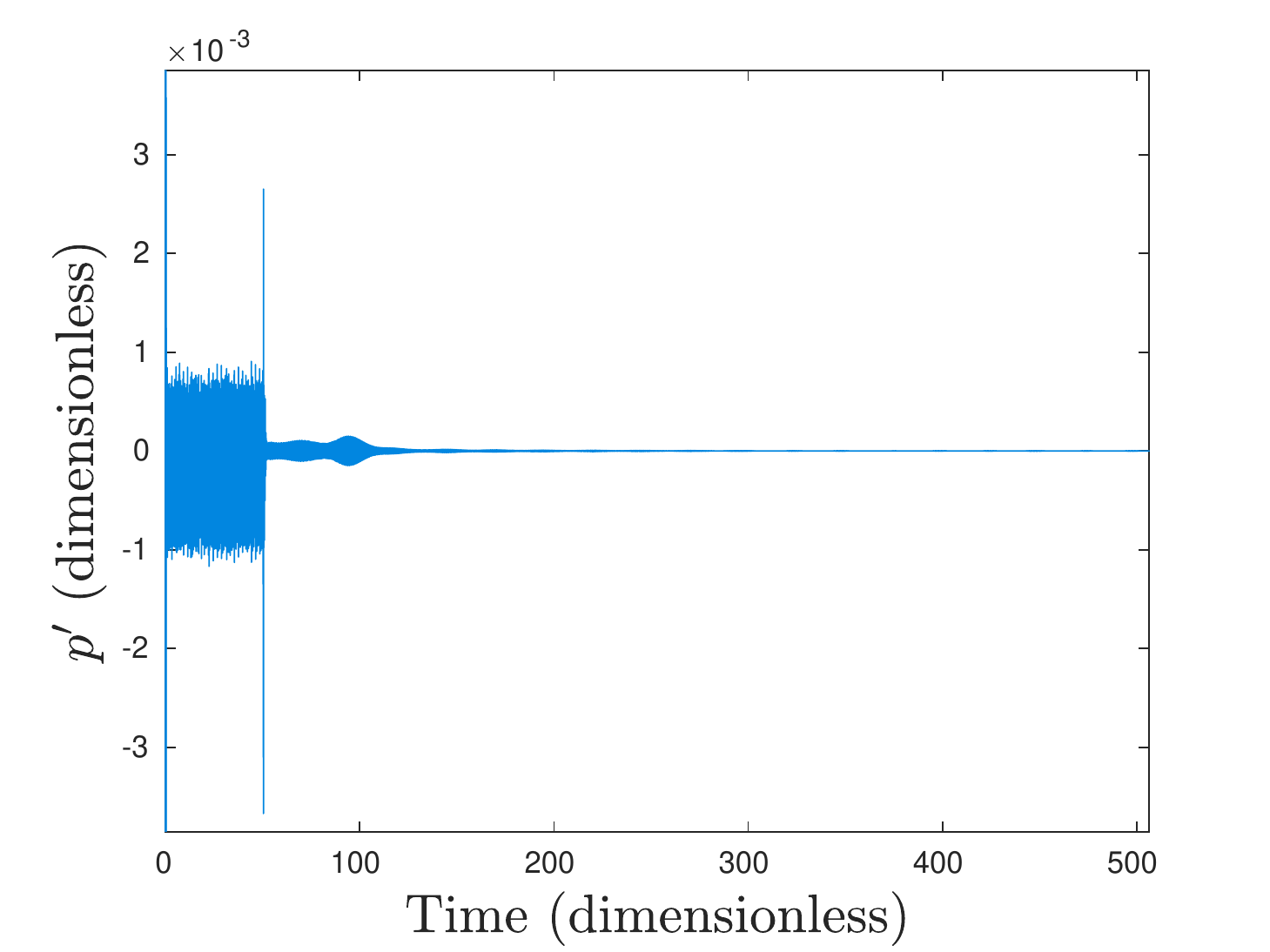}
\caption{Case \cs{109}. To avoid overshoots in the control input, a lower magnitude of the integrator gain $\eta$ is employed compared to that from Case \cs{108}.}
\label{f:e109}
\end{figure}

The motivation for this setup comes from the similarity of the present flow with the lower Reynolds number case previously studied. Although the flow topologies are similar, the blowing actuator at the trailing edge produces different results in terms of noise reduction for the two Reynolds numbers investigated. As shown by \citet{ramirez2015effects}, at low Reynolds numbers, the actuator shifts instabilities away from the airfoil surface, reducing the noise scattering mechanism. On the other hand, for the present case at $\text{Re} = 10^5$, the attempt to reduce noise with a similar control approach results in a complete flow stabilization after a critical blowing intensity and, hence, noise suppression. %One should remind that an actuator found by the ESC is placed at the suction side to suppress boundary layer instabilities.
Three simulations are presented with the parameters shown in Table \ref{tab:cases100tred}. This set of simulations is run from a restart file obtained at $t=50$ to save computational time. This effect can be seen in Figs. \ref{f:e107} - \ref{f:e109} at the moment where the oscillation amplitude changes. Since the actuators are not turned on before $t=50$, there would be no difference if it was a fresh simulation start. 
\begin{table}
  \begin{center}
\def~{\hphantom{0}}
  \begin{tabular}{lcccccccc}
     \hline
     Case    & $\delta_{ss}$  & $A_{c0}$  & $t_{c0}$ & $\alpha=\beta$ & $m$ & $2\pi/\omega$ & $\eta$    & Results\\\hline
     \cs{107} & 0.0e-00       &   00.0\%  & 80.0     & 3.0e-3         & 40  & 25.0          & -1000.0   & Fig. \ref{f:e024}\\
     \cs{108} & 5.0e-08       &   00.0\%  & 80.0     & 3.0e-3         & 40  & 25.0          & -2000.0   & Fig. \ref{f:e025}\\
     \cs{109} & 1.0e-08       &   00.0\%  & 80.0     & 3.0e-3         & 40  & 25.0          & -1300.0   & Fig. \ref{f:e026}\\\hline
  \end{tabular}
  \caption{Control parameters for cases with fixed actuator at the trailing edge with $\text{Re} = 10^5$. %, where $\delta_{ss}$ corresponds to the slope seeking compensation value. %The parameter $t_{c0}$ and the ESC wave period $2\pi/\omega$ are dimensionless temporal parameters relative to freestream velocity and chord. The ESC parameter $\alpha$ is relative to the freestream momentum.
  }
  \label{tab:cases100tred}
  \end{center}
\end{table}

Figure \ref{f:e107} shows the results for case \cs{107} with the application of the ESC ($\delta_{ss} = 0$). The loop is able to suppress the noise generation by increasing the blowing actuation. By setting the initial jet amplitude $A_{c0}=0$, convergence is reached very close to the value of stabilization. In this case, we observed that the input/output lag reaches a phase of $90^\circ$ and the ESC is unable to further change the actuator intensity.

Since after flow stabilization no noise is generated, there is no information for the controller to measure, rendering it unable to consider reductions in $A_c$ in case this value is higher than needed to keep the flow steady. This could occur due to a large initial guess, control overshoots or drifting driven by measurement noise. To overcome this issue, slope seeking compensation is used with $\delta_{ss} > 0$. The values of $\delta_{ss}$ chosen allow for the trailing edge blowing intensity to stay slightly below the stability boundary, so very small acoustic pressure fluctuations still occur. Figure \ref{f:e108} presents the results of case \cs{108} with the slope seeking parameter set as $\delta_{ss} = 5.0e-08$. The rise time is faster due the high gain $\eta$ employed, but a strong overshoot is observed. The ability to get back to a lower blowing magnitude illustrates the importance of using slope seeking in order to avoid excessively high control effort. 

\begin{figure}
\centering
\includegraphics[width=.32\textwidth]{figures/acc_vort/100k/0000.png}
\includegraphics[width=.32\textwidth]{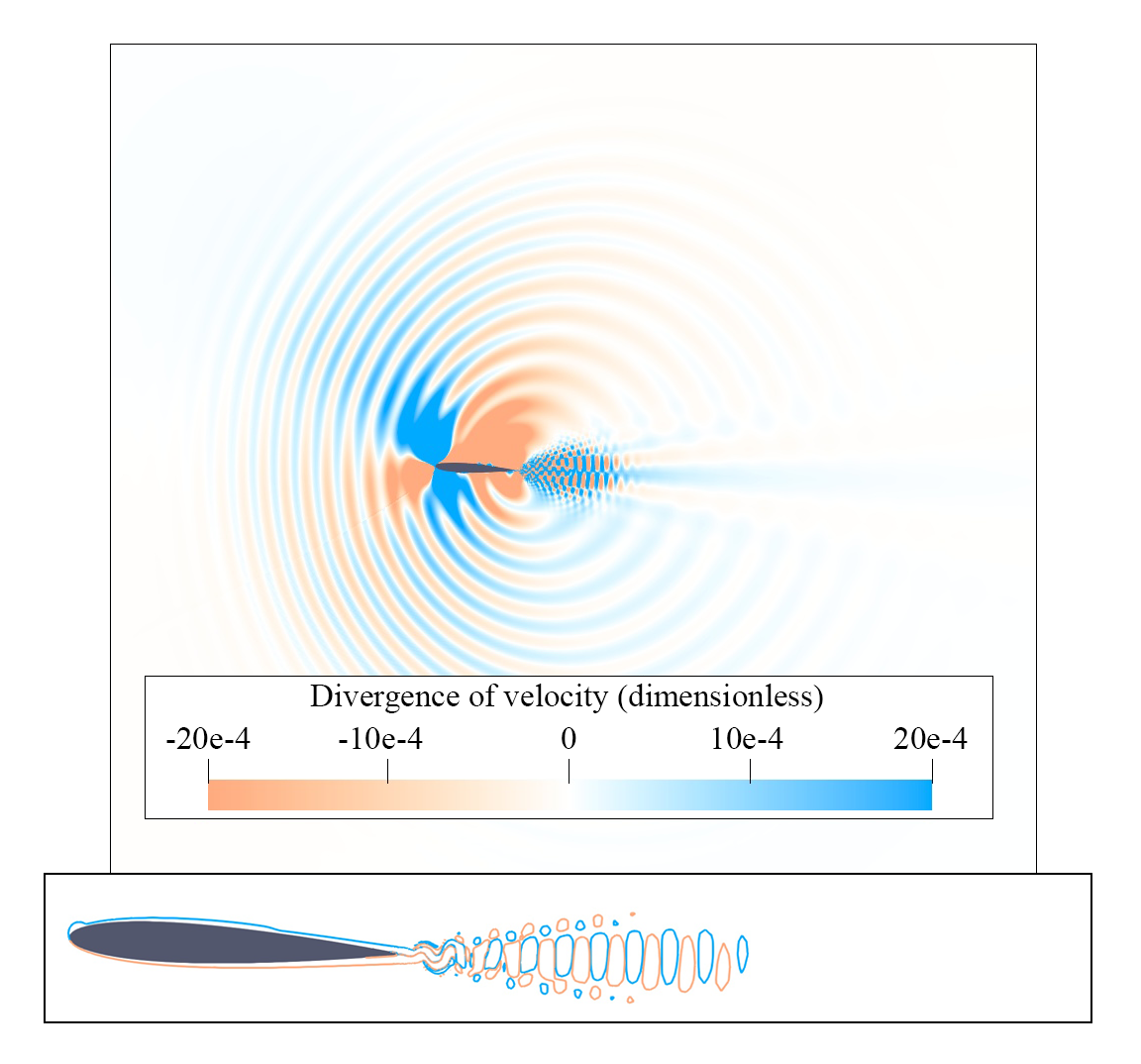}
\includegraphics[width=.32\textwidth]{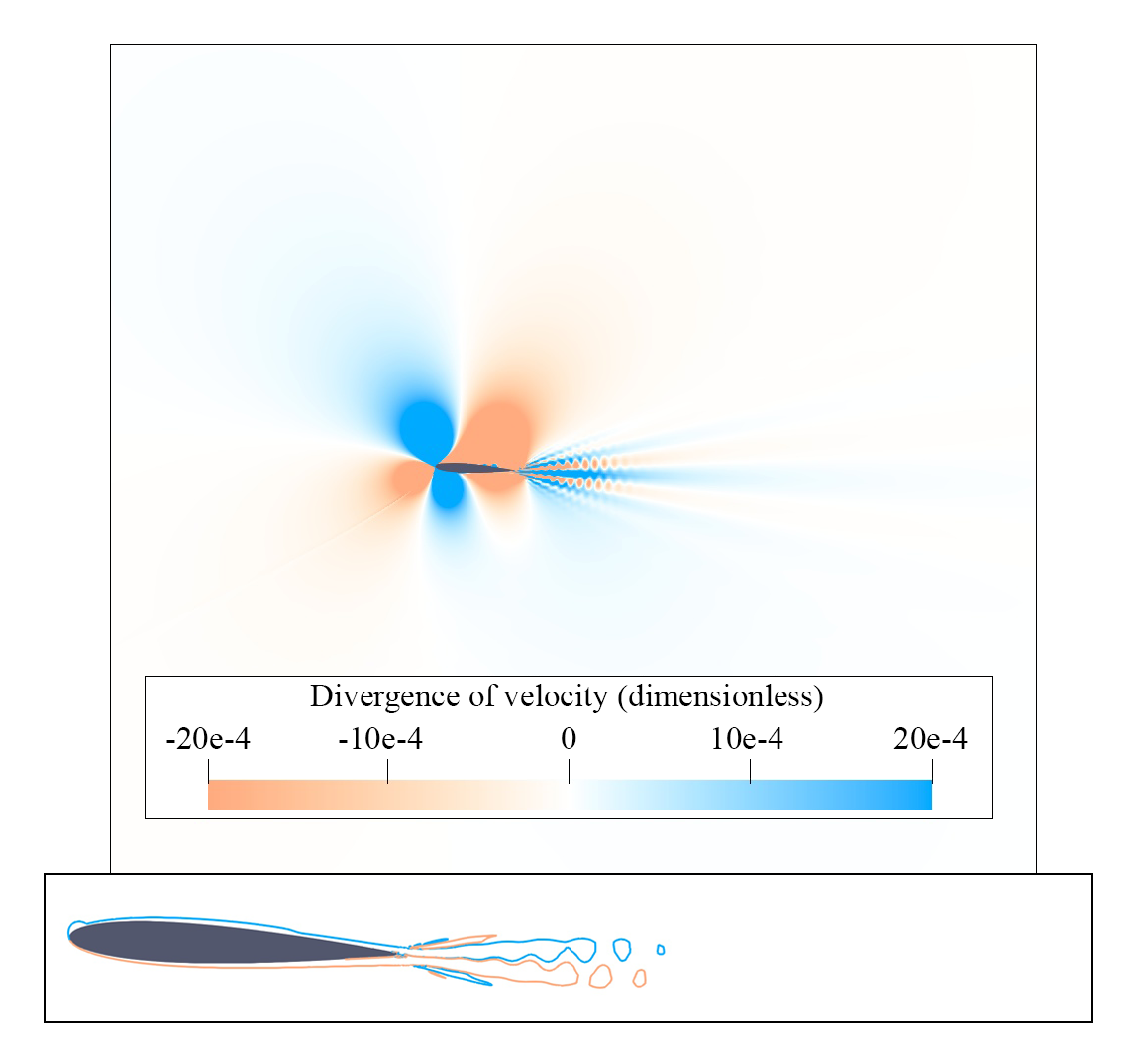}
\caption{Vorticity isolines and contours of divergence of velocity displaying the noise emission for the passive flow (left), and cases \cs{105} (center) and \cs{109} (right). By first finding an optimal actuator position at the suction side with case \cs{105}, the boundary layer instabilities are suppressed, leading to a more organized vortex shedding wake. By keeping the actuator at the ESC optimal position, slope seeking is applied using a second actuator with varying jet intensity fixed at the trailing edge. This control setup stabilizes the flow suppressing the noise generation.}
\label{f:100kall}
\end{figure}

Simulation \cs{109}, depicted in Fig. \ref{f:e109}, uses a lower integration gain $\eta$ to avoid overshooting. The slope seeking compensation is also reducd to $\delta_{ss}= 1.0e-08$, so the system settles more closely to the stability threshold. The flowfields obtained for the passive flow and the controlled cases \cs{105} and \cs{109} are illustrated in Figure \ref{f:100kall}. In the left plot, the vorticity isolines of the passive flow show instabilities along the airfoil suction side and large scale vortices along the wake. The contours of divergence of velocity display the acoustic waves generated by scattering from the previous flow structures at the trailing edge. When the ESC finds an optimal actuator position in case \cs{105}, the boundary layer instabilities are suppressed, as shown in the center plot of Fig. \ref{f:100kall}. The same figure shows a more organized wake structure composed of smaller vortices that, in turn, lead to a reduced noise emission verified in the dilatation field. Finally, the rightmost plot in Fig. \ref{f:100kall} presents the vorticity isolines and contours of divergence of velocity for case \cs{109}, which employs the slope seeking control with an actuation placed at the trailing edge. This plot shows that the flow becomes steady with the optimal actuation intensity found by the slope seeking control. 

\begin{figure}
\centering
\begin{subfigure}{.49\textwidth}
\begin{overpic}[width=0.95\textwidth, frame]{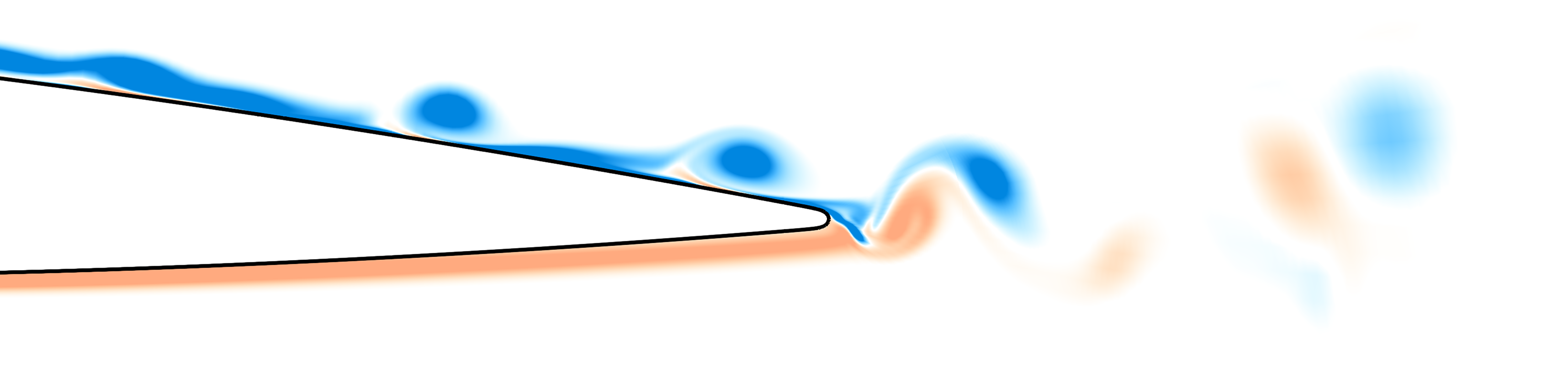}
\put(936,20){(a)}
\end{overpic}
\begin{overpic}[width=0.95\textwidth, frame]{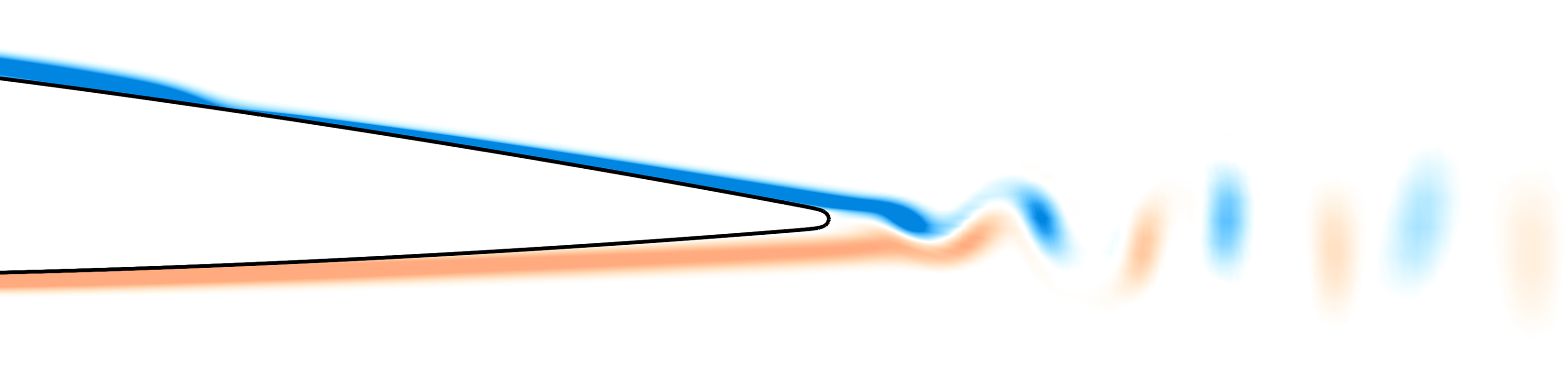}
\put(936,20){(b)}
\end{overpic}
\begin{overpic}[width=0.95\textwidth, frame]{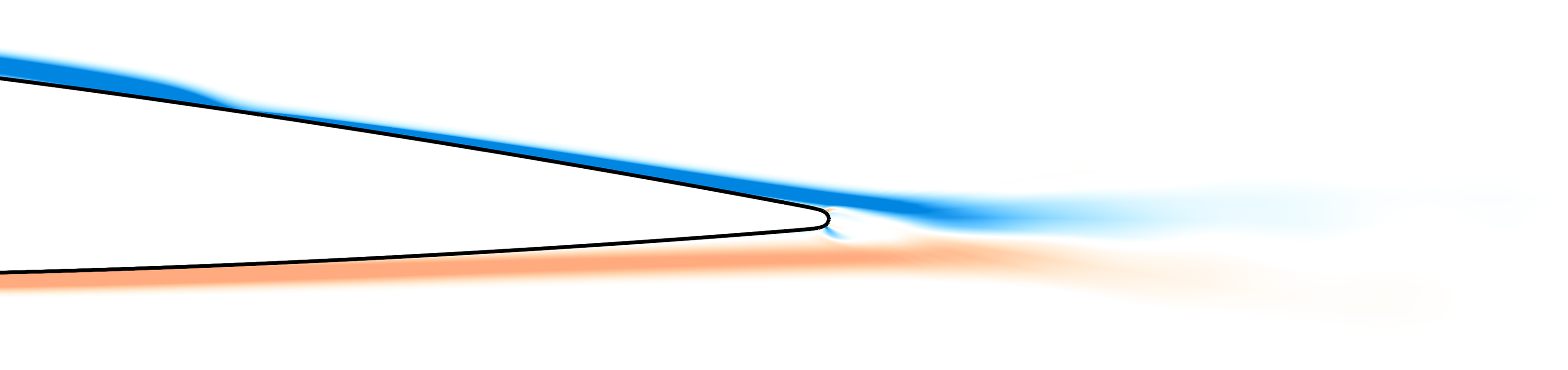}
\put(936,20){(c)}
\end{overpic}
\end{subfigure}
\begin{subfigure}{.49\textwidth}
\begin{overpic}[width=0.99\textwidth]{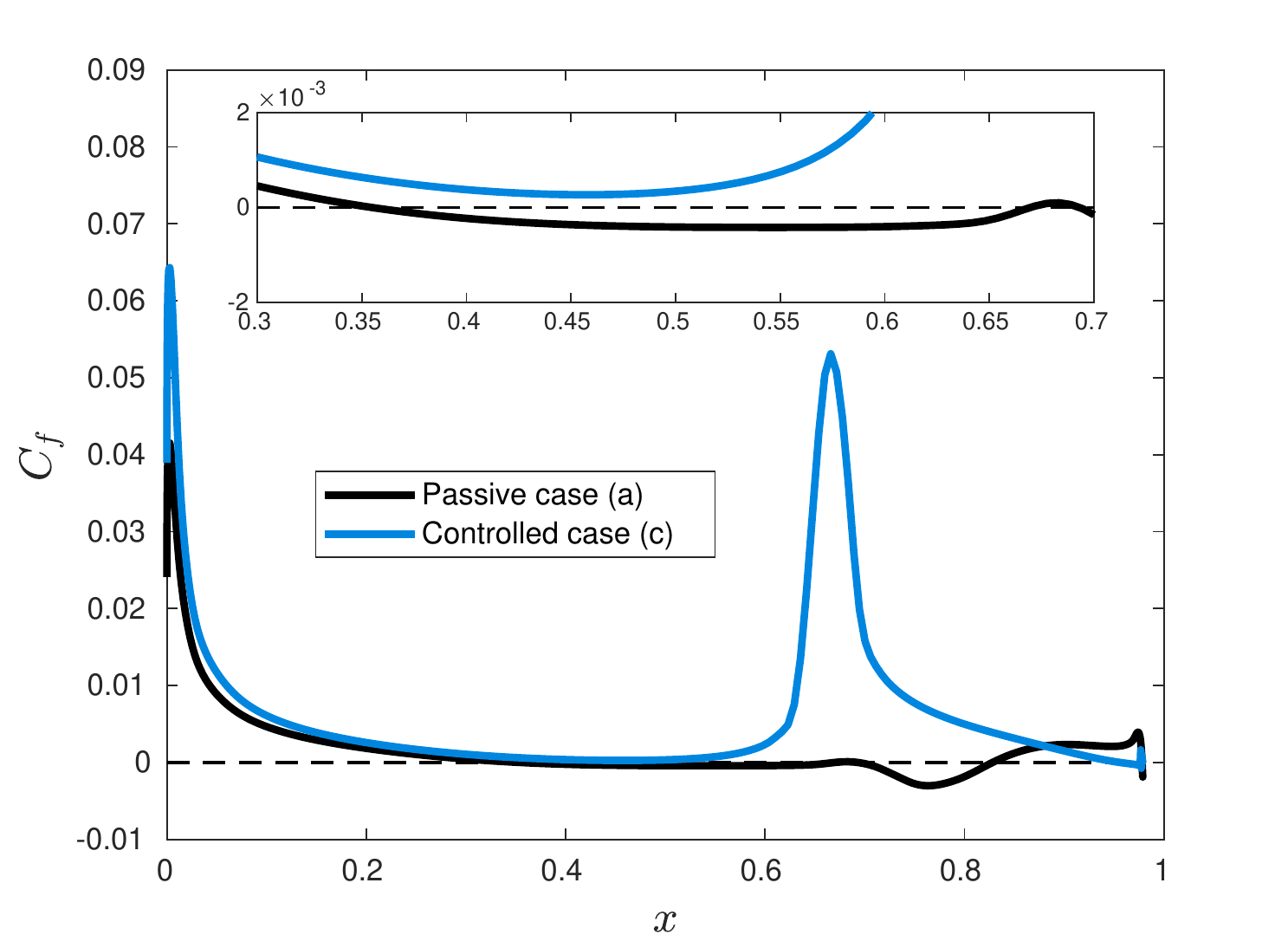}
\put(936,48){(d)}
\end{overpic}
\end{subfigure}
\caption{\green{Vorticity field and skin friction coefficient, $C_f$, at different flow conditions. Without actuation (a), instabilities grow at the suction side and the flow develops a disorganized vortex street associated with multiple tones in the noise generation. When the suction side actuator is applied (b), the vortex street becomes organized as the suction side instabilities are suppressed, reducing noise generation. The addition of an actuator at the trailing edge (c) provides further noise attenuation through suppression of wake flow instabilities. The suction side $C_f$ plot (d) shows the region of flow detachment for the passive case.}}
\label{f:100kdetail}
\end{figure}

\green{An illustration of the resulting effects observed for one controlled flow is also presented in Fig. \ref{f:100kdetail} in terms of the z-vorticity field. This figure allows a visualization of the instabilities that occur along the suction side boundary layer and the wake for the passive flow (Fig. \ref{f:100kdetail}(a)). With the actuation placed at the suction side, at $x = 0.66$, the convective instabilities are suppressed, resulting in organization of the vortex street and attenuation of sound levels as the flow reattaches (Fig. \ref{f:100kdetail}(b)). The skin friction coefficient $C_f$ along the wall at the suction side is also presented in Fig. \ref{f:100kdetail}(d). The $C_f$ profile shows that the actuation is able to prevent separation while the passive mean flow depicts a long separation region. By adding the trailing edge actuation, the vortex street is attenuated or suppressed, resulting in further noise reduction (Fig. \ref{f:100kdetail}(c)). The vorticity fields presented in the figure are obtained for simulation \cs{109} at (a) $t=0.00$, (b) $t=55.68$ and (c) $t=506.62$.}

\section{Conclusions}
\label{discussion}

This work presents the application of extremum seeking control and its slope seeking generalization for suppression of airfoil trailing-edge noise. Implementation details are discussed regarding the simulation environment and the control loop modules for several control setups that include different forms of actuation using the ESC. Special attention is given to the design of the digital filters. 
%
%\subsection{On the application of ESC and slope seeking for airfoil noise suppression}
%
Flow control for airfoil trailing-edge noise reduction is investigated for two Reynolds numbers being $\text{Re} = 10^4$ and $10^5$. For the lower Reynolds number, noise is generated at a single tone. On the other hand, for the higher Reynolds number case, the scattering of boundary layer instabilities at the trailing edge lead to noise generation including multiple tones superimposed on a broadband spectrum. Different actuation configurations are employed with the extremum seeking and slope seeking controls for noise reduction. In one approach, the control input is the actuator position, which has a fixed suction intensity. In this case, the ESC searches for an optimal actuation position to  suppress airfoil noise. In a second approach, the control input is the blowing or suction magnitude for a fixed actuator position. Both the ESC and slope seeking then search for an optimal actuation intensity. 

Results show that the ESC may not reach an optimal actuation in cases where the input/output lag reaches a phase of 90$^\circ$. This issue can be overcome by reducing the sweep frequency and, in some cases, it is shown that an increase in the integrator gain also enables convergence. For both Reynolds numbers, the  moving actuator control setup with constant suction intensity is able to find an optimal spatial range of actuation where noise is reduced. For the lower Reynolds number, flow reattachment is achieved for some suction intensities and noise is fully suppressed. For the higher Reynolds number case, this type of control setup provides noise reductions through elimination of the boundary layer instabilities. However, in these cases, vortex shedding noise remains.
Noise reduction is also observed when the actuator is fixed at the suction side or trailing edge for the lower Reynolds number flow. In the former case, abrupt flow reattachments and detachments due to the ESC control input variation may lead to discontinuities in the cost function that, in turn, increase the control effort beyond the necessary to suppress noise. To overcome this issue, an additional term is added to the cost function penalizing the control effort. In such cases, noise is suppressed and a study of the system dynamics is presented for variations in the integrator gain and digital filters. When the actuator is fixed at the trailing edge, noise reduction is achieved with both the ESC and slope seeking. The mechanisms for noise reduction are related to the downstream displacement of the vortex shedding along the wake. Slope seeking shows promising results in this case since it keeps the control effort limited. On the other hand, the ESC presents a continuous increase in the control effort seeking a further noise reduction.

The higher Reynolds number case requires a more complex control setup for noise suppression. The chordwise moving actuator is able to eliminate the boundary layer instabilities leading to a reduced tonal noise generation from vortex shedding. Due to the similarities of this flow with the lower Reynolds number one, a fixed actuator is implemented in the trailing edge while keeping a second actuator on the suction side. This second actuator is placed at the optimal position found by the ESC. Then, slope seeking and ESC are tested to find an optimal intensity of actuation at the \blue{trailing} edge. Differently from the lower Reynolds number case, the flow becomes steady and noise suppression is achieved when a threshold is reached.
%
\begin{comment}

Based on the results obtained with the present setups, one can think of two main active control applications:
%
\begin{itemize}
    \item the \textit{a priori} computation of the best location or intensity through experiments or simulations for real-world applications in open-loop configurations;
    \item the study and tuning of controller parameters for implementation of real-time control systems for real-world applications in order take advantage of some closed-loop control benefits.
\end{itemize}

The former implementation type can be useful since with a single simulation it is possible to directly find the best position for actuator placement. With the control parameters tuned, it may be possible to obtain the optimal inputs for reasonable variations in airfoil shape, Reynolds and Mach numbers and angle of attack. The latter implementation type can be useful to enable robustness to a varying optimal due to variations in the plant as it operates.

\end{comment}
%
As discussed in this work, extremum seeking control is suitable for many situations where the plant presents a variable that can be optimized by finding the best value for an input. Although, in general, fluid flows may present complex dynamics with nonlinear features and chaos, some approximate input/output relations can be found within slow timescales. It is shown that, by carefully choosing a cost function (and its parameters) capable of representing the airfoil sound emission at a given instant, there exists an almost direct correspondence between the actuator position (or intensity) and the noise generation mechanisms.

\blue{
Although in the present work the simulations are conducted with the main parameters fixed (Reynolds and Mach numbers, as well as the angle of attack), the feasibility of the ESC approach is demonstrated for flows with different sound generation mechanisms. In this sense, it is expected that the technique can provide robustness to variations in the flow parameters. In real world applications, the ESC loop could be implemented in order to optimize the desired cost function as the system characteristics change. For example, it could track the minimal possible noise generation for an airfoil flow with varying Reynolds and Mach numbers. Also, in simulations or experiments, the loop could be implemented so as to find the best location or intensity for actuators as a function of the main flow parameters without the need for a complete search within the range of possible control inputs.
}

\blue{
In this work, each simulation seeks optimization for individual parameters either by finding a position or intensity for the actuator. An ESC loop could be constructed to search for both inputs at the same time, for example, by using different values of $\omega$ for each variable to ensure that the value of $\nabla\bar{J}$ relative to one input is not affected by the other. However, it is important to mention that, if multiple variables are considered, some issues may appear in the cost function. For example, it could contain local minima or dynamics that could hinder convergence, such as hysteresis.
}

%\subsection{On the limitations of ESC and slope seeking}

One of the main limitations of the approaches implemented in this work is related to timescales. It is expected that there is an approximately static input/output function that describes the plant. When the input assumes a value, the output must respond with no relevant time delay for the operating timescale. This scale limits the ability of the control loop to quickly respond and converge to the optimal, as well as to respond to rapid plant variations that would change the optimal input value. Failure in obeying reasonable timescale separation may lead to problems as those shown in case \cs{11} analyzed.
%
%Although not present in any of the problems studied in this work, variations in the plant that would result in deviating the optimal value can also be a problem when dealing with timescales, specially for choosing the ESC frequency $\omega$. This frequency must be chosen so it disturbs the input faster than the plant changes. The convergence should also be fast enough so the controller tracks a varying optimal. This means that ESC may even not be applicable in cases where the plant parameters change at rates similar to those of the field variables themselves, given by the flow dynamics. Some examples of situations that could change the optimal input in the airfoil includes dynamic stall problems.
%
Another limitation is related to discontinuities in the cost function. In this case, the loop may not find an optimal operating condition if the system presents discontinuities, specially when the optimal input rests near the discontinuity. This issue is evident in cases \cs{18} and \cs{20} studied, where a discontinuity is present when the flow reattaches, leading to a much lower noise emission so the input related part of the cost function dominates. In this case, the best intensity for the actuator fixed at the suction side would be at the lower value to keep the flow attached, but since a small variation in this value leads to a discontinuity in the cost function, the ESC loop is unable to keep the intensity at an optimum value due to the periodic perturbation. In fact, some of the classic convergence proofs for ESC rely on an approximation of the cost (as a function of the input) to a second order polynomial near the optimal, which is not the case when discontinuities are present.

\green{The numerical solver employed in the present simulations solves the 2D compressible Navier-Stokes equations with physically accurate outputs as validated in previous work. Three-dimensional actuation effects could play an important role in several cases and the present ESC approach will be tested with turbulent flows in future work. Although experiments or real-world applications are expected to produce similar results, current technology may still present limitations that hinder some implementations. For example, actuators providing steady suction may easily accumulate dust, which can limit long-term operations. Also, moving actuators are still hard to implement; while some authors already implemented arrays of actuators with switchable elements, the limited number of actuators strongly limits the maximum resolution possible. Actuators that can promote both steady suction or blowing depending on the sign of the control input are also hard to implement.}

\appendix
\section*{Appendix A}
\label{appA}

In this Appendix, technical details about the filter parameters are reported. Table \ref{tab:filter} shows the parameters chosen for each simulation, where $\omega_s$ and $\omega_p$ refer to the stopband and passband frequencies, respectively. The stopband maximum gain allowed is $g_s$ and the maximum ripple allowed in the passband for the Chebyshev type I filter is $r$. For the control loop filters, the stopband frequency $\omega_s$ is presented divided by the $\omega$ value. For the sensor filters, $\omega_s$ is presented as a fraction of $\omega_a = 2\pi/T_a$, which approximates the angular frequency observed in the acoustic pressure signals. In the $Re=10^4$ cases, $T_a = 3.675\mbox{e-1}$ approximates the period of the tonal noise produced in the passive flow. On the other hand, for the $Re=10^5$ cases, $T_a = 1.300\mbox{e-1}$ is used. This value corresponds to an approximation of the main frequency present in the acoustic pressure computed for the passive flow, as well as for the tonal frequency that remains when the actuated flow becomes organized. %Here, $T_a$ is a dimensionless temporal parameter relative to the freestream flow velocity.
\begin{table}
  \begin{center}
\def~{\hphantom{0}}
  \resizebox{.98\textwidth}{!}{%
  \begin{tabular}{ccccccccccccc}
  \hline
    \multicolumn{4}{c}{Sensor filter} &\multicolumn{4}{c}{Loop H.P. filter} & \multicolumn{4}{c}{Loop L.P. filter} & \multicolumn{1}{c}{Cases}\\[3pt]
    r & $g_{s}$ & $\omega_{s}/\omega_a$ & $\omega_{p}/\omega_{s}$ & $r$ & $g_{s}$ & $\omega_{s}/\omega$ & $\omega_{p}/\omega_{s}$ & $r$ & $g_{s}$ & $\omega_{s}/\omega$ & $\omega_{s}/\omega_{p}$ &\\\hline
    2 & 20 & 0.4 & \multicolumn{1}{c|}{3} & 2 & 20 & 1.0 & \multicolumn{1}{c|}{3} & 2 & 20 & 0.8 & \multicolumn{1}{c|}{3} & \begin{tabular}[c]{@{}c@{}}\cs{11}, \cs{12}, \cs{13}, \cs{15}, \cs{16}, \cs{27}, \cs{17}, \cs{18}, \cs{24}, \cs{25}, \cs{26}\end{tabular}\\ [6pt]
    2 & 20 & 0.4 & \multicolumn{1}{c|}{3} & 2 & 20 & 0.3 & \multicolumn{1}{c|}{3} & 2 & 20 & 0.4 & \multicolumn{1}{c|}{3} & \cs{6}, \cs{7}, \cs{9}, \cs{10} \\ [6pt]
    2 & 20 & 0.4 & \multicolumn{1}{c|}{3} & 2 & 20 & 1.0 & \multicolumn{1}{c|}{3} & - &  - &             - & \multicolumn{1}{c|}{-} & \cs{20} \\[6pt]
    2 & 20 & 0.8 & \multicolumn{1}{c|}{3} & 2 & 20 & 1.0 & \multicolumn{1}{c|}{3} & 2 & 20 & 0.8 & \multicolumn{1}{c|}{3} & \cs{105}, \cs{106}, \cs{107}, \cs{108}, \cs{109} \\\hline       
  \end{tabular}
  }
  \caption{Filter parameters for each simulation, where $\omega_a = 2\pi/T_a$. In the $\mathrm{Re}=10^4$ cases, $T_a = 3.675\mbox{e-1}$, and for the $\mathrm{Re}=10^5$ cases, $T_a = 1.300\mbox{e-1}$.}
  \label{tab:filter}
  \end{center}
\end{table}

\section*{Appendix B}
\label{appB}

\green{In Section \ref{sec:cases}, all simulations are presented with resulting signals computed in the control loop obtained from the temporal plots. In Table \ref{tab:reduction}, a summary of the results in terms of noise reduction is presented in decibels. The approximate values of $C_\mu$ are also presented after convergence of the control inputs $x_c$ or $A_c$. These values allow for a direct evaluation of the energy consumption by the actuation. Results of simulations that did not achieve convergence are not shown. For the calculation of noise reduction, the RMS values of acoustic pressure are first computed for the passive flows. For the lower Reynolds number case, the flow produces tonal noise with $p_\mathrm{RMS}=$ 2.5825e-4, while for the higher Reynolds number flow, $p_\mathrm{RMS}=$ 4.7443e-4. A portion of the signal is selected after convergence of the control input to compute the new pressure RMS values for the controlled flows, $p_{c\mathrm{RMS}}$, which are then compared to the passive ones to obtain the noise attenuation in dB. The reduction in sound pressure level is computed as $\Delta \mbox{SPL}=20\log_{10}(p_{c\mathrm{RMS}}/p_\mathrm{RMS})$.} % according to $10\mathrm{log}(r_c^2/r_0^2) = 20\mathrm{log}(r_c/r_0)$.
\begin{table}
  \begin{center}
\def~{\hphantom{0}}
  \resizebox{.97\textwidth}{!}{%
  \begin{tabular}{lccccc}
     \hline
     Case     & $p_{RMS}$ & Attenuation  & $C_\mu$   & Noise source (passive flow)  & Noise source (controlled flow)\\\hline
     \cs{11}  & 7.2071e-06  &   -31.1dB & 1.511e-3        & Vortex shedding       & Vortex shedding        \\
     \cs{6}   & 2.1673e-08  &   -81.5dB & 1.511e-3        & Vortex shedding       & Oscillating actuation                \\
     \cs{7}   & 1.3020e-08  &   -85.9dB & 1.511e-3        & Vortex shedding       & Oscillating actuation               \\
     \cs{12}  & 2.9447e-09  &   -98.9dB & 1.511e-3        & Vortex shedding       & Oscillating actuation               \\
     \cs{13}  & 2.9447e-09  &   -98.9dB & 1.511e-3        & Vortex shedding       & Oscillating actuation               \\
     \cs{9}   & 1.9744e-08  &   -82.3dB & 3.778e-4        & Vortex shedding       & Oscillating actuation               \\
     \cs{10}  & 1.6119e-08  &   -84.1dB & 3.778e-4        & Vortex shedding       & Oscillating actuation               \\
     \cs{15}  & 7.0481e-05  &   -11.3dB & 6.045e-5        & Vortex shedding       & Vortex shedding        \\
     \cs{16}  & 6.9314e-05  &   -11.4dB & 6.045e-5        & Vortex shedding       & Vortex shedding        \\
     \cs{27}  & 6.8292e-05  &   -11.6dB & 6.045e-5        & Vortex shedding       & Vortex shedding        \\
     \cs{17}  & 8.8070e-08  &   -69.3dB & 4.103e-3        & Vortex shedding       & Oscillating actuation               \\
%     \cs{18}  & -           &            - &        -        & Vortex shedding       & -                      \\
%     \cs{20}  & -           &            - &        -        & Vortex shedding       & -                      \\
%     \cs{24}  & -           &            - &        -        & Vortex shedding       & -                      \\
     \cs{25}  & 2.4606e-05  &   -20.4dB & 9.845e-4        & Vortex shedding       & Vortex shedding        \\
     \cs{26}  & 2.1193e-05  &   -21.7dB & 9.845e-4        & Vortex shedding       & Vortex shedding        \\
     \cs{105} & 6.2671e-05  &   -17.6dB & 1.511e-3        & Boundary layer instabilities           & Vortex shedding        \\
     \cs{106} & 6.1837e-05  &   -17.7dB & 1.511e-3        & Boundary layer instabilities           & Vortex shedding        \\
%     \cs{107} & -           &            - &        -        & TS vortices           & -                      \\
     \cs{108} & 1.5067e-06  &   -49.9dB & 1.681e-3        & Boundary layer instabilities           & Vortex shedding        \\
     \cs{109} & 6.9623e-07  &   -56.7dB & 1.694e-3        & Boundary layer instabilities           & Vortex shedding \\\hline       
  \end{tabular}
  }
  \caption{\green{Noise reduction in dB for controlled cases that achieved convergence of control inputs. The $C_\mu$ values presented are obtained after convergence.}}
  \label{tab:reduction}
  \end{center}
\end{table}

\section*{Acknowledgments}
The authors acknowledge the financial support received from Fundação de Amparo
à Pesquisa do Estado de Sâo Paulo, FAPESP, under Grant No. 2013/08293-7, and from Conselho Nacional de Desenvolvimento Científico e Tecnológico, CNPq, under Grants No. 407842/2018-7 and 304335/2018-5. The first author is supported by a FAPESP PhD scholarship 2019/19179-7, which is also acknowledged. The computational resources used in this work were provided by CENAPAD-SP (Project 551), and by LNCC via the SDumont cluster (Project SimTurb).

\bibliography{sample}

\begin{thebibliography}{48}
\newcommand{\enquote}[1]{``#1''}
\providecommand{\natexlab}[1]{#1}
\providecommand{\url}[1]{\texttt{#1}}
\providecommand{\urlprefix}{URL }
\expandafter\ifx\csname urlstyle\endcsname\relax
  \providecommand{\doi}[1]{\discretionary{}{}{}https://doi.org/#1}\else
  \providecommand{\doi}[1]{\discretionary{}{}{}\urlstyle{rm}\url{https://doi.org/#1}}\fi

\bibitem[{Cattafesta~III et~al.(1997)Cattafesta~III, Garg, Choudhari, and
  Li}]{cattafesta1997active}
Cattafesta~III, L.~N., Garg, S., Choudhari, M., and Li, F., \enquote{Active
  control of flow-induced cavity resonance,} \emph{28th AIAA Fluid Dynamics
  Conference}, 1997.

\bibitem[{Alvi et~al.(2003)Alvi, Shih, Elavarasan, Garg, and
  Krothapalli}]{alvi2003control}
Alvi, F.~S., Shih, C., Elavarasan, R., Garg, G., and Krothapalli, A.,
  \enquote{Control of supersonic impinging jet flows using supersonic
  microjets,} \emph{AIAA journal}, Vol.~41, No.~7, 2003, pp. 1347--1355.

\bibitem[{Tuck and Soria(2004)}]{tuck2004active}
Tuck, A., and Soria, J., \enquote{Active flow control over a NACA 0015 airfoil
  using a ZNMF jet,} \emph{15th Australasian Fluid Mechanics Conference}, 2004.

\bibitem[{Cattafesta~III and Sheplak(2011)}]{cattafesta2011actuators}
Cattafesta~III, L.~N., and Sheplak, M., \enquote{Actuators for active flow
  control,} \emph{Annual Review of Fluid Mechanics}, Vol.~43, 2011, pp.
  247--272.

\bibitem[{Cuvier et~al.(2011)Cuvier, Braud, Foucaut, and
  Stanislas}]{cuvier2011flow}
Cuvier, C., Braud, C., Foucaut, J., and Stanislas, M., \enquote{Flow control
  over a ramp using active vortex generators,} \emph{Seventh International
  Symposium on Turbulence and Shear Flow Phenomena}, 2011.

\bibitem[{George et~al.(2015)George, Ukeiley, Cattafesta, and
  Taira}]{george2015control}
George, B., Ukeiley, L.~S., Cattafesta, L.~N., and Taira, K., \enquote{Control
  of three-dimensional cavity flow using leading-edge slot blowing,} \emph{53rd
  AIAA Aerospace Sciences Meeting}, 2015.

\bibitem[{Sinha et~al.(2018)Sinha, Towne, Colonius, Schlinker, Reba, Simonich,
  and Shannon}]{sinha2018active}
Sinha, A., Towne, A., Colonius, T., Schlinker, R.~H., Reba, R., Simonich,
  J.~C., and Shannon, D.~W., \enquote{Active control of noise from hot
  supersonic jets,} \emph{AIAA Journal}, Vol.~56, No.~3, 2018, pp. 933--948.

\bibitem[{Isfahani et~al.(2021)Isfahani, Webb, and
  Samimy}]{ghassemi2021control}
Isfahani, A.~G., Webb, N.~J., and Samimy, M., \enquote{Control of flow and
  acoustics in twin rectangular jets,} \emph{AIAA Scitech 2021 Forum}, 2021.

\bibitem[{Zigunov et~al.(2021)Zigunov, Sellappan, and
  Alvi}]{zigunov2021empirical}
Zigunov, F., Sellappan, P., and Alvi, F.~S., \enquote{An empirical platform for
  optimal placement of open-loop microjet-in-crossflow actuators,} \emph{AIAA
  Scitech Forum}, 2021.

\bibitem[{Barbagallo et~al.(2009)Barbagallo, Sipp, and
  Schmid}]{barbagallo2009closed}
Barbagallo, A., Sipp, D., and Schmid, P.~J., \enquote{Closed-loop control of an
  open cavity flow using reduced-order models,} \emph{Journal of Fluid
  Mechanics}, Vol. 641, 2009, pp. 1--50.

\bibitem[{Semeraro et~al.(2011)Semeraro, Bagheri, Brandt, and
  Henningson}]{semeraro2011feedback}
Semeraro, O., Bagheri, S., Brandt, L., and Henningson, D.~S., \enquote{Feedback
  control of three-dimensional optimal disturbances using reduced-order
  models,} \emph{Journal of Fluid Mechanics}, Vol. 677, 2011, pp. 63--102.

\bibitem[{Brunton et~al.(2014)Brunton, Dawson, and Rowley}]{brunton2014state}
Brunton, S.~L., Dawson, S. T.~M., and Rowley, C.~W., \enquote{State-space model
  identification and feedback control of unsteady aerodynamic forces,}
  \emph{Journal of Fluids and Structures}, Vol.~50, 2014, pp. 253--270.

\bibitem[{Ma et~al.(2011)Ma, Ahuja, and Rowley}]{ma2011reduced}
Ma, Z., Ahuja, S., and Rowley, C.~W., \enquote{Reduced order models for control
  of fluids using the eigensystem realization algorithm,} \emph{Theoretical and
  Computational Fluid Dynamics}, Vol.~25, 2011, pp. 233--247.

\bibitem[{Proctor et~al.(2016)Proctor, Brunton, and Kutz}]{proctor2016dynamic}
Proctor, J.~L., Brunton, S.~L., and Kutz, J.~N., \enquote{Dynamic mode
  decomposition with control,} \emph{SIAM Journal on Applied Dynamical
  Systems}, Vol.~15, No.~1, 2016, pp. 142--161.

\bibitem[{Rowley and Dawson(2017)}]{rowley2017model}
Rowley, C.~W., and Dawson, S., \enquote{Model reduction for flow analysis and
  control,} \emph{Annual Review of Fluid Mechanics}, Vol.~49, 2017, pp.
  387--417.

\bibitem[{Sasaki et~al.(2018)Sasaki, Tissot, Cavalieri, Silvestre, Jordan, and
  Biau}]{sasaki2018closed}
Sasaki, K., Tissot, G., Cavalieri, A.~V., Silvestre, F.~J., Jordan, P., and
  Biau, D., \enquote{Closed-loop control of a free shear flow: a framework
  using the parabolized stability equations,} \emph{Theoretical and
  Computational Fluid Dynamics}, Vol.~32, No.~6, 2018, pp. 765--788.

\bibitem[{Bewley(2001)}]{bewley2001flowcontrol}
Bewley, T.~R., \enquote{Flow control: new challenges for a new {R}enaissance,}
  \emph{Progress in Aerospace Sciences}, Vol.~37, 2001, pp. 21--58.

\bibitem[{Brunton and Noack(2015)}]{brunton2015closed}
Brunton, S.~L., and Noack, B.~R., \enquote{Closed-loop turbulence control:
  progress and challenges,} \emph{Applied Mechanics Reviews}, Vol.~67, No.~5,
  2015, p. 050801.

\bibitem[{You and Moin(2008)}]{you2008active}
You, D., and Moin, P., \enquote{Active control of flow separation over an
  airfoil using synthetic jets,} \emph{Journal of Fluids and Structures},
  Vol.~24, 2008, pp. 1349--1357.

\bibitem[{Avdis et~al.(2009)Avdis, Lardeau, and Lescziner}]{avdis2009large}
Avdis, A., Lardeau, S., and Lescziner, M., \enquote{Large eddy simulation of
  separated flow over a two-dimensional hump with and without control by means
  of a synthetic slot-jet,} \emph{Flow Turbulence and Combustion}, Vol.~83,
  2009, pp. 343--370.

\bibitem[{Ramirez and Wolf(2015)}]{ramirez2015effects}
Ramirez, W.~A., and Wolf, W., \enquote{The effects of suction and blowing on
  tonal noise generation by blunt trailing edges,} \emph{21st AIAA/CEAS
  Aeroacoustics Conference}, 2015.

\bibitem[{Yeh and Taira(2019)}]{yeh2018resolvent}
Yeh, C.-A., and Taira, K., \enquote{Resolvent-analysis-based design of airfoil
  separation control,} \emph{Journal of Fluid Mechanics}, Vol. 867, 2019, pp.
  572--610.

\bibitem[{Ramos et~al.(2019)Ramos, Wolf, Yeh, and Taira}]{ramos2019active}
Ramos, B.~L., Wolf, W.~R., Yeh, C.-A., and Taira, K., \enquote{Active flow
  control for drag reduction of a plunging airfoil under deep dynamic stall,}
  \emph{Physical Review Fluids}, Vol.~4, No.~7, 2019, p. 074603.

\bibitem[{Visbal and Benton(2018)}]{visbal2018exploration}
Visbal, M.~R., and Benton, S.~I., \enquote{Exploration of high-frequency
  control of dynamic stall using large-eddy simulations,} \emph{AIAA Journal},
  Vol.~56, No.~8, 2018, pp. 2974--2991.

\bibitem[{Beaudoin et~al.(2006)Beaudoin, Cadot, Aider, and
  Wesfreid}]{beaudoin2006bluff}
Beaudoin, J.-F., Cadot, O., Aider, J.-L., and Wesfreid, J.~E.,
  \enquote{Bluff-body drag reduction by extremum-seeking control,}
  \emph{Journal of Fluids and Structures}, Vol.~22, No. 6-7, 2006, pp.
  973--978.

\bibitem[{Becker et~al.(2007)Becker, King, Petz, and
  Nitsche}]{becker2007adaptive}
Becker, R., King, R., Petz, R., and Nitsche, W., \enquote{Adaptive closed-loop
  separation control on a high-lift configuration using extremum seeking,}
  \emph{AIAA journal}, Vol.~45, No.~6, 2007, pp. 1382--1392.

\bibitem[{Kim et~al.(2009)Kim, Kasnakoglu, Serrani, and
  Samimy}]{kim2009extremum}
Kim, K., Kasnakoglu, C., Serrani, A., and Samimy, M., \enquote{Extremum-seeking
  control of subsonic cavity flow,} \emph{AIAA journal}, Vol.~47, No.~1, 2009,
  pp. 195--205.

\bibitem[{Fan et~al.(2017)Fan, Wu, Yang, Li, and Zhou}]{fan2017modified}
Fan, D., Wu, Z., Yang, H., Li, J., and Zhou, Y., \enquote{Modified
  extremum-seeking closed-loop system for jet mixing enhancement,} \emph{AIAA
  Journal}, Vol.~55, No.~11, 2017, pp. 3891--3902.

\bibitem[{Brackston et~al.(2016)Brackston, Wynn, and
  Morrison}]{brackston2016extremum}
Brackston, R.~D., Wynn, A., and Morrison, J.~F., \enquote{Extremum seeking to
  control the amplitude and frequency of a pulsed jet for bluff body drag
  reduction,} \emph{Experiments in Fluids}, Vol.~57, No.~10, 2016, p. 159.

\bibitem[{Pastoor et~al.(2008)Pastoor, Henning, Noack, King, and
  Tadmor}]{pastoor2008feedback}
Pastoor, M., Henning, L., Noack, B.~R., King, R., and Tadmor, G.,
  \enquote{Feedback shear layer control for bluff body drag reduction,}
  \emph{Journal of Fluid Mechanics}, Vol. 608, 2008, p. 161.

\bibitem[{Wolf et~al.(2012)Wolf, Azevedo, and Lele}]{wolfJFM2012}
Wolf, W.~R., Azevedo, J. L.~F., and Lele, S.~K., \enquote{Convective effects
  and the role of quadrupole sources for aerofoil aeroacoustics,} \emph{Journal
  of Fluid Mechanics}, Vol. 708, 2012, p. 502–538.

\bibitem[{King et~al.(2006)King, Becker, Feuerbach, Henning, Petz, Nitsche,
  Lemke, and Neise}]{king2006adaptive}
King, R., Becker, R., Feuerbach, G., Henning, L., Petz, R., Nitsche, W., Lemke,
  O., and Neise, W., \enquote{Adaptive flow control using slope seeking,}
  \emph{2006 14th Mediterranean Conference on Control and Automation}, IEEE,
  2006.

\bibitem[{Beam and Warming(1978)}]{beam1978implicit}
Beam, R.~M., and Warming, R., \enquote{An implicit factored scheme for the
  compressible Navier-Stokes equations,} \emph{AIAA journal}, Vol.~16, No.~4,
  1978, pp. 393--402.

\bibitem[{Wray(1986)}]{wray:86}
Wray, A.~A., \enquote{Very low storage time advancement schemes,} \emph{NASA
  Technical Report 1999-209349}, 1986.

\bibitem[{Nagarajan(2004)}]{nagarajan:04}
Nagarajan, S., \enquote{Leading edge effects in bypass transition,} Ph.D.
  thesis, Stanford University, 2004.

\bibitem[{Nagarajan et~al.(2003)Nagarajan, Lele, and Ferziger}]{nagarajan:03}
Nagarajan, S., Lele, S.~K., and Ferziger, J.~H., \enquote{A robust high-order
  compact method for large eddy simulation,} \emph{Journal of Computational
  Physics}, Vol. 191, No.~2, 2003, pp. 392--419.

\bibitem[{Lele(1992)}]{lele1992compact}
Lele, S.~K., \enquote{Compact finite difference schemes with spectral-like
  resolution,} \emph{Journal of computational physics}, Vol. 103, No.~1, 1992,
  pp. 16--42.

\bibitem[{Nagarajan et~al.(2007)Nagarajan, Lele, and
  Ferziger}]{nagarajan2007leading}
Nagarajan, S., Lele, S., and Ferziger, J., \enquote{Leading-edge effects in
  bypass transition,} \emph{Journal of Fluid Mechanics}, Vol. 572, 2007, pp.
  471--504.

\bibitem[{Ricciardi et~al.(2020)Ricciardi, Arias-Ramirez, and
  Wolf}]{ricciardi2020secondary}
Ricciardi, T.~R., Arias-Ramirez, W., and Wolf, W.~R., \enquote{On secondary
  tones arising in trailing-edge noise at moderate Reynolds numbers,}
  \emph{European Journal of Mechanics-B/Fluids}, Vol.~79, 2020, pp. 54--66.

\bibitem[{Ariyur and Krsti{\'c}(2004)}]{ariyur2004slope}
Ariyur, K.~B., and Krsti{\'c}, M., \enquote{Slope seeking: a generalization of
  extremum seeking,} \emph{International Journal of Adaptive Control and Signal
  Processing}, Vol.~18, No.~1, 2004, pp. 1--22.

\bibitem[{Curle(1955)}]{curle:55}
Curle, N., \enquote{The influence of solid boundaries Upon aerodynamic sound,}
  \emph{Proceedings of the Royal Society A}, Vol. 231, 1955, pp. 505--514.

\bibitem[{Ffowcs-Williams and Hall(1970)}]{hall:70}
Ffowcs-Williams, J.~E., and Hall, L.~H., \enquote{Aerodynamic sound generation
  by turbulent flow in the vicinity of a scattering half plane,} \emph{Journal
  of Fluid Mechanics}, Vol.~40, 1970, pp. 657--670.

\bibitem[{Goodfellow et~al.(2013)Goodfellow, Yarusevych, and
  Sullivan}]{goodfellow2013momentum}
Goodfellow, S.~D., Yarusevych, S., and Sullivan, P.~E., \enquote{Momentum
  coefficient as a parameter for aerodynamic flow control with synthetic jets,}
  \emph{AIAA journal}, Vol.~51, No.~3, 2013, pp. 623--631.

\bibitem[{Benton and Visbal(2017)}]{benton2017high}
Benton, S.~I., and Visbal, M.~R., \enquote{High-frequency forcing to delay
  dynamic stall at relevant Reynolds number,} \emph{47th AIAA Fluid Dynamics
  Conference}, 2017.

\bibitem[{Benton and Visbal(2018)}]{benton2018evaluation}
Benton, S.~I., and Visbal, M.~R., \enquote{Evaluation of thermoacoustic-based
  forcing for control of dynamic stall,} \emph{2018 AIAA Flow Control
  Conference}, 2018.

\bibitem[{Tan and Jiang(2018)}]{tan2018digital}
Tan, L., and Jiang, J., \emph{Digital Signal Processing: Fundamentals and
  Applications}, Academic Press, 2018.

\bibitem[{Franklin et~al.(2015)Franklin, Powell, Emami-Naeini, and
  Sanjay}]{franklin2015feedback}
Franklin, G.~F., Powell, J.~D., Emami-Naeini, A., and Sanjay, H.,
  \emph{Feedback Control of Dynamic Systems}, Pearson London, 2015.

\bibitem[{Massarotti and Wolf(2019)}]{massarotti2019passive}
Massarotti, M.~R., and Wolf, W., \enquote{Passive and active methods to control
  the aeroacoustic noise generated by elliptical cylinders for automotive
  applications,} \emph{25th AIAA/CEAS Aeroacoustics Conference}, 2019.

\end{thebibliography}

\end{document}